\documentclass[oneside,fleqn,12pt]{book}

\usepackage[paper=a4paper,margin=2.5cm]{geometry}
\usepackage{amsmath,amssymb,amsfonts,amsthm,mathrsfs}
\usepackage{graphicx}
\usepackage{nicefrac}
\usepackage[squaren]{SIunits}
\usepackage{color}
\usepackage[utf8]{inputenc}
\usepackage{epigraph}
\usepackage[hyperfootnotes=false]{hyperref}
\usepackage{setspace}
\usepackage{pdfpages}
\usepackage{wrapfig}
\usepackage[toc,page]{appendix}

\newcommand{\var}[1]{\langle\Delta#1^2\rangle}
\newcommand{\mean}[1]{\langle#1\rangle}
\newcommand{\op}[1]{\hat{#1}}
\newcommand{\cop}[1]{\hat{#1}^\dagger}

\newcommand{\ket}[1]{\left|{#1}\right.\rangle}
\newcommand{\Gammaa}{\mathcal{G}}

\begin{document}
\pagestyle{plain}

%%%%%%%%%%%%%%%%%%%%%%%%%%%%%%%%%%%%%%%%%%%%%%%%%%%%%
\begin{titlepage}
\centering

\vspace*{1cm}

\begin{bfseries} \Large
Spectral and statistical properties\\of high-gain parametric down-conversion\\
\end{bfseries}
\noindent\makebox[\linewidth]{\rule{15cm}{1pt}}
\begin{bfseries} \Large
Spektrale und statistische Eigenschaften\\der parametrischen Fluoreszenz bei hoher\\optisch-parametrischer Verstärkung\\
\end{bfseries}
\vspace*{8cm}

Der Naturwissenschaftlichen Fakultät\\
der Friedrich-Alexander-Universität Erlangen-Nürnberg\\
zur\\
Erlangung des Doktorgrades Dr. rer. nat.\\

\vspace*{1.5cm}

vorgelegt von\\[1.3ex]
\textbf{\large Kirill Spasibko}\\[1ex]
aus Moskau

\vspace*{2.5cm}

2019

\end{titlepage}

%%%%%%%%%%%%%%%%%%%%%%%%%%%%%%%%%%%%%%%%%%%%%%%%%%%%%
\begin{titlepage}
\newpage

\begin{flushleft}
Als Dissertation genehmigt\\
von der Naturwissenschaftlichen Fakultät\\
der Friedrich-Alexander-Universität Erlangen-Nürnberg\\
Tag der mündichen Prüfung: 20.12.2019

\vspace*{18cm}

\begin{tabular}{ll}
Vorsitzender des Promotionsorgans:&Prof. Dr. Kreimer\\
Gutachter:&PD Dr. Chekhova\\
Gutachter:&Prof. Dr. Joly\\
Gutachter:&Prof. Dr. Lundeen\\
\end{tabular}
\end{flushleft}

\end{titlepage}

%%%%%%%%%%%%%%%%%%%%%%%%%%%%%%%%%%%%%%%%%%%%%%%%%%%%%
\begin{titlepage}
\begin{center}
\textit{Für Vera, Sofia und alle unsere Kuscheltiere}
\end{center}
\end{titlepage}

%%%%%%%%%%%%%%%%%%%%%%%%%%%%%%%%%%%%%%%%%%%%%%%%%%%%%
\frontmatter

\chapter*{Summary}

This thesis is devoted to high-gain parametric down-conversion (PDC). PDC is mostly known in the low-gain (spontaneous) regime, in which the correlated photon pairs are produced. Spontaneous PDC (SPDC) plays a very important role for quantum optics as a variety of quantum states is produced via SPDC, including, for instance, entangled Bell states or photon-number states. Moreover, SPDC finds its applications in metrology, cryptography, imaging, and lithography.

In the high-gain case PDC leads to generation of bright states having up to hundreds mW mean power. With such states almost any nonlinear optical interaction or light-matter interaction becomes more efficient. Even being macroscopically bright, the produced states maintain nonclassical properties as, for example, the fluctuations of electric field quadratures are squeezed below the shot-noise level.

The high-gain PDC could be used not only in the same applications as SPDC, it also can provide new ones. For example, PDC is in use in the LIGO and GEO600 gravitational-wave detectors, because squeezing increases the sensitivity of interferometry.

High-gain PDC has many remarkable spectral and statistical properties, which are in the focus of this work. The thesis discusses them in detail, both theoretically and experimentally, and shows how high-gain PDC could be used.

The description starts from the PDC generation in normal and anomalous group velocity dispersion ranges. The spectrum and mode content of high-gain PDC is considered as well as their change with the parametric gain are demonstrated.

Then, there are the interference effects emerging from the PDC correlations presented, namely the macroscopic analogue of the Hong-Ou-Mandel interference. In addition, it is shown how spatial and temporal walk-off matching could be used for the generation of giant narrowband twin beams.

Finally, the statistical properties of high-gain PDC are reviewed as well as their use for multiphoton effects is demonstrated. Photon-number fluctuations of PDC are studied via normalized correlation functions and probability distributions. These fluctuations enhance the generation efficiency for multiphoton effects by orders of magnitude and lead to tremendously fluctuating light described by heavy-tailed photon-number probability distributions.

%%%%%%%%%%%%%%%%%%%%%%%%%%%%%%%%%%%%%%%%%%%%%%%%%%%%%
\chapter*{Zusammenfassung}

Diese Arbeit befasst sich mit der experimentellen und theoretischen Untersuchung der parametrischen Fluoreszenz bei hoher optisch-parametrischer Verstärkung. Im Gegensatz zur parametrische Fluoreszenz bei geringer Verstärkung, wobei die Photonenpaare erzeugt werden, ist das Regime hoher Verstärkung noch vergleichsweise wenig erforscht. Parametrische Fluoreszenz ist bedeutsam für die Quantenoptik und wird z.B. in der Metrologie, Kryptographie, Bildgebung und Lithografie angewendet.

Bei hoher Verstärkung entstehen extrem helle Lichtzustände, die bis zu hundert Milliwatt mittlere Leistung haben können. Solchen Zustände sind viel effizienter für nicht-lineare optische Effekte und Licht-Materie-Wechselwirkungen. Obwohl die bei hoher Verstärkung erzeugten Lichtzustände makroskopisch hell sind, zeigen sie nichtklassische Eigenschaften.

Die bei hoher optisch-parametrischer Verstärkung erzeugte parametrische Fluoreszenz hat außergewöhnliche spektrale und statistische Eigenschaften, die im Mittelpunkt dieser Arbeit stehen. Die Arbeit beschreibt diese Eigenschaften nicht nur theoretisch und experimentell, sondern zeigt auch realisierbare Anwendungen.

%%%%%%%%%%%%%%%%%%%%%%%%%%%%%%%%%%%%%%%%%%%%%%%%%%%%%
\tableofcontents

%%%%%%%%%%%%%%%%%%%%%%%%%%%%%%%%%%%%%%%%%%%%%%%%%%%%%
\chapter*{Author's publications}
\label{publications}

Scopus ID: \href{https://www.scopus.com/authid/detail.uri?authorId=54421216700}{54421216700}
ORCID ID: \href{https://orcid.org/0000-0001-6667-5084}{0000-0001-6667-5084}

\section*{Discussed in the thesis}

\begin{enumerate}
\item M. Manceau, K. Yu. Spasibko, G. Leuchs, R. Filip, and M. V. Chekhova. Indefinite-mean Pareto photon distribution from amplified quantum noise. \textit{Phys. Rev. Lett.} 123, 123606 (2019).
\newline DOI: \href{https://www.doi.org/10.1103/PhysRevLett.123.123606}{10.1103/PhysRevLett.123.123606}

\item D. A. Kopylov, K. Yu. Spasibko, T. V. Murzina, and M. V. Chekhova. Study of broadband multimode light via non-phase-matched sum frequency generation. \textit{New J. Phys.} 21, 033024 (2019).
\newline DOI: \href{https://www.doi.org/10.1088/1367-2630/ab0a7c}{10.1088/1367-2630/ab0a7c}

\item K. Yu. Spasibko, D. A. Kopylov, V. L. Krutyanskiy, T. V. Murzina, G. Leuchs, and M. V. Chekhova. Multiphoton effects enhanced due to ultrafast photon-number fluctuations. \textit{Phys. Rev. Lett.} 119, 223603 (2017).
\newline DOI: \href{https://www.doi.org/10.1103/PhysRevLett.119.223603}{10.1103/PhysRevLett.119.223603}

\item K. Yu. Spasibko, D. A. Kopylov, T. V. Murzina, G. Leuchs, and M. V. Chekhova. Ring-shaped spectra of parametric downconversion and entangled photons that never meet. \textit{Opt. Lett.} 41, 2827-2830 (2016).
\newline DOI: \href{https://www.doi.org/10.1364/OL.41.002827}{10.1364/OL.41.002827}

\item A. M. P\'erez, K. Yu. Spasibko, P. R. Sharapova, O. V. Tikhonova, G. Leuchs, and M. V. Chekhova. Giant narrowband twin-beam generation along the pump-energy propagation direction. \textit{Nat. Commun.} 6, 7707 (2015).
\newline DOI: \href{https://www.doi.org/10.1038/ncomms8707}{10.1038/ncomms8707}

\item K. Yu. Spasibko, F. T\"oppel, T. Sh. Iskhakov, M. Stobi\'nska, M. V. Chekhova, and G. Leuchs. Interference of macroscopic beams on a beam splitter: phase uncertainty converted into photon-number uncertainty. \textit{New J. Phys.} 16, 013025 (2014).
\newline DOI: \href{https://www.doi.org/10.1088/1367-2630/16/1/013025}{10.1088/1367-2630/16/1/013025}

\item T. Sh. Iskhakov, K. Yu. Spasibko, M. V. Chekhova, and G. Leuchs. Macroscopic Hong-Ou-Mandel interference. \textit{New J. Phys.} 15, 093036 (2013).
\newline DOI: \href{https://www.doi.org/10.1088/1367-2630/15/9/093036}{10.1088/1367-2630/15/9/093036}

\item T. Sh. Iskhakov, A. M. P\'erez, K. Yu. Spasibko, M. V. Chekhova, and G. Leuchs. Superbunched bright squeezed vacuum state. \textit{Opt. Lett.} 37, 1919-1921 (2012). 
\newline DOI: \href{https://www.doi.org/10.1364/OL.37.001919}{10.1364/OL.37.001919}

\item K. Yu. Spasibko, T. Sh. Iskhakov, and M. V. Chekhova. Spectral properties of high-gain parametric down-conversion. \textit{Opt. Express} 20, 7507-7515 (2012).
\newline DOI: \href{https://www.doi.org/10.1364/OE.20.007507}{10.1364/OE.20.007507}
\end{enumerate}

\section*{Other publications}

\begin{enumerate}
\item O. Kovalenko, K. Yu. Spasibko, M. V. Chekhova, V. C. Usenko, and R. Filip. Feasibility of quantum key distribution with macroscopically bright coherent light. \textit{Opt. Express} 27, 36154-36163 (2019).
\newline DOI: \href{https://www.doi.org/10.1364/OE.27.036154}{10.1364/OE.27.036154}

\item E. Knyazev, K. Yu. Spasibko, M. V. Chekhova, and F. Ya. Khalili. Quantum tomography enhanced through parametric amplification. \textit{New J. Phys.} 20, 013005 (2018).
\newline DOI: \href{https://www.doi.org/10.1088/1367-2630/aa99b4}{10.1088/1367-2630/aa99b4}

\item K. Yu. Spasibko, M. V. Chekhova, and F. Ya. Khalili. Experimental demonstration of negative-valued polarization quasiprobability distribution. \textit{Phys. Rev. A} 96, 023822 (2017).
\newline DOI: \href{https://www.doi.org/10.1103/PhysRevA.96.023822}{10.1103/PhysRevA.96.023822}

\item F. Sciarrino, G. Vallone, G. Milani, A. Avella, J. Galinis, R. Machulka, A. M. Perego, K. Y. Spasibko, A. Allevi, M. Bondani, and P. Mataloni. High degree of entanglement and nonlocality of a two-photon state generated at 532 nm.
\textit{Eur. Phys. J. ST} 199, 111-125 (2011).
\newline DOI: \href{https://www.doi.org/10.1140/epjst/e2011-01507-y}{10.1140/epjst/e2011-01507-y}
\end{enumerate}

%%%%%%%%%%%%%%%%%%%%%%%%%%%%%%%%%%%%%%%%%%%%%%%%%%%%%
\mainmatter

\includepdf{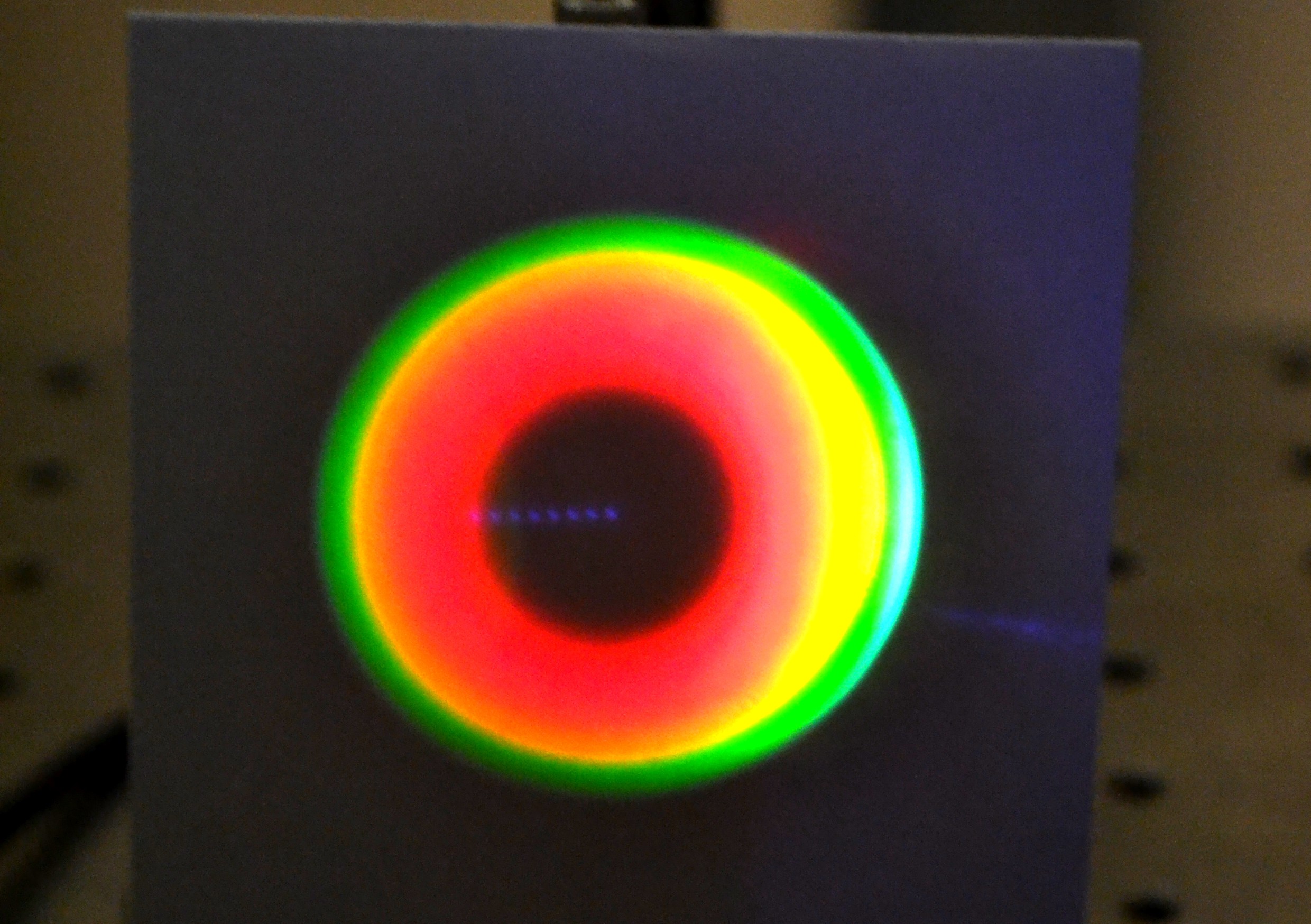}

%%%%%%%%%%%%%%%%%%%%%%%%%%%%%%%%%%%%%%%%%%%%%%%%%%%%%
\chapter{Introduction}\label{1_chapter}

\begin{center}\textit{And God said, Let there be light: and there was light.\\And God saw the light, and it was \textcolor{red}{good};\\and God divided the light from \textcolor{red}{the darkness}.}\\---The Book of Genesis
\end{center}

Parametric down conversion (PDC) naturally originates from \textcolor{red}{\it the darkness}, namely from the vacuum noise, also called zero-point vacuum fluctuations. Since its discovery in 1967~\cite{Klyshko1967,Harris1967,Magde1967,Akhmanov1967} PDC remains one of the main resources for quantum optics.

The PDC, namely spontaneous (SPDC), occupies this place, because pairs of correlated photons are generated much easier with it than, for example, with atoms in cascaded transitions~\cite{Freedman1972}. With these photon pairs, one could produce various states, which are important for quantum information including for example, single-photon~\cite{Lvovsky2001} or entangled states, bipartite~\cite{Ou1988,Shih1988} or tripartite~\cite{Bouwmeester1999} ones. The former one is a highly nonclassical state with well-defined number of quanta, whereas the latter ones have nonfactorizable wave functions for multiple particles and probe interpretations of quantum mechanics in EPR and Bell test experiments.

Furthermore, SPDC opens a variety of applications. For example, the calibration of photodetectors~\cite{Burnham1970,Malygin1981} without a reference detector, a method of photolithography~\cite{Chekhova2001}, so-called quantum lithography, providing better resolution than the classical light for the same wavelength, or ghost imaging~\cite{Pittman1995}, in which the image is retrieved even if the detector that `sees' the object has no spatial resolution. Moreover, it is also useful in cryptography, namely in quantum key distribution~\cite{Jennewein2000}, in which the key for encryption and decryption is secured by quantum mechanics.

The high-gain PDC, has potentially a huge benefit w.r.t. SPDC: instead of faint states, one gets bright ones. Thus, almost any nonlinear optical interaction or the interaction of light with the matter would be much more efficient. Since parametric amplification is accompanied by squeezing, the quantum state of light produced through high-gain PDC is called `squeezed vacuum'. The squeezing can be single-mode, quadrature~\cite{Wu1986}, or two-mode, polarization~\cite{Iskhakov2009} or photon-number one~\cite{Jedrkiewicz2004}, meaning that the fluctuations of the quadrature amplitude, one of the Stokes parameters, or the photon-number difference in two modes is below the shot-noise level. Even Bell's inequalities could be, in principle, violated~\cite{Rosolek2015}. Interestingly, these nonclassical features remain even if the state becomes macroscopical; for example, entanglement~\cite{Iskhakov2012_2} and twin-beam squeezing~\cite{Iskhakov2016} is observed for the states containing more than 10$^5$ photons. Therefore, the squeezed vacuum with the macroscopical number of photons is called bright squeezed vacuum (BSV).

The high-gain PDC could be used not only in the same applications as SPDC, like quantum key distribution~\cite{Madsen2012,Gehring2015} or absolute calibration of photodetectors~\cite{Brida2010,Vahlbruch2016}, but also it can provide new applications. For example, squeezing increases the sensitivity of optical imaging~\cite{Brida2010_2} and interferometry~\cite{Caves1981} and consequently becomes important for live cell imaging, where one should use the lowest possible photon dose~\cite{Cole2014}, and gravitational-wave detectors, where the detected signals are extremely weak. These detectors serve as a major successful example since the squeezing is used operationally at GEO600~\cite{Abadie2011}, at LIGO it has been tested~\cite{Aasi2013} and is in use starting from 1st of April 2019.

However, it still requires some hard work to show how \textcolor{red}{\it good} high-gain regime is. This thesis respectively takes a step in such a direction as it summarizes the major part of author's research work devoted to the spectral and statistical properties of high-gain PDC. Moreover, it highlights some potential applications, which could be useful for science and industry. Most essential results have already been published being presented in \nameref{publications}.

This thesis is organized as follows. Chapter~\ref{2_chapter} introduces the basic theoretical foundations of PDC and clarifies the difference between spontaneous and high-gain regimes.

Chapter~\ref{3_chapter} is devoted to the wavelength-angular spectrum of PDC and presents the results published in Refs.~\cite{Spasibko2016,Spasibko2012,Kopylov2019}. Firstly, it describes the PDC generation in the normal and anomalous group velocity dispersion (GVD) ranges. In the second case, the spectrum is restricted in both the angle and the wavelength. The shape of the spectrum suggests a new type of spatiotemporal coherence. Afterwards there is the broadening of the spectrum at high gain explained and demonstrated. Both the total spectral and the correlation widths get broader as the gain increases. Finally, the chapter presents a method for the reconstruction of the joint spectral intensity (JSI) and as well as a way to get the number of modes, namely the Fedorov ratio or the Schmidt number.

Chapter~\ref{4_chapter} to a larger extend focuses on the twin-beam correlations and summarizes the results of Refs.~\cite{Iskhakov2013,Spasibko2014,Perez2015}. It starts with the interference effects emerging from these correlations and explains the Hong-Ou-Mandel (HOM) effect and interference of Fock states on a beam splitter (BS). Besides that, the chapter deals with the macroscopic analogue of the HOM interference and the way how the Fock states interference can be accessed using twin beams. Finally, the chapter shows how spatial and temporal walk-off matching could be used for giant narrowband twin-beam generation.

Chapter~\ref{5_chapter} provides some information about the statistical properties of BSV and clarify why they are useful for multiphoton effects. Being more detailed than the other chapters, it summarizes the most recent results published in \cite{Iskhakov2012,Spasibko2017,Manceau2018}. First and foremost it discusses single-mode and multimode light, its fluctuations and measurement and presents the measurement of correlation functions (CF) and probability distributions for BSV. Afterwards, the chapter shows how BSV fluctuations enhance the efficiency of multiphoton effects and finally describes how these fluctuations lead to tremendously fluctuating light with a such heavy-tailed probability distribution that no statistical moments are defined for the photon number, not even a mean value.

%%%%%%%%%%%%%%%%%%%%%%%%%%%%%%%%%%%%%%%%%%%%%%%%%%%%%
\chapter{Parametric down-conversion (PDC)}
\label{2_chapter}

\begin{center}\textit{In which we are introduced to PDC and some features, and the stories begin.}
\end{center}

%%%%%%%%%%%%%%%%%%%%%%%%%%%%%%%%%%%%%%%%%%%%%%%%%%%%%
\section{PDC process}
Parametric down-conversion occurs in nonlinear medium with nonzero second-order susceptibility $\chi^{(2)}$. In this process each converted pump photon ($p$) produces two daughter photons, signal ($s$) and idler ($i$). This process is called \textit{parametric}, because there is no energy transfer between photons and the medium, and it is \textit{down}-conversion, because the frequencies of produced photons are below the one of the pump. The frequencies $\omega$ and wave vectors\footnote{Here and further on the bold symbols are used for vectors, e.g. $\boldsymbol{k}\equiv(k_x,k_y,k_z)$.} $\boldsymbol{k}$ are related through the energy and momentum conservation laws:
\begin{equation}\label{2_eq:energy_momentum_cons}
\omega_p = \omega_s + \omega_i, 	\qquad 	\boldsymbol{k}_p = \boldsymbol{k}_s + \boldsymbol{k}_i.
\end{equation}

PDC can be described as the optical parametric amplification (OPA) of the vacuum noise in many modes~\cite{Klyshko1988,Klyshko2011}. Although these modes differ by many parameters, like frequency, wave vector, or polarization, the most important issue is whether signal and idler photons are indistinguishable or not. In the first case, the OPA at each mode should be described as single-mode, in the second one, the two-mode picture is needed. The both cases are introduced using the quantum description of the electric field.

\subsection{Quadrature squeezing}

In the case of a single-mode OPA, the output photon creation and annihilation operators, $\cop{a}$ and $\op{a}$, are related with the input ones, $\cop{a}_0$ and $\op{a}_0$, with the Bogolyubov transformation~\cite{Klyshko1988},
\begin{equation}\label{2_eq:Bogolubov}
\op{a}=\op{a}_0\cosh G + \cop{a}_0\sinh G,
\end{equation}
arising from the Heisenberg equation of motion and the corresponding Hamiltonian, the parametric gain $G$ characterizes the amplification strength\footnote{Actually, the gain $G$ depends on many parameters, which will be discussed a bit later (section~\ref{3_sub:freq_wave_spectra}).}.

Unfortunately, the operators $\op{a}$ and $\cop{a}$ can not be measured, because they are not Hermitian, $\op{a}\ne\cop{a}$. Moreover, they do not commute, $[\op{a},\cop{a}]=1$.

Nevertheless, applying homodyne detection one can measure the quadratures $q$ and $p$, the quantum analogues of the real and imaginary parts (cosinusoidal and sinusoidal components) of the electric field,
\begin{equation}
\op{q} \equiv\frac{\op{a}+\cop{a}}{\sqrt{2}} \quad \mathrm{and} \quad	\op{p} \equiv\frac{\op{a}-\cop{a}}{i\sqrt{2}}.
\end{equation}

The OPA does not change the mean values of the quadratures and for the vacuum input state the mean values are zero, $\mean{q}=\mean{p}=\mean{q_0}=\mean{p_0}=0$. However, the noise changes: the quadrature noise becomes antisqueezed,
\begin{equation}
\var{q} = e^{2G}\var{q_0},
\end{equation}
for one quadrature while squeezed,
\begin{equation}
\var{p} =e^{-2G}\var{p_0},
\end{equation}
for the other one w.r.t. the vacuum noise\footnote{This noise is typical for a coherent state produced, for example, by a shot-noise limited laser.}, $\var{q_0}=\var{p_0}=1/2$. Therefore, the state produced by an unseeded OPA is called the \textit{squeezed vacuum}.

Remarkably, this \textit{vacuum} contains energy, i.e. photons. The mean number of photons, corresponding to the photon-number operator $\op{N}\equiv\cop{a}\op{a}$, is
\begin{equation}\label{2_eq:sinh2}
\mean{N}=\sinh^2 G.
\end{equation}
With $\mean{N}\gg1$ the \textit{vacuum} will be not only \textit{squeezed}, but also \textit{bright}. BSV can be extremely bright; in the following chapters one finds the cases with $G>15$ corresponding to $\mean{N}>3\times10^{12}$ photons! The cover for chapter~\ref{1_chapter} shows one of the examples: vacuum is amplified so strongly that it is visible on a piece of paper.

\subsection{Twin-beam squeezing}
\label{2_sub:twin-beam_sq}

Similar approach applies also to two-mode OPA. The Bogolyubov transformations for the signal and idler modes are
\begin{equation}\label{2_eq:Bogolubov2}
\op{a}_s=\op{a}_{s0}\cosh G + \cop{a}_{i0}\sinh G \quad \mathrm{and} \quad \op{a}_i=\op{a}_{i0}\cosh G + \cop{a}_{s0}\sinh G.
\end{equation}
This time the commutation relations look as follows:
\begin{equation}\label{2_eq:comm_rules}
[\op{a}_{s},\cop{a}_{s}]=1, \quad [\op{a}_{s},\op{a}_{i}]=0, \quad [\op{a}_{s},\cop{a}_{i}]=0,
\end{equation}
and the ones with replacement $s\leftrightarrow i$~\cite{Klyshko1988}. The quadrature squeezing can be observed with two-mode OPA too\footnote{The opposite applies as well: one can observe twin-beam squeezing with a single-mode OPA~\cite{Schnabel2003}.}; two-mode quadratures that depend on both $\op{a}_s$ and $\op{a}_i$ should be measured~\cite{DAuria2009}.

However, another type of squeezing is commonly in use: two-mode or twin-beam one. Instead of the quadrature noise, the one of photon-number difference between the signal and idler modes is measured. Indeed, for the signal and idler beams not only the numbers of photons are the same,
\begin{equation}\label{2_eq:N_si}
\mean{N_s} = \mean{N_i} =\sinh^2 G,
\end{equation}
but also all moments of photon-number difference, $\mean{(\op{N}_s-\op{N}_i)^n}$, are equal to zero. The same applies to the central moments, like variance. It is certainly squeezed below the shot-noise level, $\var{(\op{N}_s-\op{N}_i)}=\mean{N_s+N_i}$, which is the noise for two independent coherent beams.

Thus, the number of photons in the signal and idler beams is always the same, and therefore they are called \textit{twin} beams and the squeezing is \textit{twin-beam} one\footnote{The polarization squeezing is similar, because the Stokes operators are equal to the photon-number difference in two polarization modes, e.g. $\op{S}_1=\op{n}_H-\op{n}_V$ for horizontal and vertical modes.}.

%%%%%%%%%%%%%%%%%%%%%%%%%%%%%%%%%%%%%%%%%%%%%%%%%%%%%
\section{High-gain vs. spontaneous PDC (SPDC)}
\label{2_sec:high_vs_SPDC}

The previous section describes the properties of BSV. Apart from that, its wave function can be expressed as a series over Fock ($m$-photon) states $\ket{m}$. Hence, an unseeded single-mode OPA gives~\cite{Marian1992}
\begin{equation}\label{2_eq:BSV_state}
\ket{\Psi_1}=\frac{1}{ \cosh G}\sum_{m=0}^{\infty}\frac{(2m)!}{2^{2m} (m!)^2} (\tanh G)^{2m}\ket{2m},
\end{equation}
an unseeded two-mode one~\cite{Chekhova2015},
\begin{equation}\label{2_eq:twin_state}
\ket{\Psi_2}=\frac{1}{ \cosh G}\sum_{m=0}^{\infty}(\tanh G)^{m}\ket{m}_s\ket{m}_i.
\end{equation}
In both cases BSV is a state of light with only even photon numbers: either all photons are in the same mode, or they are equally distributed between two modes.

The case of SPDC is the one where the parametric gain $G$ is small. It is usually the case: for a continuous-wave (CW) pump laser $G\sim10^{-3}-10^{-4}$. In this case four- and higher-photon contributions are negligibly small w.r.t. two-photon one. Thus, either a two-photon state,
\begin{equation}
\ket{\Phi_1}=c_0\ket{0} + c_2\ket{2},
\end{equation}
or an entangled photon pair,
\begin{equation}\label{2_eq:SPDC_state_2}
\ket{\Phi_2}=c_0\ket{0}_s\ket{0}_i + c_1\ket{1}_s\ket{1}_i,
\end{equation}
is produced in superposition with the vacuum. Here $c_{0,1,2}$ are renormalized constants from Eqs.~\eqref{2_eq:BSV_state} and \eqref{2_eq:twin_state}. The constant $c_0$ is often omitted, because it is approximately equal to unity. As mentioned in chapter~\ref{1_chapter}, these two states are of a great importance for quantum optics. They are produced via SPDC in hundreds of experiments.

\begin{figure}[!htb]
\begin{center}
\includegraphics[width=0.7\textwidth]{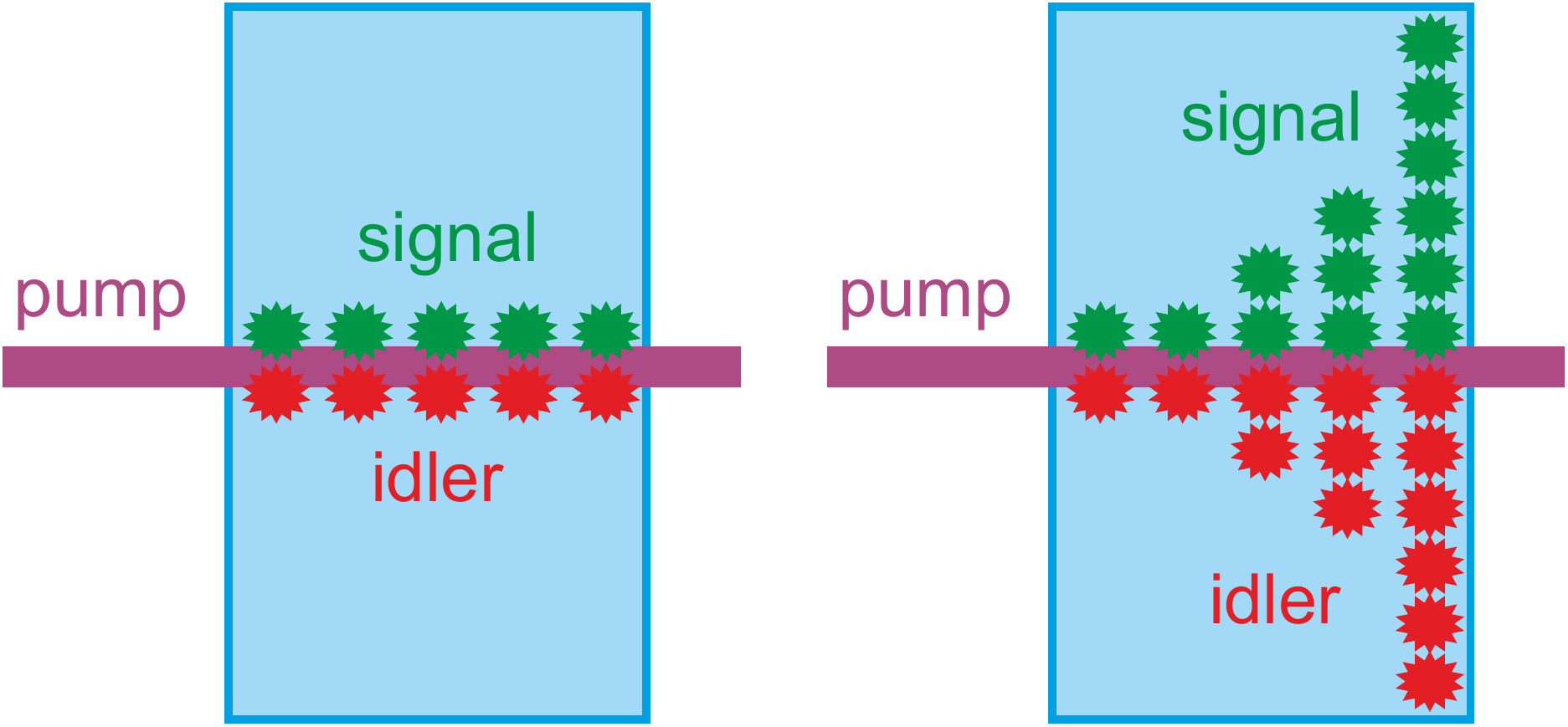}
\caption{Schematic representation of the generation rate along the nonlinear medium at low (left) and high gain (right).}
\label{2_fig:SPDC_vs_high_gain}
\end{center}
\end{figure}

Due to low gain, $\mean{N}$ is also small, $10^{-6}-10^{-8}$ photons, being much smaller than the effective brightness of zero-point vacuum fluctuations\footnote{Actually, vacuum fluctuations have no brightness but in order to explain spontaneous effects one can introduce it effectively. This effective brightness comes from the noncommutativity of $\op{a}$ and $\cop{a}$ and is equal to one photon per each mode of the field.}. The produced photons, therefore, do not enhance the generation rate and it is the same along the nonlinear medium (Fig.~\ref{2_fig:SPDC_vs_high_gain}, left).

However, this is not the case in the high-gain regime. If one reaches $G>1$, for example by using strong pulsed pump laser, $\mean{N}$ is larger than one photon. This drastically enhances the generation rate, because the photons produced at the beginning of the nonlinear medium seed the generation at the end. Therefore, the generation rate and the number of produced photons increase exponentially along the nonlinear medium (Fig.~\ref{2_fig:SPDC_vs_high_gain}, right). Hence, as long as one reaches the high-gain regime, it is extremely easy to make the state brighter and brighter.

%%%%%%%%%%%%%%%%%%%%%%%%%%%%%%%%%%%%%%%%%%%%%%%%%%%%%%%%%%%
\chapter{Wavelength-angular spectrum of PDC}
\label{3_chapter}

\begin{center}\textit{My drawing was not a picture of a hat.\\It was a picture of a boa constrictor digesting an elephant. ...\\The grown-ups always need explanations.}\\---Antoine de Saint-Exup\'ery, \textit{The Little Prince}
\end{center}

This chapter discusses the PDC spectrum, namely it shape and properties. Firstly, it describes how one can calculate the spectrum, introduces the concept of coupled spatial and temporal coherence, explains the common asymmetry of measured wavelength-angular PDC spectra, and discusses PDC correlations and eigenmodes.

Apart from that, the chapter deals with the special features that arise when PDC is generated in the anomalous GVD range. Moreover, there is the unusual type of wavelength-angular PDC spectrum presented, which suggests a new type of spatiotemporal coherence and entanglement of photon pairs. One can obtain an entangled photon pair with the photons being never at the same point at the same time.

Afterwards there is the broadening of PDC spectra at high gain regime demonstrated; both the total and the correlation spectral widths get broadened. In addition, the asymmetry of PDC spectra is discussed in more details.

Finally, the chapter presents the method for JSI reconstruction via covariance measurement, which could be directly applied for the reconstruction of BSV eigenmodes, namely Schmidt modes. Moreover, there is one more method briefly explained which gives access to the number of modes via non-phase-matched sum frequency generation (SFG).

%%%%%%%%%%%%%%%%%%%%%%%%%%%%%%%%%%%%%%%%%%%%%%%%%%%%%
\section{Introduction}
\label{3_sec:intro}

As one could have noticed, e.g. from the cover for chapter~\ref{1_chapter}, PDC is usually highly broadband in frequency and wave vector. This should be taken into account. Unfortunately, the analytical description of high-gain PDC is possible only under certain approximations. One of the ways is the assumptions of a stationary, plane-wave, and undepleted pump~\cite{Klyshko1988,Brambilla2004} and infinite transverse dimensions of a nonlinear medium. They lead to strict energy and transverse momentum conservations laws\footnote{The condition for longitudinal momentum is not so strict, because the length of nonlinear medium
is finite.},
\begin{equation}\label{3_eq:delta_corr_s_i}
\Omega_s = -\Omega_i \equiv \Omega, 	\qquad 	\boldsymbol{q}_s = -\boldsymbol{q}_i \equiv \boldsymbol{q},
\end{equation}
where $\Omega_{s,i}\equiv\omega_{s,i} - \omega_{s0,i0}$ is a frequency shift w.r.t. the central frequency $\omega_{s0,i0}$ and $\boldsymbol{q}_{s,i}$ is the transverse (w.r.t $\boldsymbol{k}_{p}$) component of $\boldsymbol{k}_{s,i}$.

Under this approximation the input-output transformations~\eqref{2_eq:Bogolubov2} look as follows~\cite{Klyshko1988,Brambilla2004}:
\begin{eqnarray}\label{3_eq:Bogolubov_q_w}
&&\op{a}_s(\boldsymbol{q},\Omega)=\op{a}_{s0}(\boldsymbol{q},\Omega)U_s(\boldsymbol{q},\Omega) + \cop{a}_{i0}(-\boldsymbol{q},-\Omega)V_s(\boldsymbol{q},\Omega),\nonumber\\
&&\op{a}_i(\boldsymbol{q},\Omega)=\op{a}_{i0}(\boldsymbol{q},\Omega)U_i(\boldsymbol{q},\Omega) + \cop{a}_{s0}(-\boldsymbol{q},-\Omega)V_i(\boldsymbol{q},\Omega),
\end{eqnarray}
where $U_{s,i}$ and $V_{s,i}$ are the functions of gain $G$. In general, the functions $U_{s,i}$ and $V_{s,i}$ have lengthy expressions, which can be greatly simplified for certain cases. Therefore, the full expressions from Ref.~\cite{Brambilla2004} are not presented, only the simplified ones for each particular case. Here it is critical that the gain functions satisfy the unitarity conditions,
\begin{eqnarray}
&&|U_{s,i}(\boldsymbol{q},\Omega)|^2 - |V_{s,i}(\boldsymbol{q},\Omega)|^2 = 1,\nonumber\\
&&U_s(\boldsymbol{q},\Omega) V_i(-\boldsymbol{q},-\Omega) = U_i(-\boldsymbol{q},-\Omega) V_s(\boldsymbol{q},\Omega),
\end{eqnarray}
which provide the commutation relations
\begin{eqnarray}
&&[\op{a}_s(\boldsymbol{q}, \Omega), \cop{a}_s(\boldsymbol{\tilde{q}}, \tilde{\Omega})] = \delta(\boldsymbol{q} - \boldsymbol{\tilde{q}}) \delta(\Omega - \tilde{\Omega}),\nonumber\\ 
&&[\op{a}_s(\boldsymbol{q}, \Omega), \op{a}_i(\boldsymbol{\tilde{q}}, \tilde{\Omega})] = 0, \qquad [\op{a}_s(\boldsymbol{q}, \Omega), \cop{a}_i(\boldsymbol{\tilde{q}}, \tilde{\Omega})] = 0,
\end{eqnarray}
and the ones with the replacement $s\leftrightarrow i$. Here $\delta(x)$ is the Dirac delta function.

\subsection{Frequency-wavevector spectrum}\label{3_sub:freq_wave_spectra}

The PDC spectrum $S(\boldsymbol{q}, \Omega)$ can be obtained using transformations~\eqref{3_eq:Bogolubov_q_w}. Indeed, the spectrum is equal to the normalized mean number $\mean{N_{s,i}}$ of emitted photons,
\begin{equation}
S(\boldsymbol{q}, \Omega) = \frac{\mean{N_{s,i}(\boldsymbol{q}, \Omega)}}{\max\left(\mean{N_{s,i}}\right)},
\end{equation}
where $\mean{N_{s,i}(\boldsymbol{q}, \Omega)}=\mean{\cop{a}_{s,i} (\boldsymbol{q}, \Omega) \op{a}_{s,i} (\boldsymbol{q}, \Omega)}$. The operators $\op{a}_{s,i}$ are taken at the end of a nonlinear medium and vacuum operators $\op{a}_{s0,i0}$ are considered at the beginning. Then,
\begin{equation}\label{3_eq:N_q_w}
\mean{N_{s,i}(\boldsymbol{q}, \Omega)} = |V_{s,i} (\boldsymbol{q}, \Omega)|^2 = \left( \frac{G}{\Gammaa(\boldsymbol{q}, \Omega)} \sinh \Gammaa(\boldsymbol{q}, \Omega )\right)^2,
\end{equation}
where
\begin{equation}\label{3_eq:Gamma_q_w}
\Gammaa(\boldsymbol{q}, \Omega) = \sqrt{G^2 - \frac{(\Delta(\boldsymbol{q}, \Omega)L)^2}{4}}
\end{equation}
and
\begin{equation}\label{3_eq:Delta_q_w}
\Delta(\boldsymbol{q}, \Omega) = k_{sz}(\boldsymbol{q}, \Omega) + k_{iz}(-\boldsymbol{q}, -\Omega) - k_p.
\end{equation}
Here $\Delta(\boldsymbol{q}, \Omega)$ is the longitudinal phase mismatch, $k_{sz}$ and $k_{iz}$ are the longitudinal components of $\boldsymbol{k}_{s,i}$, the function $\Gammaa(\boldsymbol{q}, \Omega)$ gives the parametric gain at nonzero mismatch. At zero mismatch, $\Delta(\boldsymbol{q}, \Omega)=0$, the mean number $\mean{N_{s,i}(\boldsymbol{q}, \Omega)}$ of photons reaches its maximum, the function $\Gammaa(\boldsymbol{q}, \Omega)$ equals to $G$, the parametric gain at exact phase matching, and the Eq.~\eqref{3_eq:N_q_w} transforms into Eq.~\eqref{2_eq:N_si}.

Finally, the spectrum is
\begin{equation}\label{3_eq:S_q_w}
S(\boldsymbol{q}, \Omega) = \left( \frac{G}{\Gammaa(\boldsymbol{q}, \Omega)} \frac{\sinh \Gammaa(\boldsymbol{q}, \Omega )}{\sinh G}\right)^2.
\end{equation}
At low gain, $G\to0$, $S(\boldsymbol{q}, \Omega)$ transforms into the well-known $\mathrm{sinc}$ solution\footnote{$\mathrm{sinc}(x)$ is defined as $\sin(x)/x$.}~\cite{Klyshko1988},
\begin{equation}
S(\boldsymbol{q}, \Omega) = \mathrm{sinc}^2\left(\frac{\Delta(\boldsymbol{q}, \Omega)L}{2}\right).
\end{equation}

Here and further on the gain $G$ at exact phase matching is considered to be a constant value determined by the pumping and nonlinear medium~\cite{Klyshko1988},
\begin{equation}\label{3_eq:G_eq}
G=\frac{2\pi}{c}\sqrt{\frac{\omega_{s0}\,\omega_{i0}}{n_s n_i}}\chi^{(2)}LE_p,
\end{equation}
where $\chi^{(2)}$ is the second-order susceptibility for the interaction $\omega_p\to\omega_{s0}+\omega_{i0}$, $n_{s,i}$ is the refractive index at $\omega_{s0,i0}$, $L$ is the length of nonlinear medium, $E_p$ is the pump field amplitude. This assumption is valid for a small frequency detuning, $\Omega\ll\omega_{s0,i0}$, and a small emission angle, meaning that $q\equiv|\boldsymbol{q}|\ll |\boldsymbol{k}_{s,i}|$. However, in practice Eq.~\ref{3_eq:G_eq} often gives an imprecise result, and therefore it is always better to measure the gain experimentally, for example using the simple method described later in section~\ref{3_sec:broadening}.

Let us apply Eq.~\eqref{3_eq:S_q_w} to some particular case. In the simplest case, PDC is generated in a uniaxial nonlinear crystal and both signal and idler photons are polarized ordinary, i.e. perpendicular to the principal plane\footnote{In uniaxial crystals it is the plane containing wave vector $\boldsymbol{k}$ and optic axis.}. To satisfy the phase-matching condition, $\Delta(\boldsymbol{q}, \Omega)=0$, the pump is usually polarized extraordinary, i.e. in the principal plane. This type of interaction is labeled $e\to oo$ and called type-I.

Moreover, if the pump spatial walk-off is neglected, the spectrum will depend only on the modulus $q$ of transverse wave vector and be radially symmetric w.r.t. $\boldsymbol{k}_p$ direction~\cite{Klyshko1988}. This symmetry can be seen from the cover for chapter~\ref{1_chapter}. Therefore, $S(\boldsymbol{q}, \Omega)$ can be plotted as a function of only $q_x$, $S(q_x, \Omega)$, without loosing any information\footnote{Here and further on the second variable is set to be zero, e.g. $S(q_x, \Omega)\equiv S(q_x, q_y=0, \Omega)$.}.

\begin{figure}[!htb]
\begin{center}
\includegraphics[width=0.7\textwidth]{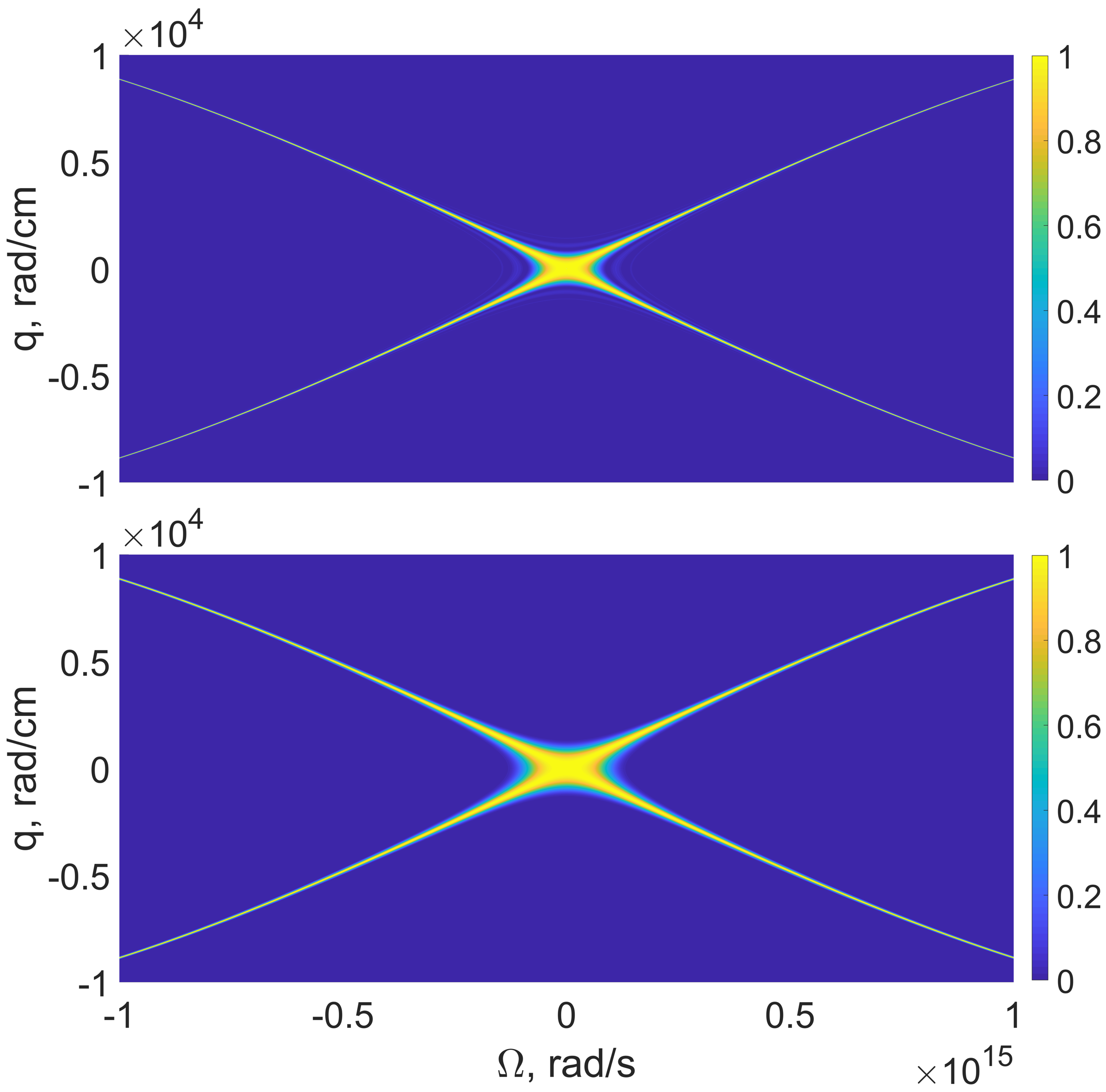}
\caption{Normalized frequency-wavevector spectrum $S(q_x, \Omega)$ for SPDC (top) and high-gain PDC with $G=10$ (bottom).}
\label{3_fig:S_q_w}
\end{center}
\end{figure}

The spectrum $S(q_x, \Omega)$ is plotted for beta barium borate (BBO) crystal with a 10~mm length and collinear frequency-degenerate phase matching, $\Delta(\boldsymbol{q}=0,\Omega=0)=0$ and $\omega_{s0}=\omega_{i0}=\omega_{p}/2$, for the pump wavelength $\lambda_p=400$~nm. The spectra for SPDC (top) and high-gain PDC (bottom) are shown in Fig.~\ref{3_fig:S_q_w}.

The spectrum is symmetric w.r.t. the central point ($q_x=0$, $\Omega=0$). It has a characteristic `X' shape and is practically unbounded, both in the frequency and in the wave vector, due to the fact that the mismatch gained by frequency detuning from degenerate collinear PDC can be compensated by noncollinear emission.

For high-gain PDC the spectrum is considerably broader. Moreover, the characteristic lobes provided by $\mathrm{sinc}$ function disappear, because at high gain the generation rate increases exponentially along the nonlinear crystal and the most photons are produced at its end. The effective length of the medium decreases, which leads to the broadening of the spectrum. This effect is demonstrated experimentally in section~\ref{3_sec:broadening}.

\subsection{X-shaped spatiotemporal coherence}

This characteristic shape has an interesting feature: the spectrum is nonfactorizable, $S(\boldsymbol{q}, \Omega)\ne S_{\boldsymbol{q}}(\boldsymbol{q})S_{\Omega}(\Omega)$. It leads to a nonfactorizable first-order CF $G^{(1)}(\boldsymbol{\xi},\tau)$, which characterizes spatial and temporal coherence of an electromagnetic field\footnote{Here and further on $\boldsymbol{\xi}$ and $\tau$ mean space and time delays, respectively.}, having an `X' shape as well~\cite{Jedrkiewicz2006}. The latter means that the coherence is neither spatial nor temporal, but has a more complicated structure, which will be clarified in this section.

Indeed, the CF $G^{(1)}(\boldsymbol{\xi},\tau)$ by virtue of the generalized Wiener–Khinchine theorem~\cite{Jedrkiewicz2007} is the Fourier transform (FT) of the spectrum\footnote{The transform in this form provides only the envelope of the CF $G^{(1)}$ without fast oscillations with an optical period in time and wavelength in space. However, mainly the envelope is important, because it defines an interferometric visibility.},
\begin{equation}
G^{(1)}(\boldsymbol{\xi},\tau)=\iint d\boldsymbol{q} d\Omega \, S(\boldsymbol{q}, \Omega) e^{i(\Omega\tau+\boldsymbol{q}\cdot\boldsymbol{\xi})}.
\end{equation}
The spectrum is radially symmetric, therefore the CF $G^{(1)}$ depends only on $\xi\equiv|\boldsymbol{\xi}|$ and can be considered as $G^{(1)}(\xi_x,\tau)$.

\begin{figure}[!htb]
\begin{center}
\includegraphics[width=0.7\textwidth]{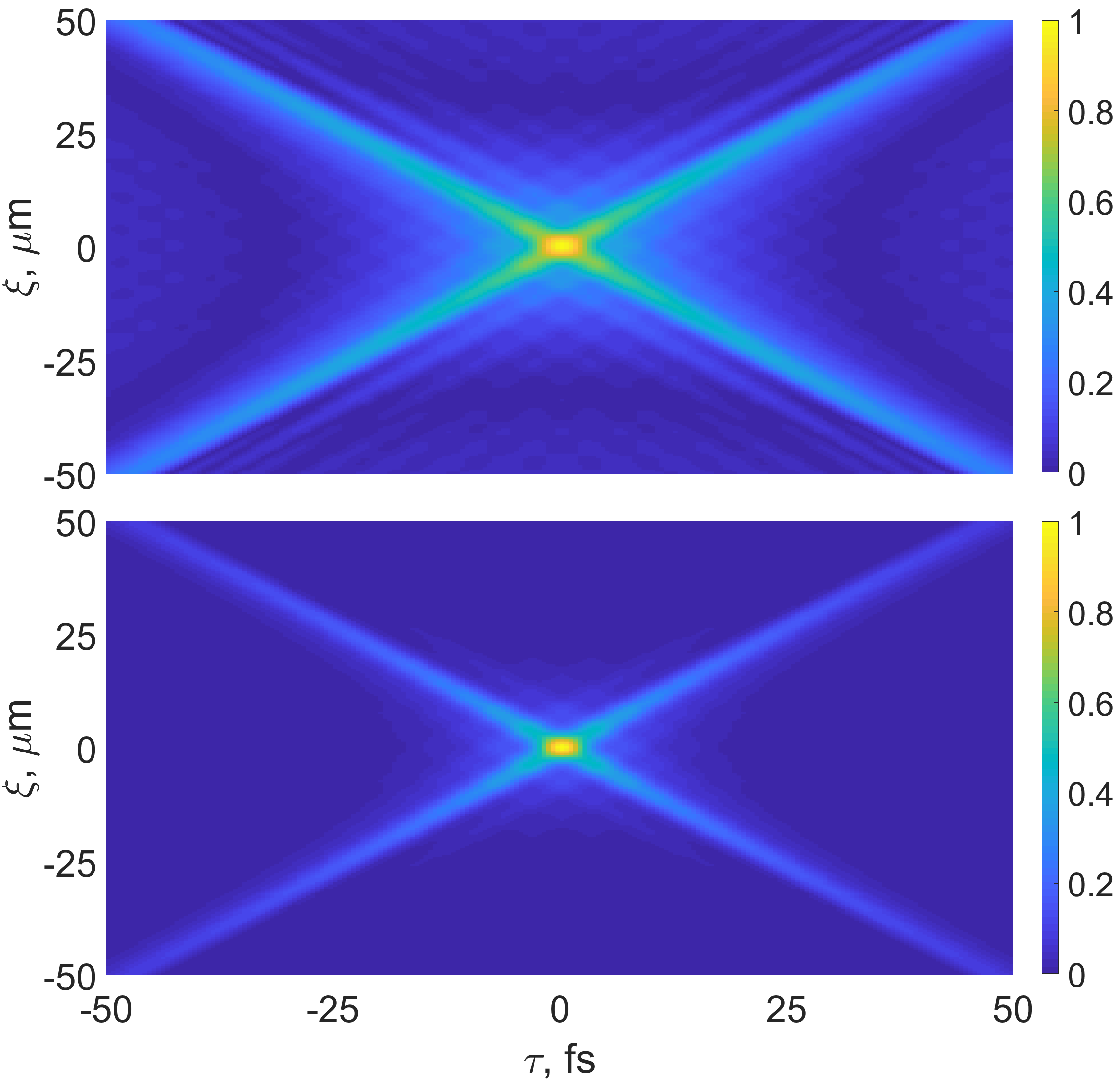}
\caption{The spatiotemporal CFs $|G^{(1)}(\xi_x,\tau)|$ (top) and $G^{(2)}(\xi_x,\tau)$ (bottom) normalized to their maximums for SPDC spectrum from Fig.~\ref{3_fig:S_q_w} (top).}
\label{3_fig:G1_G2_x_t}
\end{center}
\end{figure}

The modulus of the resulting spatiotemporal CF, $|G^{(1)}(\xi_x,\tau)|$, is shown in Fig.~\ref{3_fig:G1_G2_x_t} (top). For light with such CF the time and space coherence cannot be measured separately, on the contrary, the coherence can be still retrieved at large spatial delays by an additional temporal delay and vice versa.

This was demonstrated in the double-pinhole experiment~\cite{Jedrkiewicz2006,Jedrkiewicz2007} similar to the famous Young's double-slit one. In the Young's experiment the interference visibility is the largest at the point symmetric w.r.t. both pinholes. If the distance between the pinholes is larger than the width of the central $G^{(1)}$ maximum, the interference disappears.

On the contrary, for the X-shaped CF $G^{(1)}(\boldsymbol{\xi},\tau)$ the interference still exists having the maximum visibility at two different points. At each point there is a special time delay between light coming from different pinholes. This delay is defined by $G^{(1)}$ spatiotemporal trajectories; for example, for the CF $G^{(1)}$ from Fig.~\ref{3_fig:G1_G2_x_t} (top) the delay should be a bit less than 50~fs if the pinholes are separated by 50~$\mu$m.

A similar `X' shape and spatiotemporal behavior reveals itself in the second-order CF $G^{(2)}(\boldsymbol{\xi},\tau)$~\cite{Jedrkiewicz2009} that gives the rate of two-photon coincidences for SPDC. Indeed, for SPDC the CF $G^{(2)}(\boldsymbol{\xi},\tau)$ is given by the squared FT of the spectral amplitude $F(\boldsymbol{q}, \Omega)$~\cite{Chekhova2002,Jedrkiewicz2009},
\begin{equation}
G^{(2)}(\boldsymbol{\xi},\tau)=\left|\iint d\boldsymbol{q} d\Omega \, F(\boldsymbol{q}, \Omega) e^{i(\Omega\tau+\boldsymbol{q}\cdot\boldsymbol{\xi})}\right|^2.
\end{equation}
The square modulus, $|F(\boldsymbol{q}, \Omega)|^2$, of the spectral amplitude is equal to the spectrum $S(\boldsymbol{q}, \Omega)$. For SPDC the spectral amplitude is given by~\cite{Chekhova2002,Jedrkiewicz2009}
\begin{equation}
F(\boldsymbol{q}, \Omega)=\mathrm{sinc}\left(\frac{\Delta(\boldsymbol{q}, \Omega)L}{2}\right)\exp\left(i\frac{\Delta(\boldsymbol{q}, \Omega)L}{2}\right).
\end{equation}

Similar to the spectrum, the amplitude $F(\boldsymbol{q}, \Omega)$ has `X' shape and is nonfactorizable. Therefore, the second-order CF $G^{(2)}$ has the shape similar to the first-order CF $G^{(1)}$, the only major difference being that $G^{(2)}$ is narrower (see Fig.~\ref{3_fig:G1_G2_x_t}). In this case the `X' shape means that the spatial degrees of freedom of a photon pair are entangled to its temporal degrees of freedom; this effect is called \textit{X entanglement}~\cite{Jedrkiewicz2009}.

\subsection{Wavelength-angular spectrum and the $\lambda^4$ factor}
\label{3_sub:lambda4}

The variables $(\boldsymbol{q}, \Omega)$ are natural for the theoretical description of PDC, yet, not for the measurement techniques. In reality, one measures the radiation within a certain angular aperture $\Delta\theta_x\Delta\theta_y$ and wavelength interval $\Delta\lambda$, thereby acquiring the wavelength-angular spectrum, $S_{\theta\lambda}(\theta_x,\theta_y, \lambda)$, which differs from the change of variables in the frequency-wavevector spectrum, $S_{q\Omega}(\theta_x,\theta_y, \lambda)\equiv S[\boldsymbol{q}(\theta_x,\theta_y, \lambda), \Omega(\lambda)]$, meaning that the wavelength-dependent Jacobian should be taken into account.

Indeed, the total energy for a unit interval should be the same,
\begin{equation}
S_{\theta\lambda}(\theta_x,\theta_y, \lambda) d\theta_x d\theta_y d\lambda = S(\boldsymbol{q}, \Omega) d\boldsymbol{q} d\Omega.
\end{equation}
Under the assumptions of small angles, $\theta_{x,y}\ll1$ and a constant refractive index $n$ within the PDC band\footnote{Both assumptions are fairly valid for PDC: the angles are mostly below 0.1 radian and the change of the refractive index is below 1\%.}, the Jacobian is equal to $8\pi^3n^2c/\lambda^4$. Therefore,
\begin{equation}
S_{\theta\lambda}(\theta_x,\theta_y, \lambda) \sim \frac{S_{q\Omega}(\theta_x,\theta_y, \lambda)}{\lambda^4}.
\end{equation}
The constant factor is not important, because the spectra are usually normalized to their maximums. One can get the same $\lambda^4$ factor from considerations based on the dependence of the mode size on the wavelength~\cite{Klyshko1988,Spasibko2012,Lemieux2019}.

This $\lambda^4$ factor is often forgotten in $S_{\theta\lambda}$ theoretical plots or in experimental $S(\boldsymbol{q}, \Omega)$ ones. The asymmetry along the wavelength is either not explained at all~\cite{Jedrkiewicz2007} or attributed to losses~\cite{Baek2008}.

\begin{figure}[!htb]
\begin{center}
\includegraphics[width=0.7\textwidth]{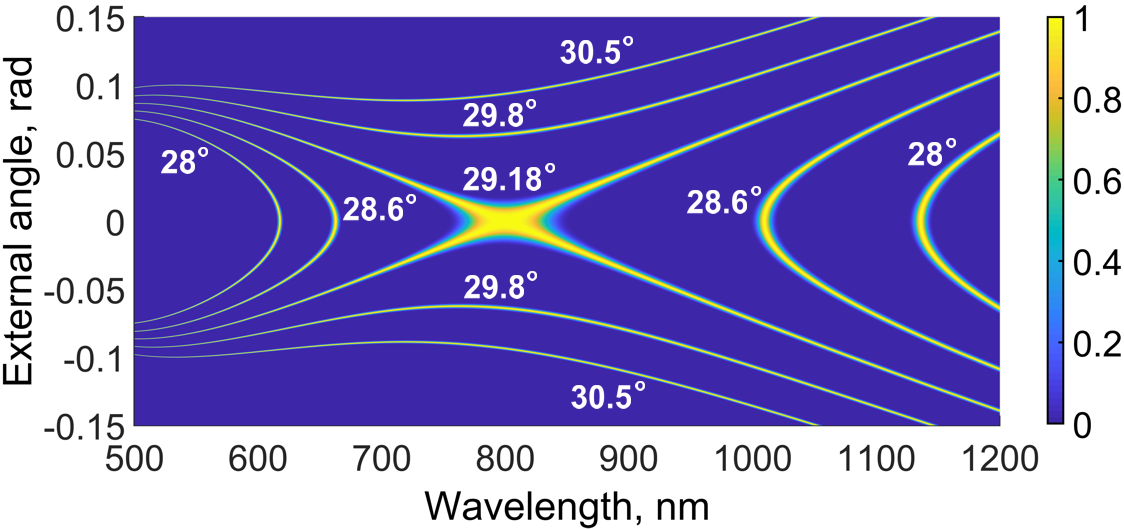}
\caption{The spectrum $S_{q\Omega}(\theta_x, \lambda)$ for collinear degenerate ($\phi=29.18^\circ$) and nondegenerate ($\phi=28^\circ$ and $28.6^\circ$), and noncollinear degenerate ($\phi=29.8^\circ$ and $30.5^\circ$) phase matching. $\theta_x$ is calculated outside the BBO crystal.}
\label{3_fig:tuning_curves}
\end{center}
\end{figure}

Nevertheless, in many cases the spectrum $S_{q\Omega}(\theta_x,\theta_y, \lambda)$ is useful without any correction; then it has the meaning of a tuning curve, which shows the phase-matched wavelengths and angles without the information on the relative brightness. As an example, $S_{q\Omega}(\theta_x, \lambda)$ is shown not only for collinear degenerate phase matching considered before, but also for nondegenerate and noncollinear degenerate ones (Fig.~\ref{3_fig:tuning_curves}). The phase matching can be changed by tuning the angle $\phi$ between $\boldsymbol{k_p}$ and the optic axis; the change in the pump refractive index changes the pump wave vector, which, in its turn, affects the phase matching.

\subsection{PDC correlations and eigenmodes}
\label{3_sub:PDC_correlations}

Previously the total PDC spectrum was discussed assuming a stationary and plane-wave pump, which leads to $\delta$-correlated signal and idler photons. However, in many cases this is not true; the correlation width is important.

Therefore, all possible frequencies and wave vectors should be considered\footnote{One can also add the polarization degree of freedom.}. For example, for the state $\ket{\Phi_2}$~\eqref{2_eq:SPDC_state_2} the main term, $\ket{\bar{\Phi}_2}=c_1\ket{1}_s\ket{1}_i$, changes to~\cite{Burlakov1997,Mikhailova2008,Perina2019}
\begin{equation}
\ket{\bar{\Phi}_2} = \iiiint d\boldsymbol{q}_s d\boldsymbol{q}_i d\Omega_s d\Omega_i \, F(\boldsymbol{q}_s,\boldsymbol{q}_i,\Omega_s,\Omega_i) \ket{1}_s\ket{1}_i.
\end{equation}
Here $F(\boldsymbol{q}_s,\boldsymbol{q}_i,\Omega_s,\Omega_i)$ is the joint spectral amplitude (JSA), which is also known as two-photon amplitude. JSA is the probability amplitude for signal and idler photons to be emitted with wave vectors $\boldsymbol{q}_{s,i}$ and frequencies $\Omega_{s,i}$.

The JSA $F(\boldsymbol{q}_s,\boldsymbol{q}_i,\Omega_s,\Omega_i)$ fully characterizes the state. Unfortunately, in the general form the JSA is too complicated, as it depends on 6 independent variables; until now there are only few attempts~\cite{Perina2019} to deal with all variables together.

At first, one usually separates spatial and temporal degrees of freedom. Experimentally it could be done by angular filtering in the collinear direction or frequency filtering at the degenerate wavelength. In the frequency domain~\cite{Mikhailova2008}, the JSA can be expressed as\footnote{The collinear emission is assumed, $\boldsymbol{q}_s=\boldsymbol{q}_i=0$.}
\begin{equation}
F(\Omega_s,\Omega_i) = \exp\left( -\frac{(\Omega_s + \Omega_i)^2 \tau_d^2}{8\ln2}\right) \mathrm{sinc}\left(\frac{\Delta(\Omega_s,\Omega_i)L}{2}\right) \exp\left(i\frac{\Delta(\Omega_s,\Omega_i)L}{2}\right),
\end{equation}
here the pump pulse duration\footnote{For non-bandwidth-limited or CW pump it is replaced by the coherence time.} $\tau_d$ is taken into account and the longitudinal mismatch $\Delta(\Omega_s,\Omega_i)$ depends on both signal and idler frequencies.

Then, one measures only the JSI, the squared modulus of the JSA, so the JSA phase is ignored. With these assumptions the spatiotemporal and phase effects are lost; however, it is still possible to describe some effects.

\begin{figure}[!htb]
\begin{center}
\includegraphics[width=0.5\textwidth]{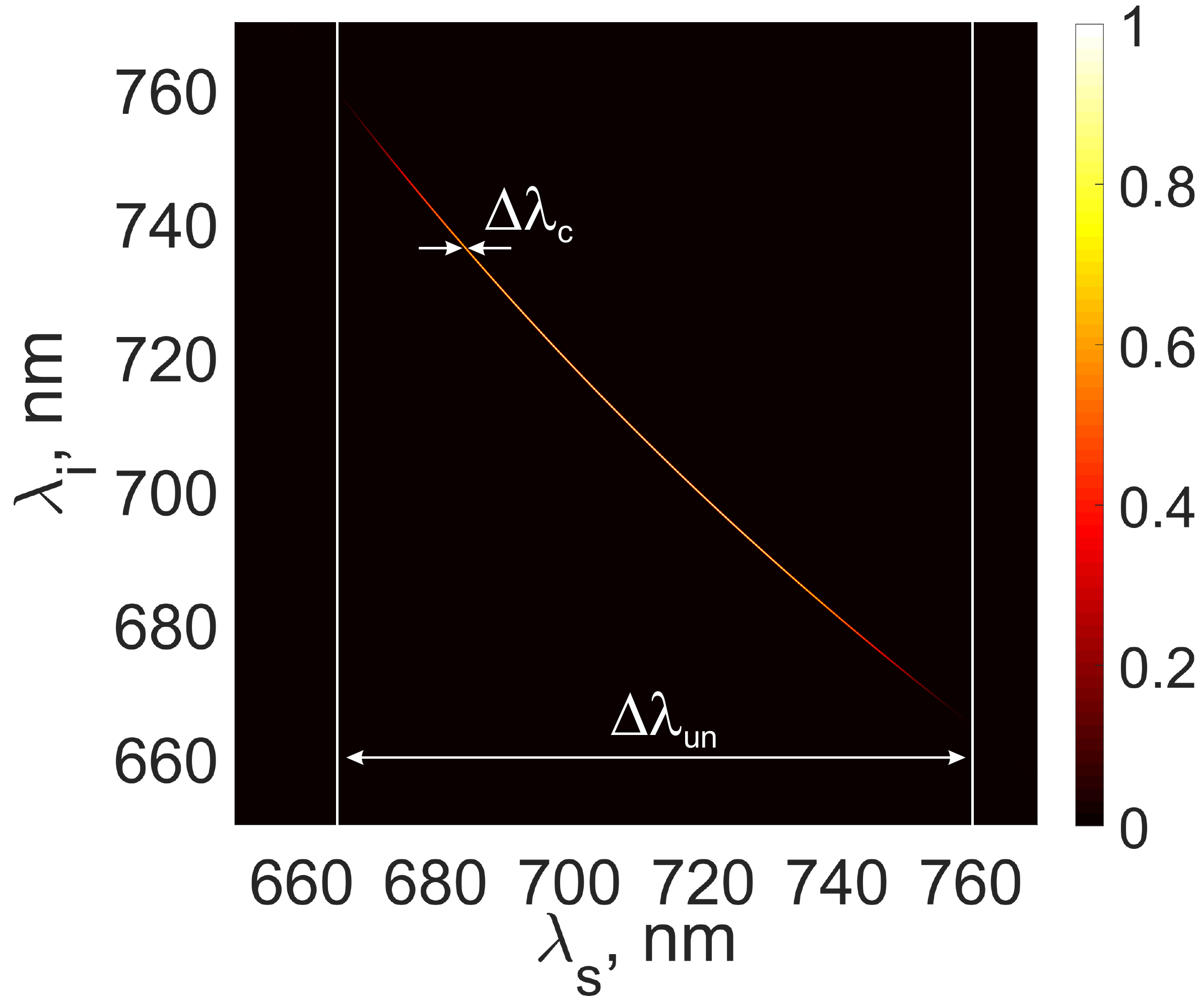}
\caption{The JSI $|F(\Omega_s,\Omega_i)|^2$ plotted for the parameters considered later in the experiment on PDC correlations (section~\ref{3_sec:correlations}). The JSI is shown in variables $\lambda_s$ and $\lambda_i$ for comparison with the experiment (Fig.~\ref{3_fig:JSI_exp}). The conditional ($\Delta\lambda_c$) and unconditional ($\Delta\lambda_{un}$) widths are marked in the figure.}
\label{3_fig:JSA}
\end{center}
\end{figure}

In Fig.~\ref{3_fig:JSA} the JSI is plotted for the two-photon analogue of the state considered later in the experiment on PDC correlations (section~\ref{3_sec:correlations}). The state is produced via PDC in a 2~mm BBO crystal with type-I collinear frequency-degenerate phase matching pumped by a pulsed laser at 354.7~nm with 5~ps coherence time.

Typically, the JSI has two characteristic parameters. The first one is the correlation width, also called the conditional width, $\Delta\Omega_c$ (or $\Delta\lambda_c$), because it is the spectral width for e.g. signal photons under the \textit{condition} $\lambda_i=\mathrm{const}$\footnote{Namely, the idler photons are measured at this wavelength.}. The other one is the total spectral width without any conditions. Therefore, it is called the unconditional width, $\Delta\Omega_{un}$ (or $\Delta\lambda_{un}$).

In order to get the eigenmodes of the state $\ket{\bar{\Phi}_2}$ and their weights one should apply the singular value (Schmidt) decomposition~\cite{Law2000} on the JSA. Then, indeed, the JSA is
\begin{equation}\label{3_eq:Schmidt_decomp}
F(\Omega_s,\Omega_i) = \sum_{l=0}^\infty \sqrt{\Lambda_l} \psi_l(\Omega_s) \phi_l(\Omega_i),
\end{equation}
where $\psi_l(\Omega_s)$ and $\phi_l(\Omega_i)$ are the orthonormal eigenmodes, i.e. the Schmidt modes, and $\Lambda_l$ are the corresponding weights, i.e the Schmidt coefficients. The latter are normalized as
\begin{equation}\label{3_eq:Schmidt_coeff_norm}
\sum_{l=0}^\infty \Lambda_l=1.
\end{equation}

In this case the effective number $K$ of modes is~\cite{Eberly2006}
\begin{equation}
K=\left(\sum_{l=0}^\infty \Lambda_l^2 \right)^{-1}.
\end{equation}
It is called the Schmidt number and provides the degree of entanglement~\cite{Law2000}, namely the dimensionality of the state, for example, for information processing in quantum computing~\cite{Lloyd1999}.

The Schmidt number can be also assessed without the JSA measurement and the Schmidt decomposition. As it is shown in Ref.~\cite{Brida2009}, in many cases the number $K$ is equal to the Fedorov ratio $R$~\cite{Fedorov2006}, the ratio of unconditional and conditional widths,
\begin{equation}\label{3_eq:Fedorov_R}
R\equiv\frac{\Delta\Omega_{un}}{\Delta\Omega_{c}}.
\end{equation}
Thus, the degree of entanglement can be estimated just from two measured JSI widths.

The Schmidt-mode description can be used not only at low gain, but also at high one. In the first approximation the Schmidt modes are the same as in the low gain\footnote{Actually, the Schmidt modes are broadened a bit with the gain~\cite{Sharapova2019}.}~\cite{Sharapova2015}, only a redistribution of the Schmidt coefficients happens,
\begin{equation}\label{3_eq:Schmidt_tilde}
\tilde{\Lambda}_l=\frac{(v_l)^2}{\sum_{l=0}^\infty(v_l)^2},
\end{equation}
since the number of photons produced by each mode, $(v_l)^2$, changes with the gain. Here $v_l=\sinh(G_l)$ and $G_l=\tilde{G}\sqrt{\Lambda_l}$ is the parametric gain for mode $l$. Due to this redistribution, the number $K$ of modes decreases and the width $\Delta\Omega_{c}$ increases~\cite{Sharapova2015}.

%%%%%%%%%%%%%%%%%%%%%%%%%%%%%%%%%%%%%%%%%%%%%%%%%%%%%%%%%%%%%%%%%%%%
\section{PDC in the anomalous group velocity dispersion (GVD) range}
\label{3_sec:O_PDC}

The previous section describes PDC with X-shaped spectrum. This is the usual case, PDC is generated in the visible range or close to it. That is the range of normal GVD in the crystal, $(\partial^2 k/\partial\omega^2)|_{\omega_p/2}>0$.

However, if the pump wavelength goes into the infrared (IR) range, GVD becomes anomalous, $(\partial^2 k/\partial\omega^2)|_{\omega_p/2}<0$. In this range the spectrum is restricted in wavelengths and angles, because the frequency detuning $\Omega$ from degenerate collinear phase matching cannot be compensated.

\begin{figure}[!htb]
\begin{center}
\includegraphics[width=0.7\textwidth]{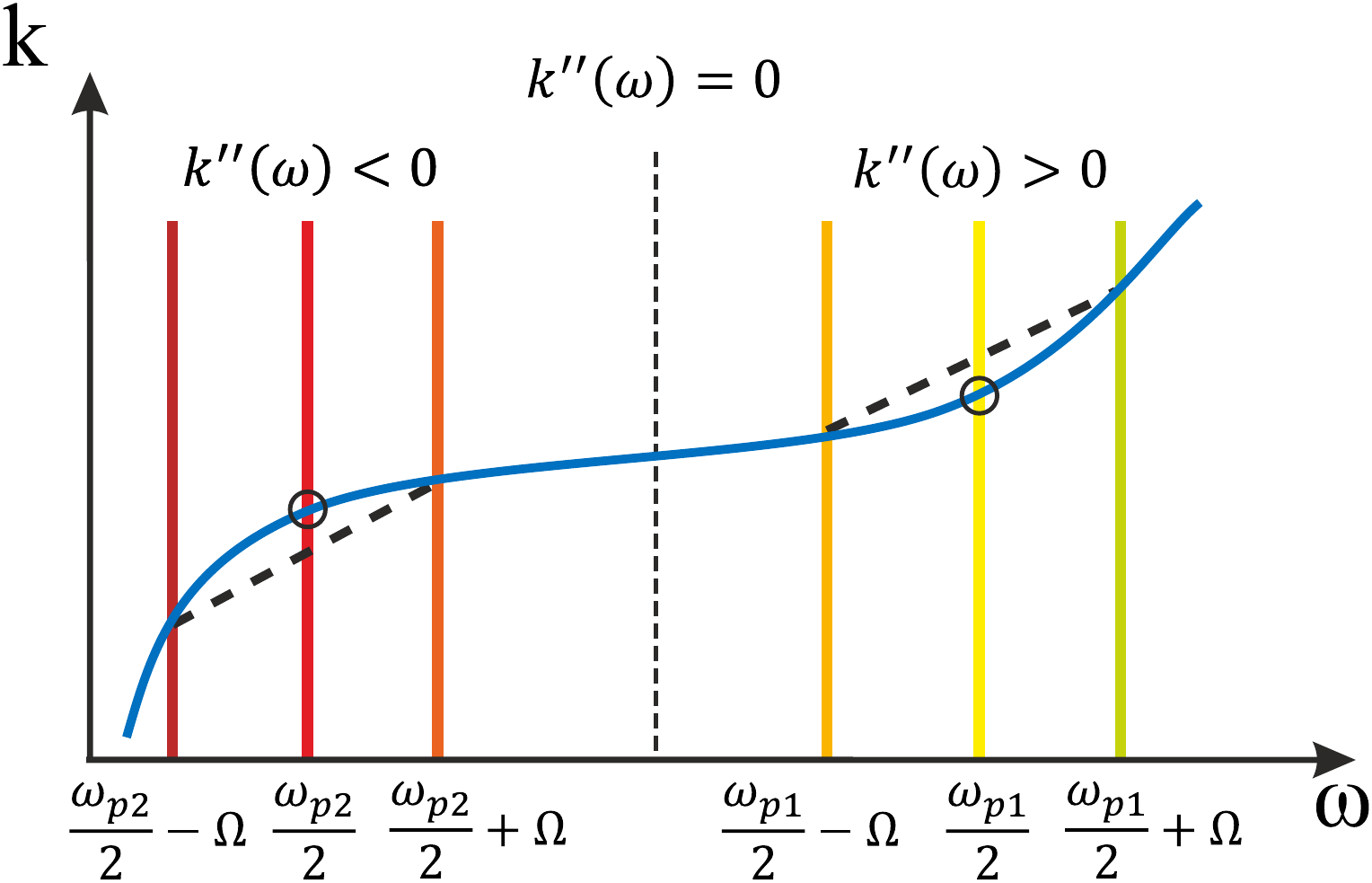}
\caption{Schematic representation of the dispersion relation $k(\omega)$ demonstrating the possibility of simultaneous collinear degenerate and noncollinear nondegenerate phase matching in normal GVD range, $k''(\omega)>0$, and impossibility in anomalous one, $k''(\omega)<0$.}
\label{3_fig:k_w_explanation}
\end{center}
\end{figure}

This difference can be explained through using the schematic representation of the dispersion relation\footnote{In reality, $k(\omega)$ is very close to straight line.} $k(\omega)$ (Fig.~\ref{3_fig:k_w_explanation}). Indeed, for degenerate collinear phase matching $k_s(\omega_p/2)=k_i(\omega_p/2)=k_p/2$ (marked by two circles). For detuned frequency the phase matching in the collinear direction is impossible, because the sum $k_s(\omega_p/2+\Omega)+k_i(\omega_p/2-\Omega)$ is either larger than $k_p/2$, $k(\omega)$ is convex at $\omega_{p1}/2$, or smaller, $k(\omega)$ is concave at $\omega_{p2}/2$. However, in the first case, normal GVD range, the phase matching is still possible for noncollinear emission, because it reduces the longitudinal components $k_{sz}$ and $k_{iz}$, whereas in the second case, anomalous GVD range, it becomes impossible since the emission being noncollinear only increases mismatch.

\subsection{PDC at zero GVD}

Moreover, it is interesting that exactly at zero GVD, $(\partial^2 k/\partial\omega^2)|_{\omega_p/2}=0$, the spectrum is extremely broadband in frequency at collinear direction~\cite{Strekalov2005}. The mystery again lies in the mismatch $\Delta(\Omega)\equiv\Delta(\boldsymbol{q}=0, \Omega)$.

In its expression~\eqref{3_eq:Delta_q_w} both\footnote{In the collinear direction $k_{sz}(\boldsymbol{q}=0, \Omega)=k_s(\Omega)$ and $k_{iz}(-\boldsymbol{q}=0, -\Omega)=k_i(-\Omega)$} $k_{s}(\Omega)$ and $k_i(-\Omega)$ can be expanded in the Taylor series around $\Omega=0$. Thus,
\begin{equation}\label{3_eq:Delta_Omega}
\Delta(\Omega) = (k_{s}(0) + k_{i}(0) - k_p) + \sum_{l=1}^\infty\left[k_{s}^{(l)}(0) + (-1)^l k_{i}^{(l)}(0)\right]\frac{\Omega^l}{l!}.
\end{equation}
The first term is equal to zero due to phase matching. In the general case $k'_{s}(0)\ne k'_{i}(0)$ and the term with $l=1$ defines the width of the spectrum. This is the case for nondegenerate phase matching (see Fig.~\ref{3_fig:tuning_curves}). Yet, in case of degenerate one, $k'_{s}(0)=k'_{i}(0)$, only the term with $l=2$ plays a role. Therefore, the spectral width in the collinear direction is much broader in the degenerate case than in the nondegenerate one; just compare the cases with $\phi=29.18^\circ$ and $28.6^\circ$ (Fig.~\ref{3_fig:tuning_curves}).

At zero GVD only the term with $l=4$ is nonzero, therefore, the phase matching is extremely broad. It is much broader than in the case of any degenerate phase matching and can cover more than 100~THz~\cite{Tiihonen2006}. This fact is used in works on parametric generation and amplification~\cite{Kuo2006,ODonnell2007,Lim2007}.

\subsection{Experimental setup}
\label{3_sub:Setup_O_PDC}

\begin{figure}[!htb]
\begin{center}
\includegraphics[width=0.7\textwidth]{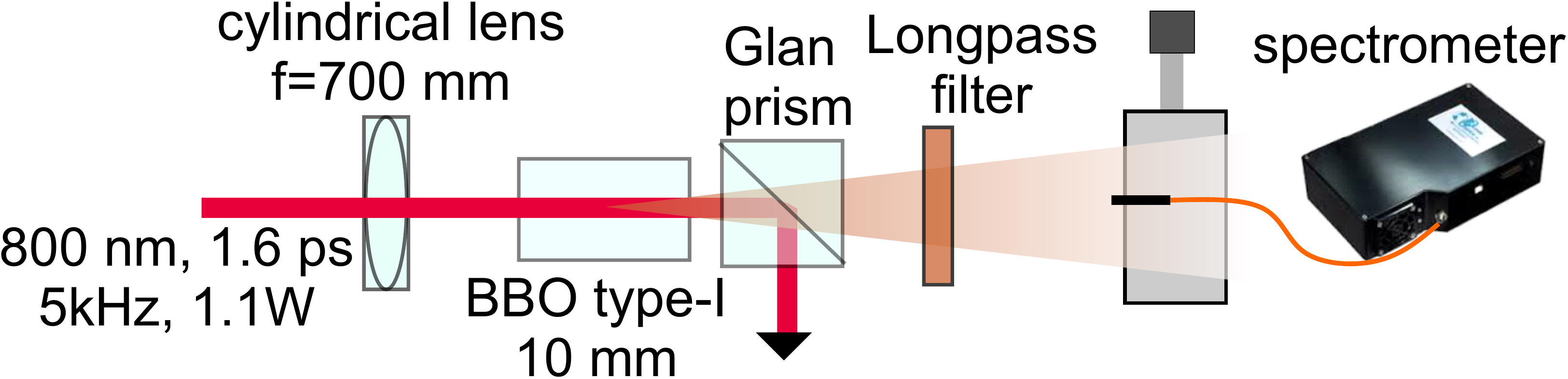}
\caption{Experimental setup for measuring PDC wavelength-angular spectra.}
\label{3_fig:Setup_O_PDC}
\end{center}
\end{figure}

The PDC is generated at the anomalous GVD range in a 10 mm BBO crystal using amplified Ti-sapphire laser with 800~nm wavelength as a pump (Fig.~\ref{3_fig:Setup_O_PDC}). The degenerate wavelength is 1600~nm, thus it is no so far from 1431~nm, which is the zero GVD wavelength for BBO. Accordingly, the spectrum is not of a record width, but it is broad enough.

The laser is a pulsed one, with 1.6~ps pulse duration and 5~kHz repetition rate. The mean power of 1.1~W is used. The crystal is cut at $19.9^{\circ}$ to the optic axis. This orientation provides nearly collinear frequency-degenerate phase matching. To reduce spatial walk-off effects, the pump radiation is focused into the crystal by means of a 700~mm cylindrical lens, so that no focusing takes place in the principal plane. The pump radiation is cut off by a Glan prism and a long-pass filter (FEL1200). The wavelength-angular spectra $S_{\theta\lambda}(\theta_x, \lambda)$ are measured with an IR fiber spectrometer (Ocean Optics NIRQuest256) with the fiber tip (600 $\mu$m) placed at 40~cm distance from the crystal. By displacing the fiber tip and each time recording the wavelength spectrum, the spectra at different angles are obtained. The angular scanning is done in the horizontal direction, in which the pump beam is not focused.

\subsection{Wavelength-angular spectra and correlation functions (CF)}

\begin{figure}[!htb]
\begin{center}
\includegraphics[width=0.7\textwidth]{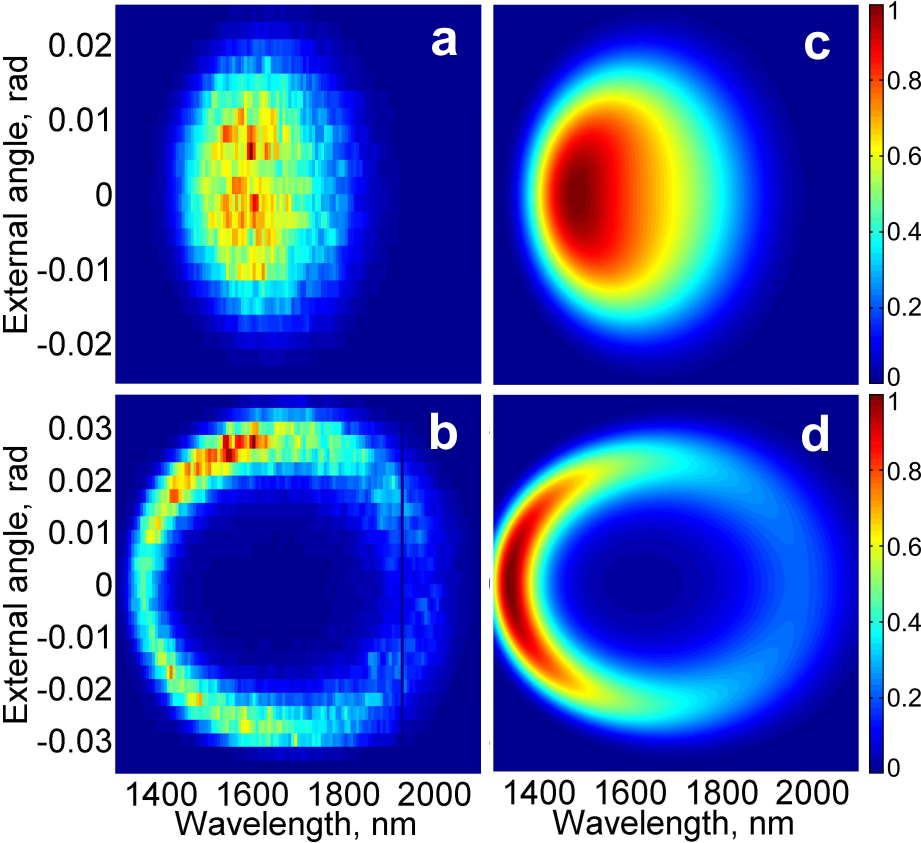}
\caption{Experimental [(a), (b)] and theoretical [(c), (d)] wavelength-angular spectra $S_{\theta\lambda}$ for crystal orientations $\phi=19.87^{\circ}$ [(a), (c)] and $\phi=19.98^{\circ}$ [(b), (d)].}
\label{3_fig:spectra_O_PDC}
\end{center}
\end{figure}

The spectra for $\phi=19.87^{\circ}$ and $19.98^{\circ}$ are obtained by combining the wavelength spectra at different angles into 2D plot. They are presented in Fig.~\ref{3_fig:spectra_O_PDC} together with the corresponding theoretical spectra calculated using the approach discussed in section~\ref{3_sub:freq_wave_spectra}.

The spectra are indeed restricted in both wavelength and angle. At exact collinear frequency-degenerate phase matching, $\phi=19.87^{\circ}$, the spectrum has the shape of a spot while, with the crystal detuned from this orientation, $\phi=19.98^{\circ}$, the shape resembles a ring. With the tuning in the other direction, the smaller angle $\phi$, the phase matching disappears. As described in section~\ref{3_sub:lambda4} the intensity decreases toward long wavelengths due to the $\lambda^4$ factor. Also one could notice that the experimental ring-shaped spectrum [panel (b)] is brighter at the large angles. This is caused by the spatial walk-off effect that will be described in section~\ref{4_sec:giant_twin_beams}.

\begin{figure}[!htb]
\begin{center}
\includegraphics[width=0.7\textwidth]{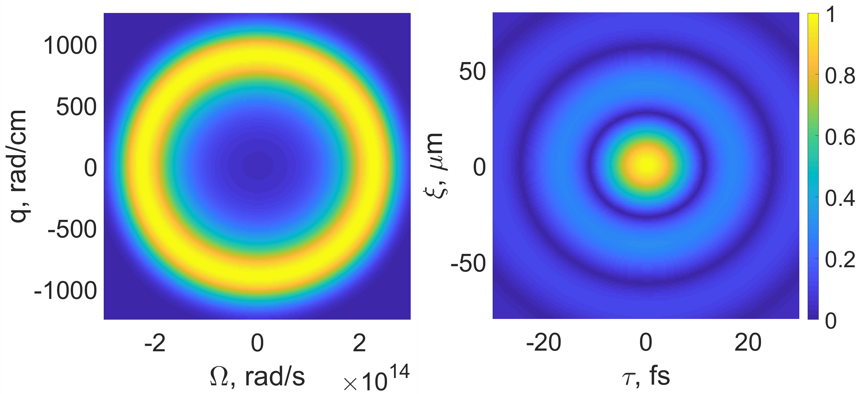}
\caption{Normalized frequency-wavevector spectrum $S(q_x, \Omega)$ (left) and the spatiotemporal CF $|G^{(1)}(\xi_x,\tau)|$ (right) for $\phi=19.98^{\circ}$.}
\label{3_fig:O_S_G1}
\end{center}
\end{figure}

The spectrum $S_{\theta\lambda}$, as shown in Fig.~\ref{3_fig:spectra_O_PDC} (b,d), converted to $S(\boldsymbol{q}, \Omega)$ acquires an ideal ring shape (Fig.~\ref{3_fig:O_S_G1}). Similarly to the X-shaped case (see Fig.~\ref{3_fig:S_q_w}) it is nonfactorizable and leads to a nonfactorizable spatiotemporal CF $G^{(1)}(\boldsymbol{\xi},\tau)$. In the rings, the lack of temporal coherence and the lack of spatial coherence can compensate for each other.

Even more interestingly, this type of phase matching leads to a new type of entanglement. Since the CF $G^{(1)}(\boldsymbol{\xi},\tau)$ at low gain has the similar shape as at high one, the CF $G^{(2)}(\boldsymbol{\xi},\tau)$ resembles ring-shaped structure. However, in contrast to the first-order CF, the second-order one is sensitive to the phase of the spectral amplitude. By imparting on the latter an axially varying phase, one can get rid of the central peak and obtain a photon pair with unusual properties: the space and time intervals between entangled photons will lie on a circle,
\begin{equation}\label{3_eq:ring_formula}
\frac{\xi_x^2}{\xi_0^2} + \frac{\tau_x^2}{\tau_0^2} = 1,
\end{equation}
where $\xi_0$ and $\tau_0$ are the correlation distance and time, which depend on the phase matching condition and the applied phase. This means that the two photons are never at the same point at the same time.

\begin{figure}[!htb]
\begin{center}
\includegraphics[width=1\textwidth]{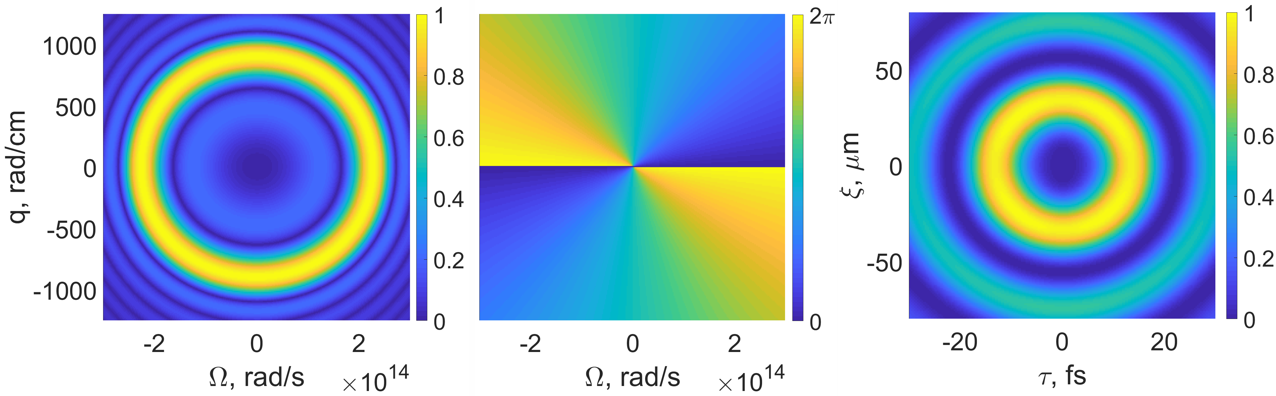}
\caption{Absolute value of the spectral amplitude $|F(q_x, \Omega)|$ (left), the phase distribution $e^{2i\psi}$ imparted on it (center), and the resulting CF $G^{(2)}(\xi_x,\tau)$ for $\phi=19.98^{\circ}$ (right).}
\label{3_fig:O_S_G2_phase}
\end{center}
\end{figure}

The axially varying phase can be imparted on the spectral amplitude by a spatial light modulator at the output of crossed-dispersion scheme~\cite{Burlakov1997,Jedrkiewicz2007} where the 2D distribution as in Fig.~\ref{3_fig:spectra_O_PDC} is formed. Figure~\ref{3_fig:O_S_G2_phase} shows an example of a such CF. The left panel presents the absolute value of a ring-shaped spectral amplitude $|F(q_x, \Omega)|$ for low-gain PDC for $\phi=19.98^{\circ}$. The azimuthally varying phase in the frequency-wavevector space replaces the spectral amplitude $F(q_x, \Omega)$ by
\begin{equation}
\tilde{F}(q_x, \Omega)=F(q_x, \Omega)e^{il\psi},
\end{equation}
where $\psi$ is the azimuthal angle and integer $l=2$ in the example (central panel). The resulting CF indeed has the shape of a ring given by Eq.~\eqref{3_eq:ring_formula}, with $\xi_0=38$~$\mu$m and  $\tau_0=13$~fs. In general, any phase modulation with $l\ne0$ will lead to a similar result; a larger integer $l$ leads to a bigger size of the ring.

\subsection{Possible applications of PDC in the anomalous GVD range}

PDC in the anomalous GVD range is useful from two viewpoints. Firstly, a spectrum restricted in angle and wavelength provides a possibility to collect all radiation without losses, which is an advantage in experiments with squeezed light. Indeed, to use the photon-number correlations, one usually has to filter the spectrum, which is usually accompanied by losses. For lossless filtering one should use a projection technique~\cite{Perez2015_2} that is difficult to realize, especially for the frequency domain. If the spectrum is restricted, no filtering is necessary.

The unusual type of correlation, with two photons always separated in space or in time, can be useful if one needs to avoid two-photon effects such as absorption. On the other hand, the original single-peaked CF can be restored by an appropriate phase modulation. This fact can be used in time and space-resolved nonlinear spectroscopy with two-photon light. For instance, by varying the size of the ring and observing a two-photon effect, one can measure the nonlinear response time of the material.

%%%%%%%%%%%%%%%%%%%%%%%%%%%%%%%%%%%%%%%%%%%%%%%%%%%%%%%%%%%%%%%%%%%%%%%%%%%%%
\section{Broadening of PDC spectrum}\label{3_sec:broadening}

In section~\ref{3_sub:freq_wave_spectra} it was mentioned that PDC spectrum exhibits broadening at high-gain regime. Here the first experimental measurement of the effect is presented and it is discussed in more details.

\subsection{Experimental setup}

\begin{figure}[!htb]
\begin{center}
\includegraphics[width=0.7\textwidth]{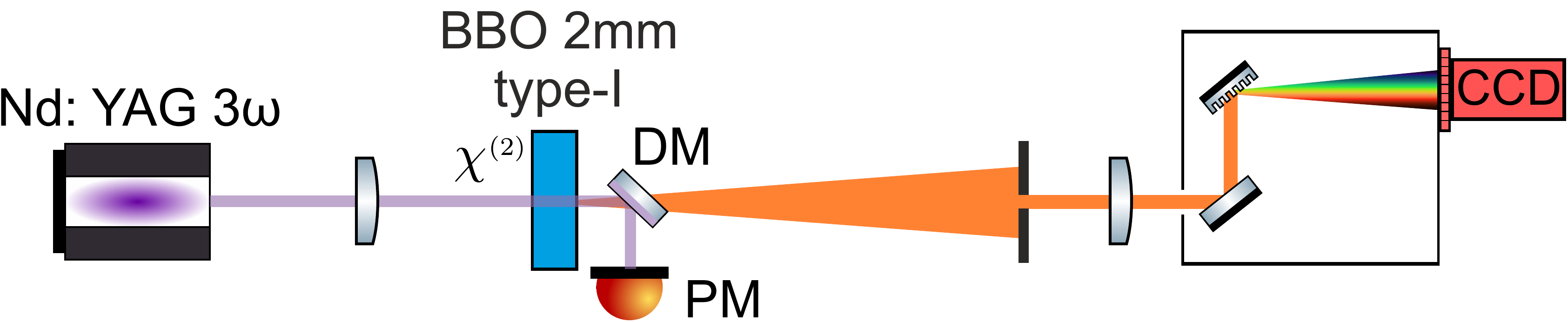}
\caption{Experimental setup for measuring the spectral broadening.}
\label{3_fig:Setup_broadening}
\end{center}
\end{figure}

In this experiment PDC is obtained in a 2 mm BBO crystal cut for type-I collinear frequency-degenerate phase matching (Fig.~\ref{3_fig:Setup_broadening}). As the pump, the third harmonic of a~Nd:YAG laser with the wavelength $\lambda_p = 354.7$~nm, pulse duration 18~ps, repetition rate 1~kHz, and the energy per pulse up to 0.1~mJ is used. The coherence time of the pump is less than the pulse duration and equal to 5~ps. The pump is focused by a 1~m lens into the crystal and reflected after it by a dichroic mirror (DM). The pump power is controlled with a power meter (PM).

This experiment is devoted to the frequency domain, therefore the nonfactorazibility of $S(\boldsymbol{q}, \Omega)$ is removed via the filtering in the collinear direction, being done with an aperture placed in the far-field zone. The emission angle selected this way is $0.17^\circ$ (inside the crystal).

All radiation passing through the aperture is focused on the input slit of a HORIBA Jobin Yvon Micro HR monochromator equipped with a CCD array. The spectral measurements are performed with the best possible resolution 0.2~nm.

\begin{figure}[!htb]
\begin{center}
\includegraphics[width=0.7\textwidth]{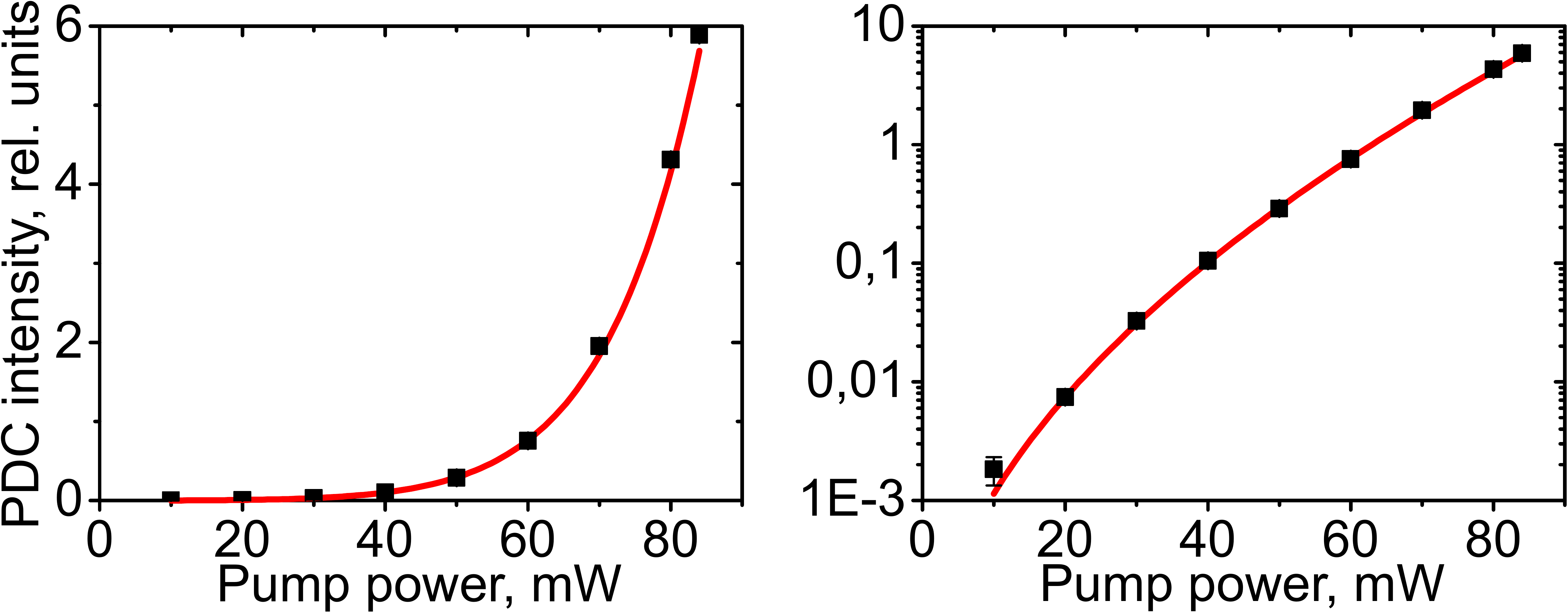}
\caption{The dependence of the PDC intensity on the pump power (points) and its fit (line) with Eq.~\eqref{3_eq:gain_fit} in linear (left) and log-linear (right) scales.}
\label{3_fig:gain_dependence}
\end{center}
\end{figure}

In all measurements, the spectral properties are studied depending on the parametric gain $G$, which is found from the nonlinear dependence of PDC intensity $I$ on the pump power $P$ (see Fig.~\ref{3_fig:gain_dependence}). The fitting function~\cite{Ivanova2006},
\begin{equation}\label{3_eq:gain_fit}
I = I_0\sinh^2(B\sqrt{P}),
\end{equation}
is based on Eqs.~\eqref{2_eq:sinh2} and \eqref{3_eq:N_q_w} with the fitting parameters $I_0$ and $B$. The gain should be measured around the zero mismatch, otherwise instead of $G$ one gets some averaged value of $\Gammaa$, see Eqs.~\eqref{3_eq:N_q_w} and \eqref{3_eq:Gamma_q_w}. In this experiment the PDC should be well filtered in $\boldsymbol{q}$ and $\Omega$ near the collinear frequency-degenerate point. After the fitting, each pump power corresponds to a certain gain value, $G = B\sqrt{P}$.

\subsection{Measurement of the spectral broadening}

\begin{figure}[!htb]
\begin{center}
\includegraphics[width=0.5\textwidth]{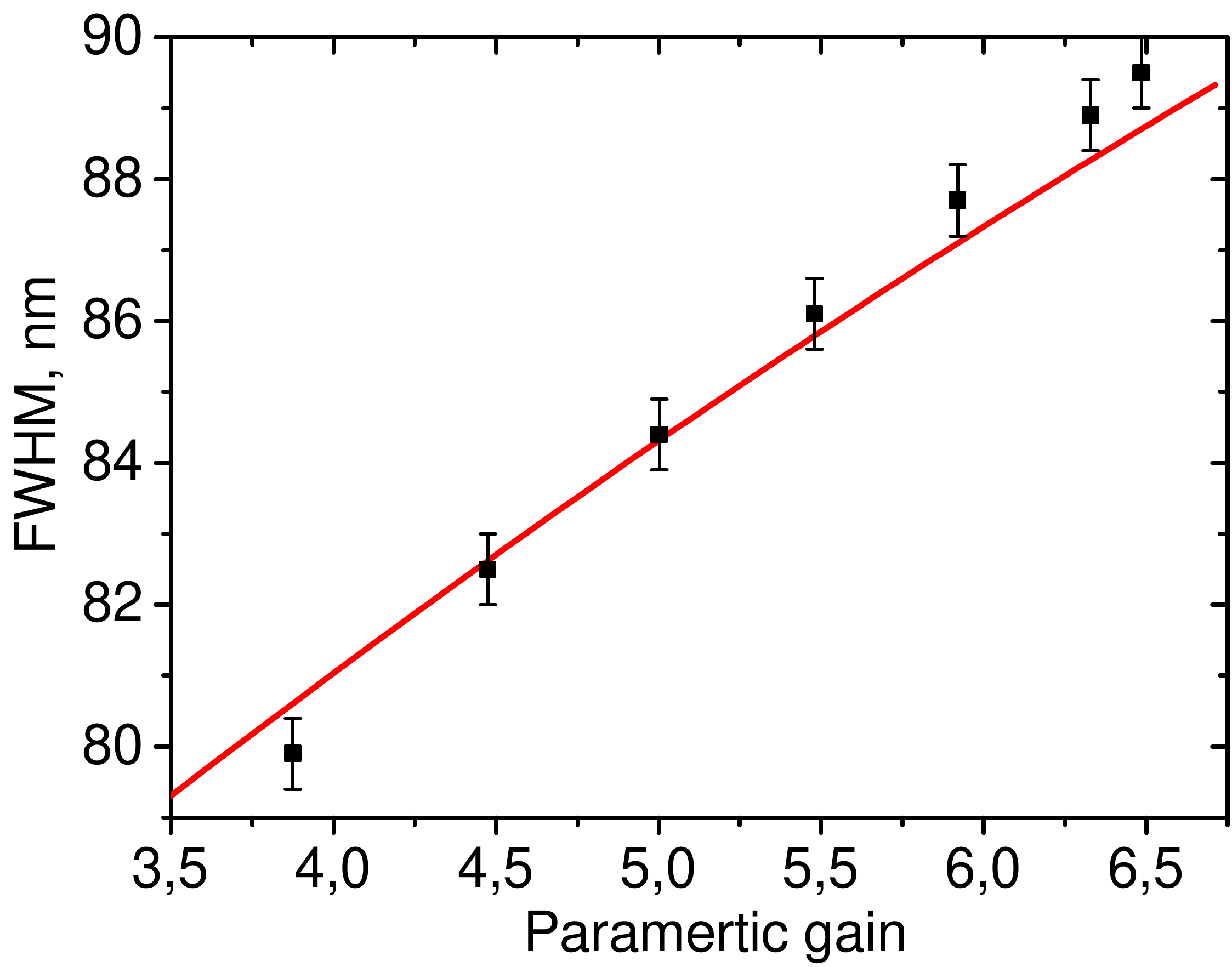}
\caption{The experimental dependence of the PDC spectral width on the parametric gain $G$ (points). The line is the calculation using the theory discussed in section~\ref{3_sub:freq_wave_spectra}.}
\label{3_fig:width_vs_gain}
\end{center}
\end{figure}

Indeed, the dependence of the PDC spectral width on the parametric gain $G$ measured for the range $3.9 < G < 6.5$ (Fig.~\ref{3_fig:width_vs_gain}) shows the broadening. At lower gain values, the PDC signal is too low to make spectral measurements while higher gain values could not be achieved with available pump powers.

In the chosen range the broadening is not large, the spectral width changes by 12\%; the increase compared to the spectral width at low gain is 27\%. However, the broadening should be much stronger in the case of type-II ($e\to oe$ interaction) or frequency-nondegenerate phase matching~\cite{Klyshko1988}.

\begin{figure}[!htb]
\begin{center}
\includegraphics[width=0.9\textwidth]{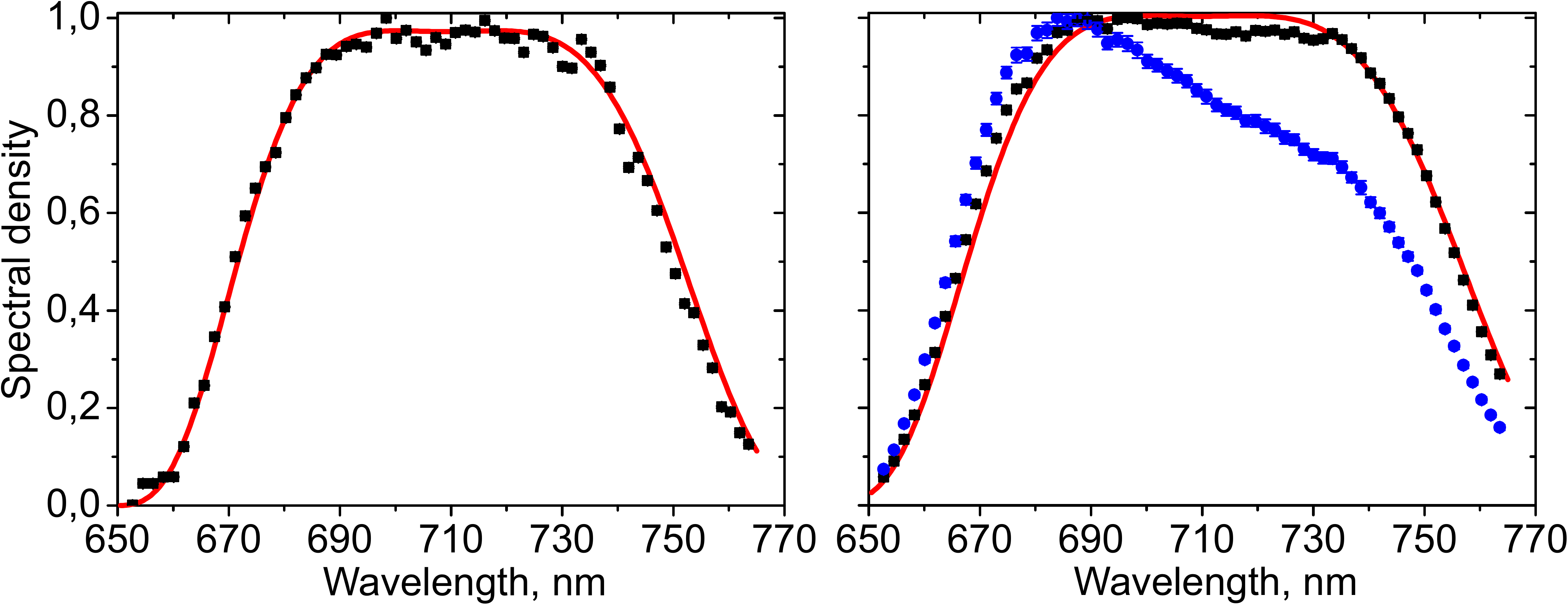}
\caption{The experimental (black squares) and theoretical (lines) spectra $S_{\Omega}(\lambda)$ at the gain $G=3.9$ (left) and 6.5 (right). The right panel also shows the measured spectrum $S_{\lambda}(\lambda)$ before the correction by the $\lambda^4$ factor (blue circles).}
\label{3_fig:broadening_spectrums}
\end{center}
\end{figure}

Figure~\ref{3_fig:broadening_spectrums} shows the spectra $S_{\Omega}(\lambda)$ recorded at the lowest and highest gain values. The shapes are well fitted with the theory discussed in section~\ref{3_sub:freq_wave_spectra}. The only fitting parameter is the crystal orientation, which turned out to differ by $0.0025^\circ$ from the one for exact collinear frequency-degenerate phase matching. This leads to about 2 nm shift for the whole dependence. Such an error in the crystal orientation is difficult to notice in experiment.

As it has already been discussed in section~\ref{3_sub:lambda4} the ‘raw’ spectra $S_{\lambda}(\lambda)$ obtained from the spectrometer are always asymmetric, while the long-wavelength part having smaller intensity (Fig.~\ref{3_fig:broadening_spectrums}, right).

This shape resembles the `Drawing~No.~1' from `The Little Prince', especially, if the phase matching goes a bit out of the exact frequency degeneracy like in the Fig.~7a from Refs.~\cite{Baek2008}.

Thus, the conversion to the experimental $S_{\Omega}$ requires the $\lambda^4$ factor. Moreover, about 10\% variation of the wavelength resolution along the CCD array is taken into account. This variation appears because the wavelength interval selected by one pixel slightly depended on wavelength.

The $\lambda^4$ factor is needed to understand the asymmetry to the same extent as the `Drawing~No.~2' is needed to grasp the hidden elephant.

%%%%%%%%%%%%%%%%%%%%%%%%%%%%%%%%%%%%%%%%%%%%%%%%%%%%%%%%%%%%%%
\section{Measurement of PDC correlations}
\label{3_sec:correlations}

This section shows how to retrieve the correlation width and reconstruct the JSI as well as it explains how to get the BSV eigenmodes from the JSI.

\subsection{PDC correlations in the noise of intensity difference}

The correlations between signal and idler beams can be studied by measuring the noise of the intensity difference for various pairs of wavelengths in the spectrum. Ideally, for correlated wavelengths this noise is zero, see section~\ref{2_sub:twin-beam_sq}, while for uncorrelated ones the noise has independent contributions from both beams.

\begin{figure}[!htb]
\begin{center}
\includegraphics[width=0.9\textwidth]{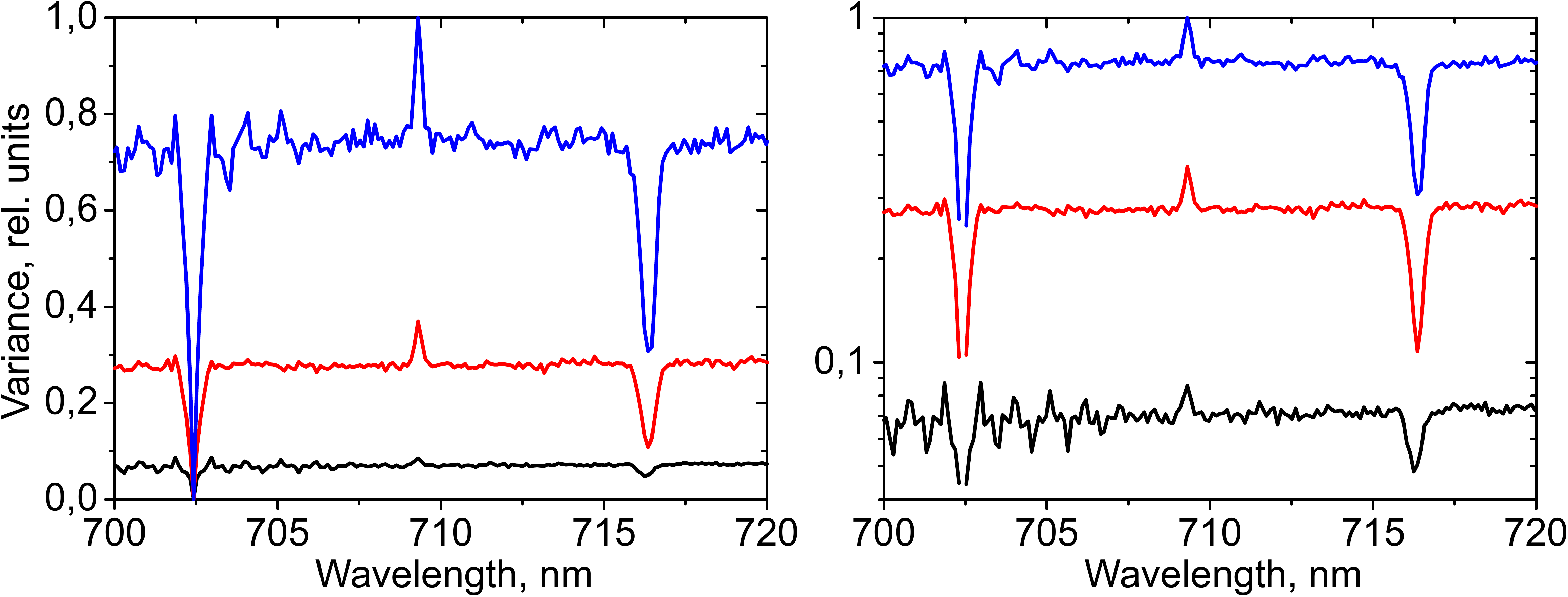}
\caption{Variance of the difference between PDC intensities at given wavelength, 702.4~nm, and scanned wavelengths, 700-720~nm, for the parametric gain $G=6.4$ (black), 6.5 (red), and 6.8 (blue) in linear (left) and log-linear (right) scales. The variance is normalized to the maximum of the last curve.}
\label{3_fig:var_diff}
\end{center}
\end{figure}

The PDC correlations are measured in the setup from the previous section (Fig.~\ref{3_fig:Setup_broadening}). The measurement procedure is the following: one wavelength is fixed, by fixing a single pixel of the CCD array, and the other one is scanned. The variance of the difference between the signals from these wavelengths is calculated via the ensemble averaging over the acquired spectra (Fig.~\ref{3_fig:var_diff}). For different gain values, 10-20 thousands spectra are acquired, each collecting the intensity from 500-1000 pulses. In all spectra, the fixed wavelength, signal, is chosen to be $\lambda_s = 702.4$~nm, while the other one, idler $\lambda_i$, is scanned around degenerate wavelength (709.3~nm), from 700~nm to 720~nm.

Each variance distribution contains two ‘dips’ and one `peak'. The left-hand (auto-correlation) dip is caused by correlations existing within a single mode. The dip goes down to zero because the variance is identically zero when the reading of a pixel is subtracted from itself. The dip is surrounded by the oscillating noise, which in its respect is an artifact and should therefore not be considered.

The right-hand (cross-correlation) dip is caused by correlations between signal and idler beams, with the wavelengths related as $\lambda_s^{-1} + \lambda_i^{-1} = \lambda_p^{-1}$, i.e. in accordance with energy conservation law [Eq.~\eqref{2_eq:energy_momentum_cons}]. Respectively, the peak at degenerate wavelength, 709.3~nm, occurs due to superbunching that is to be discussed in chapter~\ref{5_chapter}.

In more detail, the variance of photon-number difference is \cite{Iskhakov2011}
\begin{equation}\label{3_eq:var_diff}
\var{(\op{N}_s-\op{N}_i)} = (g^{(2)}_{ss} - 1) \mean{N_s}^2 + (g^{(2)}_{ii} - 1) \mean{N_i}^2 - 2 (g^{(2)}_{si} - 1) \mean{N_s} \mean{N_i} + \mean{N_s+N_i},
\end{equation}
where $g^{(2)}_{12}$ is the second-order normalized CF,
\begin{equation}\label{3_eq:g2_12}
g^{(2)}_{12}\equiv\frac{\mean{:\op{N}_1\op{N}_2:}}{\mean{\op{N}_1}\mean{\op{N}_2}},
\end{equation}
where $\{1,2\}$ means $\{s,i\}$ and $\mean{:\dots:}$ denotes normal ordering.

Outside of the peak PDC has thermal statistics\footnote{The statistical properties will be discussed in more details in chapter~\ref{5_chapter}.}, $g^{(2)}_{ss}=g^{(2)}_{ii}=2$, and outside of the dips the correlations are absent, $g^{(2)}_{si}=1$. Therefore, at the background $\var{(\op{N}_s-\op{N}_i)}=2\mean{N}^2$, if $\mean{N_i}=\mean{N_s}\equiv \mean{N}\gg1$. In the peak $g^{(2)}_{ii}=3 + 1/\mean{N}$, therefore the variance is equal to $3\mean{N}^2$ for large $\mean{N}$. In the cross-correlation dip $g^{(2)}_{si}=2 + 1/\mean{N}$ and the variance should be zero. However, this complete noise suppression can be observed only under strict requirements~\cite{Iskhakov2008}, e.g. perfect mode matching and 100\% detection efficiency, that are not satisfied in the experiment.

The Eq.~\eqref{3_eq:var_diff} is not valid in the auto-correlation dip, because it is obtained under the assumption of commuting signal $\op{a}_{s}$ and idler $\op{a}_{i}$ photon creation (or annihilation) operators, see Eq.~\eqref{2_eq:comm_rules}. However, it is clear that the variance of the number of photons subtracted from itself is equal to zero, $\var{(\op{N}_s-\op{N}_s)}=0$.

According to Eq.~\eqref{3_eq:var_diff}, the peak in Fig.~\ref{3_fig:var_diff} should be 50\% higher than the background value, independently of the gain $G$. In all dependences, only a $30\pm5$\% increase of the variance is seen. This difference can be attributed to the insufficient frequency selection. As to the cross-correlation dip, the imperfect mode matching reduces its depth; the value at the dip is $40\pm5$\% of background one. At smaller gain values, the measurement is not accurate enough to reveal the correlations.

Another observation from Fig.~\ref{3_fig:var_diff} is that the peak width is smaller than the width of the dips. Accurate fitting by Gaussian functions yields, in the case of $G=6.5$, the values of $0.26\pm0.06$~nm, $0.45\pm0.06$~nm, and $0.52\pm0.06$~nm for the peak, auto- and cross-correlation dips, respectively.

The spectral widths of the dips coincide with the spectral widths of the CF $g^{(2)}$ and at low gain should be given by the pump bandwidth; this results in $\Delta\lambda = 0.15$~nm at wavelengths around 709~nm. The larger width at a higher gain is explained by the redistribution of the Schmidt coefficients, see Eq.~\eqref{3_eq:Schmidt_tilde}. From simplified argumentation it can be explained by nonlinear dependence of PDC intensity on the pump one. At the `top' of the pump pulse the gain is larger than on the slopes; so the most photons are produced there. For this reason, the effective pulse width decreases and the effective pump `bandwidth' increases. The same should happen in the spatial domain; indeed, it is shown experimentally~\cite{Brida2009} that the angular correlation width (speckle size) increases with the gain.

As to the peak, its width, theoretically, should be narrower than the correlation width by a factor of 2. Simply it can be understood as follows: for the dip one wavelength ($\lambda_s$) is fixed, the other one ($\lambda_i$) is scanned, whereas for the peak both wavelengths are kind of scanned together due to the superbunching requirement\footnote{Signal and idler photons should have exactly the same wavelength, because superbunching is observed only if each pixel of a CCD array used in spectrometer detect both photons.}, $\lambda_s=\lambda_i$. The observed widths of $0.26\pm0.06$ nm for the peak and $0.52\pm0.06$ nm for the dip are in agreement with these considerations.

\subsection{Measurement of the joint spectral intensity (JSI)}

As it was discussed in section~\ref{3_sub:PDC_correlations}, the more complete information about correlations can be obtained by measuring the 2D distribution of some quantity, as a function of both wavelengths $\lambda_s$ and $\lambda_i$. As such quantity, one can choose the variance of the difference signal as in Fig.~\ref{3_fig:var_diff}, but it is more natural to measure the covariance of the signals at two wavelengths,
\begin{equation}\label{3_eq:covariance}
\mathrm{cov}(N_s, N_i) \equiv \mean{\op{N}_s\op{N}_i} - \mean{\op{N}_s}\mean{\op{N}_i} = (g^{(2)}_{si} - 1)\mean{N_s} \mean{N_i}.
\end{equation}
A bit latter it will be shown that it is equal, up to a constant factor, to the JSI.

\begin{figure}[!htb]
\begin{center}
\includegraphics[width=0.8\textwidth]{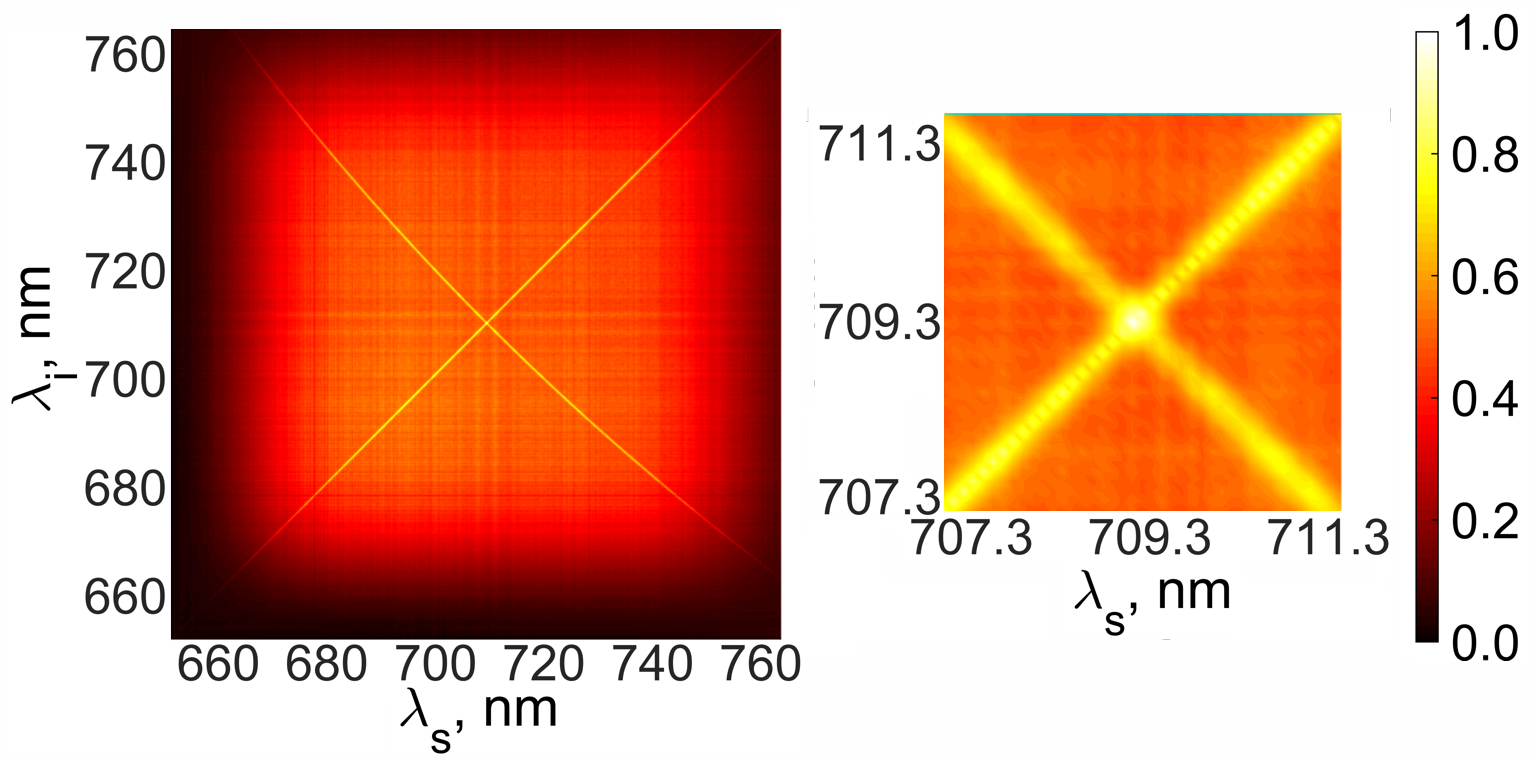}
\caption{2D distribution of covariance $\mathrm{cov}(N_s, N_i)$ (left) and its zoomed central part (right) for the gain $G=6.5$. It is normalized to its maximum value. Note that only cross-correlation peaks correspond to the JSI.}
\label{3_fig:JSI_exp}
\end{center}
\end{figure}

The covariance gives a full account of pairwise correlations, which now manifest themselves as peaks (Fig.~\ref{3_fig:JSI_exp}). As for Fig.~\ref{3_fig:var_diff}, the auto-correlation peaks occur on the straight line $\lambda_s=\lambda_i$ and the cross-correlation ones are on the line $\lambda_s^{-1} + \lambda_i^{-1} = \lambda_p^{-1}$. The superbunched peak also reveals itself.

The cross-correlation peaks represent the JSI, similar to the two-photon case (see Fig.~\ref{3_fig:JSA}): its cross-section gives the conditional width, while its projections on the axes $\lambda_s$ and $\lambda_i$ provide the unconditional ones. As it was discussed in the previous section, these widths are different from the ones for the SPDC case, and therefore the number of modes is different. Namely, in the low-gain case the full spectral width is estimated to be 70.2~nm and the correlation width is to be 0.15~nm, which results in the Fedorov ratio $R\approx470$, while at the parametric gain $G=6.5$ the corresponding widths are $89.5\pm0.5$~nm and $0.52\pm0.06$~nm, respectively, which gives the ratio $R=172\pm20$.

The analogy between the covariance and the JSI can be understood using the Schmidt modes, see section~\ref{3_sub:PDC_correlations}. The covariance is equal to~\cite{Finger2017}
\begin{equation}
\mathrm{cov}[N_s(\Omega_s), N_i(\Omega_i)] = \left| \sum_{l=0}^\infty u_l v_l \psi_l(\Omega_s) \phi_l(\Omega_i) \right|^2,
\end{equation}
where $u_l=\cosh(G_l)$.

At low gain $u_lv_l\sim G_l$ and the covariance is equal to JSI up to a factor $\tilde{G}$. This measurement is similar to the registration of coincidences ($g^{(2)}$ measurement)~\cite{Baek2008}; indeed, at low gain the CF $g^{(2)}_{si}\gg1$ and $\mathrm{cov}(N_s, N_i)\propto g^{(2)}_{si}$, see Eq.~\eqref{3_eq:covariance}. At high gain the covariance again provides JSI, because $u_l\approx v_l$ and the Schmidt coefficients are redistributed, see Eq.~\eqref{3_eq:Schmidt_tilde}.

Accordingly, the singular value decomposition [Eq.~\eqref{3_eq:Schmidt_decomp}] of $\sqrt{\mathrm{cov}[N_s(\Omega_s), N_i(\Omega_i)]}$ gives the shapes of reconstructed Schmidt modes, $\psi_l(\Omega_s)$ and $\phi_l(\Omega_i)$, and the corresponding Schmidt coefficients, $\Lambda_l$. The latter should be normalized according to Eq.~\eqref{3_eq:Schmidt_coeff_norm}.

In Fig.~\ref{3_fig:JSI_exp} the JSI `sits' on the background\footnote{It looks red in the figure.} that repeats the total PDC spectrum. This background could degrade the quality of the reconstructed modes; in experiment, it could be eliminated by single-shot measurements synchronized with the laser pulses~\cite{Finger2017,Beltran2017}. 

Interestingly, a similar method has been applied to characterize the intensity correlations in fiber optical solitons~\cite{Spalter1998} and to get the spatiotemporal correlations~\cite{Allevi2014,Haderka2015}. Finally, there is another, quite elegant, method for the JSI reconstruction called the stimulated emission tomography~\cite{Liscidini2013,Fang2014}.

%%%%%%%%%%%%%%%%%%%%%%%%%%%%%%%%%%%%%%%%%%%%%%%%%%%%%%%%%%%%%%55
\section{Schmidt number and non-phase-matched sum frequency generation}
\label{3_sec:schmidt_SFG}

This section explains how one can assess the number of modes using non-phase-matched SFG. Despite the fact that the main part of work was done by Denis Kopylov, the author's contribution in the experimental and theoretical work was significant. Therefore, the experiment is briefly discussed below.

\subsection{The pump reconstruction}

As it has been already discussed, BSV has two characteristic parameters: the total spectral width and the correlation width. What will happen if we add the second nonlinear crystal and generate the sum frequency from BSV? These two characteristic parameters should manifest themselves.

In our experiment the BSV is produced via PDC, the same way as in the experiment on PDC generation in the anomalous GVD range (section~\ref{3_sec:O_PDC}). The produced BSV is filtered to a single spatial mode in the collinear direction and then sharply focused into a 1~mm lithium niobate (LiNbO$_3$) crystal where SFG occurs through $ee\to e$ interaction without phase matching. The generated light is separated from the BSV and measured with a visible spectrometer with 1.3~nm resolution.

\begin{figure}[!htb]
\begin{center}
\includegraphics[width=0.6\textwidth]{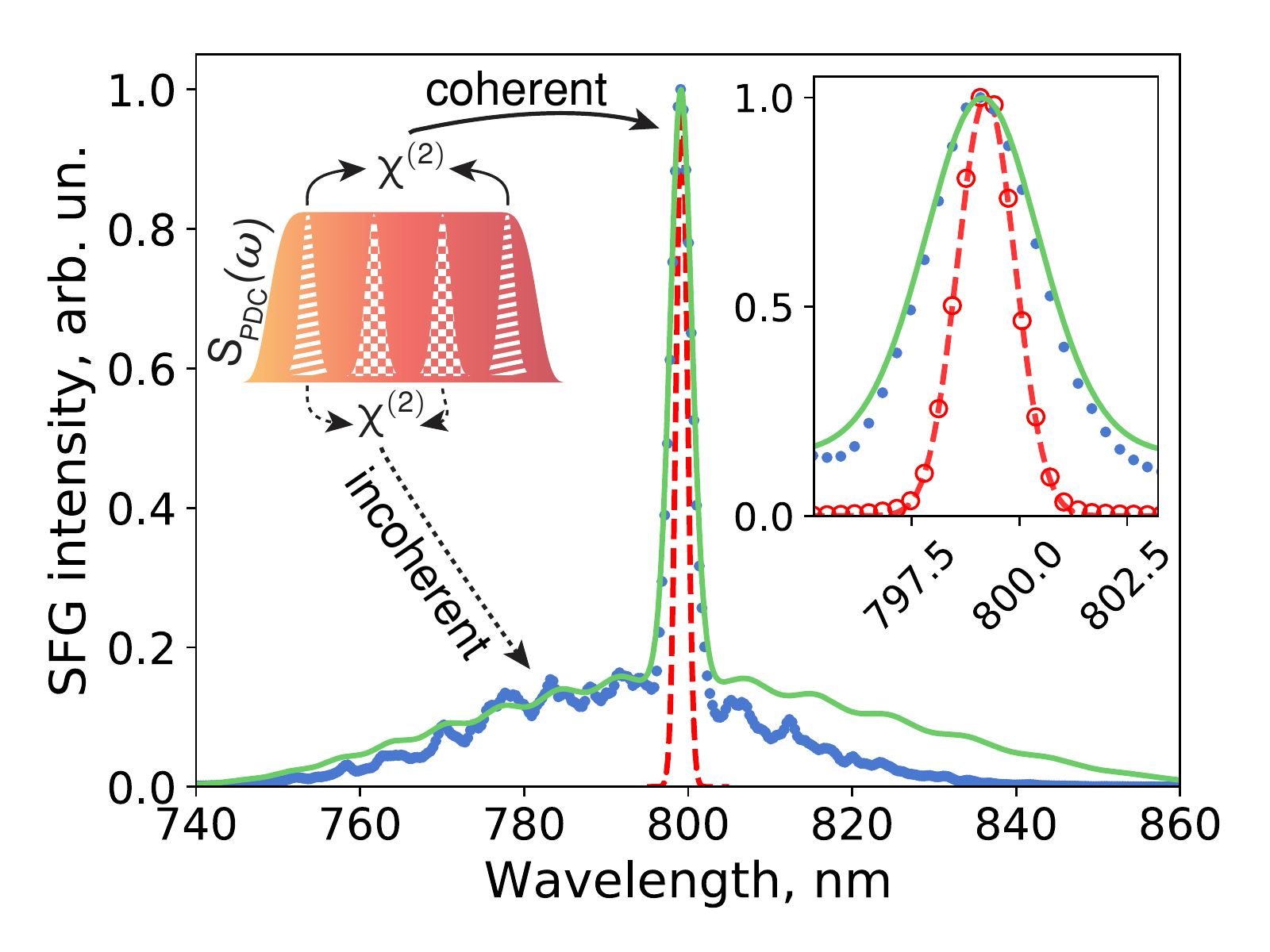}
\caption{The experimental (blue dots) and theoretical (green line) SFG spectra from BSV. Red dashed line shows the spectrum of the PDC pump laser (Gaussian fit). The right inset is a zoom into the central part of the SFG spectrum showing additionally the experimental points for the laser spectrum (red circles). The left inset schematically shows how the peak and the pedestal are formed.}
\label{3_fig:SFG_PDC}
\end{center}
\end{figure}

Indeed, the spectrum of SFG from BSV contains a narrow spectral peak and a broad pedestal (Fig.~\ref{3_fig:SFG_PDC}), see also Refs~\cite{Abram1986,Jedrkiewicz2011}. The width of the pedestal is determined by the whole PDC spectrum, whereas the width of the peak is comparable to the one of the PDC pump. Therefore, the peak is often interpreted as a result of the `pump reconstruction'.

These features could be quite well interpreted within the Schmidt-mode formalism. We show~\cite{Kopylov2019} that the peak, the so-called coherent contribution, corresponds to the SFG from each Schmidt mode with itself, i.e. the process reverse to PDC, while the pedestal, the incoherent contribution, is caused by the SFG from different (uncorrelated) Schmidt modes. 

\begin{figure}[!htb]
\begin{center}
\includegraphics[width=0.5\textwidth]{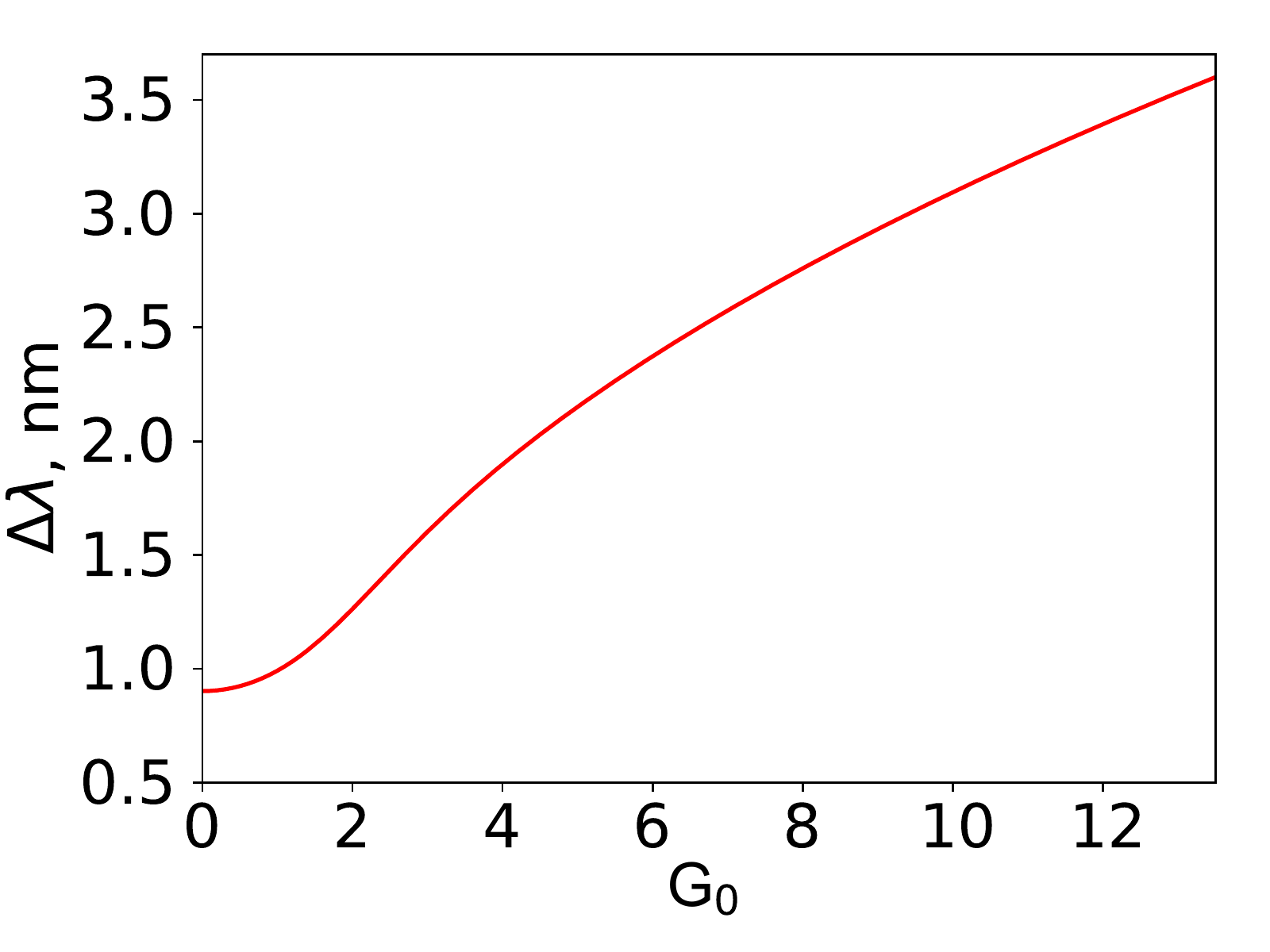}
\caption{Calculated bandwidth of the SFG peak vs. the PDC gain $G_0$ for the first Schmidt mode; at low gain, $G_0\ll1$, the peak bandwidth is equal to the PDC pump bandwidth.}
\label{3_fig:peak_width_gain}
\end{center}
\end{figure}

Actually, it is not fully correct to consider the peak as the result of the pump reconstruction. In the Schmidt mode formalism it is possible to show that the width of the `coherent' peak at high gain is not the same as the pump one; it is somewhat broader and increases with the gain (Fig.~\ref{3_fig:peak_width_gain}). It happens because the Schmidt number $K$ decreases with the gain and this effect is noticeable only when $K$ is not large. In order to observe this effect one should go beyond the narrowband pump approximation, used e.g. in Ref.~\cite{Dayan2007}, and take into account the finite width of photon correlations.

\subsection{Schmidt number measurement}
\label{3_sub:Schmidt_SFG}

The SFG can be also interpreted as a fast correlator or coincidence circuit~\cite{Sensarn2010}. Expanding this analogy further, one can set the correspondence between the SFG spectrum and the conditional and unconditional bandwidths. The width of the peak, $\Delta\Omega_{coh}$, corresponds to the first one, while the width of the pedestal, $\Delta\Omega_{incoh}$, to the second one.

\begin{figure}[!htb]
\begin{center}
\includegraphics[width=0.5\textwidth]{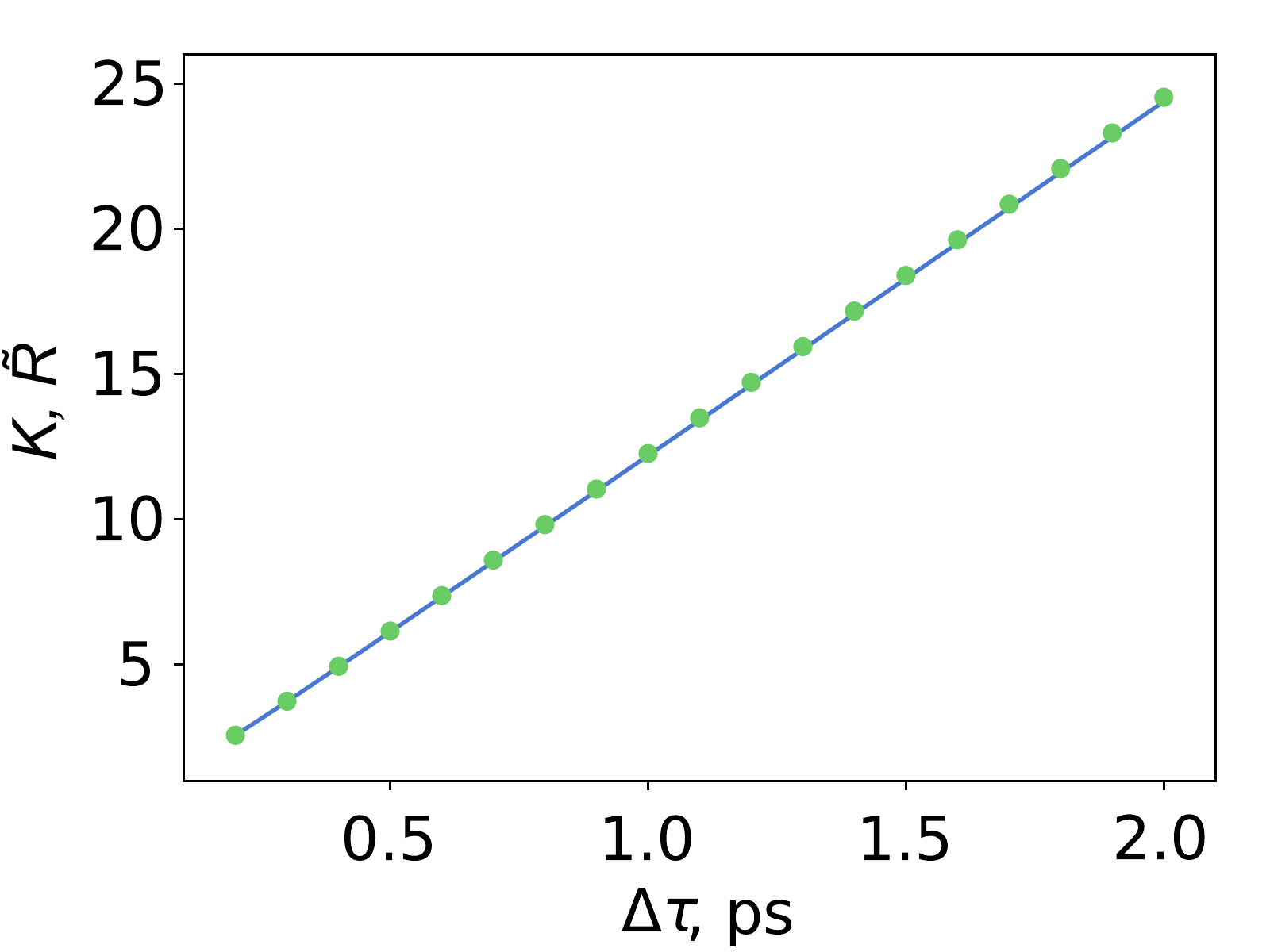}
\caption{The Schmidt number $K$ (blue line) and the ratio $\tilde{R}$ (green dots) calculated for the experimental parameters from section~\ref{3_sec:O_PDC} as functions of the PDC pump pulse duration.}
\label{3_fig:K_R_vs_tau}
\end{center}
\end{figure}

It implies that the ratio of the widths, $\tilde{R} = \Delta\Omega_{incoh}/\Delta\Omega_{coh}$, has a similar meaning as the Fedorov ratio [Eq.~\eqref{3_eq:Fedorov_R}]. Similarly, the ratio $\tilde{R}$ should be equal to the Schmidt number $K$. To test this statement, we calculate $K$ and $\tilde{R}$ for different values of pump pulse duration (Fig.~\ref{3_fig:K_R_vs_tau}); indeed, the result is that $\tilde{R}=K$.

In experiment we get the ratio $\tilde{R}_{exp}\approx13$, which is smaller than the expected one $\tilde{R}_{theor}=19.5$. It happens because the BSV spectrum is partially cut due to experimental imperfections, which leads to a reduced bandwidth $\Delta\Omega_{incoh}$. The estimate for the Schmidt number is correct only for frequency-independent losses, the ones that depends on the frequency change the result. The same applies if the SFG phase matching is not broad enough; that is why it is important to make the SFG broadband, e.g. by avoiding the phase matching.

The major possible limitation is that the method requires a presence of the coherent peak, which can be observed as long as signal-idler correlations are not completely lost. The peak is pronounced if its spectral intensity is higher than the pedestal one. Therefore, the ratio of the peak and pedestal spectral intensities quantifies the quality of signal-idler correlations; in Fig.~\ref{3_fig:SFG_PDC} this ratio is equal to $7.3\pm0.9$.

\begin{figure}[!htb]
\begin{center}
\includegraphics[width=0.5\textwidth]{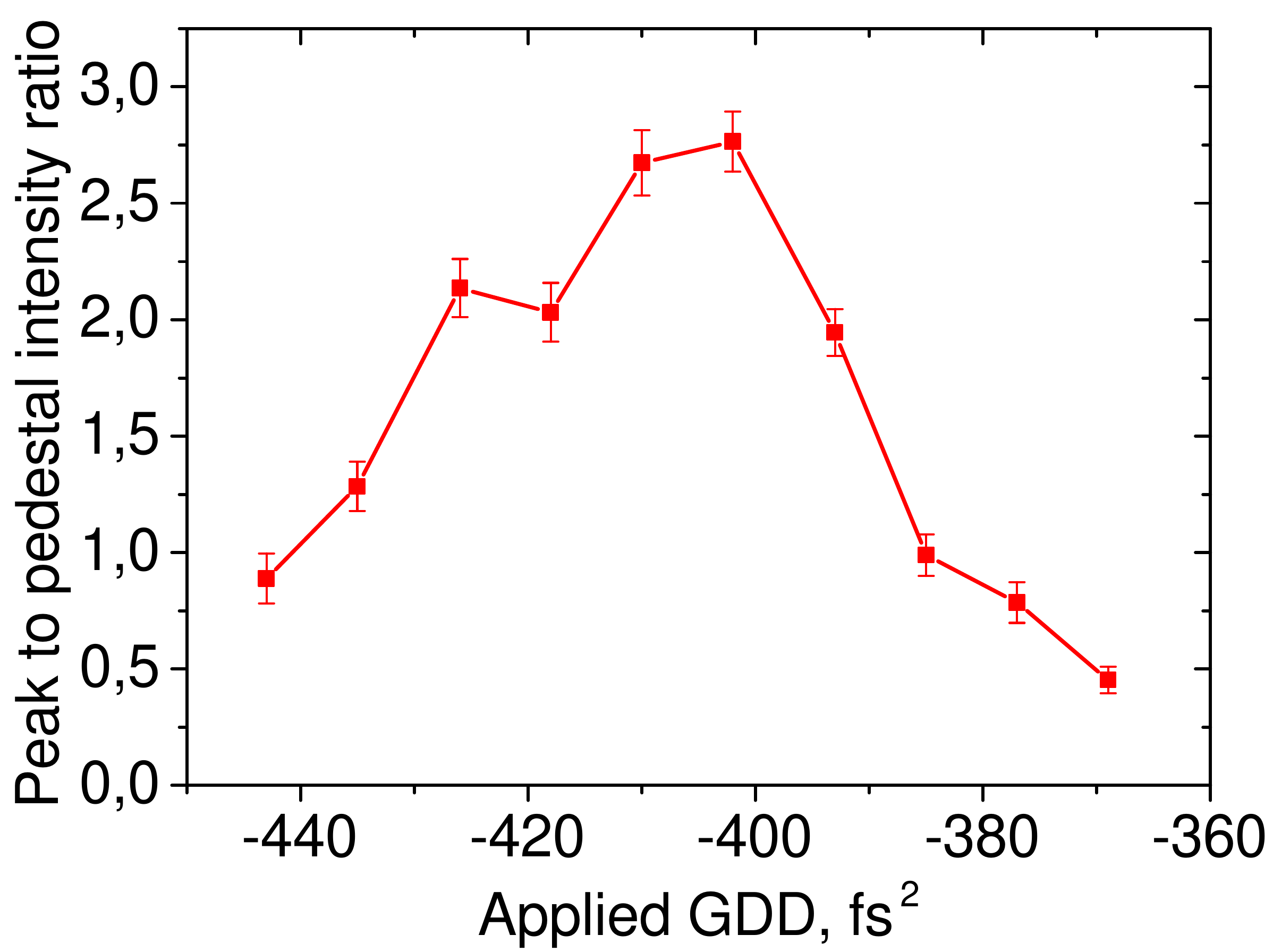}
\caption{The ratio of the peak and pedestal spectral intensities vs. applied GDD.}
\label{3_fig:peak_GDD}
\end{center}
\end{figure}

In particular, the loss of these correlations can occur due to the group delay dispersion (GDD), which leads to the time delay between the signal and idler photons. Sometimes one can prevent it with dispersion-compensating methods. In the other setup we successfully demonstrate this together with Hani Abou Hadba~\cite{Hadba2018}. Indeed, as shown in Fig.~\ref{3_fig:peak_GDD}, the peak is retrieved and the peak to pedestal ratio reaches its maximum, if the correct GDD (-400~fs$^2$) is applied.

%%%%%%%%%%%%%%%%%%%%%%%%%%%%%%%%%%%%%%%%%%%%%%%%%%%%%%%%%%%%%%%%
\includepdf{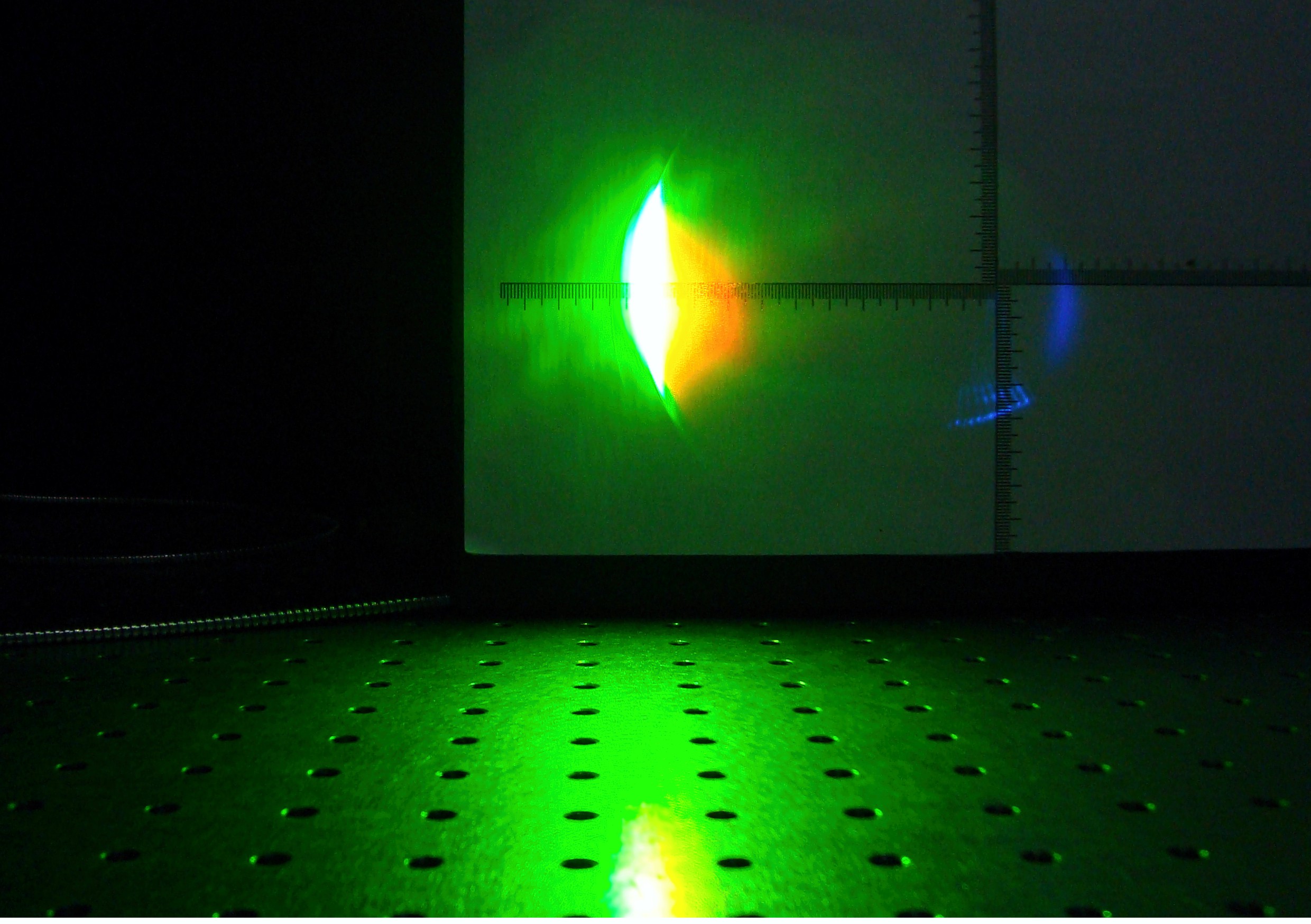}

\chapter{Twin-beam correlations in PDC}
\label{4_chapter}

\begin{center}\textit{The Signal and the Idler are twin brothers.\\Who is more valuable for Physics?\\We say the Idler and we mean the Signal,\\We say the Signal and we mean the Idler.}\\---Derivative on Vladimir Mayakovsky's \href{https://en.wikipedia.org/wiki/Vladimir_Ilyich_Lenin_(poem)}{poem}
\end{center}

This chapter focuses on some interesting effects emerging from signal-idler correlations. It starts with the interference of signal and idler beams on a 50\% BS, the HOM effect, and the Fock states interference, which are greatly considerable for quantum optics.

Further, there is the macroscopic HOM interference, the case of bright twin beams, demonstrated. Although the standard measurement technique leads to a very low visibility, a modified technique provides a visibility close to unity. Applying this technique the interference manifests itself as a peak, which additionally provides the information about the PDC spectral properties, including the number of temporal modes.

Afterwards, the chapter focuses on the statistics at the BS outputs, which is similar to the one obtained for the Fock states interference. The phase fluctuations are converted into photon-number fluctuations manifesting themselves in a U-shaped photon-number distribution.

Finally, it is shown how spatial and temporal walk-off matching can be used for giant narrowband twin-beam generation. If one of the twin beams is emitted along the pump Poynting vector or its group velocity matches that of the pump, the generation of both beams will be drastically enhanced; the effect is exemplified on the chapter's cover. This enhancement leads to a considerable narrowing of the wavelength and angular spectrum.

%%%%%%%%%%%%%%%%%%%%%%%%%%%%%%%%%%%%%%%%%%%%%%
\section{Introduction}

\subsection{Hong-Ou-Mandel (HOM) interference}

In 1987 Hong, Ou, and Mandel suggested a method for measuring subpicosecond time intervals between photons using the interference on a 50\% BS~\cite{Hong1987}. If signal and idler photons come to BS exactly at the same time, they always leave it from the same output port, never from the different ones.

In experiment one measures the number of coincidences at the BS outputs, namely the number of events, when the photons are detected at the same time, i.e. in coincidence. This measurement, up to a constant factor, gives the second-order CF $g^{(2)}_{12}$~\cite{Ivanova2006}.  If the photons come to the BS at the same time, the number of coincidences drops drastically; so, a narrow `dip' in the coincidence rate is observed, if the delay between signal and idler photons is scanned. Therefore, the HOM effect is often called `the HOM dip', although under certain conditions it manifests itself as a peak~\cite{Walborn2003}.

\begin{figure}[!htb]
\begin{center}
\includegraphics[width=0.8\textwidth]{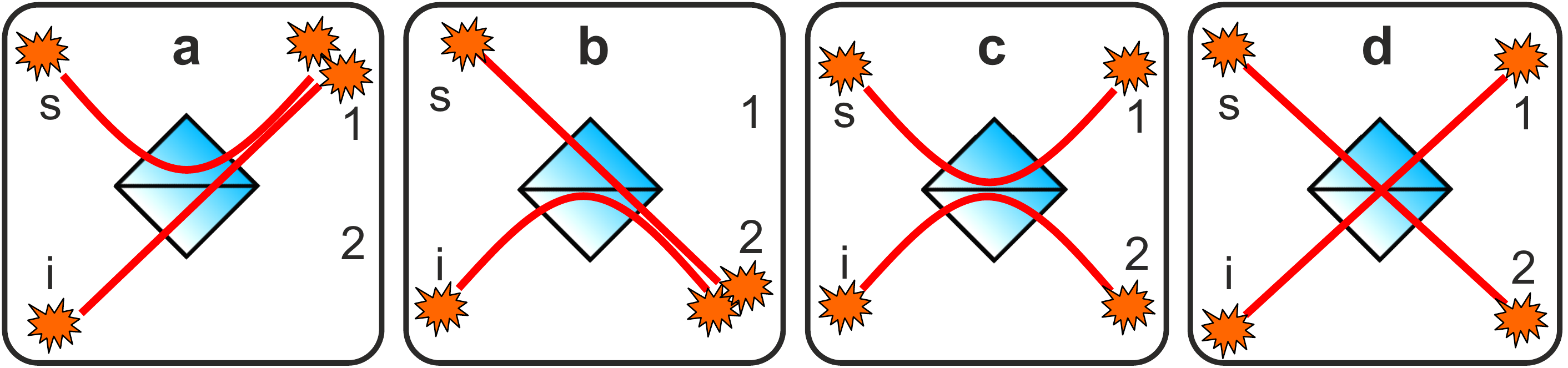}
\caption{Four possible two-photon incidents on a BS: one photon is transmitted and the second one is reflected (a,b), both photons are reflected (c), and both photons are transmitted (d). For indistinguishable photons the last two cases interfere destructively.}
\label{4_fig:BS_inter}
\end{center}
\end{figure}

The effect happens due to the destructive interference of the probability amplitudes for both photons to be transmitted and to be reflected (Fig.~\ref{4_fig:BS_inter}). The key point here is that the photons should be indistinguishable in all characteristics such as wavelength, polarization, wave vector, etc.; any distinguishability degrades the HOM effect visibility.

This feature allows to use the HOM interference for characterizing distinguishability~\cite{Toppel2012} and as an identity test for single-photon states~\cite{Santori2002}, including ones generated from two different sources, like crystals~\cite{Mosley2008} or molecules~\cite{Lettow2010}.

As it was suggested in the pioneering work~\cite{Hong1987}, the effect can be used to measure time intervals with high accuracy, because the width of the dip could be extremely small, up to a few femtoseconds~\cite{Nasr2008}; for example, the HOM effect was used to measure the tunneling time~\cite{Steinberg1993}.

The width of the dip is twice smaller~\cite{Burlakov2001} than the one of the first-order CF $G^{(1)}(\tau)$ and is not affected by GVD~\cite{Steinberg1992}. Due to these features the HOM effect is proposed to use in the quantum version of optical coherence tomography (OCT)~\cite{Abouraddy2002,Nasr2003}.
 
In the quantum OCT the signal photons travel though the sample before the BS, while the idler ones are passing the delay line, which introduces a variable time delay. The positions of the HOM dips give the positions of reflecting/scattering layers in the sample. The resolution of the quantum OCT is twice higher than the one of the standard OCT for the same bandwidth of light. Furthermore, the method is not sensitive to the acquired GDD.

\subsection{Fock states interference}
\label{4_sub:Fock_interf}

\begin{figure}[!htb]
\begin{center}
\includegraphics[width=1\textwidth]{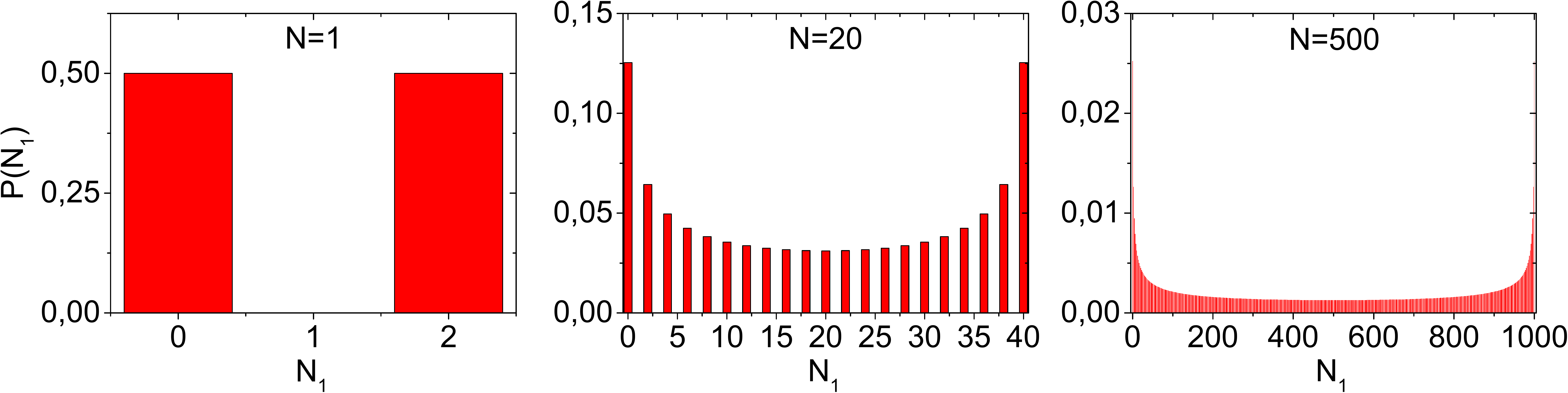}
\caption{Probability distributions $P(N_1)$ at one of the BS outputs for the interference of Fock states $\ket{N}$: $N=1$ (left), 20 (center), and 500 (right).}
\label{4_fig:Fock_states_interf}
\end{center}
\end{figure}

The HOM interference can be considered as a special case of Fock states $\ket{N}$ interference on a 50\% BS. Indeed, if the vacuum and higher photon contributions to the PDC state [Eq.~\eqref{2_eq:twin_state}] are neglected, the interference just transforms the state $\ket{1}_s\ket{1}_i$ into $\ket{2}_1\ket{0}_2+\ket{0}_1\ket{2}_2$. Therefore, at each BS output, e.g. the output number $1$, the probability $P(N_1)$ to get $N_1=1$ photon is zero, see Fig.~\ref{4_fig:Fock_states_interf} (left).

The increase of $N$ leads to a more complicated distribution~\cite{Campos1989},
\begin{equation}\label{4_eq:U_shape}
P(N_1=2m)=\frac{1}{2^{2N}}\binom{2m}{m}\binom{2N-2m}{N-m}\quad \mathrm{and} \quad P(N_1=2m+1)=0,
\end{equation}
where $\binom{k}{l}$ is the binomial coefficient. The examples for $N=20$ and 500 are shown in Fig.~\ref{4_fig:Fock_states_interf} (center and right). The maximal probabilities are at $N_1=0$ and $2N$, meaning that all $2N$ photons go together to one output, and the envelope manifests an `U' shape. Moreover, the odd photon numbers do not occur at all; unfortunately, at high $N$ this is hardly distinguishable.

The distribution~\eqref{4_eq:U_shape} is called the discrete arcsine one~\cite{Campos1989}. This type of distribution occurs also in a classical random walk~\cite{Campos1989}. The distribution~\eqref{4_eq:U_shape} provides the probability that a random walker visited the initial point for the last time on step $N_1$ after $2N$ steps have been made. However, the position of the same walker after $2N$ steps is given by binomial distribution. Exactly the same binomial law describes the probability distribution for distinguishable, non-interfering, Fock states at the BS inputs~\cite{Campos1989}.

Such a U-shaped distribution appears due to the general property of a BS: it converts phase fluctuations into amplitude ones and vice versa~\cite{Klyshko1994,Kim1998}. Indeed, the photon-number difference is zero and the phase difference is uncertain for Fock states at the inputs, at the same time after the BS the photon-number difference fluctuates a lot, whereas the phase difference is certain up to the Heisenberg limit, $1/2N$~\cite{Holland1993}.

Differently it can be understood from `the conservation law for the sum of correlations and fluctuations'~\cite{Klyshko1994}. The sum of the second-order auto-CFs $G_{11}^{(2)}=\mean{:\op{N}_1^2:}$ and $G_{22}^{(2)}=\mean{:\op{N}_2^2:}$ for outputs 1 and 2, respectively, and the doubled cross-CF $G_{12}^{(2)}=\mean{:\op{N}_1\op{N}_2:}$  is the same before and after BS,
\begin{equation}
G_{11}^{(2)} + G_{22}^{(2)} + 2G_{12}^{(2)} = \mathrm{const}.
\end{equation}
For the Fock states the photons numbers at the inputs do not fluctuate and are highly correlated; hence, at the output the drop of correlations is accompanied by an increase in fluctuations.

%%%%%%%%%%%%%%%%%%%%%%%%%%%%%%%%%%%%%%%%%%%%%%%%%%%%%%%%%%%%%%%%%%%%%%
\section{Macroscopic HOM interference}
\label{4_sec:Macro_HOM}

The introduction described the HOM interference for a pair of photons neglecting the higher photon contributions in SPDC. However, these contributions exist and degrade the HOM effect visibility~\cite{Cosme2008}. Moreover, the visibility reduces even more if many modes are involved. Here for observing the HOM effect with bright beams an alternative approach is suggested, which is based on the measurement of the normalized variance of the difference signal, which in its turn is robust against multimode detection~\cite{Iskhakov2008}.

\subsection{Theoretical description: CF and variance of the difference signal}

The theoretical treatment is based on the model described in section~\ref{3_sec:intro}. Like in the several previous sections, only the frequency domain, $\boldsymbol{q}=0$, and one spatial mode are considered. Assuming that the signal ($s$) and idler ($i$) beams arrive at a 50\% BS with a relative delay $\tau$, the BS transformations for the output ($1,2$) take the form  
\begin{equation}
\op{a}_1(t)=\frac{\op{a}_s(t-\tau)+ \op{a}_i(t)}{\sqrt{2}},	\qquad 	\op{a}_2(t)=\frac{-\op{a}_s(t-\tau) + \op{a}_i(t)}{\sqrt{2}},
\end{equation}
where the operators $\op{a}_{s,i}(t)$ are related with $\op{a}_{s,i}(\Omega)$, see Eq.~\eqref{3_eq:Bogolubov_q_w}, via the inverse FT.

At this stage three important aspects should be discussed. Firstly, the problem of multimode detection appearing if the detector is not fast enough. In this case the fluctuations and correlations are averaged over the detection time $T$. Thus, the photon numbers,
\begin{equation}
\op{N}_{j}=\frac{1}{T}\int_{-T/2}^{T/2}dt\,\cop{a}_{j}(t)\op{a}_{j}(t),
\end{equation}
and the correlators,
\begin{equation}
\op{N}_l\op{N}_j=\frac{1}{T^2}\iint_{-T/2}^{T/2}dtdt'\,\cop{a}_{l}(t)\op{a}_{l}(t)\cop{a}_{j}(t')\op{a}_{j}(t'),
\end{equation}
are averaged over $T$; $l,j={1,2}$. Moreover, the time $T$ is assumed to be much larger than the width of the HOM dip. Note that this time averaging should not be mixed with the ensemble one, $\mean{...}$, which will be applied later.

Secondly, in the pulsed regime the parametric gain $G$, see Eq.~\eqref{3_eq:G_eq}, is not the same along the pulse; it depends on the pump field envelope $E_p(t)$,
\begin{equation}\label{4_eq:gain_envelope}
G(t)=\sigma LE_p(t),
\end{equation}
where $\sigma$ takes into account all other parameters. In the subsequent calculations the pump pulse duration $\tau_d$ is assumed to be much larger than the width of the dip.

Finally, only the first nonzero term in the mismatch expansion, see Eq.~\eqref{3_eq:Delta_Omega}, is considered and, namely, for type-I phase matching it is the term with $l=2$ and for the type-II case it is the one with $l=1$. Then, the mismatch $\Delta(\Omega)$ becomes an even or odd function of frequency detuning $\Omega$; this property is used in the calculations. 

With an account for all these assumptions, for the CF $g^{(2)}_{12}$, see Eq.~\eqref{3_eq:g2_12}, the calculation gives
\begin{equation}
\label{4_eq:g2_final_type_I}
g^{(2)}_{12}(\tau)=1+\frac{\int_0^{\infty}d\Omega\,|V(\Omega, 0)|^2\left[|V(\Omega, 0)|^2-|V(\Omega, \tau)|^2+|U(\Omega, 0)|^2\left(1-\cos(2\Omega\tau)\right)\right]}{\frac{T}{\pi}\left(\int_0^{\infty}d\Omega\,|V(\Omega, 0)|^2\right)^2}
\end{equation}
for type-I phase matching and
\begin{equation}
\label{4_eq:g2_final_type_II}
g^{(2)}_{12}(\tau)=1+\frac{\int_0^{\infty}d\Omega\,|V(\Omega, 0)|^2\left[|V(\Omega, 0)|^2-|V(\Omega, \tau)|^2+|U(\Omega, 0)|^2-\Re\left(\left(U^*(\Omega, 0)\right)^2\,e^{2i\Omega\tau}\right)\right]}{\frac{T}{\pi}\left(\int_0^{\infty}\,d\Omega\,|V\left(\Omega, 0\right)|^2\right)^2}
\end{equation}
for type-II case, where
\begin{equation}
\label{U_def}
U(\Omega,t)=\cosh\Gammaa(\Omega,t)+i\frac{\Delta(\Omega)L}{2\Gammaa(\Omega,t)}\sinh\Gammaa(\Omega,t),
\end{equation}
\begin{equation}
\label{V_def}
V(\Omega,t)=\frac{G(t)}{\Gammaa(\Omega,t)}\sinh\Gammaa(\Omega,t),
\end{equation} 
$\Gammaa(\Omega,t)$ is given by Eq.~\eqref{3_eq:Gamma_q_w} taking into account the pump field [Eq.~\eqref{4_eq:gain_envelope}] envelope and $\Re$ means the real part.

\begin{figure}[!htb]
\begin{center}
\includegraphics[width=1\textwidth]{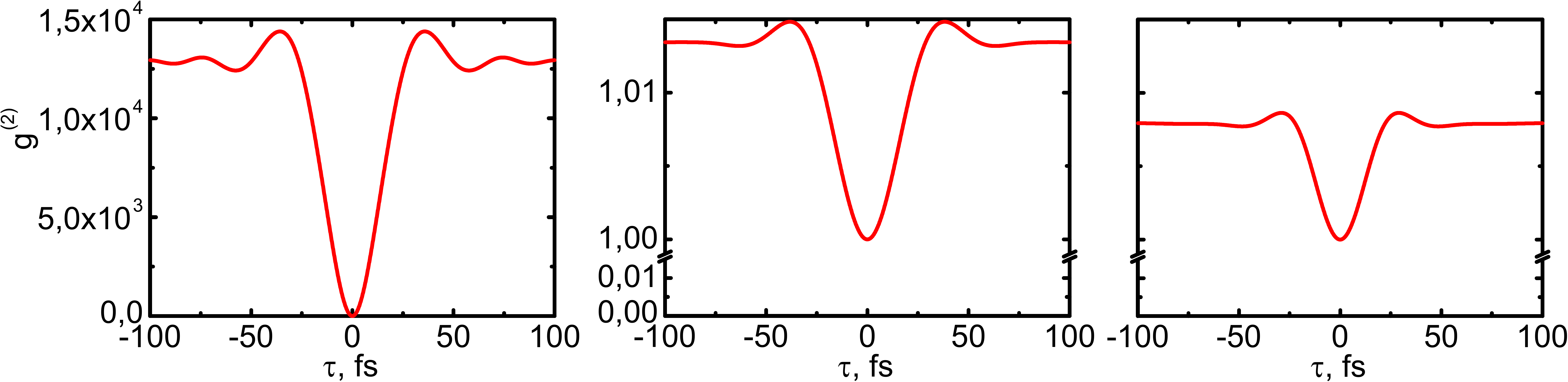}
\caption{The HOM effect obtained via the $g^{(2)}_{12}$ measurement for type-I phase matching and gain $G=0.001$ (left), 1.5 (center), and 10 (right).}
\label{4_fig:g2_type_I}
\end{center}
\end{figure}

\begin{figure}[!htb]
\begin{center}
\includegraphics[width=1\textwidth]{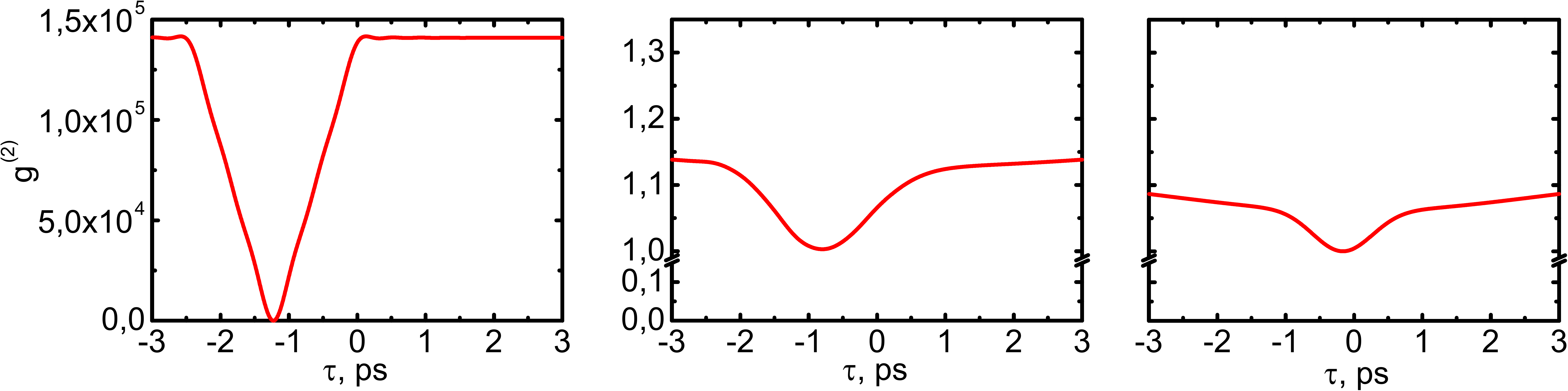}
\caption{The HOM effect obtained via the $g^{(2)}_{12}$ measurement for type-II phase matching and gain $G=0.001$ (left), 1.5 (center), and 7.5 (right).}
\label{4_fig:g2_type_II}
\end{center}
\end{figure}

The CF $g^{(2)}_{12}(\tau)$ is calculated for both types of phase matching, see Figs.~\ref{4_fig:g2_type_I} and \ref{4_fig:g2_type_II}. For type-I case the same configuration as for the calculation of frequency-angular spectrum (section~\ref{3_sub:freq_wave_spectra}) is used, a 10~mm BBO crystal pumped by 400~nm laser; for type-II case, the experimental configuration considered later in the experiment on HOM interference (section~\ref{4_sub:HOM_exp}). In this case not only the phase matching, but also the pumping wavelength is different, 354.7~nm. 

As expected, at low gain ($G=0.001$) the HOM dip has very good visibility, because the CF $g^{(2)}$ is high outside of the dip. As it is known, the shape of the dip is similar to the shape of the CF $G^{(1)}(\tau)$~\cite{Burlakov2001}. For type-I case both shapes are smooth, whereas for type-II one they are triangular. For type-II case the dip is shifted from zero, because the delay between the signal and idler photons exists already after the crystal due to the difference in the group velocities for an ordinary and extraordinary waves. The idler photons always arrive earlier than the signal ones~\cite{Brambilla2010}.

As the gain increases, the HOM visibility~\cite{Ou1999,Cosme2008},
\begin{equation}\label{4_eq:HOM_visibility}
\mathcal{V}\equiv\frac{\left|g^{(2)}_{12}(\infty)-g^{(2)}_{12}(0)\right|}{g^{(2)}_{12}(\infty)},
\end{equation}
drops drastically: from unity to almost zero. It happens because the CF $g^{(2)}$ outside of the dip decreases and at the same time at $\tau=0$ it is always equal to unity. This value, $g^{(2)}_{12}(0)=1$, is given by four- and higher-photon contributions in PDC and it is absent for ideal pairs of indistinguishable photons~\cite{Lettow2010}. However, for SPDC it could be ignored, because the CF $g^{(2)}$ is high outside of the dip.

The width and shape of the HOM dip change with the gain due to the fact that the spectrum changes, see section~\ref{3_sub:freq_wave_spectra}. For the considered type-I case (Fig.~\ref{4_fig:g2_type_I}) the width first increases from 28.2~fs ($G=0.001$) to its maximum of 31.6~fs ($G=1.5$) and then decreases, for $G=10$ it is equal to 24.6~fs; the same happens for the type-II case (Fig.~\ref{4_fig:g2_type_I}): 1.27~ps ($G=0.001$), 1.53~ps ($G=1.5$), and 0.99~ps ($G=7.5$). Moreover, for the type-II case the dip shifts closer to zero, because most photons are produced at the end of the crystal; the correlations demonstrate the same behavior~\cite{Brambilla2010}.

At the same time, the HOM effect can be observed with a high visibility by measuring, instead of the CF, the variance of photon-number difference $\var{\op{N}_-}$, $\op{N}_-\equiv\op{N}_1-\op{N}_2$, which is convenient to normalize to the mean sum of photon numbers, $\mean{\op{N}_+}\equiv\mean{\op{N}_1+\op{N}_2}$. In this case it has a meaning of so-called noise reduction factor, which is widely used for twin beams, see e.g. Ref.~\cite{Brida2010_2}.

\begin{figure}[!htb]
\begin{center}
\includegraphics[width=1\textwidth]{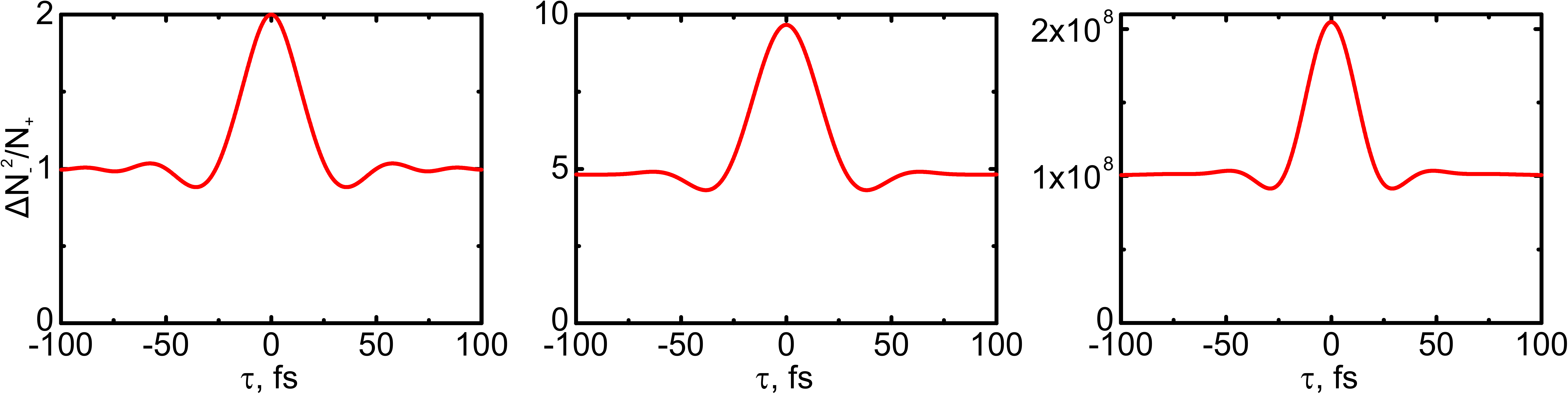}
\caption{The HOM effect in the dependence of $\var{\op{N}_-}/\mean{\op{N}_+}$ on the time delay for type-I phase matching and gain $G=0.001$ (left), 1.5 (center), and 10 (right).}
\label{4_fig:NRF_type_I}
\end{center}
\end{figure}

\begin{figure}[!htb]
\begin{center}
\includegraphics[width=1\textwidth]{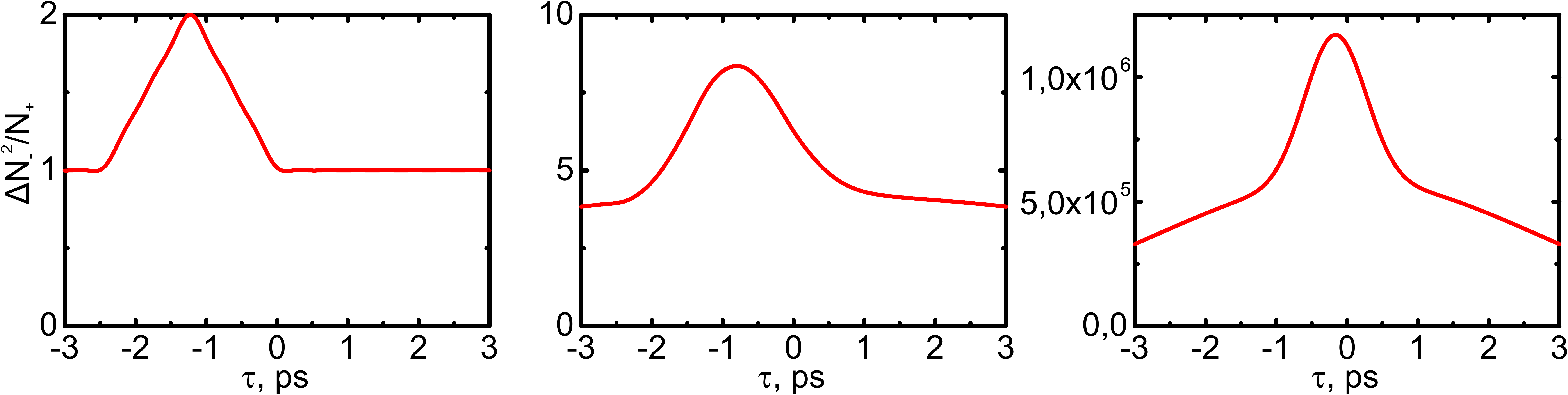}
\caption{The HOM effect in the dependence of $\var{\op{N}_-}/\mean{\op{N}_+}$ on the time delay for type-II phase matching and gain $G=0.001$ (left), 1.5 (center), and 7.5 (right).}
\label{4_fig:NRF_type_II}
\end{center}
\end{figure}

Indeed, the calculation gives
\begin{equation}
\label{4_eq:NRF_final_type_I}
\frac{\var{\op{N}_-}(\tau)}{\mean{\op{N}_+}}=1+\frac{\int_0^{\infty}d\Omega\,|V(\Omega, 0)|^2\left[|V(\Omega, \tau)|^2+|U(\Omega, 0)|^2\cos(2\Omega\tau)\right]}{\int_0^{\infty}d\Omega\,|V(\Omega, 0)|^2}
\end{equation}
for type-I phase matching and
\begin{equation}
\label{4_eq:NRF_final_type_II}
\frac{\var{\op{N}_-}(\tau)}{\mean{\op{N}_+}}=1+\frac{\int_0^{\infty}d\Omega\,|V(\Omega, 0)|^2\left[|V(\Omega, \tau)|^2+\Re\left(\left(U^*(\Omega, 0)\right)^2\,e^{2i\Omega\tau}\right)\right]}{\int_0^{\infty}d\Omega\,|V(\Omega, 0)|^2}
\end{equation}
for type-II one. If the variance of the photon-number difference is measured versus the delay between the signal and idler beams, a peak is observed instead of the dip (Figs.~\ref{4_fig:NRF_type_I} and \ref{4_fig:NRF_type_II}). The HOM visibility, defined in the same way as in Eq.~\eqref{4_eq:HOM_visibility}, is 100\%. The rest of the behavior is exactly the same as in the case of $g^{(2)}$ measurement.

As long as only small, dip-size, time delays are considered, the pump pulse duration is not important unless it is femto- or picosecond one. However, one might notice a variation of the background level in the right panels of Figs.~\ref{4_fig:g2_type_II} and \ref{4_fig:NRF_type_II}, because the calculations are made with a finite pulse duration (18 ps) of the pump. The envelope is assumed to be Gaussian,
\begin{equation}
E_p(t)=E_{p0}e^{-2\ln2(\nicefrac{t}{\tau_{d}})^2},
\end{equation}
with the duration of the pulse $\tau_{d}$.

\begin{figure}[!htb]
\begin{center}
\includegraphics[width=1\textwidth]{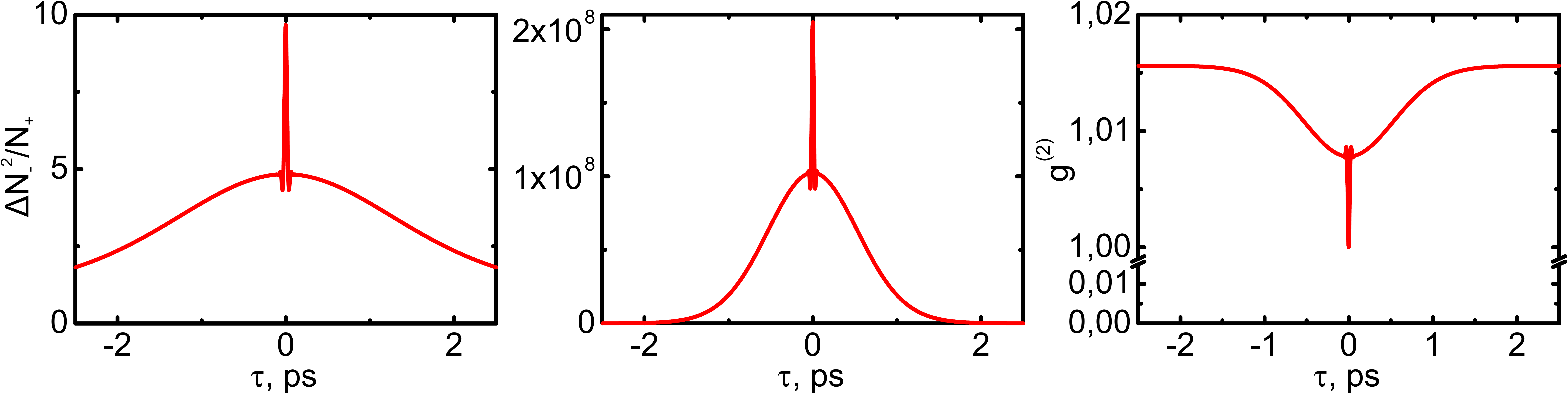}
\caption{The normalized variance of the difference $\var{\op{N}_-}/\mean{\op{N}_+}$ within a broad range of delays for type-I phase matching and gain $G_{\mathrm{max}}\equiv\sigma LE_{p0}=1.5$ (left) and 10 (center). For the last case also the CF $g^{(2)}_{12}$ (right) is presented.}
\label{4_fig:NRF_g2_large_delay}
\end{center}
\end{figure}

The normalized variance of the difference $\var{\op{N}_-}/\mean{\op{N}_+}$ is plotted within a broad range of delays for the case of type-I phase matching and $\tau_{d}=4$~ps, see Fig.~\ref{4_fig:NRF_g2_large_delay}. The background (pedestal) drops to unity, which is the shot-noise level, for delays larger than $\tau_{d}$. At higher gain the pedestal drops faster, because the PDC pulse shrinks; this effect is similar to the increase of the correlation width discussed in section~\ref{3_sec:correlations}. Therefore, the ratio between the peak and pedestal widths, similarly to the SFG case (section~\ref{3_sub:Schmidt_SFG}), should provide the number modes, namely the number $M_{temp}$ of temporal modes, which is equal to the number of frequency modes and also often called the number of longitudinal modes.

The CF $g^{(2)}_{12}$ shows the same behavior, but instead of decrease the background increases. It does not go to unity, because both detectors still measure signal and idler beams, the signal and idler photons just do not overlap on the BS. Therefore, the CF $g^{(2)}_{12}$ at large delay reflects the total number, $M_{tot}$, of the detected modes. The latter could be obtained from BSV statistical properties, namely from Eq.~\eqref{5_eq:g2_multimode} and the statistics of signal and idler beams, which will be discussed in chapter~\ref{5_chapter}.

\subsection{Experimental results: HOM dip and HOM peak}
\label{4_sub:HOM_exp}

Here the author's theory is presented as well as the experimental results, mainly obtained by Timur Iskhakov, showing a nice agreement between the theory and experiment.

The interfering twin beams are obtained in two 5~mm BBO crystals with type-II phase matching in the collinear frequency-degenerate regime. The crystals are placed close to each other, so that they are equivalent to a single 10~mm crystal, and pumped by 18 ps pulses of the third harmonic of a Nd:YAG laser at wavelength 354.7~nm. The HOM effect is observed by changing the time delay between signal and idler photons. This delay is introduced by the delay line made of a polarizing BS (PBS) and two quarter-wave plates oriented at $45^\circ$ w.r.t. PBS.

\begin{figure}[!htb]
\begin{center}
\includegraphics[width=0.9\textwidth]{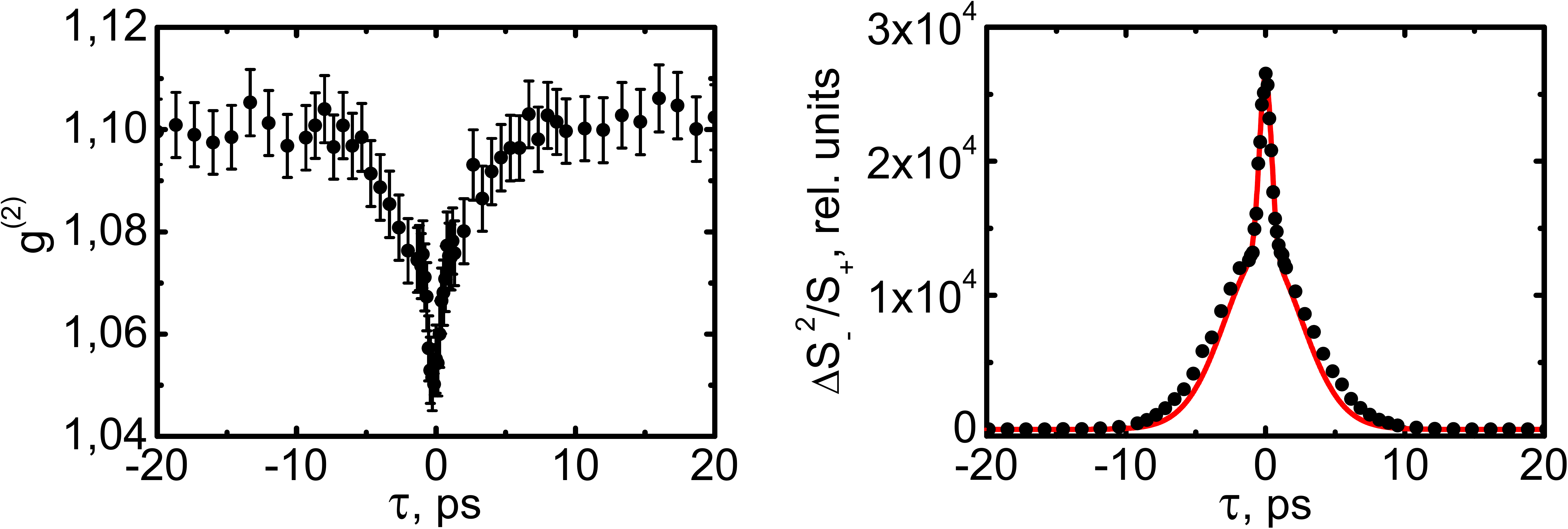}
\caption{The HOM effect in the $g^{(2)}_{12}$ (left) and $\var{S_-}/\mean{S_+}$ (right) dependence on the delay (black points) for gain $G=7.5$. For the latter also the theoretical dependence (red line) according to Eq.~\eqref{4_eq:NRF_final_type_II} is presented.}
\label{4_fig:HOM_experiment}
\end{center}
\end{figure}

The experimental dependences perfectly agree with the theory, see Fig.~\ref{4_fig:HOM_experiment}. In the $g^{(2)}_{12}$ measurement the visibility of the HOM effect is small; furthermore, a complicated structure of the dip could be hardly resolved. The value of $g^{(2)}_{12}(0)$ does not reach unity due to experimental imperfections.

On the contrary, the variance of the difference $\var{S_-}/\mean{S_+}$ measured simultaneously presents all features ideally detailed with almost 100\% visibility. Here $S_{1,2}\propto N_{1,2}$ are the signals from the detectors\footnote{More details follow in section~\ref{4_sub:setup_BS_interf}.}. There is only one fitting parameter, the height of the peak, used in the theoretical dependence (red line). The pedestal in the theory is narrower than in the experiment by about 15\%, which could be caused by the fact that the pump pulses are a bit longer than 18~ps.

From the $g^{(2)}$ value outside of the dip we get the total number of modes, $M_{tot}\approx10$. At the same time from the widths ratio in $\var{S_-}/\mean{S_+}$ we get the number of temporal modes, $M_{temp}\approx8$. Hence, there are about 1.25 spatial modes, which is a reasonable value, because the spatial filtering is not perfect.

\begin{figure}[!htb]
\begin{center}
\includegraphics[width=0.45\textwidth]{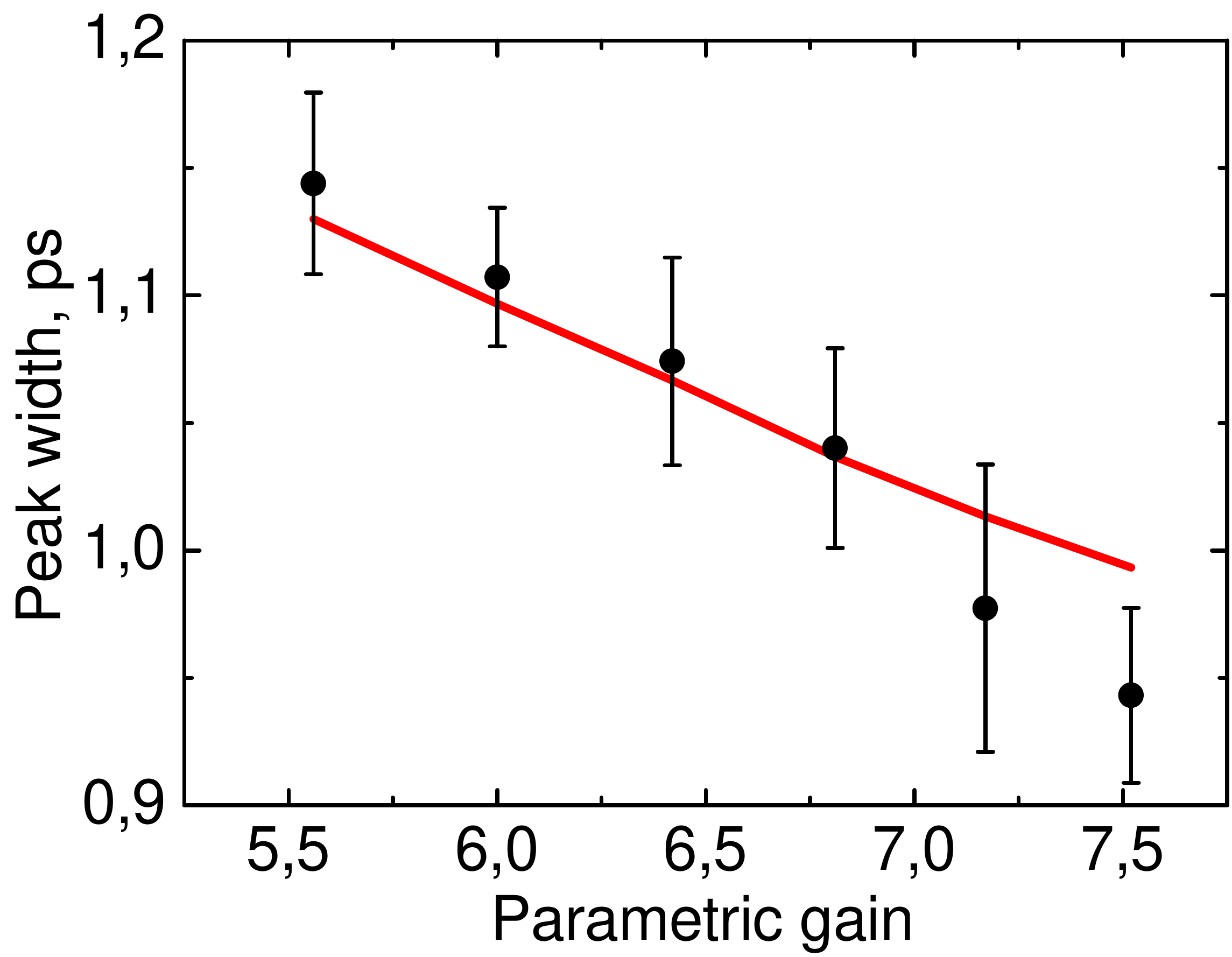}
\caption{Theoretical (red line) and experimental (black points) widths of the central peak in the $\var{S_-}/\mean{S_+}$ dependence presented versus the parametric gain $G$.}
\label{4_fig:HOM_peak_gain}
\end{center}
\end{figure}

\begin{figure}[!htb]
\begin{center}
\includegraphics[width=0.45\textwidth]{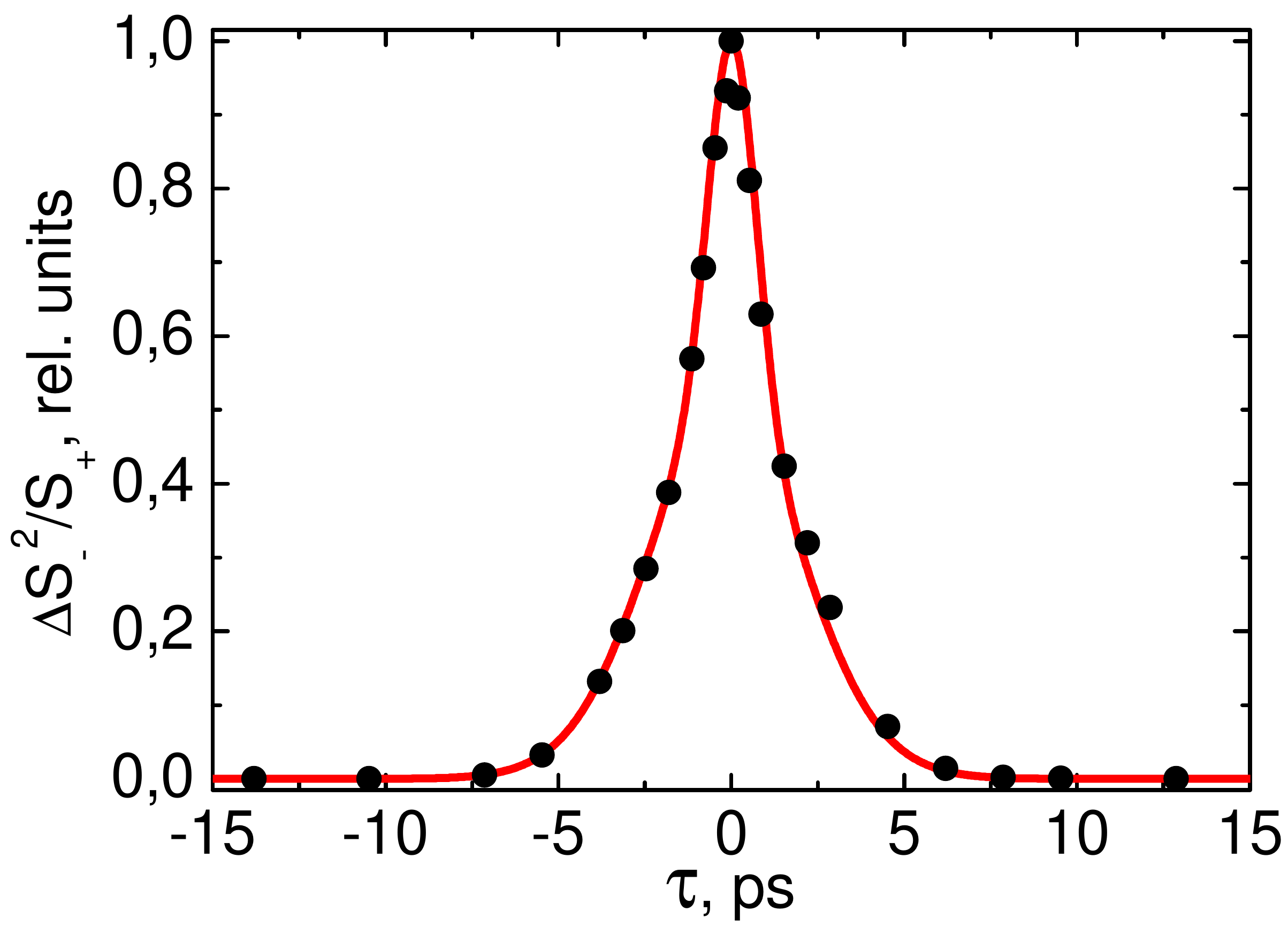}
\caption{The HOM effect for a 20~mm of BBO and $G=12.5$: theoretical (red line) and experimental (black points) $\var{S_-}/\mean{S_+}$ dependences.}
\label{4_fig:HOM_peak_2}
\end{center}
\end{figure}

The peak width changes with the gain (Fig.~\ref{4_fig:HOM_peak_gain}) due to the spectral broadening (see section~\ref{3_sec:broadening}). We observe about 20\% narrowing by changing the gain $G$ from 5.5 to 7.5, which agrees with the theoretical expectations.

Moreover, in the setup described in section~\ref{4_sub:setup_BS_interf} the author also measures the HOM effect in order to overlap the signal and idler pulses on the BS. Currently a 20~mm of BBO and $G=12.5$ are used  (Fig.~\ref{4_fig:HOM_peak_2}). Although the number of temporal modes is small and therefore Eq.~\eqref{4_eq:NRF_final_type_II} is not fully reliable for this case, there is still a nice agreement between the theory and experiment.

%%%%%%%%%%%%%%%%%%%%%%%%%%%%%%%%%%%%%%%%%%%%%%%%%%%%%%%%%%%%%%%%%5
\section{Interference on a beam splitter (BS)}
\label{4_sec:interf_on_BS}

This section considers in details the photon statistics at the BS output in the case of twin beams at the inputs and, namely, how the phase fluctuations are converted into photon-number ones. It is also shown how this effect could be mimicked by classical beams with an artificially mixed phase.

\subsection{Twin beams and phase fluctuations}

As it was discussed in section~\ref{4_sub:Fock_interf}, the interference of Fock states leads to a broad U-shaped photon-number distribution at the BS outputs. The photon-number difference does not fluctuate at the input and is completely uncertain at the output, whereas the phase difference is completely uncertain at the input and does not fluctuate at the output, suggesting applications in phase super-resolution. However, the Fock states are extremely difficult to obtain.

Fortunately, twin beams have similar features. As it was mentioned in section~\ref{2_sub:twin-beam_sq}, the photon-number difference is zero and does not fluctuate. For example, in Ref.~\cite{Iskhakov2016_2} the photon-number difference is squeezed 7.8~dB below the shot-noise level. Similar is the situation with the phase difference: it is completely uncertain\footnote{Note that the sum of phases is certain and equal to the pump phase.}.

Accordingly, similar to the Fock states interference, the twin-beam one should provide a phase super-resolution. However, this is difficult to observe; although, the photon-number difference is zero, the total number of photons could be very different, see Eq.~\eqref{2_eq:twin_state}, because each of the beams is in a thermal state. Fortunately, these fluctuations could be suppressed using a postselection on the sum of photon numbers. This could be done using an additional detector~\cite{Iskhakov2016, Manceau2017} or with the same ones as used for the main measurement~\cite{Iskhakov2016_2}.

Interestingly, similar has been predicted~\cite{Bouyer1997} and discovered~\cite{Lucke2011} for Bose-Einstein condensates with pair correlations. They also have strong fluctuations in the total number of correlated atoms and ultralow ones in the difference of populations with the spins up and down. By implementing a `beam splitter' an interferometric sensitivity 1.6~dB below the shot noise limit was shown. Moreover, the problem of strong fluctuations is solved in the same way: the condition on the total number of correlated atoms is applied, namely this number should be within the certain boundaries.

\begin{figure}[!htb]
\begin{center}
\includegraphics[width=0.9\textwidth]{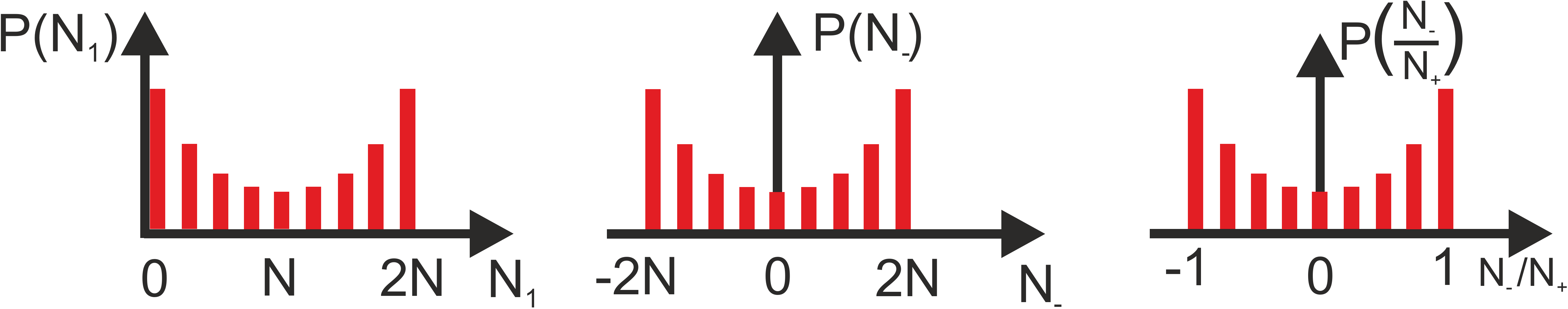}
\caption{Schematic probability distributions $P(N_1)$ (left), $P(N_-)$ (center), and $P(N_-/N_+)$ (right) at the BS output for the case of interfering Fock states $\ket{N}$.}
\label{4_fig:P_N_diff_expl}
\end{center}
\end{figure}

Furthermore, one can use an additional trick\footnote{Interestingly, a similar one is used also in Ref.~\cite{Lucke2011} for atoms.} for observing a U-shaped distribution. For the interfering Fock states, see Eq.~\ref{4_eq:U_shape}, the shape of the distribution remains the same regardless of whether $P(N_1)$, $P(N_-)$, or $P(N_-/N_+)$ is plotted (Fig.~\ref{4_fig:P_N_diff_expl}). However, there is a big difference in the case of twin beams: for $P(N_1)$ and $P(N_-)$ distributions the `U' shape smears out, because the positions of maxima are different for the different Fock states. Therefore, some postselection discussed above is necessary. Yet, for $P(N_-/N_+)$ distribution the positions are the same, and therefore the selection is not necessary, instead of postselection one can use the whole dataset.

\subsection{Experimental setup}
\label{4_sub:setup_BS_interf}

\begin{figure}[!htb]
\begin{center}
\includegraphics[width=0.65\textwidth]{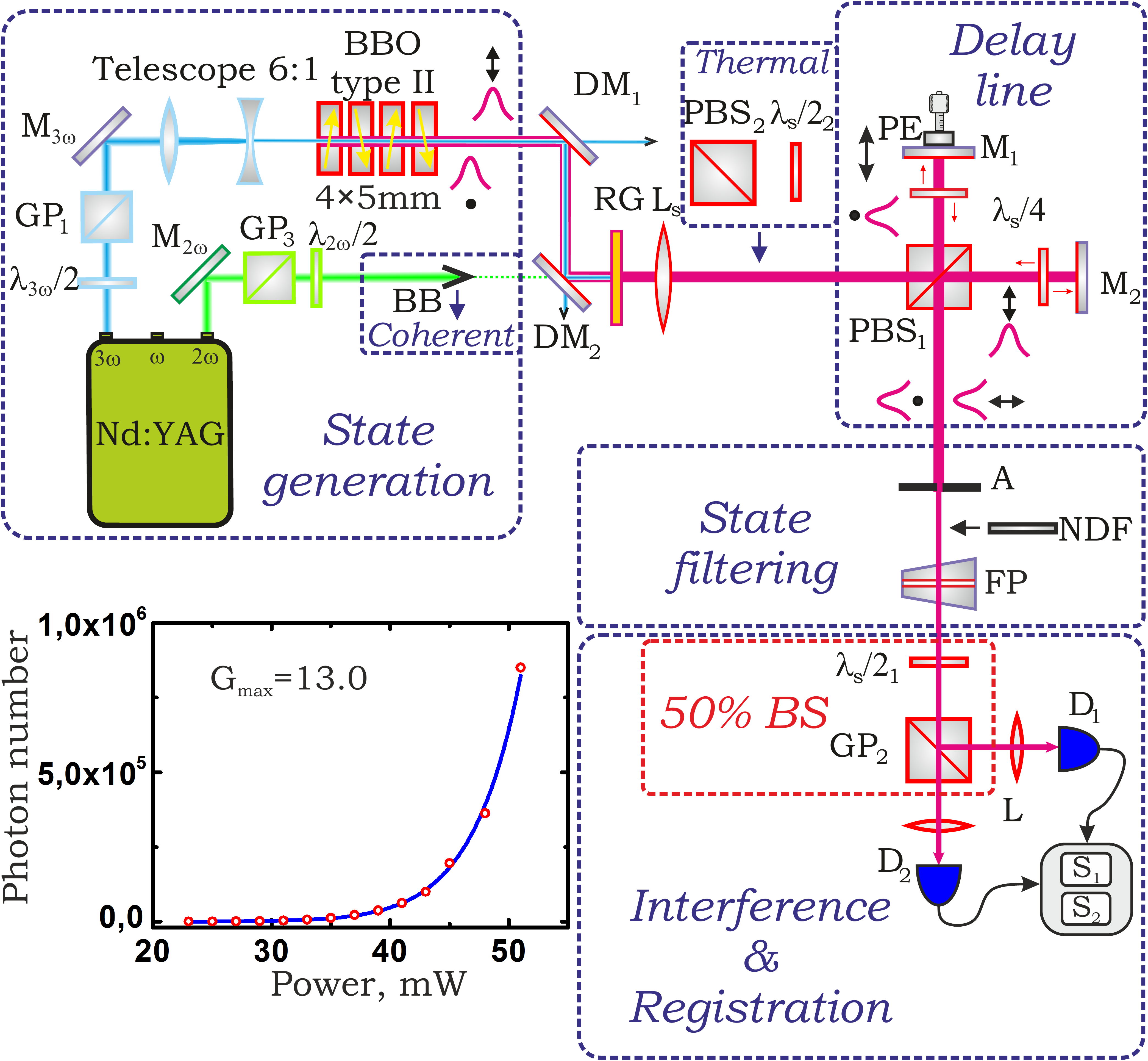}
\caption{Experimental setup for measuring twin-beam interference. Inset: the dependence of the mean photon number in twin beams on the pump power (points) and its fit with Eq.~\eqref{3_eq:gain_fit}.}
\label{4_fig:setup_BS_interf}
\end{center}
\end{figure}

The experimental setup is presented in Fig.~\ref{4_fig:setup_BS_interf}. The twin beams are generated in four 5 mm BBO crystals with the collinear frequency-degenerate type-II phase matching pumped by the same laser as 
in the experiment on spectral broadening (section~\ref{3_sec:broadening}). The crystals have optic axes tilted in opposite directions to reduce the effect of spatial walk-off\footnote{More details about this effect will follow in section~\ref{4_sec:giant_twin_beams}.}~\cite{Slusher1987}. The pump is focused using a 6:1 telescopic system. After the crystals the pump is cut off by two dichroic mirrors (DM$_{1,2}$) and a long-pass filter (red glass, RG).

As it was discussed, for type-II phase matching the signal and idler photons are always delayed w.r.t. each other; a delay line is used to compensate for that. The beams are split on a PBS (PBS$_1$) and then propagate along different paths to mirrors M$_1$ and M$_2$, respectively. The position of mirror M$_1$ can be changed roughly using a micrometer and accurately with a piezoelectric actuator (PE). Polarization of the beams is then rotated by $90^\circ$ after a double pass through quarter-wave plates ($\lambda_s/4$) oriented at $45^\circ$ w.r.t. PBS$_1$. Therefore, both beams exit through the same output port of PBS$_1$. The delay is controlled by observing the HOM effect (Fig.~\ref{4_fig:HOM_peak_2}), for which the simultaneous arrival of signal and idler photons is critical.

In order to obtain nearly single-mode beams, the spatial and spectral filtering is applied. The spatial filtering is performed in a focal plane of a lens (L$_s$) by an aperture (A) selecting an angle of $1$ mrad. The spectral filtering is done by a Fabry-P\'erot interferometer (FP) with a $250$ $\mu$m base, which is adjusted to achieve a spectrum with only one transmission maximum. To make the signal and idler beams indistinguishable, this maximum is shifted to the degenerate wavelength, 709.3~nm, by slightly rotating the FP. The spectrum is monitored by a spectrometer (HORIBA Jobin Yvon Micro HR). After the filtering, the spectral width is only $0.013$ nm being estimated from the free spectral range and finesse of the FP~\cite{Born1980}.

A combination of a Glan prism (GP$_2$) and a half-wave plate ($\lambda_s/2_1$) oriented at $22.5^\circ$ serves as a 50\% BS for signal and idler beams. The reflected and transmitted beams are focused onto charge-integrating detectors (D$_1$ and D$_2$) based on Si PIN photodiodes (Hamamatsu S3883). The electronic signals $S_{1,2}$ from the detectors are proportional to the numbers $N_{1,2}$ of photons per pulse, namely $S_{1,2}=A_{1,2}N_{1,2}$, where $A_1=10.0$~pV$\times$s and $A_2=11.1$~pV$\times$s per photon~\cite{Iskhakov2009}. The signals are digitized by a fast analog-to-digital converter card. If the light is too bright for the detectors, the intensity is reduced using neutral density filters (NDF). 

Like above the effective number $M$ of modes in the beams is estimated from the $g^{(2)}$ measurement. With the spectral filtering a nearly single-mode case, $M = 1.2$, is produced, while without it, a few modes one, $M = 3.4$.

The parametric gain is measured in a similar way as in the experiment on spectral broadening (section~\ref{3_sec:broadening}), see inset in Fig.~\ref{4_fig:setup_BS_interf}; maximally the gain $G_{max}=13$ is obtained, which corresponds to $4.9\times 10^{10}$ photons per mode.

\subsection{Measurement of the twin-beam interference}

\begin{figure}[!htb]
\begin{center}
\includegraphics[width=0.85\textwidth]{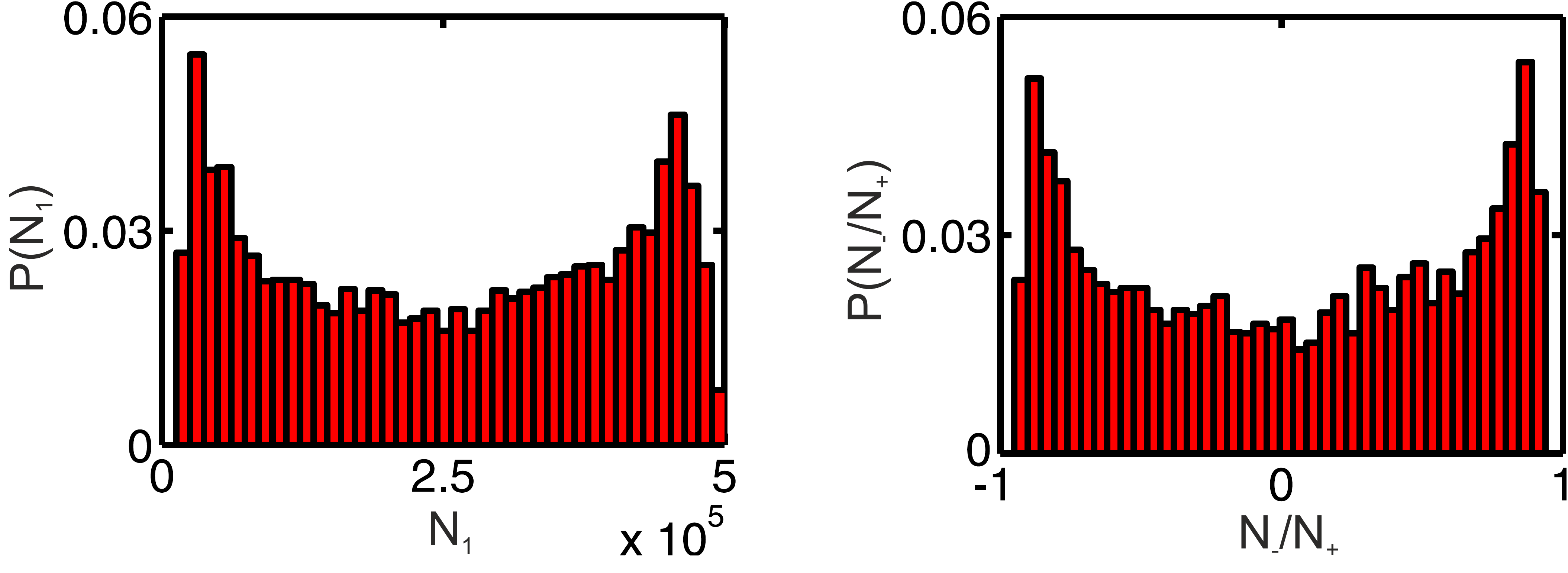}
\caption{Probability distributions $P(N_1)$ (left) and $P(N_-/N_+)$ (right) for the number of modes $M=1.2$. The total photon number $\mean{N_{+}}$ is fixed at $(5\pm0.2)\times10^5$ photons.}
\label{4_fig:4_P_N_exp_M_1_2}
\end{center}
\end{figure}

The experimental probability distributions $P(N_1)$ and $P(N_-/N_+)$ for the nearly single-mode case are shown in Fig.~\ref{4_fig:4_P_N_exp_M_1_2}. Both distributions clearly demonstrate a `U' shape, the case of almost all photons at one of the outputs has the highest probability. As it was discussed above, this reflects the relative phase uncertainty between twin beams. The odd/even structure expected for Fock states, see Fig.~\ref{4_fig:Fock_states_interf}, should not be observed here due to losses and lack of photon-number resolution. 

However, it is clear that the distribution $P(N_-/N_+)$ is more useful than $P(N_1)$, which is blurred, mainly on the right side, despite the rather strict condition on the total photon number, $\mean{N_{+}}=5\times10^5$ photons with 4\% tolerance; the distribution $P(N_-/N_+)$ does not have this problem.

The distribution $P(N_-/N_+)$ does not occupy all range from -1 to 1 because the BS is unbalanced, the GP always introduces 5\% losses into one polarization. This we confirm theoretically~\cite{Spasibko2014}: the higher the bias of the BS, the closer the maxima get to each other, ultimately converging into one central peak.

Both theoretical and experimental results show that the `U' shape is not destroyed even in the presence of high losses in both beams, because the losses do not reduce the phase uncertainty. In experiment they are higher than 99\% due to the spatial and spectral filtering.

\begin{figure}[!htb]
\begin{center}
\includegraphics[width=0.4\textwidth]{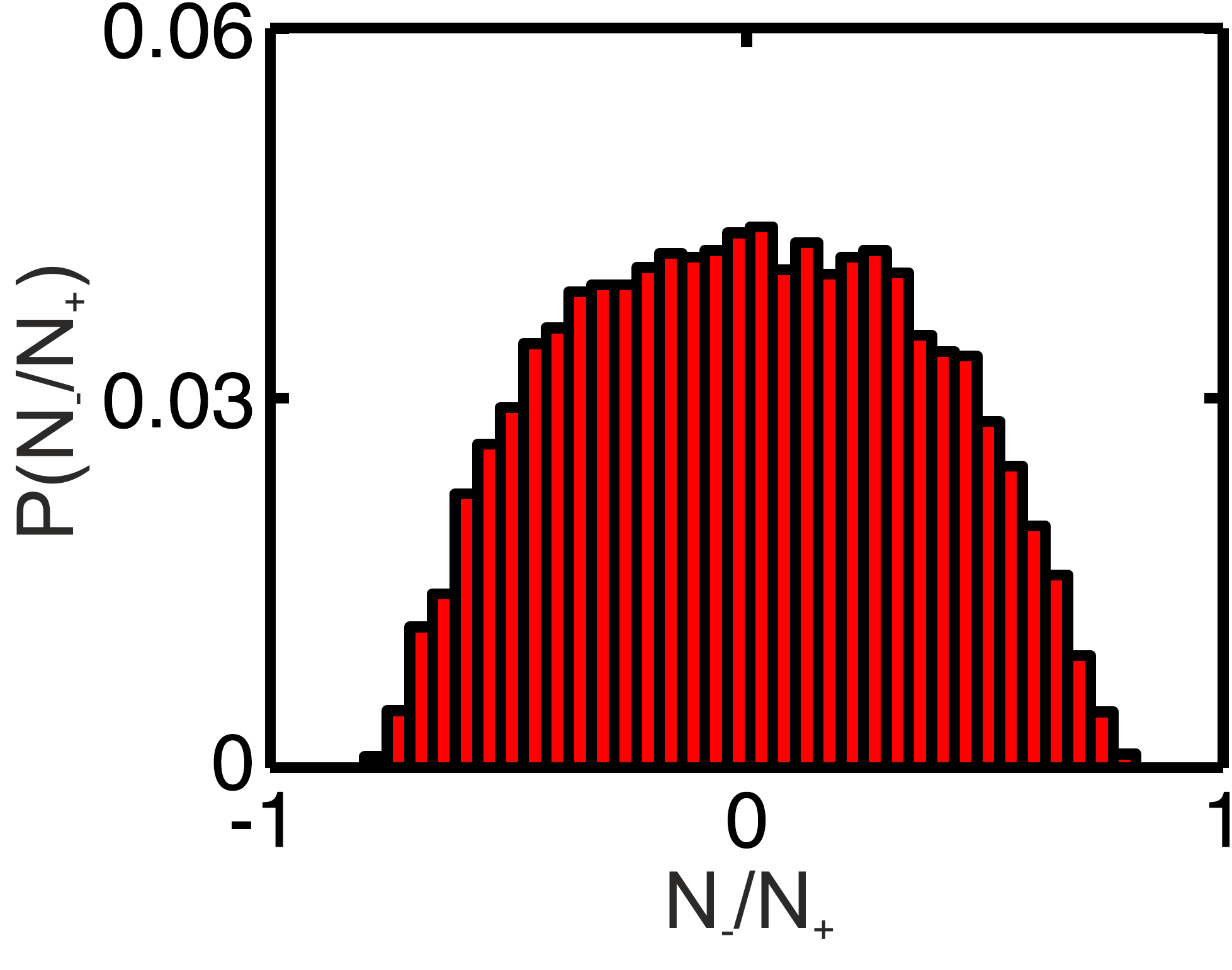}
\caption{Probability distribution $P(N_-/N_+)$ for the number of modes $M=3.4$.}
\label{4_fig:4_P_N_exp_M_3_4}
\end{center}
\end{figure}

The theory also shows that the `U' shape disappears for the number of modes $M>2$. Without the frequency filtering, $M=3.4$, indeed the experiment yields a peaked distribution (Fig.~\ref{4_fig:4_P_N_exp_M_3_4}). In the multimode case the `U' shape smears out due to independent photon-number fluctuations in different modes.
 
\begin{figure}[!htb]
\begin{center}
\includegraphics[width=1\textwidth]{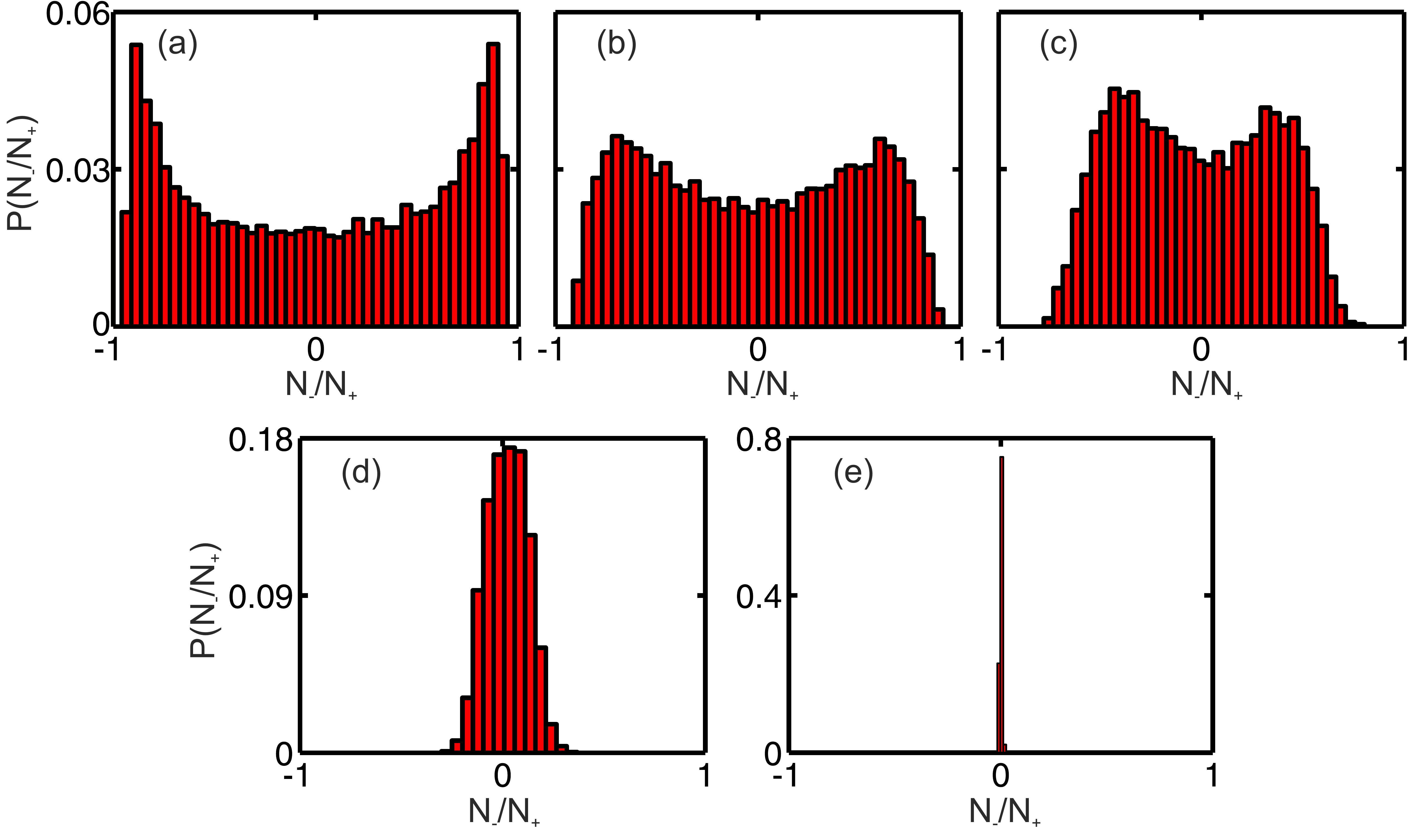}
\caption{Probability distributions $P(N_-/N_+)$ for the number of modes $M=1.2$ and different ratios of the input photon numbers: 1 : 1 (a), 1 : 3 (b), 1 : 5 (c), 1 : 17 (d), and 1 : $\infty$ (e).}
\label{4_fig:P_N_exp_diff_ratio}
\end{center}
\end{figure}

The interference of non-equally populated signal and idler beams is also studied by introducing the losses in one of the beams (Fig.~\ref{4_fig:P_N_exp_diff_ratio}). Notably, the shape remains concave up to a considerable, 5:1, ratio between the input photon numbers (panel c). This feature is mentioned in Ref.~\cite{Bouyer1997} in connection with the robustness of phase resolution to an imbalance between twin beams. For a higher imbalance, the `U' shape transforms into a single peak (panels d,e), similarly to the case of Fock states~\cite{Campos1989}. Note that a single peak also appears for the case of two independent (distinguishable) Fock states at the BS inputs.

\subsection{Interference of the classical beams}

The `U' shape can be also observed for classical beams with fluctuations in the relative phase increased artificially; this classical analogy was pointed out in Ref.~\cite{Campos1989}. To demonstrate this analogy a thermal or coherent beam is used with the following strategy applied to it: the beam is split in two, the phase of one of the beams is varied, both beams are overlapped on the BS, and the distribution $P(N_-/N_+)$ is measured at the output.

\begin{figure}[!htb]
\begin{center}
\includegraphics[width=0.78\textwidth]{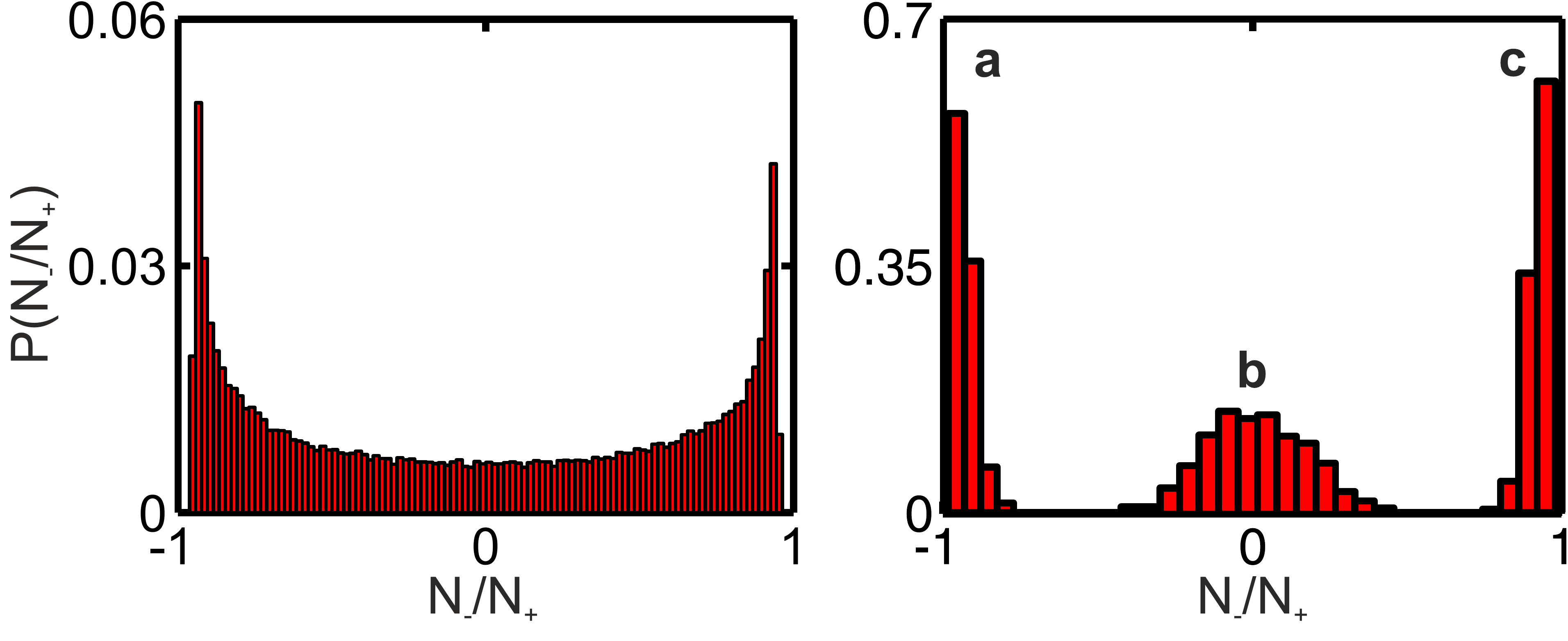}
\caption{Probability distributions $P(N_-/N_+)$ for the case of two input thermal states with a randomized (left) and fixed (right) relative phase. For the latter, three typical distributions for the different phases are shown: all photons are divided equally between the detectors (b) or go together to the detectors D$_1$ (c) and D$_2$ (a).}
\label{4_fig:P_N_exp_thermal}
\end{center}
\end{figure}

As a thermal radiation, one of the twin beams is used. To separate the signal beam from the idler one another PBS (PBS$_2$, see Fig.~\ref{4_fig:setup_BS_interf}) is introduced. Using a half-wave plate ($\lambda_s/2_2$) at $22.5^\circ$ the signal beam is split on the PBS$_1$ and the latter one interferes at the 50\% BS. To randomize the phase, a varying voltage is applied to the PE actuator. The resulting distribution demonstrates the `U' shape if the phase is randomized and it becomes peaked if the phase is fixed (Fig.~\ref{4_fig:P_N_exp_thermal}).

\begin{figure}[!htb]
\begin{center}
\includegraphics[width=0.85\textwidth]{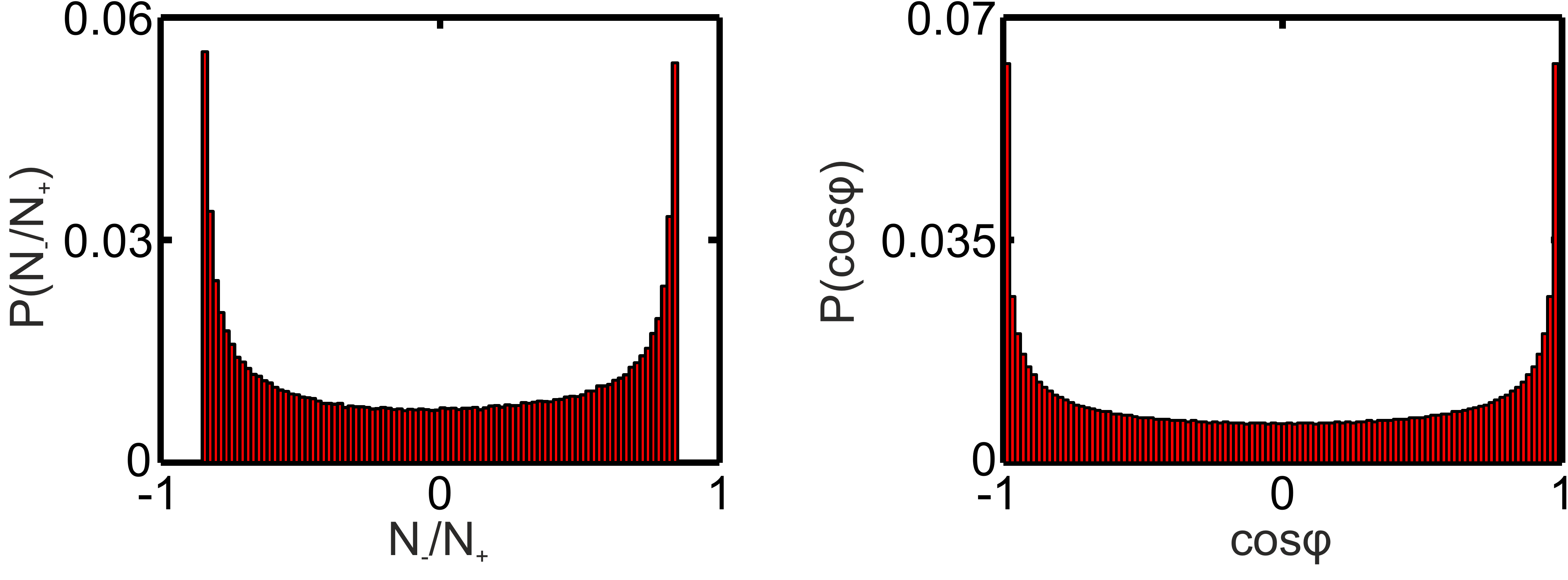}
\caption{Probability distributions $P(N_-/N_+)$ for the case of two coherent states with a randomized phase at the inputs (left). As an analogy also the distribution $P(\cos\varphi)$ for the randomized phase $\varphi$ is shown (right).}
\label{4_fig:P_N_exp_coherent}
\end{center}
\end{figure}

As a coherent beam, the second harmonic of a Nd:YAG laser (Nd:YAG 2$\omega$) at wavelength 532 nm is used (see Fig.~\ref{4_fig:setup_BS_interf}). As in the case of thermal light, a half-wave plate ($\lambda_{2\omega}/2$) oriented at $22.5^\circ$ followed by PBS$_1$ is used to split the beam. The beam block (BB), the long-pass filter (RG), and the FP interferometer are removed in this case. The probability distribution $P(N_-/N_+)$ shows the same `U' shape if the phase is randomized (Fig.~\ref{4_fig:P_N_exp_coherent}, left). It is narrower than the one for the thermal case due to the nonideal interference caused by the fact that optical elements are not optimized for 532~nm.

The origin of the `U' shape here can be simply understood from the distribution $P(\cos\varphi)$ for the phase $\varphi$ distributed uniformly at $[0~2\pi]$ (Fig.~\ref{4_fig:P_N_exp_coherent}, right). Indeed, if two beams with equal intensities and the phase difference $\varphi$ interfere on a 50\% BS, the intensity difference between the outputs is proportional to $\cos\varphi$.

It should be stressed that in contrast to a pure twin-beam state~\cite{Anisimov2010}, U-shaped distributions observed for classical light do not lead to any phase super-resolution, because the phase fluctuations are increased artificially by ‘mixing’ the state.

\subsection{Possible applications of the BS interference}

The BS interference could have an important application for quantum state engineering. The distributions shown in Fig.~\ref{4_fig:P_N_exp_diff_ratio} clearly demonstrate the following tendency: the more symmetric are photon numbers at the input, the more asymmetric they are at the output, and vice versa. This suggests that the BS interference could be used for filtering two-mode macroscopically populated states of light with either close or much different photon numbers~\cite{Stobinska2012}.

Indeed, from the initial two-mode state one can tap off a small portion, which is directed to a BS, where these two modes interfere. The measurement at the BS outputs will show that the photon numbers in the initial modes are likely equal or different. Depending on the result and desired filtering, the main part of the state will be used or not used later, for example one can perform a postselection or open/close some special shutter. Since the measurement does not provide information about the photon numbers at the input ports of BS, this filter would select only the modulus of the photon-number difference but not the sign of this difference.

This feature could be useful for macroscopic qubits, $\alpha\ket{\Phi}+\beta\ket{\Phi_\perp}$, considered in Ref.~\cite{DeMartini2008}, the states obtained after amplifying a single-photon polarization qubit, $\alpha\ket{1}+\beta\ket{1}_\perp$, in an OPA. The amplification keeps the information encoded in a qubit, but it is difficult to extract it due to a big effective overlap in the photon number distributions for the basis qubits $\ket{\Phi}$ and $\ket{\Phi_\perp}$. The filter described above reduces this overlap and at the same time maintains quantum superpositions.

%%%%%%%%%%%%%%%%%%%%%%%%%%%%%%%%%%%%%%%%%%%%%%%%%%%%%%%%%
\section{Giant twin-beam generation}\label{4_sec:giant_twin_beams}

This section is devoted to the amplification of twin beams and, in particular, it considers an application of the walk-off matching, spatial and temporal, to the generation of giant narrowband twin beams. Such an amplifier could be considered as a highly tunable optical parametric generator (OPG) without any cavity.

\subsection{Walk-off effects in SPDC and high-gain PDC}

Walk-off effects are caused by the difference between the velocities and directions of energy and phase propagation. The most famous manifestation is the phenomenon of double refraction caused by the birefringence of anisotropic materials. In this case the Poynting vector is non-collinear to the wave vector. For example, in uniaxial crystals this is the case for an extraordinarily polarized light~\cite{Dmitriev1999}. Similarly, the longitudinal (temporal) walk-off appears because the energy of a pulse travels at a different speed than its phase. 

Since in nonlinear optical processes the phase matching determines the frequency-wavevector spectrum, interacting pulses are often group mismatched and separate in space and time during the propagation. This limits the length of nonlinear interactions for short pulses and focused beams. 

This is definitely the case for high-gain PDC: the exponential amplification is provided only if the produced photons are overlapped in space and time with the pump ones. For example, for the case of type-I phase matching the pump is polarized extraordinary whereas signal and idler beams, ordinary, therefore only the former exhibits the spatial walk-off. The situation with the temporal walk-off is similar: the group velocities of the pump and the PDC beams are different, as they have different frequencies.

The spatial walk-off can be, in principle, eliminated by a using non-critical phase-matching, double-crystal configuration, used in the experiment on BS interference (section~\ref{4_sub:setup_BS_interf}), or periodically poled materials. Unfortunately, the temporal walk-off is almost inevitable.

For SPDC in its respect these effects are less important, because the produced photons do not enhance the generation rate. Therefore, the walk-off effects lead only to an asymmetry of the frequency-angular spectrum~\cite{Fedorov2007,Cavanna2014}, rather than to a beamlike emission shown below.

\subsection{Twin-beam generation along the pump Poynting vector}

This section is intended to show that the twin-beam generation can be drastically enhanced, if one of the beams is emitted along the pump Poynting vector. This experiment was done by Angela P\'erez, the author's contribution was the explanation of some important properties of the effect and the data processing.

Twin beams are produced in a 5~mm BBO crystal with the collinear frequency-degenerate type-I phase matching, which corresponds to the optic axis angle $\phi=32.97^\circ$. The pump, the same laser as in the experiment on spectral broadening (section~\ref{3_sec:broadening}), is focused into the crystal to $60~\mu$m size.

\begin{figure}[!htb]
\begin{center}
\includegraphics[width=1\textwidth]{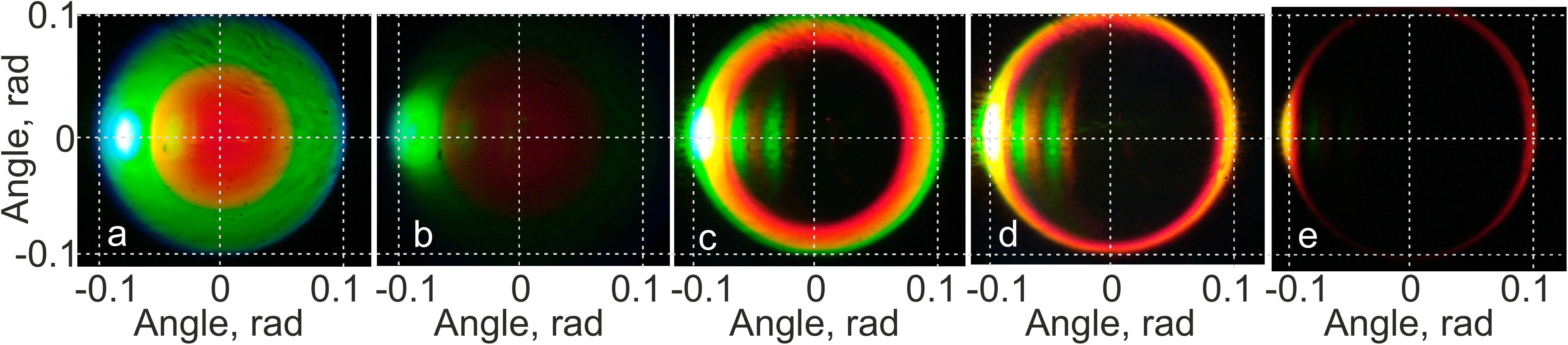}
\caption{Snapshots with a photographic camera show the spectra of high-gain PDC at different crystal orientations: $32.97^\circ$ (a,b), $33.4^\circ$ (c), $33.7^\circ$ (d), and $34.0^\circ$ (e). Additional stripes in panels d,e appear due to reflections in the crystal. A snapshot with a neutral density filter (b) shows colors without saturation.}
\label{4_fig:spat_walk_off_camera}
\end{center}
\end{figure}

\begin{figure}[!htb]
\begin{center}
\includegraphics[width=0.65\textwidth]{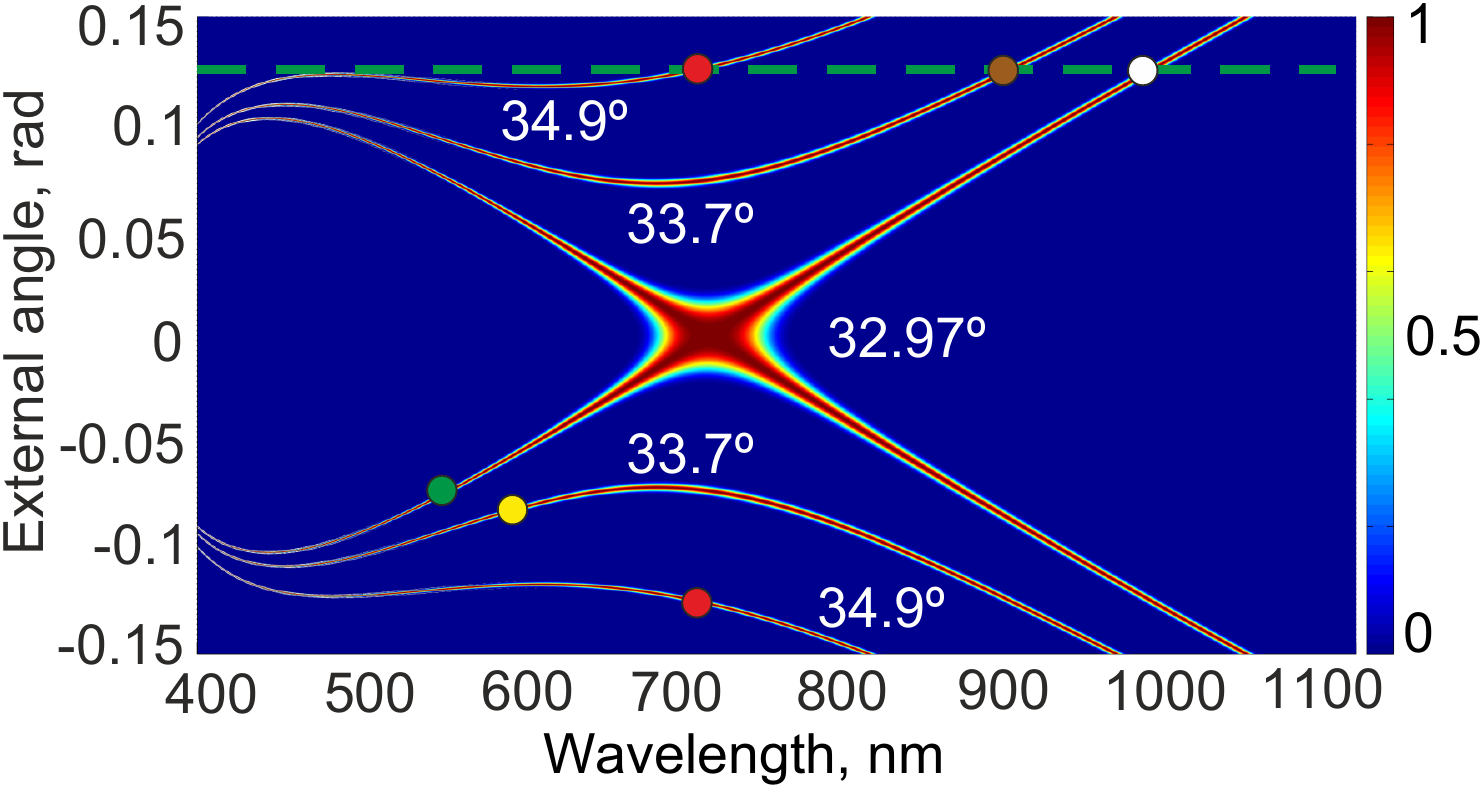}
\caption{The spectrum $S_{q\Omega}(\theta_x, \lambda)$ for the degenerate collinear ($\phi=32.97^\circ$) and noncollinear ($\phi=33.7^\circ$ and $34.9^\circ$) phase matching. By green line the walk-off direction is shown. The phase matching for the idler beam is shown by white, brown, and red dots; the corresponding signal beams are shown by green, yellow, and red dots, respectively.}
\label{4_fig:spat_walk_off_tuning}
\end{center}
\end{figure}

The angular spectra captured by using a digital photographic camera are shown in Fig.~\ref{4_fig:spat_walk_off_camera}. One could see strongly enhanced beamlike emission in green spectral range (panel a). The color is modified due to saturation of the camera; a snapshot with a neutral density filter (panel b) clearly demonstrates this. The green beam is the only one of the twins, the signal one. Moreover, it is the idler beam that propagates along the pump Poynting vector; it belongs to the IR range and is not visible (see Fig.~\ref{4_fig:spat_walk_off_tuning}, white dot).

By tilting the crystal we change the phase matching and the spectral range for the beamlike emission (Fig.~\ref{4_fig:spat_walk_off_camera}, panels c-d). As the idler wavelength approaches the visible range, both left and right parts of the ring become pronounced (panel e).

\begin{figure}[!htb]
\begin{center}
\includegraphics[width=0.9\textwidth]{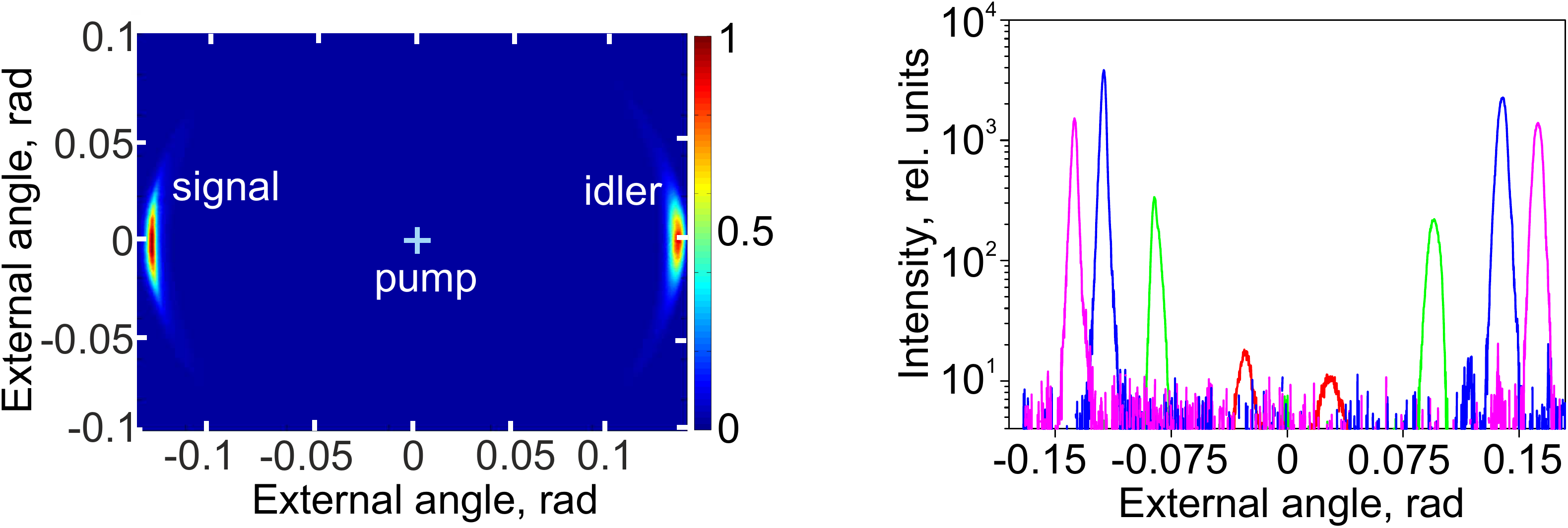}
\caption{2D angular spectrum for the optic axis angle $\phi=34.9^\circ$ at 710~nm (left) and the cross-sections of the 2D spectra in the walk-off plane (right) for $\phi=33^\circ$ (red), $34^\circ$ (green), $34.9^\circ$ (blue), and $36^\circ$ (magenta). The asymmetry in the positions of the peaks is caused by large tilt of the crystal.}
\label{4_fig:spat_walk_off_CCD}
\end{center}
\end{figure}

In order to show both enhanced beams we install a bandpass filter centered at the degenerate wavelength and orient the crystal to the angle $\phi=34.9^\circ$ (Fig.~\ref{4_fig:spat_walk_off_tuning}, red dots). The angular distribution is recorded by a CCD camera (Fig.~\ref{4_fig:spat_walk_off_CCD}, left panel). One can hardly see the whole ring of emission, yet, there is a strong peak in the walk-off direction as well as in the symmetrical one w.r.t. to the pump. At the other orientations of the crystal, the emission at 710~nm is much weaker (right panel): twin beams observed in the walk-off direction are brighter than the near-collinear emission by more than two orders in magnitude. The theory provides the same result~\cite{Perez2015}.

The angular divergence of both beams is small because the amplification occurs within a small angle. The pump divergence is on the same order, therefore the twin beams should be nearly single-mode spatially. Indeed, from the $g^{(2)}$ measurement we get $1.4$ spatial modes. This decrease in the number of modes is similar to the one which occurs in the PDC generation in two separated crystals~\cite{Perez2014}.

\subsection{Experimental setup: temporal walk-off matching}
\label{4_sub:setup_temp_walk}

\begin{figure}[!htb]
\begin{center}
\includegraphics[width=0.55\textwidth]{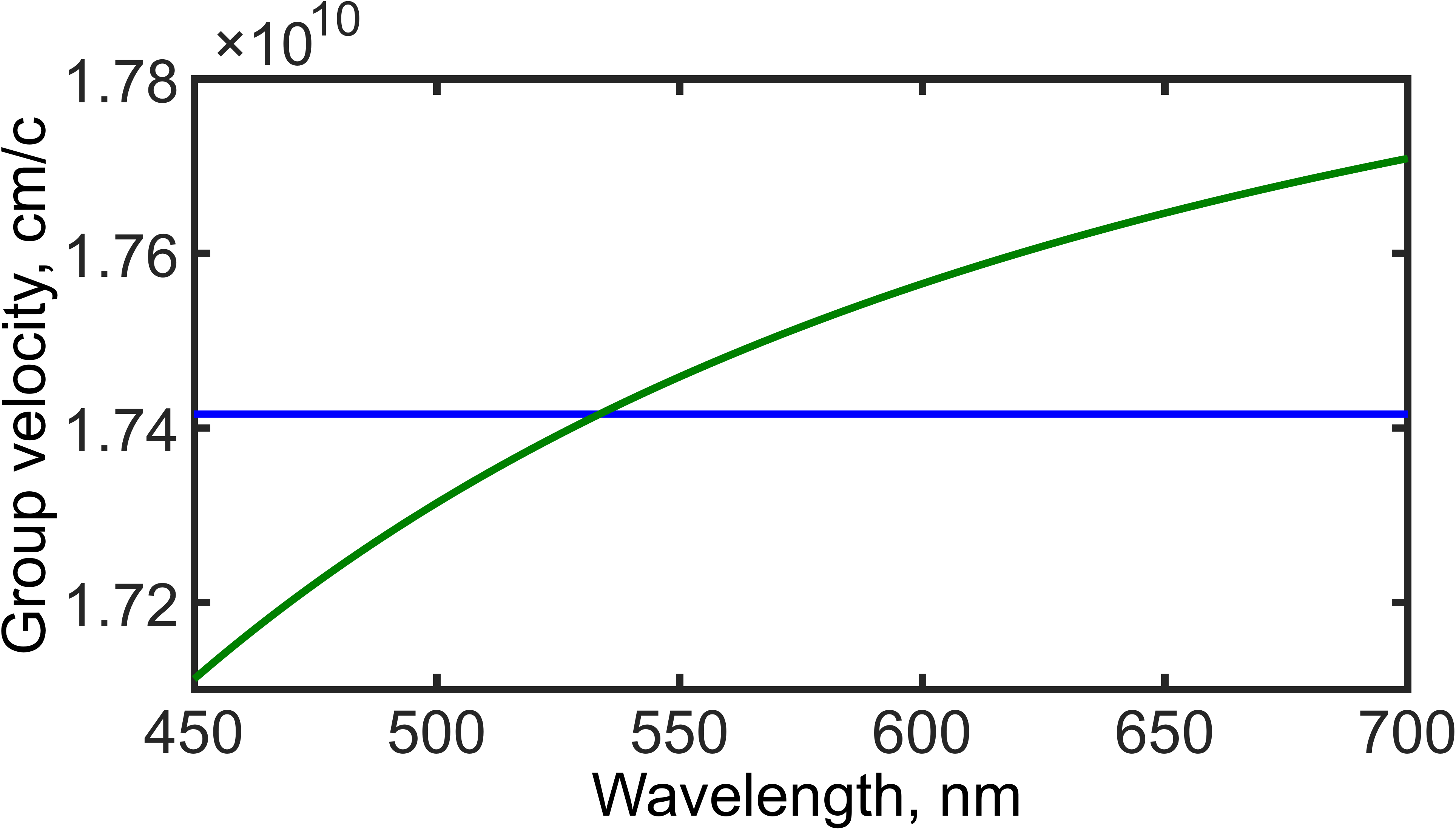}
\caption{Group velocity in a BBO crystal for the extraordinary polarized light at 400~nm (blue line) and the ordinary polarized one at different wavelengths (green line) for the angle $\phi=29.18^\circ$.}
\label{4_fig:group_vel}
\end{center}
\end{figure}

The same effect exists also in temporal domain; moreover, it is almost inevitable for ultrashort pulses. For instance, for the frequency-degenerate type-I PDC generated in a BBO crystal from a 400~nm pump, a 180~fs delay between the pump and PDC pulses emerges in each 1~mm of the crystal. However, at some other wavelength of the signal beam, its group velocity can be equal to that of the pump (Fig.~\ref{4_fig:group_vel}). The phase matching can then be fulfilled using noncollinear emission. Note that only one pulse, signal or idler, has to propagate together with the pump one, its twin will be amplified as well, similarly to the spatial walk-off matching case.

\begin{figure}[!htb]
\begin{center}
\includegraphics[width=1\textwidth]{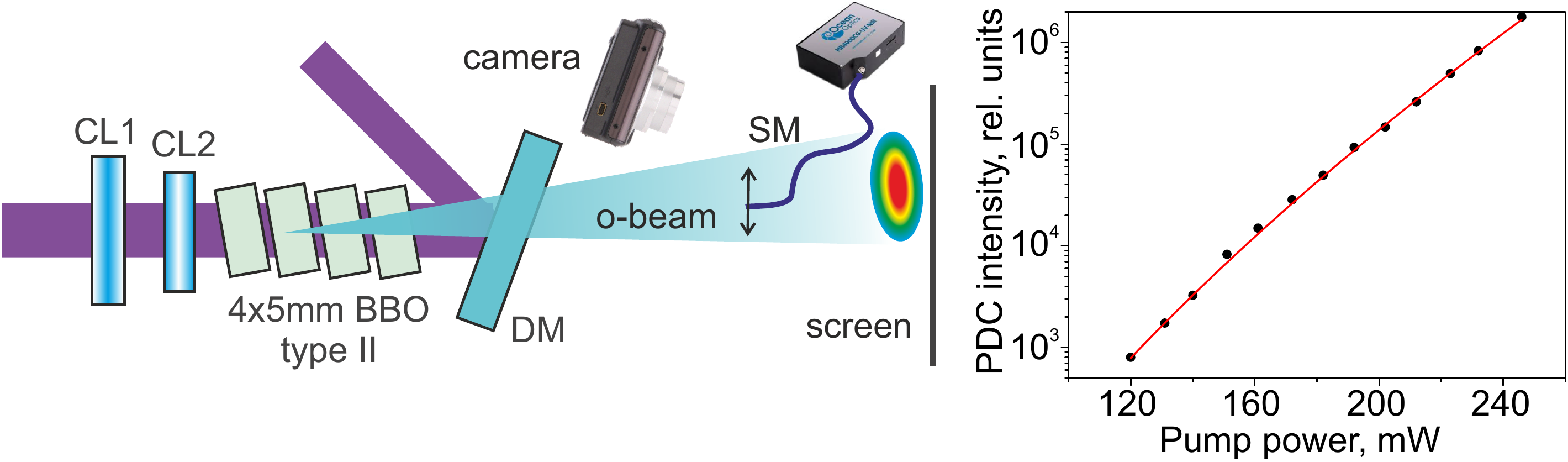}
\caption{Left: experimental setup for observing the effect of temporal walk-off matching (left). Right: the dependence of the PDC intensity on the pump power (points) and its fit with Eq.~\eqref{3_eq:gain_fit} (line).}
\label{4_fig:setup_temp_walk_off}
\end{center}
\end{figure}

The group velocity matching is realized for type-II PDC (Fig.~\ref{4_fig:setup_temp_walk_off}). The frequency-doubled radiation of the Ti-sapphire laser, which is the same as used in the experiment on PDC generation in the anomalous GVD range (section~\ref{3_sec:O_PDC}), is used as a pump. The pump has 400~nm wavelength, 1.4~ps pulse duration, and up to 700~mW mean power.

The length of the nonlinear medium is changed by using either one 5 mm BBO crystal or four ones stacked together with 3~mm spaces between them. In the last case each crystal is adjusted to the phase matching individually keeping the same direction of the optic axis for all crystals. Therefore, they work in the same way as a single 20~mm crystal.

The pump is focused into the crystals differently for both cases, with a 12:1 cylindrical telescope (CL1 and CL2) and with a 500~mm cylindrical lens, respectively. Accordingly, it leads to 680~$\mu$m and 130~$\mu$m width in the plane orthogonal to principal one. The tighter focusing is required in a shorter crystal, which, in its turn, leads to a Rayleigh length too small for the four-crystal stack. The pump beam is kept wide enough (3.4~mm) in the principal plane to minimize the effect of spatial walk-off discussed above.

After the crystals, the pump radiation is cut off by a dichroic mirror (DM) and the ordinary, signal, beam (o-beam) is observed on a screen placed at $76$~cm distance from the crystals. The signal beam is studied without overlapping with the idler one, because in type-II PDC the ordinary beam is shifted in angle w.r.t. to the extraordinary one.

The screen has a ruler pasted to it, so the external angles of emission can be determined from the coordinate on the screen. The example was already presented on the chapter's cover. This time the full PDC ring is not observed, because the apertures of the crystals restrict the beam in vertical direction. The spectra on the screen are captured by a photographic camera. 

Alternatively, the spectra is measured at different angles in the same way as in the experiment on PDC generation in the anomalous GVD range (section~\ref{3_sec:O_PDC}). A 400~$\mu$m multimode fiber connected to a spectrometer (Ocean Optics HR4000) is placed at 87~cm distance from the crystals and scanned in the direction orthogonal to the pump propagation. The parametric gain is measured in a similar way as in the experiment on spectral broadening (section~\ref{3_sec:broadening}), see Fig.~\ref{4_fig:setup_temp_walk_off} (right panel).

\subsection{Twin-beam generation using temporal walk-off matching}

\begin{figure}[!htb]
\begin{center}
\includegraphics[width=0.85\textwidth]{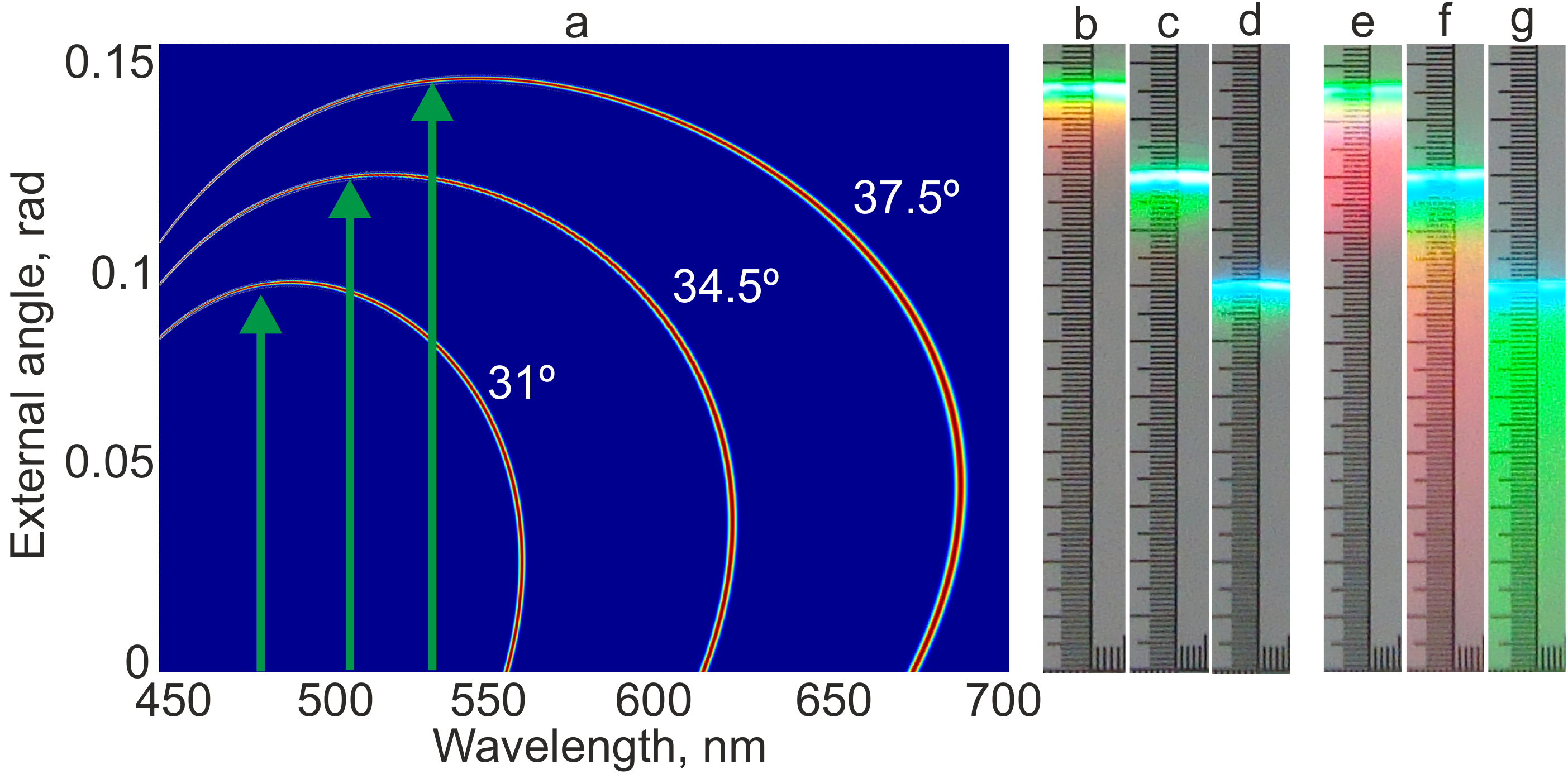}
\caption{The spectra $S_{q\Omega}(\theta_x, \lambda)$ for the ordinary beam shown for positive angles and $\phi=31^\circ$, $34.5^\circ$, and $37.5^\circ$ (a). Green arrows indicate the group-velocity-matching wavelengths. Snapshots of intensity distributions on the screen for a 20~mm (b-d) and a 5~mm (e-g) of BBO captured for the same orientations, $31^\circ$ (d,g), $34.5^\circ$ (c,f), and $37.5^\circ$ (b,e). }
\label{4_fig:temp_walk_off_camera}
\end{center}
\end{figure}

The spectra $S_{q\Omega}(\theta_x, \lambda)$ of the ordinary beam for positive angles are shown in Fig.~\ref{4_fig:temp_walk_off_camera}a for $\phi=31^\circ$, $34.5^\circ$, and $37.5^\circ$. Green arrows mark the angles corresponding to the wavelengths at which the group velocity of the signal beam matches that of the pump. One should expect a considerable amplification at these wavelengths in the case of long crystal and high gain.

Indeed, the snapshots of intensity distributions on the screen (Fig.~\ref{4_fig:temp_walk_off_camera}b-g) clearly demonstrate the effect. For twin beams generated in a 20~mm of BBO (panels b–d), there is huge amplification within a relatively narrow spectral range, while for 5~mm crystal (panels e–g) the spectrum is more uniform over frequency. The angle, at which amplification occurs, depends on the orientation; this clearly shows that it is not related to the spatial walk-off.

\begin{figure}[!htb]
\begin{center}
\includegraphics[width=0.55\textwidth]{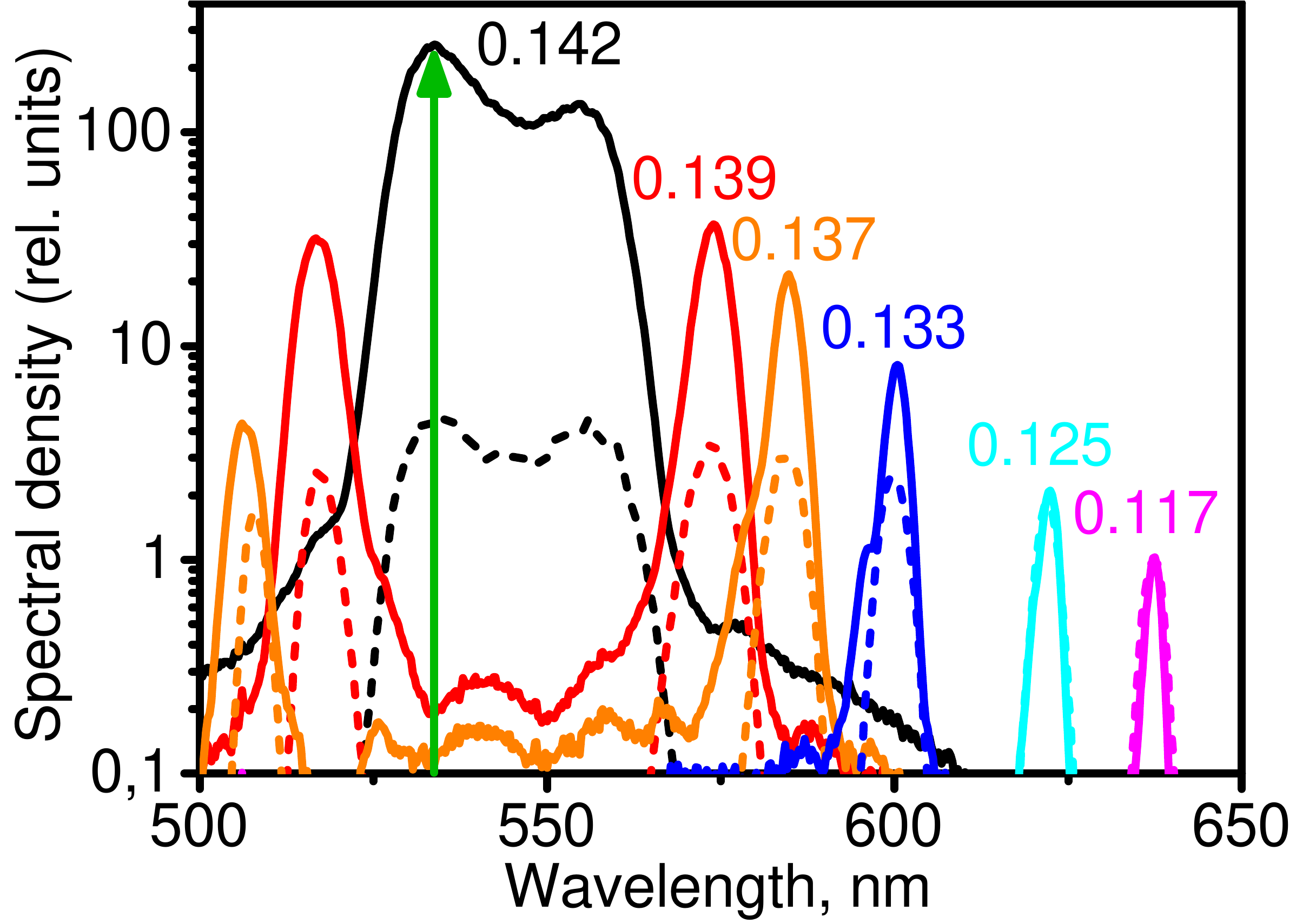}
\caption{Frequency spectra recorded at external angles 0.142, 0.139, 0.137, 0.133, 0.125, and 0.117~rad for a 5 mm (dashed line) and a 20 mm (solid line) of BBO oriented at $\phi=37.5^\circ$.}
\label{4_fig:temp_walk_off_spectr}
\end{center}
\end{figure}

For the quantitative characterization of the effect, the frequency spectra are recorded at different angles for the crystal orientation $\phi=37.5^\circ$ (Fig.~\ref{4_fig:temp_walk_off_spectr}). Each angle corresponds to a peak at a different wavelength, see Fig.~\ref{4_fig:temp_walk_off_camera}a. The measurements are performed for a 20~mm and a 5~mm of BBO choosing 637.5~nm wavelength as a reference one. At this wavelength the parametric gain is kept the same, $G = 8.6$, for both cases.

For twin beams generated in four crystals the peak at the group-velocity-matching wavelength (533.5~nm) exceeds the one at 637.5~nm for 250 times. Because this difference is influenced by many frequency-dependent factors~\cite{Klyshko1988}, the spectra are also recorded for one crystal. Although the temporal walk-off is not negligible in this case too, the spectral distribution is much more uniform.

\subsection{Possible applications of the spatial and temporal walk-off matching}

Although, the spatial walk-off matching is known and used in parametric oscillators~\cite{Gale1995}, we demonstrate the directionality of the twin beams achieved without using a cavity. Moreover, the wavelength of the enhanced emission is tunable within a broad range~\cite{Cavanna2016}. The possibility for building of such a OPG has been stated by David Klyshko long ago~\cite{Klyshko1988}: `Given powerful pulse pumping, PDC passes into a fairly intense parametric superluminescence\footnote{In other words, high-gain PDC.}, which provides a source of smoothly frequency-tunable short (up to $10^{-12}$~sec) light pulses'.

Furthermore, the idler beam is a source of tunable broadband IR radiation, which is especially convenient because its twin is maintained in the visible range.

The temporal walk-off matching could be especially useful for generating twin beams in cavities and whispering-gallery mode resonators~\cite{Furst2011}, where, due to the large quality factor, nonlinear interaction occurs on a very large distance. If the group velocity of down-converted radiation coincides with that of the pump, strong twin-beam generation can be realized even from short pulses.  Moreover, near resonances in medium the pump group velocity could differ from the signal or idler ones significantly, thereby providing additional possibilities for engineering the spectrum.

%%%%%%%%%%%%%%%%%%%%%%%%%%%%%%%%%%%%%%%%%%%%%%%%%%%%%
\chapter{High-gain PDC as a pump for multiphoton effects}
\label{5_chapter}

\begin{center}\textit{Money in a bank account is something important, but certainly not physical.\\As such it can take any value without necessitating the expenditure of energy.\\It is just a number!}\\---Nassim Taleb, \textit{The Black Swan}
\end{center}

BSV has an interesting statistics and serves as a useful source for multiphoton effects. Both aspects are discussed in detail in this chapter. Firstly, the fluctuations of light and their measurement are considered, in particular, how they change in the multimode case and after the spectral filtering.

Then, it is shown that BSV has very strong photon-number fluctuations and, namely, there are the measured CFs and probability distributions presented. This makes BSV useful for pumping nonlinear effects, because their efficiency increases dramatically, what is demonstrated using the generation of the second (SH), third (TH), and fourth (FH) harmonics as an example.

Finally, it is shown that the generated harmonics fluctuate even more than BSV and have unusual statistics. They have a lot of extreme events and are described by heavy-tailed photon-number distributions. Furthermore, if supercontinuum is generated from BSV, even more exceptional photon-number distributions appear. We manage to observe power-law (Pareto) probability distributions with power exponents less than 2 leading to indefinite mean values and higher-order moments.

%%%%%%%%%%%%%%%%%%%%%%%%%%%%%%%%%%%%%%%%%%%%%%%
\section{Introduction}

Although the fluctuations of light have been previously discussed a bit, they still are to be analyzed deeper.

\subsection{Fluctuations of light: coherent, thermal, and superbunched}\label{5_sub:fluctuations}

The fluctuations of photon number is a stochastic process, and therefore they can be described using different statistical characteristics. Although moments and CFs were mostly used in the previous chapters, probability distributions provide the most complete information.

They can be obtained by applying different approaches as, e.g., the quantum mechanical approach, the most rigorous one, used before. Unfortunately, in many cases it is quite complicated, and therefore in these cases the semi-classical or even classical approach becomes more preferable as long as it gives reasonable outcomes. Let us explain here the difference between approaches and justify their applicability.

In the semi-classical approach the electric field is considered classically, whereas the detection process as a quantum. Using such a treatment and, namely, by describing the probability of photoelectric emission in a detector exposed by fluctuating electric field, one obtains the Mandel formula \cite{Mandel1995,Klyshko2011},
\begin{equation}
p(m)=\frac{1}{m!}\int_0^\infty N^me^{-N}P(N)dN,
\label{5_eq:Mandel}
\end{equation}
which links the semi-classical approach with the classical one. Here $P(N)$ is the photon-number distribution\footnote{Rigorously it is an \textit{intensity} distribution, since `photons' do not appear in the classical picture. Nevertheless, it is called \textit{photon-number} distribution to be consistent in the thesis.} in the classical approach, $p(m)$, in the semi-classical one, and $N, m$ are the corresponding photon numbers.

Using Eq.\eqref{5_eq:Mandel} the mean values and the variances are related as
\begin{equation}
\langle m\rangle = \langle N\rangle, \qquad \langle\Delta m^2\rangle = \langle N\rangle + \langle\Delta N^2\rangle.
\end{equation}
The main difference is caused by shot-noise fluctuations, which arise in the semi-classical case and do not appear in the classical one. Of course, the quantum mechanical approach provides them too, moreover, they can be suppressed as well, which does not happen in semi-classical description, $\langle\Delta m^2\rangle\ge\langle N\rangle$.

This difference can be exemplified by coherent light. In the semi-classical approach $p(m)$ has the Poissonian distribution~\cite{Klyshko2011},
\begin{equation}
p_{coh}(m)=\frac{1}{m!}\mean{N}^me^{-\mean{N}},
\end{equation}
and $\langle\Delta m^2\rangle=\mean{N}$, while in the classical case, the Dirac delta one, $P_{coh}(N)=\delta(N-\mean{N})$, and $\langle\Delta N^2\rangle=0$. The shot-noise contribution is essential here, because it leads to a nonzero variance.

Yet, the situation is different for more fluctuating light. For example, chaotic or thermal light has the Bose-Einstein 
photon-number distribution~\cite{Klyshko2011},
\begin{equation}
p_{th}(m) = \frac{\langle N\rangle^m}{(\langle N\rangle+1)^{m+1}},
\end{equation}
in the semi-classical approach and a negative-exponential distribution,
\begin{equation}
P_{th}(N) = \frac{1}{\langle N\rangle}e^{-\frac{N}{\langle N\rangle}},
\label{5_eq:P_th}
\end{equation}
in the classical one. The variances are $\langle\Delta m^2\rangle=\mean{N}^2+\mean{N}$ and $\langle\Delta N^2\rangle=\mean{N}^2$, respectively. They are equal at $\mean{N}\gg1$, in this case the classical approach provides a correct result. This is the case in general: if $\langle\Delta N^2\rangle\gg\langle N\rangle$, the classical approach (often) can be used.

The CFs $g^{(n)}$ are defined differently in all three approaches~\cite{Klyshko1996}. In the classical one, it is done through moments,
\begin{equation}
g^{(n)} \equiv \frac{\mean{N^n}}{\mean{N}^n},
\end{equation}
in the semi-classical approach via factorial moments,
\begin{equation}
g^{(n)} \equiv \frac{\mean{m(m-1)\dots(m-n+1)}}{\mean{m}^n},
\end{equation}
and in the quantum approach with normally-ordered operators,   
\begin{equation}
g^{(n)} \equiv \frac{\mean{:\op{N}^n:}}{\mean{\op{N}}^n}.
\end{equation}  
The CFs defined this way take the same values for the same state of light in all three approaches as long as the distributions $P(N)$ and $p(m)$ exist at all.

Following the definitions, the $n$-th order CF $g_{coh}^{(n)}=1$ for coherent light and $g_{th}^{(n)}=n!$ for thermal one. In particular, for the latter $g_{th}^{(2)}=2$, which is larger than for coherent light. In this case the photons tend to bunch together, therefore thermal light is referred to as a `bunched' one.

Thermal light is produced by various effects and sources, which can be either `true' chaotic sources, like stars~\cite{Hanbury1956} and light emitting diodes~\cite{Jechow2013}, or their analogues, like multimode lasers~\cite{Lecompte1975} and rotating ground glass discs~\cite{Ferri2005}.

Particularly noteworthy that for BSV generated in the twin-beam configuration, each beam, signal or idler, is in a true thermal state~\cite{Tapster1998}. In order to observe this thermal state, the two beams should be distinguishable, namely some detector\footnote{The detector is meant in a broad sense. For example, the generation of optical harmonics can be also considered as such a `detector'.} should distinguish them. 

For degenerate BSV, i.e. in the case of indistinguishable signal and idler beams, the photon-number distribution is different. It contains only even numbers of photons~\cite{Klyshko1996,Iskhakov2012},
\begin{equation}\label{5_eq:p_m_sb}
p_{sb}(2m) = \frac{(2m)!}{2^{2m}(m!)^2}\frac{\langle N\rangle^m}{(\langle N\rangle+1)^{m+1/2}},\qquad p_{sb}(2m+1)=0,
\end{equation}
and its variance is twice larger than for thermal light, $\langle\Delta m^2\rangle=2\mean{N}^2+2\mean{N}$. For this strongly nonclassical state, see Eq.~\eqref{2_eq:BSV_state}, the quantum mechanical approach should be used.

However, if one neglects the fine structure\footnote{For example, if the detector has no photon-number resolution.} of distribution~\eqref{5_eq:p_m_sb}, at $\mean{N}\gg1$ its envelope is well-described by the gamma distribution~\cite{Hogg1978} with the shape parameter $1/2$ and the scale parameter $1/(2\langle N\rangle)$~\cite{Akhmanov1981, Leuchs2015},
\begin{equation}
P_{sb}(N) = \frac{1}{\sqrt{2\pi\langle N\rangle N}}e^{-\frac{N}{2\langle N\rangle}}.
\label{5_eq:P_sb}
\end{equation}
The variance $\langle\Delta N^2\rangle=2\mean{N}^2$ and the CF $g_{sb}^{(n)}=(2n-1)!!$ provided by this distribution are justified for large $\mean{N}$. The CFs are higher than for thermal light; for this reason the degenerate BSV is often called `superbunched', `extrabunched', or `superthermal' light~\cite{Boitier2011,Iskhakov2012,Zhou2017,Alves2019}.

\begin{figure}[!htb]
\begin{center}
\includegraphics[width=0.45\textwidth]{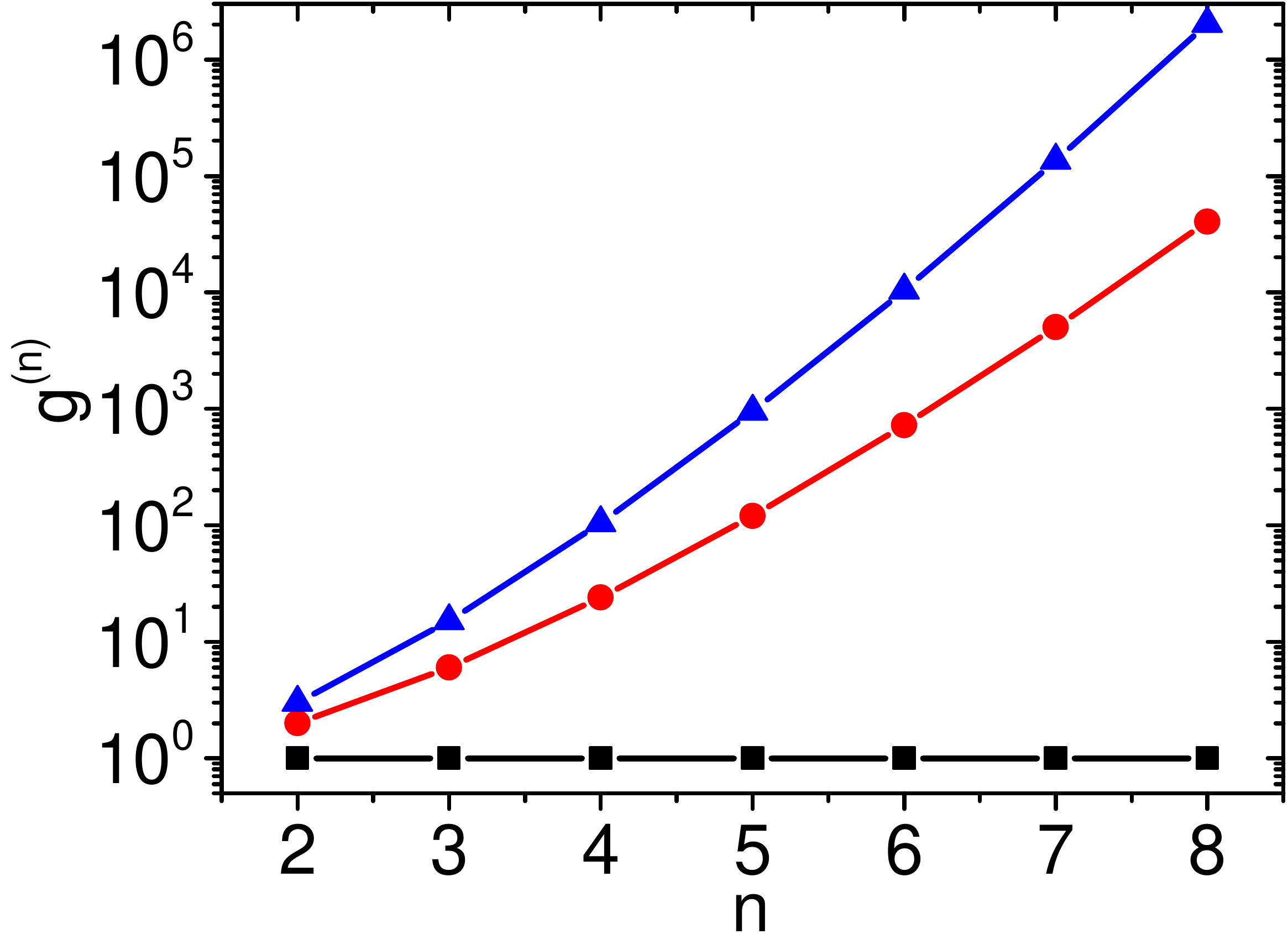}
\caption{The CF $g^{(n)}$ on the order $n$ for coherent (black) and thermal light (red) and superbunched BSV (blue).}
\label{5_fig:g_n}
\end{center}
\end{figure}

To compare the CFs for coherent and thermal light and superbunched BSV the dependence of $g^{(n)}$ on the order $n$ is shown in Fig.~\ref{5_fig:g_n}. The difference increases fast with $n$; due to this fact, the fluctuating light is useful for multiphoton processes, an example will be shown in section~\ref{5_sec:stat_enhancement}.

The coherent light can be also described by means of the classical approach, but this description is not strict. In this case the Dirac delta distribtion should be replaced by the Gaussian one,
\begin{equation}
P(N)=\frac{1}{\sigma\sqrt{2\pi}}e^{-\frac{(N-\mean{N})^2}{2\sigma^2}},
\label{5_eq:P_gauss}
\end{equation}
following the fact that the Poissonian distribution tends to the Gaussian one for large $\mean{N}$. The variance $\langle\Delta N^2\rangle=\sigma^2$ should be set `by hand' to the shot-noise contribution $\mean{N}$.

\subsection{Multimode light and its measurement}
\label{5_sub:multimode}

The distributions~\eqref{5_eq:P_th} and \eqref{5_eq:P_sb} and the expressions for CFs $g_{th}^{(n)}$ and $g_{sb}^{(n)}$ are obtained for single-mode light. If multimode light is measured they will be valid only provided that the detector traces all photon-number fluctuations and this will be the case only under the condition of spatially and temporary single-mode detection~\cite{Ivanova2006,Klyshko2011}. The latter means that the detection area\footnote{In the case of a multipixel detector, e.g. a CCD camera, each pixel is a separate detector.} and time are (much) smaller than the coherence area and time, namely smaller than the spatial and temporal size of the detected mode. Otherwise several modes are detected and both the probability distributions and the CFs are different from the single-mode ones.

In the case of $M$-mode detection, the probability distributions change. For independent modes the multimode distributions can be obtained using the probability theory and the fact that the probability distribution of a sum of two or more independent variables is given by the convolution of their probability distributions~\cite{Hogg1978}. 

Therefore, for $M$ equally-populated modes the probability distribution~\eqref{5_eq:P_th} for thermal light changes to~\cite{Mandel1995,Allevi2015}
\begin{equation}\label{5_eq:P_th_M}
P_{th,M}(N)=\frac{N^{M-1}}{(M-1)!}\left(\frac{M}{\langle N\rangle}\right)^{M}e^{-\frac{MN}{\langle N\rangle}},
\end{equation}
and the distribution~\eqref{5_eq:P_sb} for superbunched BSV becomes
\begin{equation}\label{5_eq:P_sb_M}
P_{sb,M}(N)=\frac{N^{M/2-1}}{\Gamma\left(\frac{M}{2}\right)}\left(\frac{M}{2\langle N\rangle}\right)^{M/2}e^{-\frac{MN}{2\langle N\rangle}},
\end{equation}
where $\Gamma(x)$ is the gamma function. Both distributions approach the Gaussian one~\eqref{5_eq:P_gauss} for a large $M$.

At the same time, the CFs tend to unity for large $M$. In particular, the second-order CF for $M$-mode light is~\cite{Ivanova2006,Klyshko2011} 
\begin{equation}\label{5_eq:g2_multimode}
g_M^{(2)}=1+\frac{g_1^{(2)}-1}{M},
\end{equation}
where $g_1^{(2)}$ is the single-mode CF.

The last expression provides a possibility to estimate the total number $M$ of detected modes from the $g^{(2)}$ measurement if the single-mode statistics is known; exactly this method is applied in all chapters of this thesis.

\subsection{Filtering of the multimode light}
\label{5_sub:filtering}

The condition of single-mode detection can be fulfilled for multimode radiation not only by means of a small and fast detector, but also through Fourier filtering. Such filtering can be considered in two different ways: either as a method to shape the light itself or as a part of detection process. In the first picture, the light is considered to be strongly filtered to a single mode and then detected. In the second one, the filtering `stretches' the time and area of intensity fluctuations, so that a big and slow detector can resolve them.

Although the first picture is quite obvious, the second one may cause objections. Below it is shown that the filtering does not change the normalized intensity moments for thermal light and superbunched BSV, which are further considered experimentally. Only the temporal domain is considered, the spatial one can be taken into account similarly.

Indeed, the normalized CF describing intensity correlations is~\cite{Klyshko2011}
\begin{equation}
g^{(n)}(t_1,\dots,t_n) \equiv \frac{G^{(n)}(t_1,\dots,t_n)}{\prod_{i=1}^nG^{(1)}(t_i)},
\end{equation} 
where
\begin{equation}
G^{(n)}(t_1,\dots,t_n) \equiv \langle \prod_{i=1}^n E^{(-)}(t_i)E^{(+)}(t_i)\rangle.
\end{equation}
Here $E^{(+)}(t)=[E^{(-)}(t)]^{*}$ is the analytical signal, defined as
\begin{equation}
E^{(+)}(t) \equiv \int_0^{\infty}d\omega\,e^{-i\omega t}E(\omega),
\label{5_eq:an_signal}
\end{equation}
and $E(\omega)$ is the FT of the real electric field $E(t)$.

The field for thermal light and BSV can be described as a Gaussian random process with zero mean value~\cite{Akhmanov1981}. The higher-order CFs can be therefore represented as combinations of the first-order ones~\cite{Akhmanov1981,Mandel1995,Klyshko2011}. For example, 
\begin{equation}
G^{(2)}(t,t+\tau)=G^{(1)}(t)G^{(1)}(t+\tau)+|\mathbb{G}(t,t+\tau)|^2+|\langle E^{(+)}(t)E^{(+)}(t+\tau)\rangle|^2,
\label{5_eq:G2_t_tau}
\end{equation}
where $\mathbb{G}(t,t+\tau) \equiv \langle E^{(-)}(t)E^{(+)}(t+\tau)\rangle$ is the first-order CF of the electric field. It is related with the spectrum $|E(\omega)|^2$ via Wiener-Khinchin theorem~\cite{Akhmanov1981,Mandel1995}, namely via inverse FT, and reaches the maximum value at $\tau=0$.

Keeping only the dependence on the time difference (stationarity), $g^{(2)}(\tau)\equiv g^{(2)}(t,t+\tau)$, one obtains
\begin{equation}
g^{(2)}(\tau) = 1+|\gamma(\tau)|^2+\frac{|\langle E^{(+)}(t)E^{(+)}(t+\tau)\rangle|^2}{[G^{(1)}(t)]^2},
\label{5_eq:g2_tau}
\end{equation}
where $\gamma(\tau) \equiv \mathbb{G}(t,t+\tau)/G^{(1)}(t)$.

As far as thermal light is considered, only the first two terms in Eq.~\eqref{5_eq:g2_tau} are nonzero; one obtains the Siegert relation~\cite{Lemieux1999,Klyshko2011}. Although the filtering changes the spectrum and therefore changes the correlation time and the $\gamma(\tau)$ shape, $\gamma(0)$ is always unity.

The generalized Siegert relation is more complicated, but the tendency is the same. So, the CFs $g^{(3)}(\tau_1,\tau_2)\equiv g^{(3)}(t,t+\tau_1,t+\tau_2)$ and  $g^{(4)}(\tau_1,\tau_2,\tau_3)\equiv g^{(4)}(t,t+\tau_1,t+\tau_2,t+\tau_3)$ depend only on the combination of $\gamma(\tau_j-\tau_i)$ ($i,j=0,1,2$ or $0,1,2,3$ with $\tau_0=0$)~\cite{Lemieux1999}. Therefore, the CFs $g^{(3)}(0,0)$ and $g^{(4)}(0,0,0)$ do not depend on filtering.

The same can be shown also for superbunched BSV. In this case the last term in Eq.~\eqref{5_eq:g2_tau} can be nonzero as well~\cite{Klyshko1995}. In the simplest case of $\delta$-correlated signal and idler beams, see section~\ref{3_sec:intro}, detuned from the central frequencies by $\Omega_s$ and $\Omega_i$~\cite{Klyshko1988} one gets
\begin{equation}
\langle E(\Omega_s)E(\Omega_i)\rangle=\delta(\Omega_s+\Omega_i)\langle E(\Omega_s)E(\Omega_i)\rangle.
\end{equation}
Thus, using Eq.~\eqref{5_eq:an_signal} the last term in Eq.~\eqref{5_eq:G2_t_tau} at $\tau=0$ is
\begin{equation}
\bigl|\langle[E^{(+)}(t)]^2\rangle\bigr|^2 = \left|\int_{-\Omega_{\max}}^{\,\Omega_{\max}}d\Omega_s\langle E(\Omega_s)E(-\Omega_s)\rangle\right|^2,
\end{equation}
where $2\Omega_{\max}$ is the spectral width determined by the filtering. This filtering should be around the central BSV frequency, $\omega_{s0}=\omega_{i0}=\omega_p/2$.

Signal and idler fields in BSV have the same amplitude and opposite phases\footnote{If one assumes that the pump phase is equal to zero, see section~\ref{4_sec:interf_on_BS} for more details.}, $E(-\Omega_s)=[E(\Omega_s)]^*$; thus,
\begin{equation}
\bigl|\langle[E^{(+)}(t)]^2\rangle\bigr|^2 = \left(\int_{-\Omega_{\max}}^{\,\Omega_{\max}}d\Omega_s\langle|E(\Omega_s)|^2\rangle\right)^2.
\end{equation}
By virtue of the Wiener-Khinchin theorem~\cite{Akhmanov1981,Mandel1995}, the last expression equals exactly\footnote{Note that the exponent in the Wiener-Khinchin theorem disappears, because the delay is zero, $\tau=0$.} to $[\mathbb{G}(t,t)]^2=[G^{(1)}(t)]^2$. Therefore, $g^{(2)}(0)=3$, regardless of the filtering.

%%%%%%%%%%%%%%%%%%%%%%%%%%%%%%%%%%%%%%%%%%%%%%%%%%%%%%%%%%%
\section{Measurement of correlation functions and probability distributions}

Here it is demonstrated that BSV has high photon-number fluctuations and, namely, the measurement of probability distributions and CFs for thermal and superbunched BSV is presented.

\subsection{Experimental setup}
\label{5_sub:setup_CF}

\begin{figure}[!htb]
\begin{center}
\includegraphics[width=0.6\textwidth]{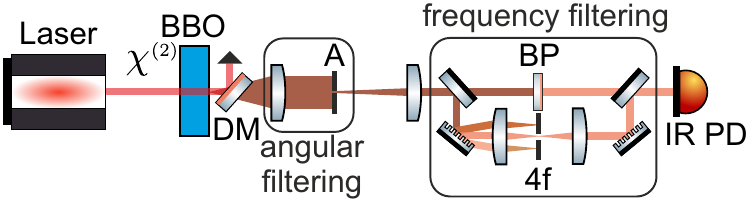}
\caption{Experimental setup for measuring the statistics of thermal and superbunched BSV.}
\label{5_fig:setup_BSV_stat}
\end{center}
\end{figure}

The experimental setup is shown in Fig.~\ref{5_fig:setup_BSV_stat}. BSV is produced in the same BBO crystal and with the same Ti-sapphire pump laser as in the experiment on PDC generation in the anomalous GVD range (section~\ref{3_sec:O_PDC}). The collinear frequency-degenerate configuration is used, see Fig.~\ref{3_fig:spectra_O_PDC}a. This time the pump is brighter, 2.5~W mean power, and is cut off with dichroic mirrors (DM) and a longpass filter. With such pumping the gain $G=15.3\pm0.5$ is reached\footnote{For the crystal orientation that maintains the BSV fluctuations, see below Fig.~\ref{5_fig:g2_crystal_power}.}.

The CFs and probability distributions are measured using a PIN diode-based IR charge-integrating detector (IR PD) with neither spatial nor temporal resolution within a pulse, therefore, as discussed in the previous section, filtering is needed in order to trace intensity fluctuations. Angular filtering is made with an aperture (A) in the focal plane of a 200~mm lens and frequency filtering, either with bandpass filters (BP) or with a $4f$-monochromator (4f) containing two diffraction gratings, two lenses, and a slit. After each filtering the beam is collimated.

The detector has the same design as the one used in the visible range, see section~\ref{4_sub:setup_BS_interf}, but is based on an InGaAs photodiode (Hamamatsu G12180-020A). It has a dynamic range from few thousands to several millions of photons per pulse, a quantum efficiency of 85\% at 1600 nm, and an electronic noise equivalent to $1600$~photons per pulse. Similarly to the visible one, it produces electronic pulses with the area $S=A_{IR}N$, where $N$ is the input number of photons and $A_{IR}=5.0\pm0.4$~pV$\times$s is the calibration coefficient at 1600~nm. 

This coefficient is obtained from the calibration measurement, which ideally should be done with a low-noise laser. Such a laser at 1600~nm was not available, therefore for calibration a radiation at 1064~nm of the first harmonic of the Nd:YAG laser is used, which has been already used in the experiment on spectral broadening (section~\ref{3_sec:broadening}). The pulse energy is measured with a power meter\footnote{The measured power should be divided by a repetition rate.} and then attenuated in a controlled way using neutral density filters and two polarizers. The pulse energy is converted thereafter to the number of photons per pulse. Therefore, each signal from the detector corresponds to a certain photon number.

\begin{figure}[!htb]
\begin{center}
\includegraphics[width=1\textwidth]{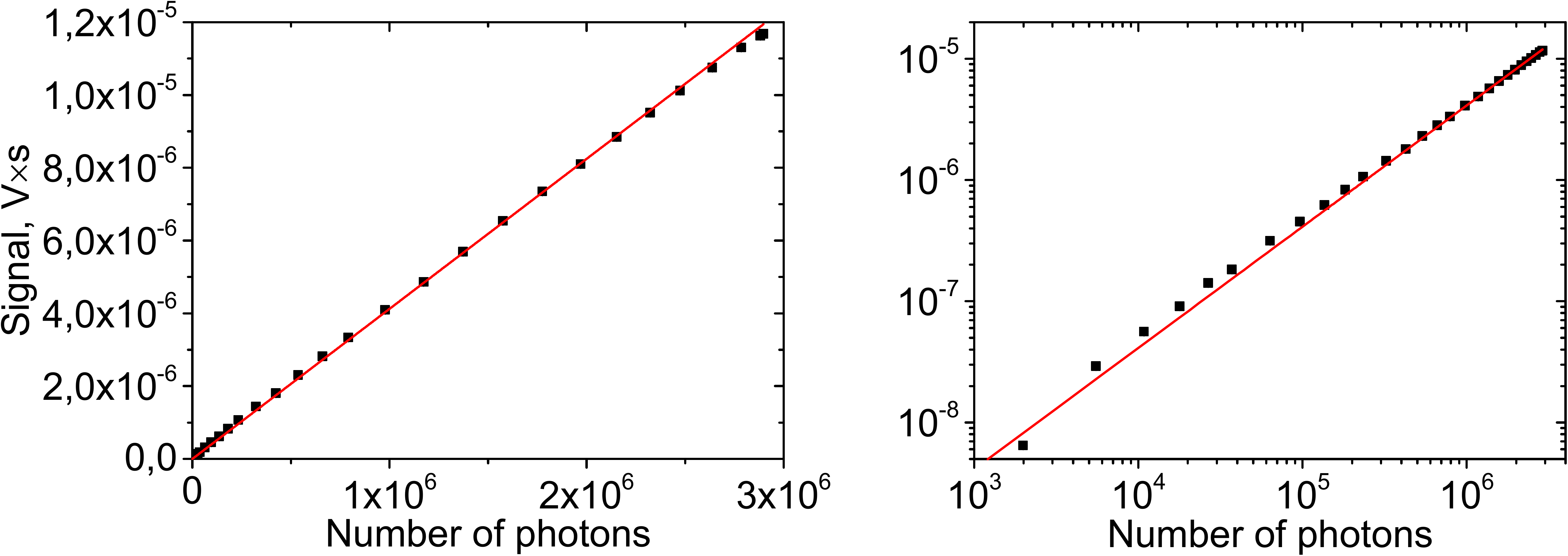}
\caption{The dependence of the signal from IR detector on the input number of photons at 1064~nm (points) and its linear fit (line) in linear (left) and log-log (right) scales.}
\label{5_fig:IR_det_calib}
\end{center}
\end{figure}

\begin{figure}[!htb]
\begin{center}
\includegraphics[width=0.45\textwidth]{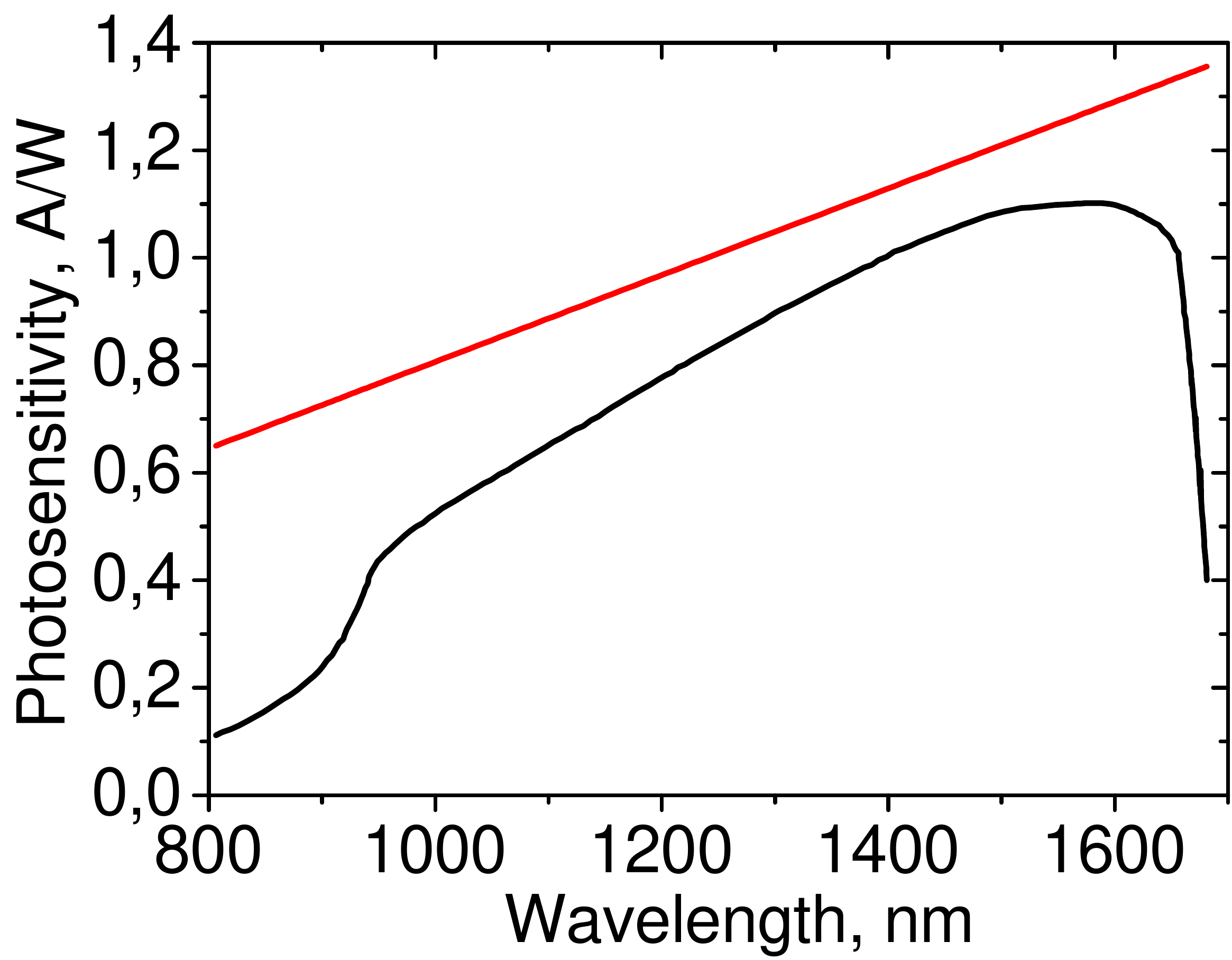}
\caption{Photosensitivity in A/W for Hamamatsu G12180-020A photodiode (black) and the dependence corresponding to a 100\% quantum efficiency (red).}
\label{5_fig:IR_det_photosens}
\end{center}
\end{figure}

Indeed, the detector shows nearly linear response on the input signal (Fig.~\ref{5_fig:IR_det_calib}). From the linear fit of the dependence one gets the calibration coefficient at wavelength 1064~nm, for different wavelengths the detector's response is different, and, therefore, the coefficient is recalculated using the photosensitivity dependence provided by the manufacturer (Fig.~\ref{5_fig:IR_det_photosens}).

The photosensitivity is usually provided in amperes per watt; however, these units are not so convenient to use, because we are more interested in a quantum efficiency, which shows with what probability an incoming photon produces an electron. When working with such dependencies the author notices one helpful rule of thumb: a 100\% quantum efficiency in A/W equal to an inverse energy of incoming photon in eV.

\subsection{Thermal and superbunched BSV: correlation functions}
\label{5_sub:CFs_exp}

\begin{figure}[!htb]
\begin{center}
\includegraphics[width=1.0\textwidth]{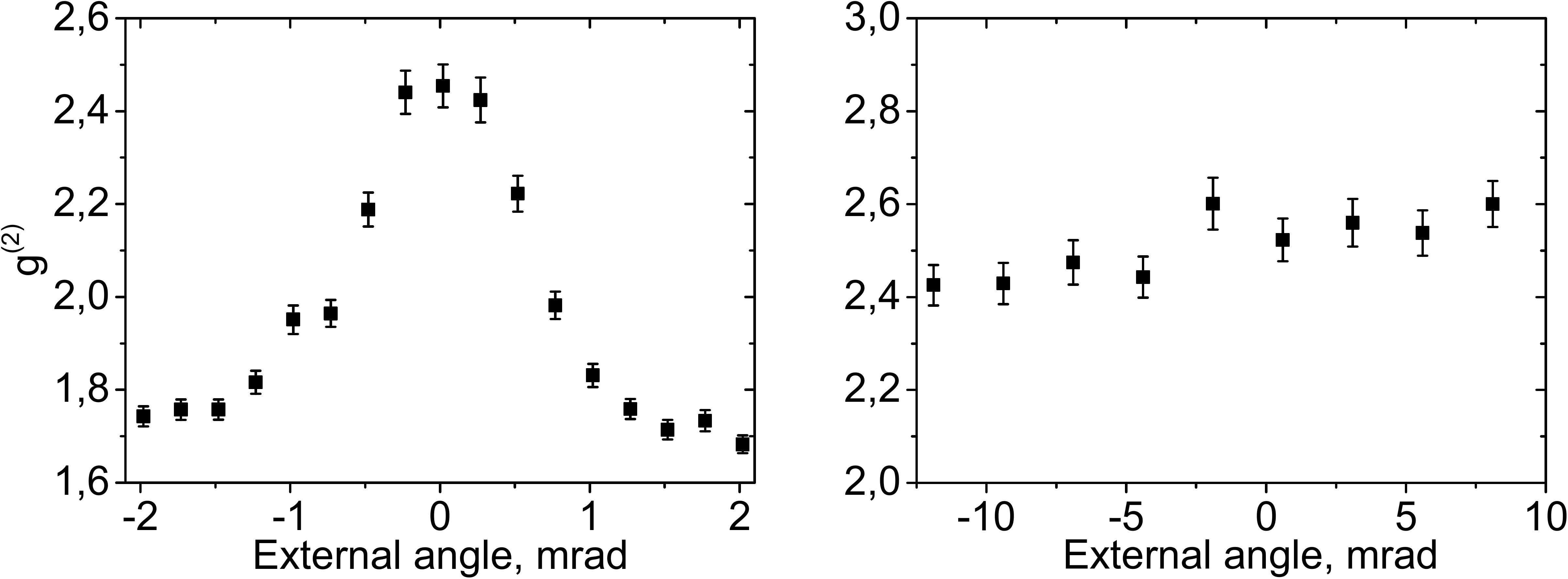}
\caption{The CF $g^{(2)}$ vs. the external angle in the horizontal (left) and vertical (right) directions.}
\label{5_fig:g2_ang_scan}
\end{center}
\end{figure}

The CF $g^{(2)}$ angular dependence is measured by scanning an aperture in the horizontal and vertical directions with an angular resolution of 0.5 mrad (Fig.~\ref{5_fig:g2_ang_scan}). The scanning is performed with 7~nm filtering around 1600~nm, which is done by using two 10~nm bandpass filters with a Gaussian transmission profile placed one after another; the transmission band in this case is given by the convolution of transmission profiles and, therefore, it decreases by 30\%.

Indeed, the CF $g^{(2)}$ reveals the fluctuations: its value is far above unity. It manifests a peak with 1.3~mrad width in the horizontal direction (left panel) and remains constant along all spectrum in the vertical one (right). This is largely due to the pump focusing and the restricted angular spectrum.

The angular dependence of CF $g^{(2)}$ shows that the BSV is single-mode in the vertical direction and multimode in the horizontal one. In this experimental configuration the superbunching, $g^{(2)}>2$, could be observed only for the collinear selection of modes, because both signal and idler photons should be detected at the same time. If the superbunching is observed for all possible angles, it means that the BSV is single-mode in this direction. 

In both cases the value $g^{(2)}=3$ is not achieved due to insufficient frequency filtering. The same problem we had quite a while ago and, therefore, we got similar result with the CF $g^{(2)}=2.4-2.5$ at the maximum~\cite{Iskhakov2012}.

\begin{figure}[!htb]
\begin{center}
\includegraphics[width=1\textwidth]{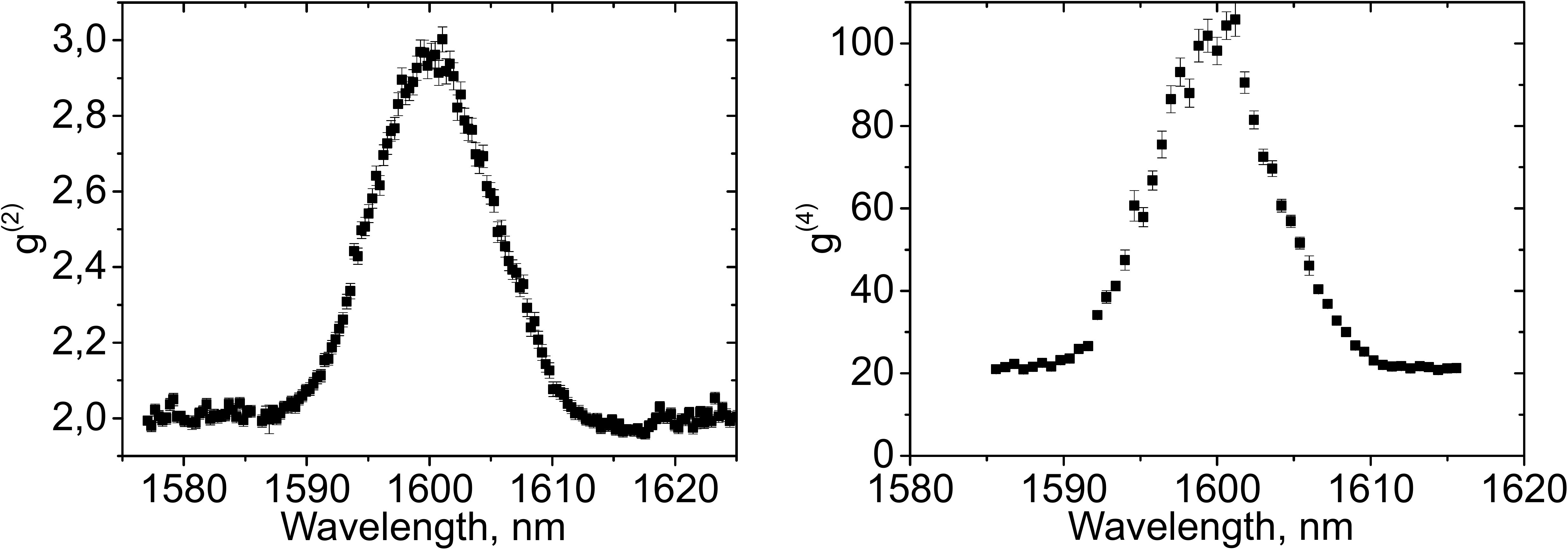}
\caption{The second-order $g^{(2)}$ (left) and fourth-order $g^{(4)}$ (right) CFs vs. the wavelength.}
\label{5_fig:g2_g4_wave_scan}
\end{center}
\end{figure}

In order to improve the measurement a $4f$-monochromator with a 150~$\mu$m slit, providing a 2.3~nm resolution, is implemented (Fig.~\ref{5_fig:g2_g4_wave_scan}, left). In this case the peak reveals the expected statistics: for degenerate BSV the CF $g^{(2)}=2.96\pm0.05$, for nondegenerate one, $2.01\pm0.03$. For the other CFs the situations is the same, one gets the peaks with similar widths; the CF $g^{(4)}$ is shown as an example (right panel). It is equal to $102\pm6$ and $21.6\pm0.9$ for degenerate and nondegenerate BSV, respectively, meeting well the theoretical expectations.

However, the BSV is so bright that it experiences further SH generation (SHG) in the BBO crystal, $1440-1780$~nm~$\to720-890$~nm. This process acts on the BSV as a nonlinear absorber, it reduces fluctuations by absorbing more photons from higher energy bursts~\cite{KRASINSKI1976, Leuchs1986}.

\begin{figure}[!htb]
\begin{center}
\includegraphics[width=0.9\textwidth]{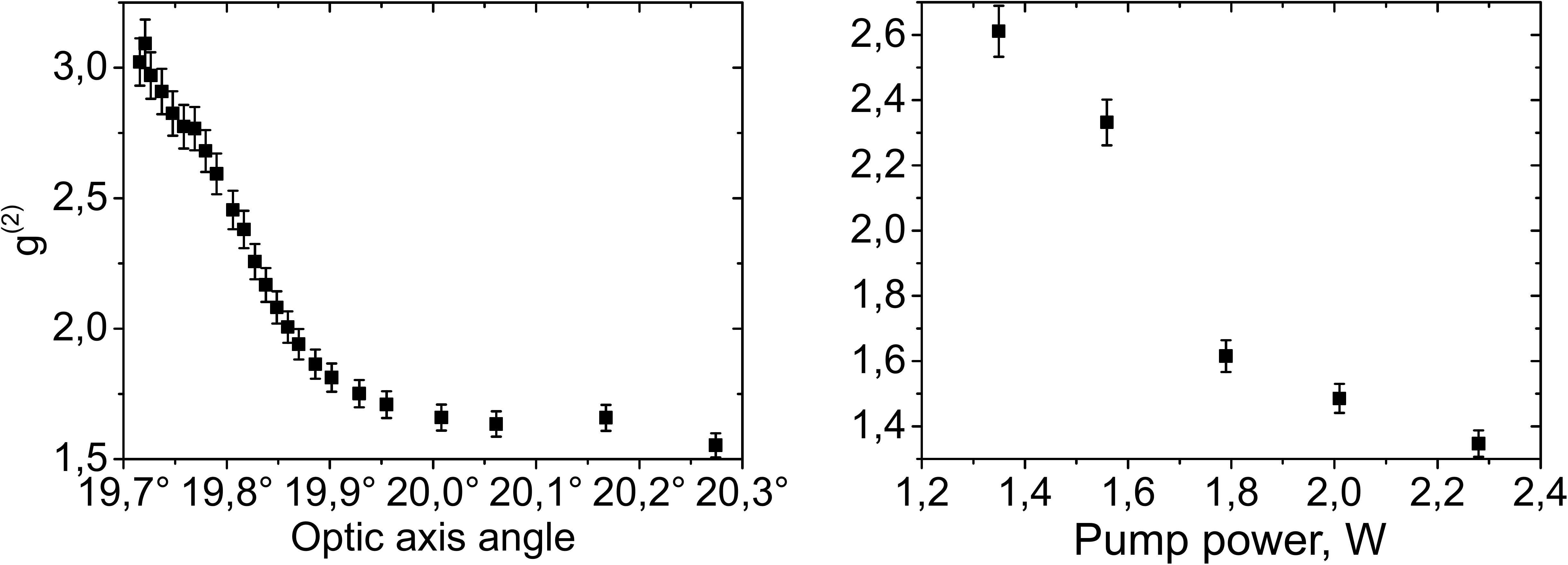}
\caption{The CF $g^{(2)}$ for the collinear degenerate BSV depending on the crystal orientation $\phi$ (left) and the pump power (right).}
\label{5_fig:g2_crystal_power}
\end{center}
\end{figure}

The contribution of this process can be tuned by tilting the crystal closer or further from the exact phase-matching position, because the SHG efficiency depends on the phase mismatch. The CF $g^{(2)}$ is measured depending on the crystal orientation $\phi$ (Fig.~\ref{5_fig:g2_crystal_power}, left). With the crystal oriented at the angle $\phi$ smaller than required for the exact phase matching, $19.87^\circ$, strong photon-number fluctuations are maintained and the CF $g^{(2)}$ reaches its theoretical value, while for the larger $\phi$ the CF keeps decreasing. This occurs because at the smaller angles there is no phase matching whereas at the larger ones, it exists, see section~\ref{3_sec:O_PDC}. This effect will be used to vary the BSV statistics in section~\ref{5_sec:stat_enhancement}.

For the same reason, the CF $g^{(2)}$ also depends on the pump power (Fig.~\ref{5_fig:g2_crystal_power}, right), because the brighter is the BSV, the more efficient is the unwanted SHG. This dependence is obtained without sufficient frequency filtering, therefore the CF $g^{(2)}$ does not reach the value 3 even at low powers.

\subsection{Thermal and superbunched BSV: probability distributions}
\label{5_sub:BSV_prob}

\begin{figure}[!htb]
\begin{center}
\includegraphics[width=0.7\textwidth]{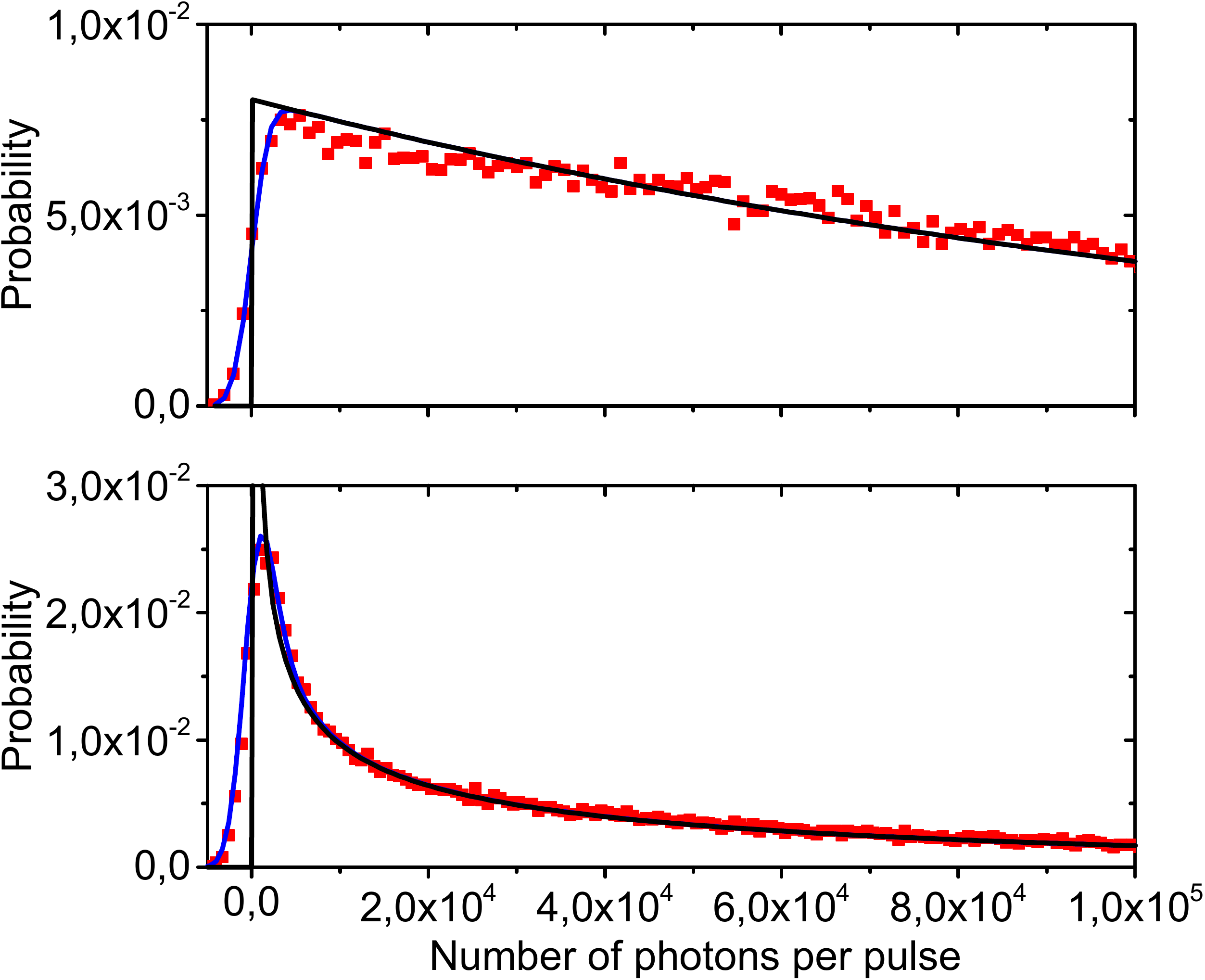}
\caption{Experimental photon-number histograms (red points) for the thermal (top) and superbunched (bottom) BSV. Theoretical distributions [\eqref{5_eq:P_th} and \eqref{5_eq:P_sb}] are shown by black lines. Blue lines show the convolution of theoretical distributions with the noise distribution~\eqref{5_eq:P_gauss}}
\label{5_fig:P_th_sb_linear}
\end{center}
\end{figure}

\begin{figure}[!htb]
\begin{center}
\includegraphics[width=0.65\textwidth]{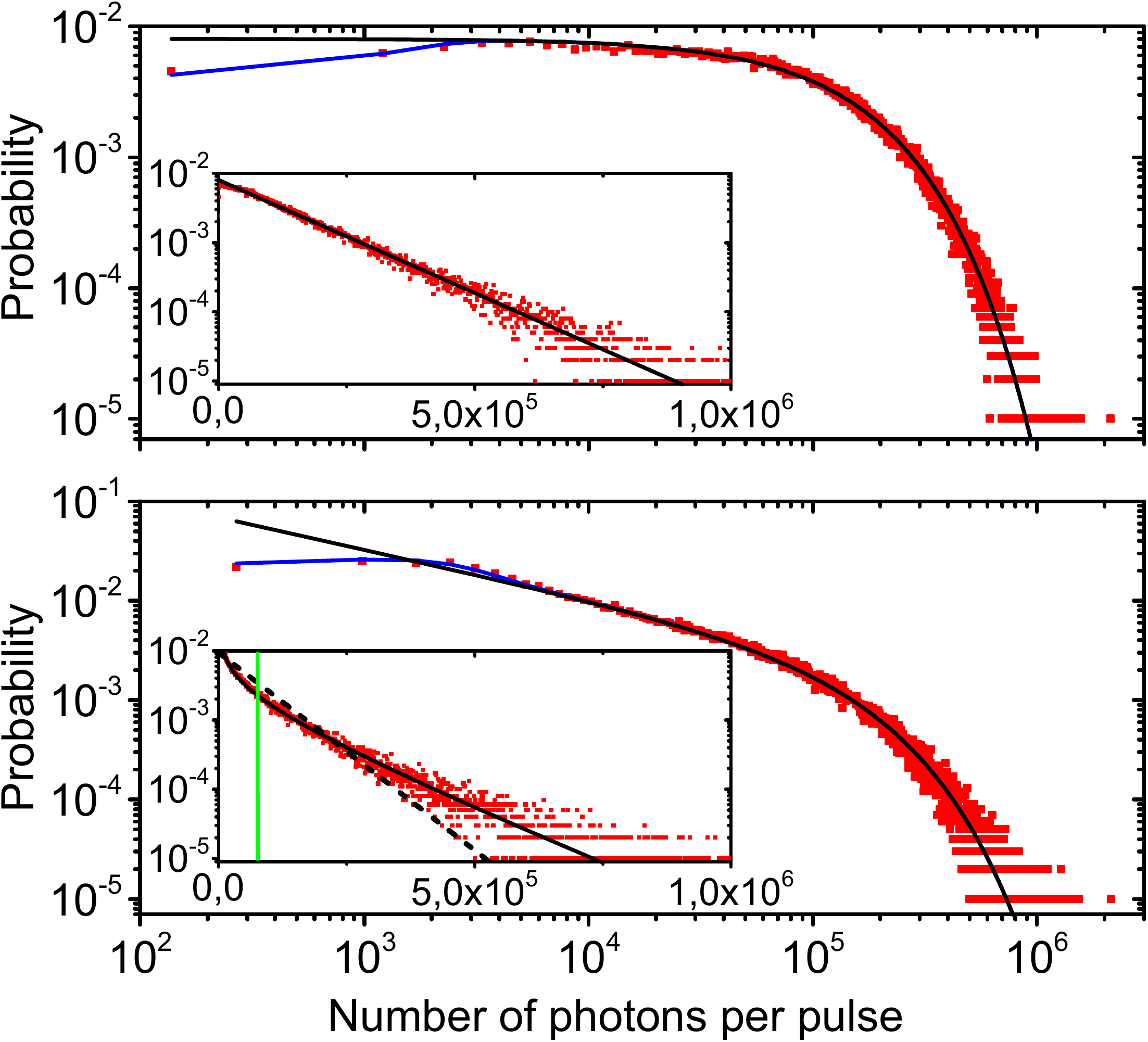}
\caption{The same data as in Fig.~\ref{5_fig:P_th_sb_linear}, but in log-log and log-linear (insets) scales. For the superbunched BSV the thermal (black dashed line) and Poissonian (green line) distributions are also shown with the same mean value.}
\label{5_fig:P_th_sb_log}
\end{center}
\end{figure}

The probability distributions of photon number reveal the fluctuations as well. In Fig.~\ref{5_fig:P_th_sb_linear} the experimental probability distributions are plotted for the thermal BSV filtered at 1585~nm (top) and for the superbunched one at 1600~nm (bottom). Both distributions are very broad, they decay slowly in linear scale, therefore it is better to present them in log-log or log-linear ones (Fig.~\ref{5_fig:P_th_sb_log}). In these scales one can see that the superbunched distribution decays slower than the thermal one; to highlight the difference the thermal distribution~\eqref{5_eq:P_th} with the same mean value (dashed line) is added in the inset in the bottom panel.

The distributions are much broader than the one for coherent light. The same inset also shows the Poissonian distribution (green line) with the same mean value, which looks rather like the Dirac delta distribution under such binning. 

Both distributions are in a good agreement with the theoretical ones, \eqref{5_eq:P_th} and \eqref{5_eq:P_sb}, with no fitting parameters (black lines). The mean values $\mean{N}$ are taken from the experimental data, $1.33\times10^5$ and $7.6\times10^4$ photons per pulse for the thermal and superbunched BSV, respectively.

\begin{figure}[!htb]
\begin{center}
\includegraphics[width=0.65\textwidth]{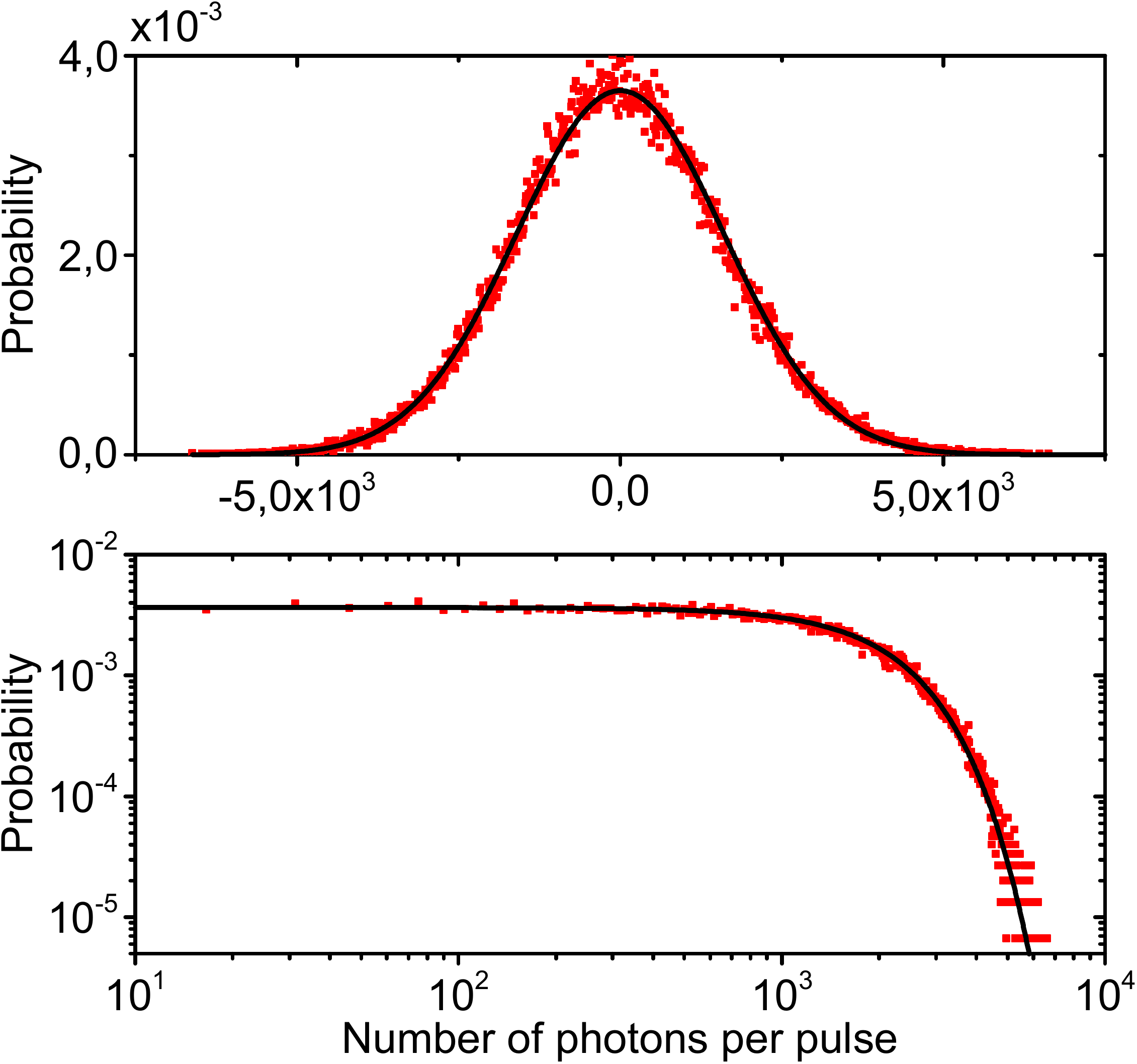}
\caption{Experimental photon-number histograms (red points) for the IR detector noise in linear (top) and log-log (bottom) scales. The Gaussian distribution~\eqref{5_eq:P_gauss} is shown by black line.}
\label{5_fig:P_noise}
\end{center}
\end{figure}

As one can see from Figs.~\ref{5_fig:P_th_sb_linear} and \ref{5_fig:P_th_sb_log}, the experimental histograms deviate from the theoretical distributions at low photon numbers. For this the dark noise of the detector is largely responsible, its distribution for the IR detector shown in Fig.~\ref{5_fig:P_noise}.

This noise is well described by the Gaussian distribution~\eqref{5_eq:P_gauss} with zero mean value and standard deviation $\sigma=1600$ photons taken from the experimental data. Being independent from the fluctuations of detected light, the dark noise just add to the light fluctuations, and therefore the detected light should be described by the convolution of light and noise probability distributions.

The convolution of Eq.~\eqref{5_eq:P_gauss} with Eq.~\eqref{5_eq:P_th} perfectly coincides with the experimental histogram for the thermal BSV without any fitting parameters, see Figs.~\ref{5_fig:P_th_sb_linear} and \ref{5_fig:P_th_sb_log} (top, blue lines); for the superbunched BSV the situation is the same, the convolution of Eq.~\eqref{5_eq:P_gauss} with Eq.~\eqref{5_eq:P_sb} is presented in the bottom panels (blue lines).

%%%%%%%%%%%%%%%%%%%%%%%%%%%%%%%%%%%%%%%%%%%%%%%%%%%%%%%
\section{Statistical enhancement of multiphoton processes}
\label{5_sec:stat_enhancement}

In the previous section it was demonstrated that BSV has very strong photon-number fluctuations. Here it is to show how these fluctuations increase the efficiency of miltiphoton effects.

Indeed, the rate $R^{(n)}$ of an $n$-photon effect generally scales as the $n$-th order CF $g^{(n)}$ of the incident light~\cite{Ducuing1964,Lambropoulos1966,Mollow1968,Agarwal1970},
\begin{equation}
R^{(n)}\sim  g^{(n)}\mean{N}^n.
\label{5_eq:multisignal}
\end{equation}
Accordingly, the \textit{statistical efficiency} $\xi^{(n)}$ of an $n$-photon effect~\cite{Lecompte1975,Delone1980,Qu1992,Qu1993,Qu1995}, defined as
\begin{equation}
\xi^{(n)}\equiv \frac{R^{(n)}}{\mean{N}^n},
\label{5_eq:stat_eff}
\end{equation}
should scale with the CF $g^{(n)}$.

This enhancement has been demonstrated for the SHG~\cite{Qu1992,Qu1993,Qu1995}, two-photon absorption~\cite{Smirnova1977,Jechow2013}, and multiphoton ionization~\cite{Lecompte1975,Arslanbekov1977} from different sources with thermal-like statistics, namely a light emitting diode~\cite{Jechow2013}, lasers below threshold~\cite{Qu1992,Qu1995}, and multimode lasers~\cite{Lecompte1975,Smirnova1977,Arslanbekov1977,Qu1993}. The latter ones have intensity fluctuations due to the contributions of different temporal modes. 

Ref.~\cite{Lecompte1975} deserves a special consideration since in this work the authors used a highly nonlinear process, 11-photon ionization of Xe atoms, leading to a dramatic enhancement. $11!\approx10^{7.6}$ factor is expected and, indeed, about $10^{6.9}$ was obtained in the experiment! The statistics was varied by changing the number of modes in nanosecond laser pulses.

Unfortunately, fluctuations of thermal-like sources are usually either slow or the sources themselves are not so bright. As it will be shown below, BSV is notable here, because its fluctuations are not only stronger than the thermal ones, see Fig.~\ref{5_fig:P_th_sb_log}, but they can be extremely fast as well.

\subsection{Experimental setup}
\label{5_sub:setup_HG}

\begin{figure}[!htb]
\begin{center}
\includegraphics[width=0.8\textwidth]{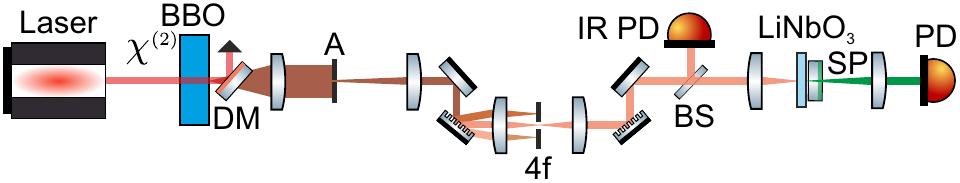}
\caption{Experimental setup for optical harmonics generation from BSV.}
\label{5_fig:setup_HG}
\end{center}
\end{figure}

The experimental setup is shown in Fig.~\ref{5_fig:setup_HG}. The first part of the experimental setup is the same as in the experiment on CF measurement (section~\ref{5_sub:setup_CF}), but now the IR detector is used for monitoring the mean photon number and the statistics of BSV. A beam sampler (BS) reflects 0.6\% of the BSV power to it.

\begin{figure}[!htb]
\begin{center}
\includegraphics[width=1\textwidth]{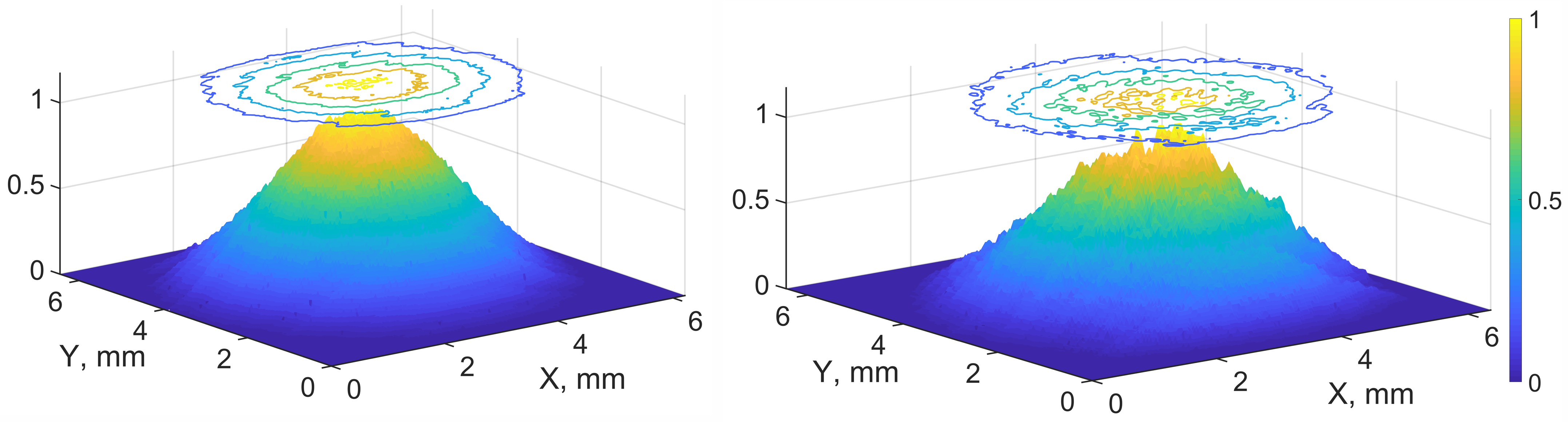}
\caption{BSV beam before (left) and after (right) the $4f$ monochromator.}
\label{5_fig:setup_beam}
\end{center}
\end{figure}

In order to get a spatially single-mode beam, the BSV is filtered only in the horizontal direction with a 140~$\mu$m slit, because the BSV is already single-mode in the vertical one (see Fig.~\ref{5_fig:g2_ang_scan}, right). The BSV beam is collimated with a 300~mm cylindrical lens thereafter and shows a good quality~(Fig.~\ref{5_fig:setup_beam}, left).

As before the $4f$-monochromator is used for frequency filtering; unfortunately, it transmits maximally $75$~nm bandwidth, therefore the full BSV spectrum (Fig.~\ref{3_fig:spectra_O_PDC})  cannot be used. Furthermore, a larger bandwidth would lead to problems with chromatic aberrations. After the spectral filtering the beam quality remains quite good (Fig.~\ref{5_fig:setup_beam}, right). 

\begin{figure}[!htb]
\begin{center}
\includegraphics[width=1\textwidth]{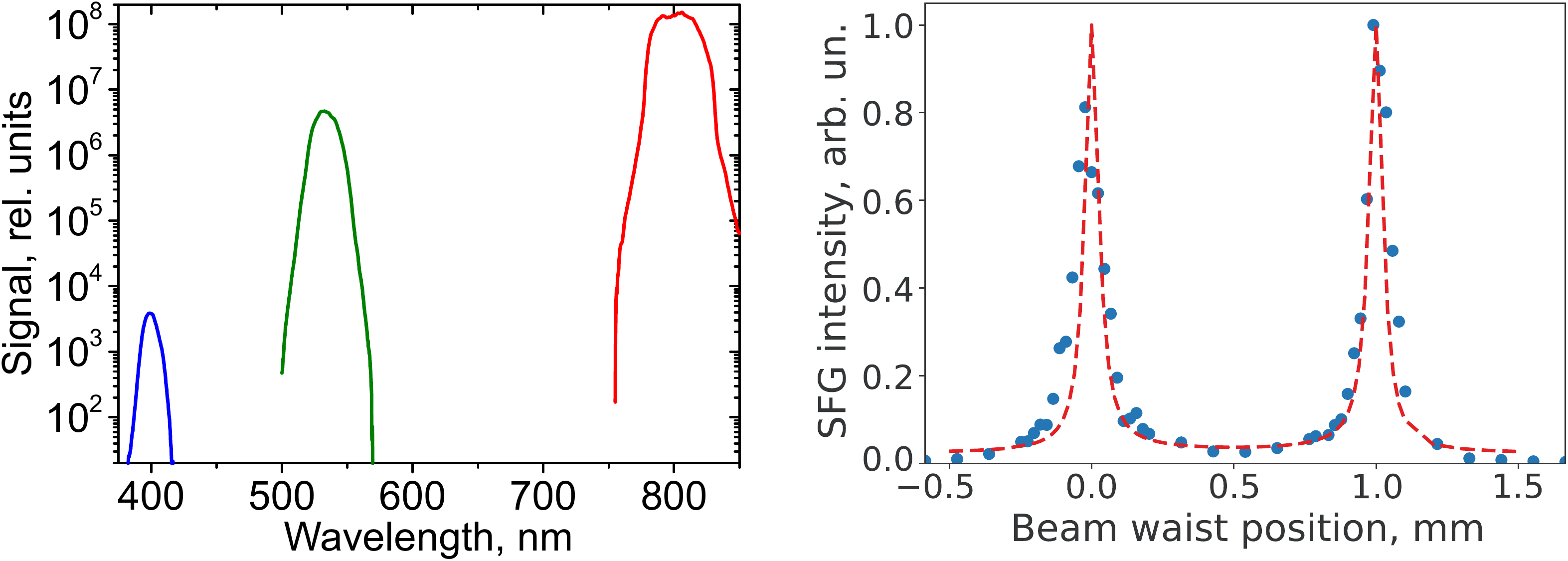}
\caption{Left: SH (red), TH (green), and FH (blue) spectrum from the broadband BSV. Right: Experimental (points) and theoretical (line) SFG intensity vs. the beam waist position for the experiment on Schmidt number measurement (section~\ref{3_sec:schmidt_SFG}).}
\label{5_fig:broad_and_surf}
\end{center}
\end{figure}

The main BSV beam is tightly focused by a $3.1$~mm lens on the surface of a 1~mm LiNbO$_3$ crystal doped with 5.1\% of MgO~\cite{Kitaeva2000}, which has been already used in the experiment on Schmidt number measurement (section~\ref{3_sec:schmidt_SFG}). The crystal has its optic axis parallel to the facet, therefore the SH, TH, and FH are generated together without phase matching though $ee\to e$, $eee\to e$, and $eeee\to e$ interactions, respectively. The coherence length of these interactions is small; on the other hand, they occur by means of very large susceptibility components, for example the second-order one, $\chi^{(2)}\sim60$~pm/V, exceeds the one for BBO for more than an order in magnitude~\cite{Dmitriev1999}. Moreover, the absence of the phase matching leads to the broadband generation of optical harmonics (Fig.~\ref{5_fig:broad_and_surf}, left).

For more efficient harmonic generation, the beam waist should be placed on the one of the crystal surfaces~\cite{Rostovtseva1980}. This could be exemplified by the dependence of the SFG intensity from the beam waist position (Fig.~\ref{5_fig:broad_and_surf}, right) for the experiment on Schmidt number measurement (section~\ref{3_sec:schmidt_SFG}) --- the intensity is the highest when the BSV is focused on the crystal facets and the lowest if the focusing happens exactly between them.

Such behavior is caused by the Gouy phase for the fundamental radiation\footnote{In our case for the BSV.} and pronounced only in the case of strong focusing, namely if the crystal length is comparable or larger than the Rayleigh length of focused beam. The phase shifts by $\pi$ before and after the waist leading to the corresponding phase shift in the nonlinear polarization. Thus, the harmonic radiation generated before and after the waist interferes destructively, which results into a low generation efficiency if the waist is placed in the bulk material.

The optical harmonics are separated from the BSV and from each other by short-pass (SP) and bandpass filters. The harmonic photon numbers are measured with a fiber-coupled avalanche photodiode (APD, Perkin\&Elmer SPCM-AQRH-16) or with the visible detector used in the experiment on BS interference (section~\ref{4_sub:setup_BS_interf}).

It is important to keep the detectors in the linear range defined by the dark noise and the saturation level. For APD the 'saturation' occurs because the detector cannot detect more than one photon per pulse. Therefore, the probability of single-photon events should be much higher than the one of two-photon events; the detector's response is nonlinear otherwise.

However, a utilization of fluctuating light is not so straightforward. The probability of high-energy spikes is relatively high and the nonlinear response sometimes affected the measurements despite author's attempts to avoid it. Note that the `noisier' light is detected, the more this issue comes into play.

In order to get the harmonics generation efficiency in absolute units the power ratio between the tapped off and the main BSV beams is estimated; thereby the detection and transmission losses for the generated harmonics are taken into account. The transmission losses in the generated harmonics are caused by the filters used for the harmonics separation and the quantum efficiency of the APD. The detection efficiencies are equal to 33\% for the SH, 37\% for the TH, and 3.6\% for the FH. The internal losses (reflection and absorption) of the LiNbO$_3$ crystal are not taken into account, because they are part of the generation efficiency.

The power ratio between the tapped off and the main BSV beams for the generation of different harmonics is controlled by means of various neutral density filters and a film polarizer to avoid the saturation of the detectors. The resulting ratio between the tapped off and the main BSV beams is $1.4\times 10^{-2}$ for the SHG, $6.0\times 10^{-5}$ for the TH generation (THG), and $3.6\times 10^{-6}$ for the FH generation (FHG). Note that the calculated ratio and losses affect only the generation efficiency in absolute units, not its enhancement due to statistics.

\subsection{Second, third, and forth harmonic generation from BSV}

\begin{figure}[!htb]
\begin{center}
\includegraphics[width=1\textwidth]{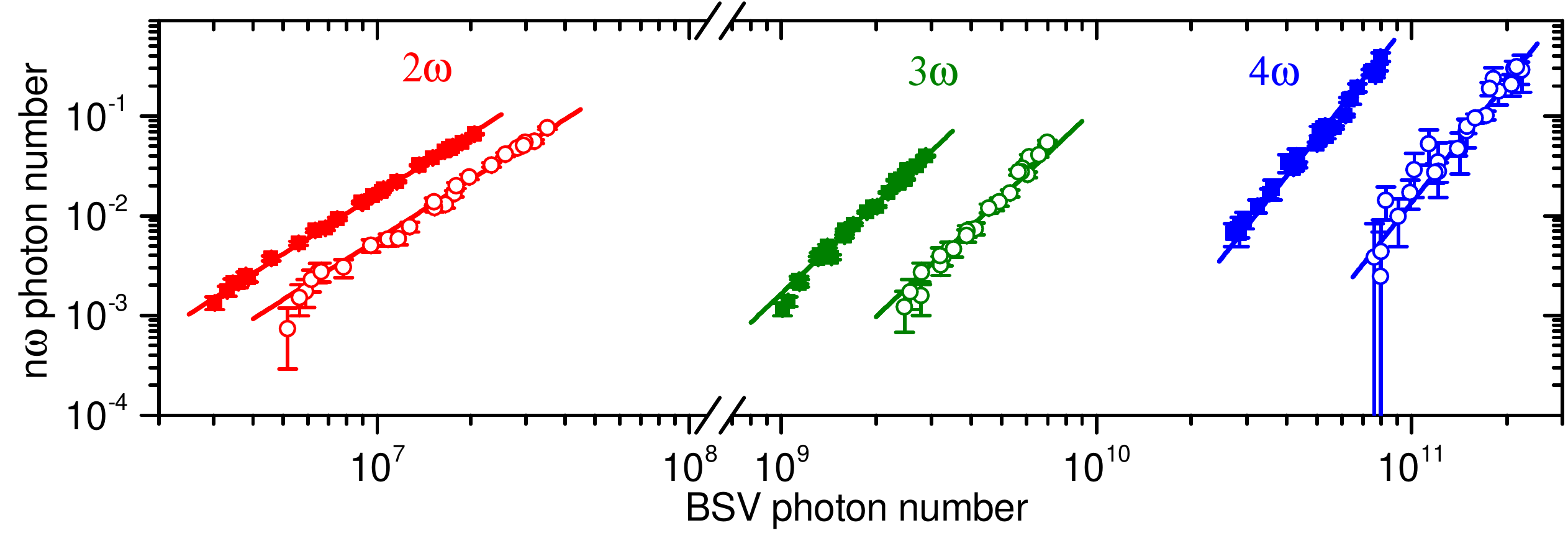}
\caption{Measured dependence of the SH (red), TH (green) and FH (blue) photon number on the pump photon number with the pump being BSV (filled squares) and pseudo-coherent light (empty circles). Theoretical fits are shown by lines.}
\label{5_fig:nw_vs_BSV}
\end{center}
\end{figure}

Firstly, the optical harmonics are measured from the BSV spectrally filtered to $3.3$~nm. The dependence of the output mean photon number $\mean{N_{n\omega}}$, corresponding to the rate $R^{(n)}$, on the BSV photon number $\mean{N}$ is shown in Fig.~\ref{5_fig:nw_vs_BSV} (filled squares). Meanwhile, the detection and transmission losses are taken into account.

All harmonics show correct power dependences, which are obtained from the linear fit in log-log scale (Table~\ref{5_tab:charact}, second column). A bit lower value for the FHG comes from the APD nonlinearity discussed above, because the FH radiation has exceptionally high fluctuations; these fluctuations will be discussed in the next section.

\begin{table}[!htb]
\centering
\renewcommand{\arraystretch}{1.25}
\begin{tabular}{|c|c|c|c|c|}
\hline
n&Exponent&$A^{(n)}_{BSV}/A^{(n)}_{pc}$&$g^{(n)}_{BSV}/g^{(n)}_{pc}$&$\eta^{(n)}_{\max}\times100\%$\\
\hline
2&$1.98\pm0.01$&$2.86\pm0.08$&$2.94\pm0.06$&$(3.2\pm0.3)\times10^{-7}$\\
\hline
3&$3.05\pm0.06$&$13.6\pm0.8$&$14.5\pm0.6$&$(1.38\pm0.12)\times10^{-9}$\\
\hline
4&$3.59\pm0.07$&$71\pm6$&$63\pm2$&$(4.9\pm0.7)\times10^{-10}$\\
\hline
\end{tabular}
\caption{The power exponents, the statistical enhancement values ($A^{(n)}_{BSV}/A^{(n)}_{pc}$), the ratios of the CFs $g^{(n)}_{BSV}/g^{(n)}_{pc}$, and the maximal generation efficiencies $\eta^{(n)}_{\max}$ from Fig.~\ref{5_fig:nw_vs_BSV}.}
\label{5_tab:charact}
\renewcommand{\arraystretch}{1}
\end{table}

The statistical efficiencies [Eq.~\eqref{5_eq:stat_eff}] can be obtained from the fits with the corresponding dependencies, $f_n(x)=A^{(n)}x^n$, $n=2,3,4$ (Fig.~\ref{5_fig:nw_vs_BSV}, solid lines). The maximal harmonic generation efficiencies $\eta^{(n)}_{\max}\equiv \mean{N_{n\omega}}/\mean{N}$ are shown in Table~\ref{5_tab:charact} (fifth column).

This has to be compared with the usual situation in which the harmonics are generated from coherent light. In the absence of a coherent source with exactly the same wavelength and pulse duration as the BSV, a pseudo-coherent source is created by reducing the photon-number fluctuations through postselection.

\begin{figure}[!htb]
\begin{center}
\includegraphics[width=1\textwidth]{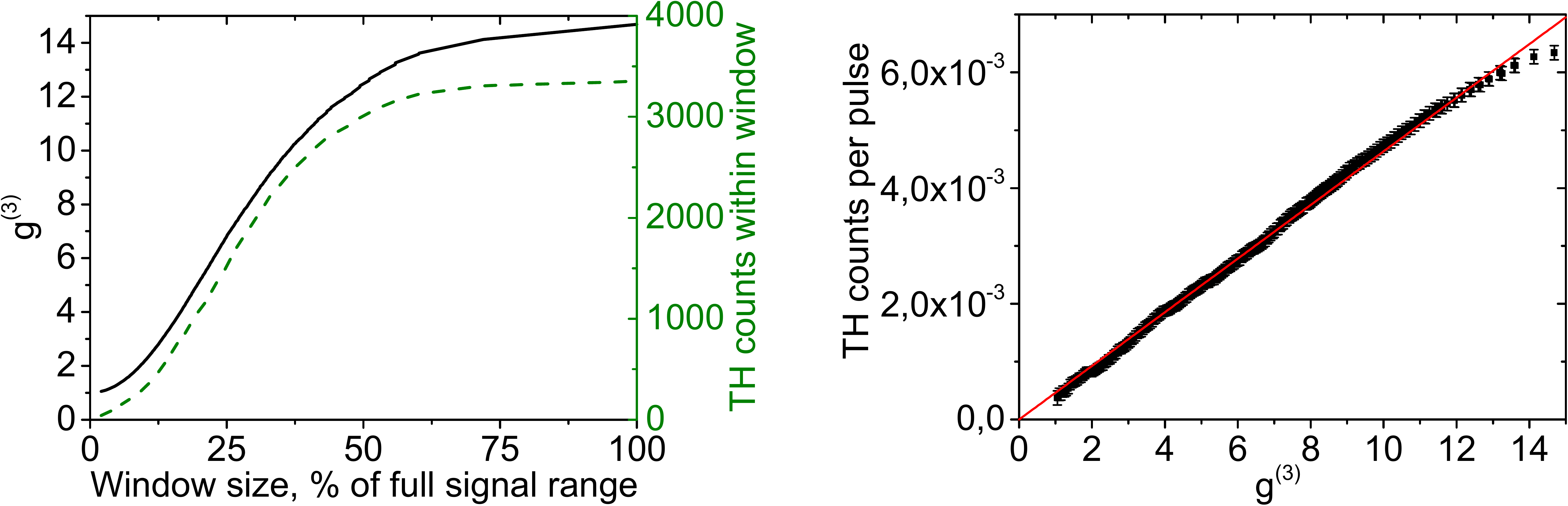}
\caption{Left: the CF $g^{(3)}$ (black solid line) and the number of TH counts (green dashed line) for the postselected data vs. the size of postselection window. Right: TH counts per pulse (points) vs. the CF $g^{(3)}$; the linear fit is shown by red line.}
\label{5_fig:postsel_expl}
\end{center}
\end{figure}

From the measured dataset only pulses, in which the number of photons in the BSV lies within the certain boundaries, the postselection window, are chosen, namely only for those pulses the data on the harmonics are further processed. The CFs for the postselected data decrease almost to unity when the postselection window for the BSV signal becomes narrower. Fig.~\ref{5_fig:postsel_expl} (left panel) shows the third-order CF $g^{(3)}$ (black solid line) as a function of the window size as an example. At the same time, the number of TH counts (green dashed line) for the chosen pulses within this window decreases.

Thus, the post-selection window is chosen to be as small as possible, but simultaneously large enough for the photon number to be measurable with a reasonably small uncertainty after the post-selection. The minimal values of CFs obtained this way for our data are $g^{(2)}=1.010\pm0.002$, $g^{(3)}=1.020\pm0.003$, $g^{(4)}=1.10\pm0.007$. After the post-selection, the number of counts per pulse is linear in the corresponding CF (Fig.~\ref{5_fig:postsel_expl}, right).

The power dependences for different harmonics generated from this pseudo-coherent source are plotted in Fig.~\ref{5_fig:nw_vs_BSV} by empty circles. One can see that the statistical efficiency is indeed considerably higher for the BSV. The enhancement factors are shown in Table~\ref{5_tab:charact} (third column) together with the ratios of CFs for the BSV and the pseudo-coherent source (forth column). As expected, there is an agreement between the two.

The FH is produced with a reduced value of $g^{(4)}=69\pm2$ due to the unwanted SHG in the BBO crystal discussed above. A brighter pump is needed for the FHG, because its efficiency is lower. At the same time, the brighter pump leads to the brighter parasite SH and more suppressed BSV fluctuations, see Fig.~\ref{5_fig:g2_crystal_power}.

\begin{figure}[!htb]
\begin{center}
\includegraphics[width=1\textwidth]{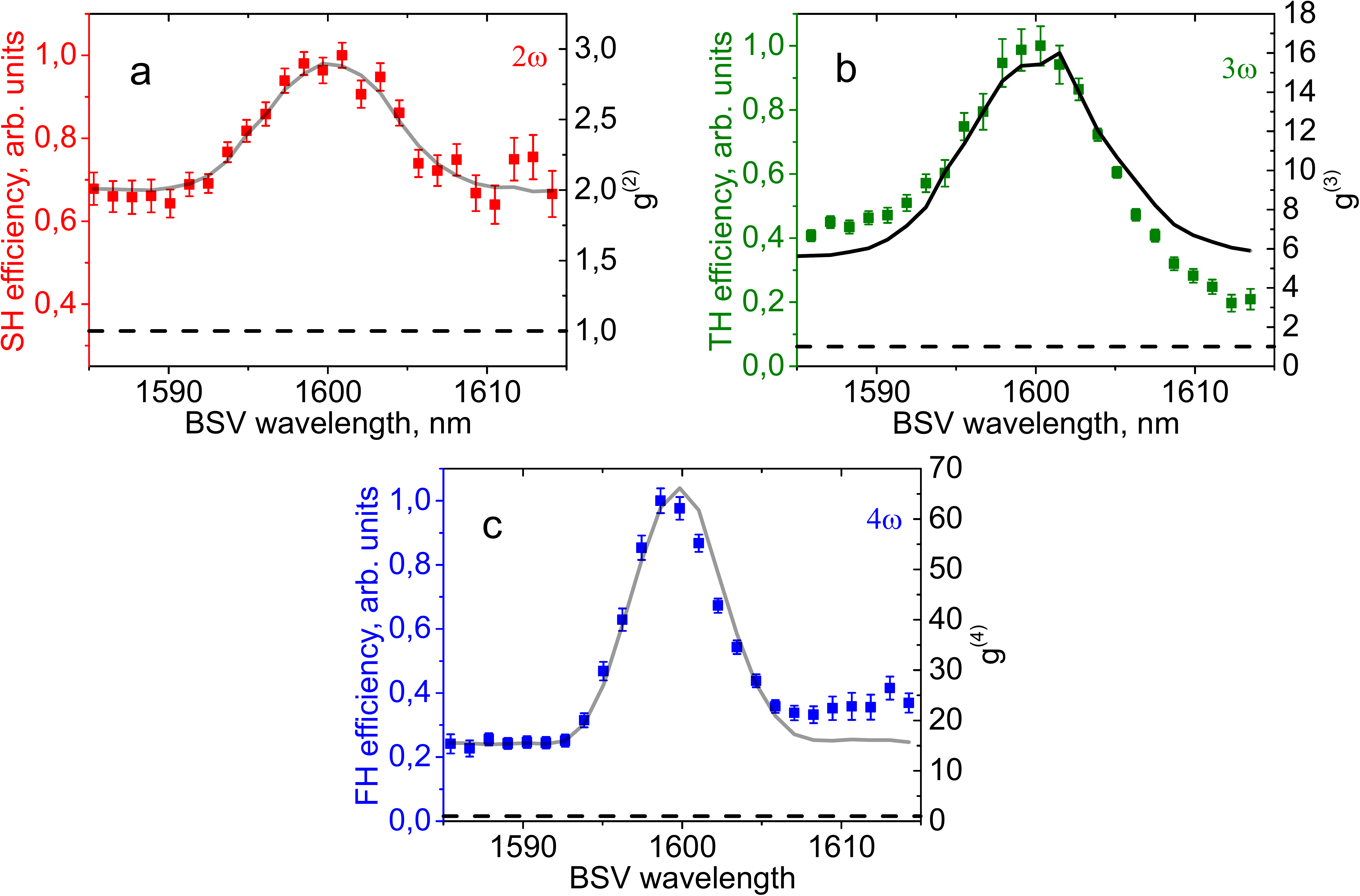}
\caption{SH (a), TH (b) and FH (c) statistical efficiencies $\xi^{(n)}$ (points) as well as the corresponding $g^{(n)}$ values (solid lines) measured vs. the BSV wavelength. For comparison, the expected efficiencies for coherent light are shown by dashed lines.}
\label{5_fig:HG_wave_scan}
\end{center}
\end{figure}

To compare the efficiency of harmonics generation from the superbunched BSV with the one for the thermal BSV, the BSV wavelength is scanned around the degeneracy point, $1600$ nm, and measure the rates of the SH around $800$~nm, the TH around $533$~nm, and the FH around $400$~nm. The obtained statistical efficiencies $\xi^{(n)}$ are plotted in Fig.~\ref{5_fig:HG_wave_scan}, together with the corresponding BSV CFs. By scanning the wavelength, the thermal BSV turns into the superbunched one and the value of $\xi^{(n)}$ follows the one of $g^{(n)}$.

Next, it is shown that the generation efficiency scales with the CF value also in the case of broadband BSV. As discussed, the generation of optical harmonics can be considered as a fast multiphoton detector~\cite{Peer2005,Sensarn2010}, which follows the fast fluctuations; in other words, the condition of single-mode detection discussed in section~\ref{5_sub:filtering} is realized despite the fact that the BSV has many modes.

\begin{figure}[!htb]
\begin{center}
\includegraphics[width=1\textwidth]{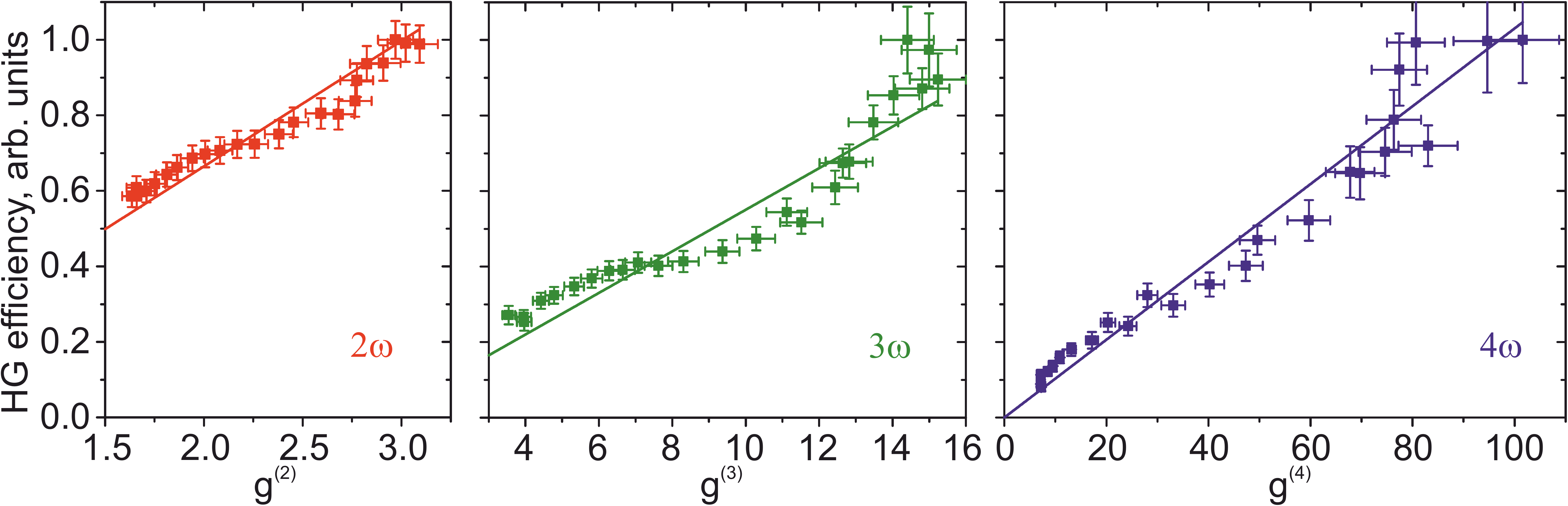}
\caption{Statistical efficiencies $\xi^{(n)}$ (points) of the SH (left), TH (middle) and FH (right) from the broadband BSV show linear dependences on the corresponding CFs $g^{(n)}$.  Solid lines show linear fits, $\xi^{(n)}\sim g^{(n)}$.}
\label{5_fig:HG_broad}
\end{center}
\end{figure}

In order to change the statistics the unwanted SHG (Fig.~\ref{5_fig:g2_crystal_power}) is used. As discussed, this effect reduces the fluctuations and CFs; by tilting the BBO crystal the latter can be reduced from their original values down to $g^{(2)}=1.55\pm0.05$, $g^{(3)}=3.5\pm0.2$, and $g^{(4)}=7.1\pm0.5$. 

Figure~\ref{5_fig:HG_broad} shows that even for 75~nm broad BSV, the statistical efficiencies $\xi^{(n)}$ of the SHG (left), THG (middle), and FHG (right) follow the CFs. As already mentioned, our detector resolves only pulse-to-pulse fluctuations and the measured CFs are reduced due to multimode detection, see section~\ref{5_sub:multimode}. Therefore, the CFs $g^{(n)}$ are measured under the narrowband, $2.5$~nm, filtering. The filtering allows one to retrieve the correct values of CFs, see section~\ref{5_sub:filtering}. Therefore, for each orientation of the BBO crystal first the CF $g^{(n)}$ is measured with the narrowband filtering and immediately after that, the efficiency $\xi^{(n)}$ is measured. Due to this fact, rather large error bars appear in the axis of abscissas.

\subsection{BSV fluctuations: possible applications}

The results for optical harmonic generation illustrate that BSV is a useful source for multiphoton effects in general. For example, BSV as a pump will be beneficial for the multiphoton microscopy of fragile structures~\cite{Valev2011} including biological objects~\cite{Helmchen2005,Cole2014}. Indeed, the harmonics are generated without phase matching from low mean powers (few nW -- tens of $\mu$W) and, according to Fig.~\ref{5_fig:nw_vs_BSV}, a certain rate of a four-photon effect is achieved with BSV having the mean power about three times less than in the case of coherent pumping. This can increase the sensitivity while not overcoming the damage threshold, what becomes critical in the case of live cell imaging, where one should use the lowest possible photon dose~\cite{Cole2014}.

As mentioned above, BSV has ultrafast fluctuations. Here a bandwidth of $9$~THz is used, what corresponds to the correlation time of $80$~fs, but one can achieve times an order of magnitude shorter, see chapter~\ref{2_chapter}. Such behavior cannot be mimicked by any external intensity modulation~\cite{Straka2018}. With such timescales BSV is useful for ultrafast spectroscopy, for example for the testing of materials response time. Although here a single spatial mode is used, the multimode spatial structure will not change the noisy behavior of BSV. One can use the whole frequency and angular spectrum of BSV for exciting multiphoton effects~\cite{Jedrkiewicz2011}.

Finally, unlike a pulsed coherent source, BSV has two typical times, the pulse duration and the correlation time, see section~\ref{4_sec:Macro_HOM}. Both parameters can be tuned in a broad range, which can be used in the two-dimensional fluorescence spectroscopy~\cite{Raymer2013,Dorfman2016} exploiting a two-photon absorption effect. Moreover, the photon fluxes are drastically higher than in the case of photon pairs.

%%%%%%%%%%%%%%%%%%%%%%%%%%%%%%%%%%%%%%%%%%%%%%%%%%%%%%%%%%%%%%%%%%%%%%%
\section{Extreme events, extreme bunching, and heavy-tailed distributions}
\label{5_sec:extreme}

In the previous section it was shown that BSV fluctuations increase the efficiency of multiphoton effects. Here it is demonstrated that the light produced with BSV as a pump fluctuates even more. It has an exceptional statistics with a lot of extreme events.

The existence of such events is a fascinating phenomenon in natural and social sciences, because they could lead to catastrophic changes in the system despite being quite rare.

One famous example is `rogue waves' in ocean~\cite{Kharif2003} that comes from sailor's fairy-tails. For centuries sailors have been telling about giant waves that appear from nowhere and disappear immediately. These stories were considered as a myth until first time such a wave has been detected instrumentally on the New Year Eve 1995 at the Draupner oil platform located in the North Sea. Importantly, these rogue waves have their analogues in nonlinear optics~\cite{Akhmediev2016}, mainly for supercontinuum generation in optical fibers~\cite{Solli2007}.

What is the definition for a \textit{rogue wave}? There are several ones, but the most general one is the following~\cite{Akhmediev2010}: `a rogue wave is a wave that is much higher than others around it, and which has a habit of appearing unpredictably'. It means that the process is chaotic or random and its fluctuations are higher than the usual, namely equilibrium, ones. For light the latter indicates a deviation from thermal statistics [Eq.~\eqref{5_eq:P_th}].

The corresponding probability distributions are called `heavy-tailed'~\cite{Foss2013}. The `heaviest' is the power-law (Pareto) distribution,
\begin{equation}\label{5_eq:P_pareto}
P(x)\propto \frac{1}{x^{(1+\alpha)}},
\end{equation}
with the Pareto index $\alpha>0$. 

Originally it arises from the observations of Italian sociologist Vilfredo Pareto, who at the beginning of the 20th century noted that 80\% of land in Italy belongs to the 20\% of population. In its respect, such spreading leads to the power-law probability distribution. Pareto's observations give rise to the famous `80/20 rule' or Pareto principle: 20\% of the input, e.g. resources or efforts, accounts for 80\% of the output, e.g. rewards or results.

The Pareto distributions are often employed for the social quantities~\cite{Clauset2009}, for example for the frequency of words and surnames, number of received telephone calls and hits on websites, intensity of wars and terrorist attracts, and the impact of academic papers, of course. The most outrageous is the distribution of income and wealth, the richest 1\% accumulate almost the same as the rest 99\% of the population~\cite{Wealth2018}, being attributable to the very non-uniform behavior of different people~\cite{Yakovenko2009}, namely due to the difference between the ones who live on salary and the ones whose sources of income are investments and bonuses.

Mostly the social quantities are just numbers, they could take any value, whereas in the natural sciences it is not so straightforward. Nevertheless, the Pareto distributions appear in biology~\cite{Zhao2013}, astrophysics~\cite{Lu1991}, geophysics~\cite{Taubert2018}, and condensed matter physics~\cite{Stefani2009,Campi2015}. In optics they appear in scattering~\cite{Barthelemy2008,Mercadier2009} or highly nonlinear phenomena~\cite{Borlaug2009,Kasparian2009,Alves2019}. In the latter, the Pareto distributions appear due to the exponential amplification of an initially broad distribution, which is one of the main origins of power-law scalings~\cite{Newman2005}.

However, the power laws can be very different. If the power exponent, $1+\alpha$, is less than~2, the mean value and higher moments are indefinite; such an example for supercontinuum pumped by BSV is presented below. Furthermore, other heavy-tailed distributions appear if the optical harmonics are generated from BSV.

\subsection{Theoretical description: nonlinear processes from fluctuating light}

Here a theoretical description of harmonic and four-wave mixing (FWM) generation from fluctuating light is presented. The description is based on probability distributions and despite its simplicity it wonderfully explains the obtained experimental results on the harmonic and supercontinuum generation discussed in sections~\ref{5_sub:HG_extreme} and  \ref{5_sub:SC_extreme}, respectively. Despite complexity of supercontinuum generation the model can be applied to it as well since the elementary physical process behind is FWM. Notably, a quite similar approach was also used in Ref.~\cite{Borlaug2009}.

In the absence of pump depletion, the number $N_{n\omega}$ of generated photons for the $n$th harmonic scales as the $n$th power of the number $N_\omega$ of photons in the fundamental radiation,
\begin{equation}
N_{n\omega} = \mathcal{K}N_\omega^n,
\label{5_eq:N_nw}
\end{equation}
where $\mathcal{K}$ is related to the conversion efficiency.

The distribution $P_{n\omega} (N_{n\omega})$ for the harmonic radiation is obtained from the one $P_{\omega}(N_\omega)$ of the fundamental radiation~\cite{Hogg1978,Akhmanov1981},
as $P_{n\omega}(N_{n\omega})dN_{n\omega}=P_{\omega}(N_\omega)dN_\omega$. Thus, using Eq.~\eqref{5_eq:N_nw} one can get
\begin{equation}
P_{n\omega}(N_{n\omega}) = \frac{P_{\omega}\left(\sqrt[n]{N_{n\omega}/\mathcal{K}}\right)}{n\sqrt[n]{\mathcal{K}}N_{n\omega}^{1-\nicefrac{1}{n}}}.
\label{5_eq:P_nw}
\end{equation}
It follows that the $m$th moment of the harmonic radiation is proportional to the $nm$th moment of the fundamental radiation, $\langle N_{n\omega}^m\rangle = \mathcal{K}^m\langle N_{\omega}^{nm}\rangle$. As a result, the CFs $g_{n\omega}^{(m)}$ can be obtained from the ones for the fundamental radiation~\cite{Teich1966,Qu1995}, 
\begin{equation}
g_{n\omega}^{(m)}=\frac{g_\omega^{(mn)}}{\left(g_\omega^{(n)}\right)^m}.
\label{5_eq:g_m_nw}
\end{equation}

At first, let us consider the simplest case of bright coherent pump. In this case the variance increases for the harmonic radiation, $\var{N_{n\omega}}\sim\mean{N_{n\omega}}^{2-\nicefrac{1}{n}}$ at $\mean{N_\omega}\gg1$, and as a result the harmonic radiation is not shot-noise limited. However, all CFs $g_{n\omega}^{(m)}$ are equal to unity since the CF $g_\omega^{(n)}=1$. Moreover, like the original distribution~\eqref{5_eq:P_gauss}, the probability distribution $P_{n\omega}(N_{n\omega})$ remains `bell-shaped'. Therefore, the probability of extreme events with $N_{n\omega}\gg\mean{N_{n\omega}}$ remains negligible. Note that since the distribution~\eqref{5_eq:P_gauss} is used, the description is not fully strict.

However, the situation is very different for input light with larger photon-number fluctuations. For the $n$th harmonic of thermal light, see Eq.~\eqref{5_eq:P_th}, one can get
\begin{equation}
P_{n\omega}(N_{n\omega}) = \frac{\sqrt[n]{n!}}{n\sqrt[n]{\langle N_{n\omega}\rangle}N_{n\omega}^{1-\nicefrac{1}{n}}}e^{-\sqrt[n]{n!\frac{N_{n\omega}}{\langle N_{n\omega}\rangle}}},
\label{5_eq:P_nw_th}
\end{equation}
while pumping with superbunched BSV [Eq.~\eqref{5_eq:P_sb}] gives
\begin{equation}
P_{n\omega}(N_{n\omega})=\frac{\sqrt[2n]{(2n-1)!!}}{n\sqrt{2\pi}\sqrt[2n]{\langle N_{n\omega}\rangle}N_{n\omega}^{1-\nicefrac{1}{2n}}}e^{-\frac{1}{2}\sqrt[n]{(2n-1)!!\frac{N_{n\omega}}{\langle N_{n\omega}\rangle}}}.
\label{5_eq:P_nw_sb}
\end{equation}
The distribution~\eqref{5_eq:P_nw_th} is the Weibull distribution~\cite{Beirlant2005} with the shape parameter $1/n$ and the scale parameter $\langle N_{n\omega}\rangle/n!$, the distribution~\eqref{5_eq:P_nw_sb} is the generalized Gamma distribution~\cite{Bell1988} with the shape parameters $1/2$ and $1/n$ and the scale parameter $2^n\langle N_{n\omega}\rangle/(2n-1)!!$. Both distributions are heavy-tailed.

Indeed, to characterize the tail of a probability distribution, one can use the complementary cumulative distribution function (CCDF)~\cite{Beirlant2005},
\begin{equation}
\bar{C}(N)=\int_{N}^{\infty} P(N')dN',
\label{5_eq:CCDF}
\end{equation}
also known as the survival function. The probability distribution is called heavy-tailed if its tail index, defined as
\begin{equation}
\beta=\lim_{N\rightarrow\infty} H(N)/N,
\label{alpha}
\end{equation}
is equal to zero~\cite{Foss2013}. Here $H(N)=-\ln\bar{C}(N)$ is the hazard function.

For thermal light [Eq.~\eqref{5_eq:P_th}], the CCDF is
\begin{equation}\label{5_eq:CCDF_th}
\bar{C}_{th}(N)=e^{-\frac{N}{\langle N\rangle}},
\end{equation}
for superbunched BSV [Eq.~\eqref{5_eq:P_sb}],
\begin{equation}\label{5_eq:CCDF_sb}
\bar{C}_{sb}(N)=\mbox{Erfc}\left[\sqrt{\frac{N}{2\langle N\rangle}}\right],
\end{equation}
where $\mbox{Erfc}$ is the complementary error function. Then, the index $\beta$ is equal to $1/\langle N\rangle$ and $1/(2\langle N\rangle)$, respectively. Hence, both distributions are not heavy-tailed. Furthermore, now it is clear that the tail index $\beta=0$ means that the distribution decays slower than the exponential distribution~\eqref{5_eq:P_th}.

\begin{figure}[!htb]
\begin{center}
\includegraphics[width=0.55\textwidth]{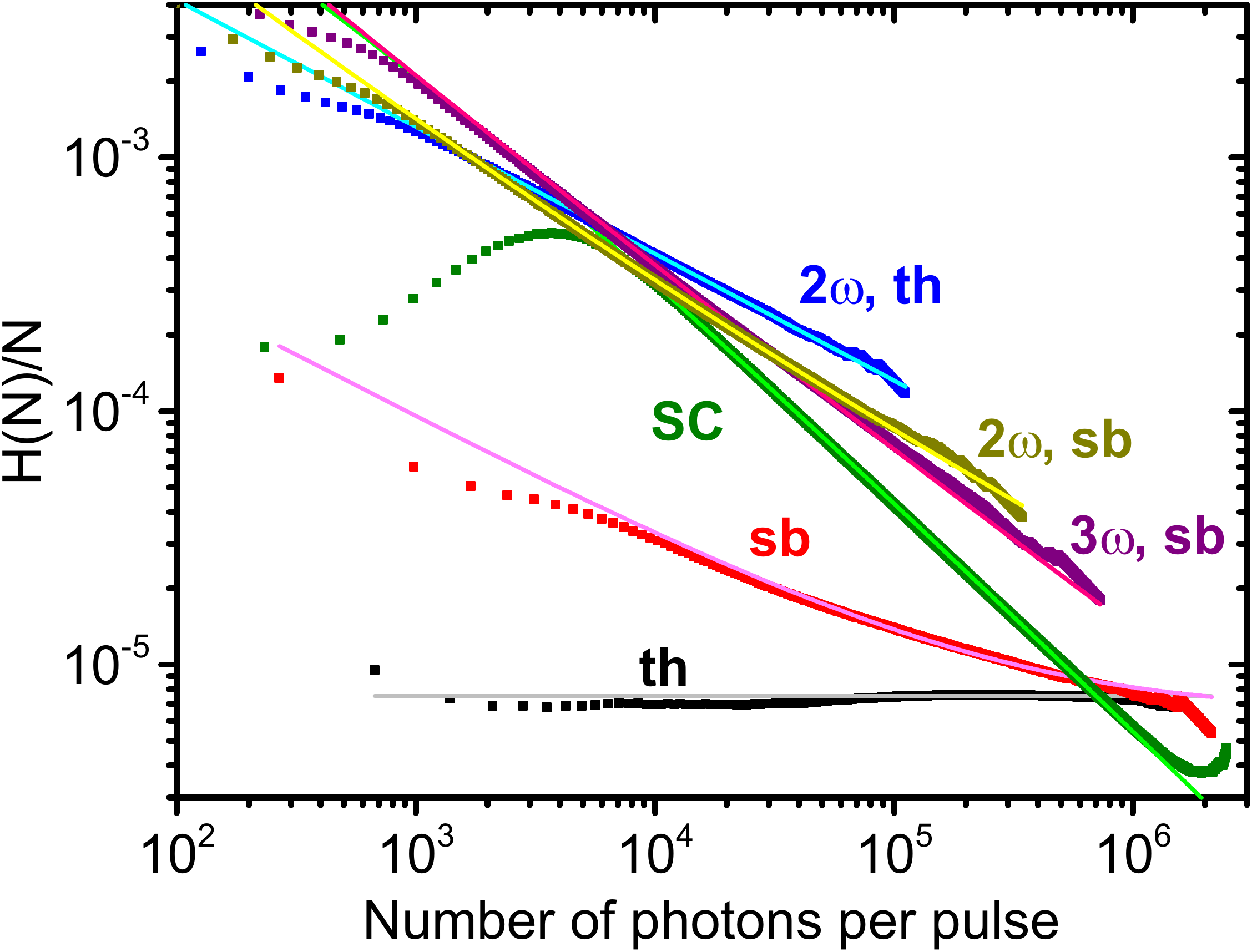}
\caption{Experimental (points) and theoretical (lines) $H(N)/N$ for the thermal (th) and superbunched (sb) BSV presented in Fig.~\ref{5_fig:P_th_sb_log}, the second ($2\omega$) and third ($3\omega$) harmonics presented in Fig.~\ref{5_fig:P_nw_log}, and the supercontinuum (SC) presented in Fig.~\ref{5_fig:CCDF_sc}. For the SC, the theoretical values are calculated from a fit with the Pareto distribution.}
\label{5_fig:tail_index}
\end{center}
\end{figure}

The experimental $H(N)/N$ with the data taken from Fig.~\ref{5_fig:P_th_sb_log} are presented in~Fig.~\ref{5_fig:tail_index} together with the theoretical curves calculated from Eqs.~\eqref{5_eq:CCDF_th} and \eqref{5_eq:CCDF_sb}. The curves fully demonstrate this behavior: at large $N$ they tend to the corresponding theoretical indices $\beta$. The thermal (black points and gray line) and superbunched BSV (red points and pink line) distributions have the same tail index if the mean photon number for the first one is twice as large as for the second one.

For optical harmonics, the CCDF is
\begin{equation}
\bar{C}_{n\omega}(N_{n\omega})=e^{-\sqrt[n]{n!\frac{N_{n\omega}}{\langle N_{n\omega}\rangle}}}
\label{5_eq:CCDF_nw_th}
\end{equation}
for thermal pumping [Eq.~\eqref{5_eq:P_th}] and
\begin{equation}
\bar{C}_{n\omega}(N_{n\omega})=\mbox{Erfc}\left[\frac{\sqrt[2n]{(2n-1)!!\frac{N_{n\omega}}{\langle N_{n\omega}\rangle}}}{\sqrt{2}}\right]
\label{5_eq:CCDF_nw_sb}
\end{equation}
for superbunched BSV one [Eq.~\ref{5_eq:P_sb}]. In both cases the index $\beta$ tends to zero, therefore both distributions are heavy-tailed. In Fig.~\ref{5_fig:tail_index} the theoretical $H(N)/N$ (cyan, yellow, and magenta lines) are plotted from Eqs.~\eqref{5_eq:CCDF_nw_th} and \eqref{5_eq:CCDF_nw_sb} together with the experimental data from Fig.~\ref{5_fig:P_nw_log} (blue, dark yellow, and purple points). All curves show fast tendency to zero, which increases with the harmonic number $n$ and is more pronounced for the fundamental radiation with larger fluctuations. In other words, superbunched BSV pump makes the tail heavier than the thermal one.

In the case of FWM the number $N_F$ of photons is given by the equation similar to the one, see Eq.~\eqref{2_eq:sinh2}, describing PDC,
\begin{equation}
N_F=\sinh^2\kappa N_\omega,
\label{5_eq:FWM_sinh2}
\end{equation} 
but the gain scales linearly with the pump photon number $N_\omega$~\cite{Finger2015}. The value $\kappa$ characterizes the interaction strength and as before the pump is supposed to be undepleted.

\begin{figure}[h]
\begin{center}
\includegraphics[width=0.55\textwidth]{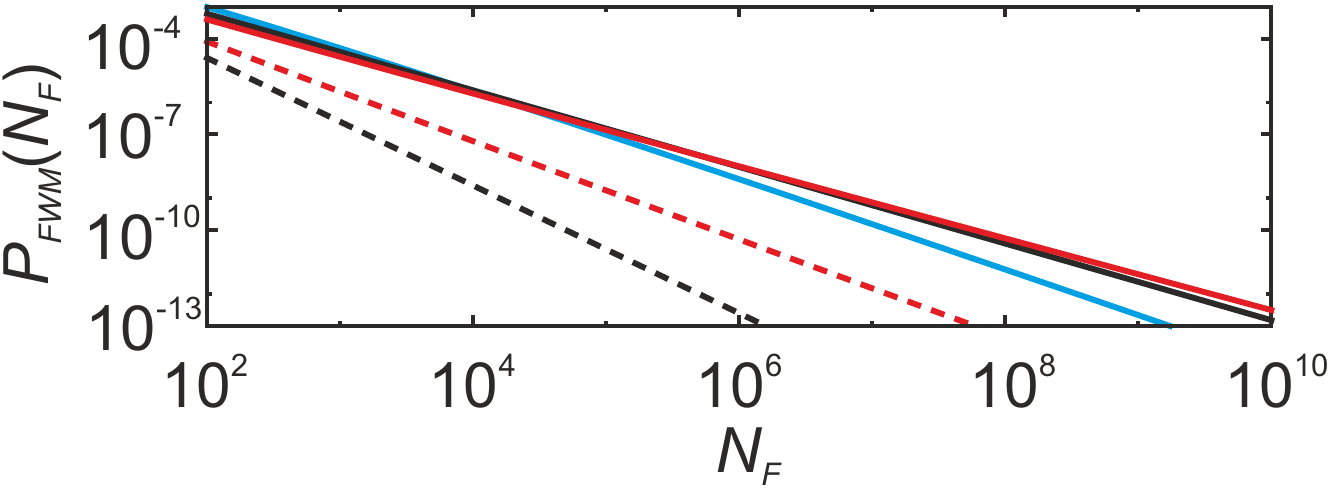}
\caption{Calculated photon-number distribution for FWM from thermal [Eq.~\eqref{5_eq:P_fwm_th}, black] and superbunched [Eq.~\eqref{5_eq:P_fwm_sb}, red] 
BSV with the gain $\kappa\langle N_\omega\rangle=0.5$ (dashed line) and $2.5$ (solid line). The blue line shows the distribution for multimode superbunched BSV pumping [Eq.~\eqref{5_eq:P_fwm_sb_M}] with $\kappa\langle N_\omega\rangle=2.5$ and $M=5$.} 
\label{5_fig:P_FWM}
\end{center}
\end{figure}

Similar to the harmonics case, from the pump photon-number distribution $P(N_\omega)$ the FWM ones are obtained:
\begin{equation}\label{5_eq:P_fwm_th}
P_{FWM}(N_F)=\frac{e^{-\frac{\mathrm{arcsinh} \sqrt{N_F}}{\kappa\langle N_\omega\rangle}}}{2\kappa\langle N_\omega\rangle\sqrt{N_F(1+N_F)}}
\end{equation}
for thermal pumping [Eq.~\eqref{5_eq:P_th}] and
\begin{equation}\label{5_eq:P_fwm_sb}
P_{FWM}(N_F)=\frac{e^{-\frac{\mathrm{arcsinh}\sqrt{N_F}}{2\kappa\langle N_\omega\rangle}}}{\sqrt{8\pi \kappa\langle N_\omega\rangle
N_F(1+N_F)\mathrm{arcsinh}\sqrt{N_F}}}
\end{equation}
for superbunched BSV one [Eq.~\eqref{5_eq:P_sb}]. Both distributions are shown in Fig.~\ref{5_fig:P_FWM}.

The corresponding CCDFs are
\begin{equation}
\bar{C}_{FWM}(N_F)=e^{-\frac{\mathrm{arcsinh}\sqrt{N_F}}{\kappa\mean{N_\omega}}}
\label{5_eq:CCDF_fwm_th}
\end{equation}
and
\begin{equation}
\bar{C}_{FWM}(N_F)=\mbox{Erfc}\left[\sqrt{\frac{\mathrm{arcsinh}\sqrt{N_F}}{2\kappa\mean{N_\omega}}}\right]
\label{5_eq:CCDF_fwm_sb}
\end{equation}
for thermal and superbunched BSV pumping, respectively.

Both CCDFs are of the form $x^{-\alpha}L(x)$, where $L(x)$ is a slowly varying function~\cite{Foss2013}, i.e. $\lim_{x\rightarrow \infty} \nicefrac{L(tx)}{L(x)}=1$, for any $t>1$. They both have the same tail as the Pareto distribution~\eqref{5_eq:P_pareto},
\begin{equation}\label{5_eq:CCDF_Pareto}
\bar{C}(N_F) \propto \frac{1}{(N_F)^{\alpha}},
\end{equation}
with the same exponent $\alpha$, i.e. they are tail equivalent\footnote{If $\lim_{x\rightarrow \infty}\nicefrac{\bar{C}_1(x)}{\bar{C}_2(x)}=1$, then $\bar{C}_1(x)$ and $\bar{C}_2(x)$ are tail equivalent~\cite{Foss2013}.} to it.

The tail exponents tend to $1/(2\kappa\langle N_\omega\rangle)$ for the distribution~\eqref{5_eq:CCDF_fwm_th} and $1/(4\kappa\langle N_\omega\rangle)$ for the distribution~\eqref{5_eq:CCDF_fwm_sb}. In other words, BSV pumping leads to twice smaller exponent $\alpha$ than pumping with thermal light with the same power. In both cases the exponent goes below unity for sufficiently bright pump and then all moments are indefinite.

The first scaling is straightforward; since $\mathrm{arcsinh}(x)=\ln(x+\sqrt{x^2+1})$,
\begin{equation}
\ln\left[\bar{C}_{FWM}(N_F)\right]	=	-\frac{\mathrm{arcsinh}\sqrt{N_F}}{\kappa\mean{N_\omega}} \approx -\frac{\ln(N_F)}{2\kappa\langle N_\omega\rangle} = \ln \left(N_F^{-\nicefrac{1}{2\kappa\langle N_\omega\rangle}} \right)	\,\,\, \mathrm{at} \,\,\, N_F\rightarrow \infty.
\end{equation}
The second scaling can be found similarly by using the asymptotic for the error function, $\mbox{Erfc}(\sqrt{x})\sim \exp(-x)/(\sqrt{\pi x})$.

These Pareto-like distributions are surely heavy-tailed; the tail index $\beta$ tends to zero as the photon number tends to infinity. Moreover, the tendency is faster than for any harmonic; the experimental $H(N)/N$ (green points in Fig.~\ref{5_fig:tail_index}) fitted with the Pareto distribution illustrate this behavior.

As discussed in section~\ref{5_sub:multimode}, the probability distributions are different for multimode light. Here the single-mode detection with the multimode pumping is assumed. This will be the experimental case considered below; we are able to filter the supercontinuum very well, but we cannot do it with the pump at the same time.

Similarly to the single-mode case, from the pump multimode distributions~\eqref{5_eq:P_th_M} and \eqref{5_eq:P_sb_M} the FWM photon-number distributions are obtained:
\begin{equation}\label{5_eq:P_fwm_th_M}
P_{FWM,M}(N_F)=\frac{e^{-\frac{\mathrm{arcsinh}\sqrt{N_F}}{\kappa\langle N_\omega\rangle/M}}\;\mathrm{(arcsinh}\sqrt{N_F})^{M-1}}{2\sqrt{N_F(N_F+1)}\;(M-1)!\;(\kappa\langle N_\omega\rangle/M)^M}
\end{equation}
for $M$-mode thermal pumping and
\begin{equation}\label{5_eq:P_fwm_sb_M}
P_{FWM,M}(N_F)=\frac{e^{-\frac{\mathrm{arcsinh}\sqrt{N_F}}{2\kappa\langle N_\omega\rangle/M}}\;\mathrm{(arcsinh}\sqrt{N_F})^{M/2-1}}{2\sqrt{N_F(N_F+1)}\;\Gamma\left(\frac{M}{2}\right)(2\kappa\langle N_\omega\rangle/M)^M}
\end{equation}
for $M$-mode superbunched BSV one. The corresponding CCDFs are
\begin{equation}\label{5_eq:CCDF_fwm_th_M}
\bar{C}_{FWM,M}(N_F)=\frac{\Gamma\left(M,\frac{\mathrm{arcsinh}\sqrt{N_F}}{\kappa\langle N_\omega\rangle/M}\right)}{(M-1)!}
\end{equation}
and
\begin{equation}\label{5_eq:CCDF_fwm_sb_M}
\bar{C}_{FWM,M}(N_F)=\frac{\Gamma\left(\frac{M}{2},\frac{\mathrm{arcsinh}\sqrt{N_F}}{2\kappa\langle N_\omega\rangle/M}\right)}{\Gamma\left(\frac{M}{2}\right)},
\end{equation}
respectively. Here $\Gamma(s,x)$ is the upper incomplete Gamma function.

Since $\Gamma(s,x)\sim x^{s-1}e^{-x}$ at $x\rightarrow \infty$, one can get the tail exponents
\begin{equation}\label{5_eq:alpha_scaling}
\alpha=\frac{M}{2\kappa\langle N_\omega\rangle} \quad \mathrm{and} \quad \alpha=\frac{M}{4\kappa\langle N_\omega\rangle}
\end{equation}
for distributions~\eqref{5_eq:CCDF_fwm_th_M} and \eqref{5_eq:CCDF_fwm_sb_M}, respectively. Thus, the exponents linearly increase with the number $M$ of modes for the same mean number $\langle N_\omega\rangle$ of pump photons. This can be seen from Fig.~\ref{5_fig:P_FWM}: the distribution~\eqref{5_eq:P_fwm_sb_M} for the gain $\kappa\langle N_\omega\rangle=2.5$ and $M=5$ (blue line) has nearly the same scaling as the single-mode one~\eqref{5_eq:P_fwm_sb} for the gain $\kappa\langle N_\omega\rangle=0.5$ (red dashed line).

\subsection{Optical harmonic generation: extreme events}
\label{5_sub:HG_extreme}

The optical harmonics are generated in the same setup as in the experiment on statistical enhancement (section~\ref{5_sub:setup_HG}). In order to demonstrate that the generated harmonic becomes extremely `noisy' the CFs $g^{(2)}_{n\omega}$ are measured for each harmonic in a Hanbury Brown -- Twiss setup~\cite{Hanbury1956}. After the LiNbO$_3$ crystal the harmonic radiation is collimated by a $3.1$~mm lens, splitted using an achromatic half-wave plate and a Glan prism, and fed in two APDs. The photocount pulses are sent to a coincidence circuit and the correlation function is found from the numbers of coincidences and single counts in detectors~\cite{Ivanova2006}.

\begin{figure}[!htb]
\begin{center}
\includegraphics[width=0.5\textwidth]{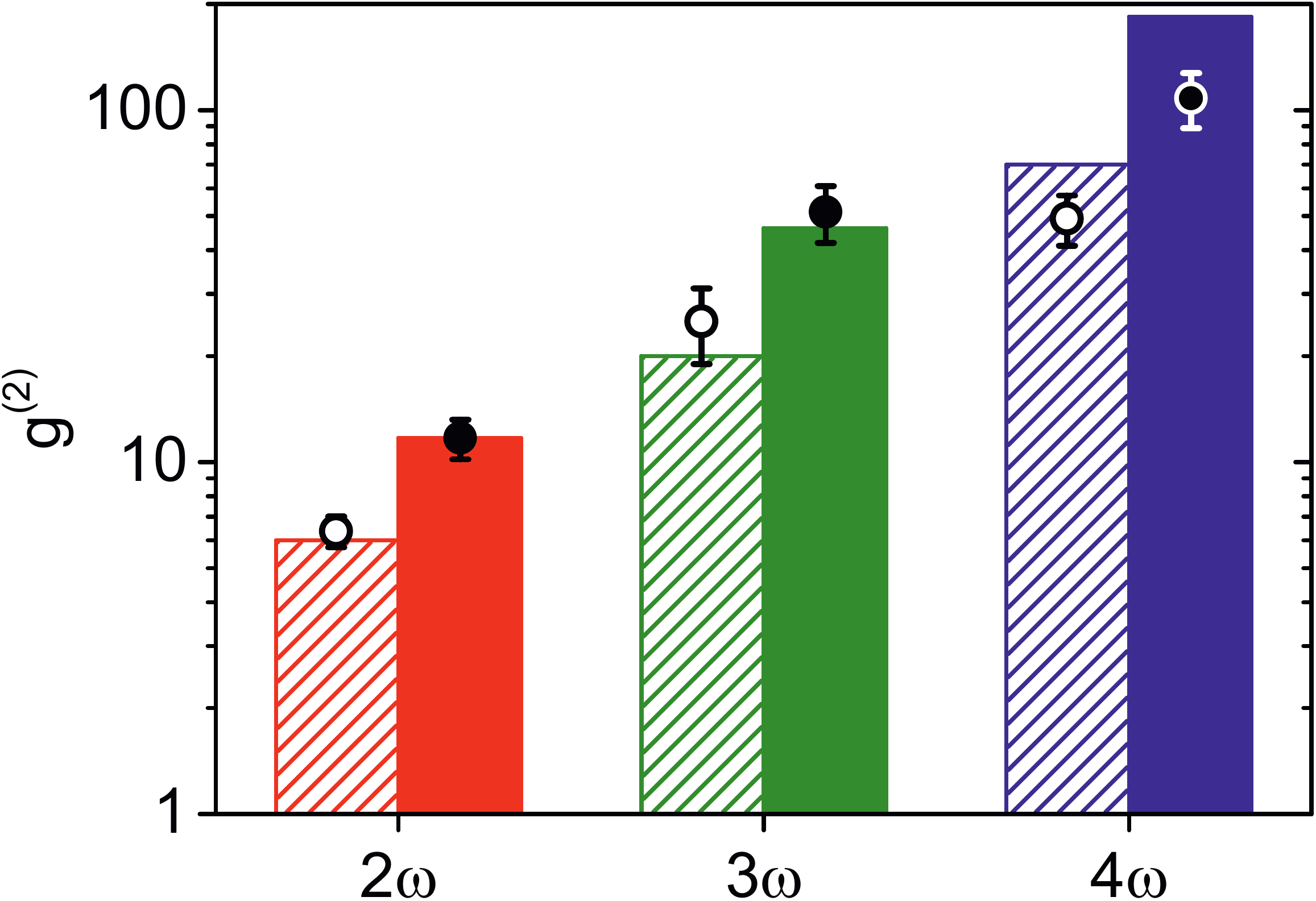}
\caption{Measured second-order CFs $g^{(2)}_{n\omega}$ for the harmonics generated from the superbunched (filled circles) and thermal (empty circles) BSV. The theoretical values are shown by filled and shaded color bars, respectively.}
\label{5_fig:g2_harm}
\end{center}
\end{figure}

Fig.~\ref{5_fig:g2_harm} shows the second-order CF $g^{(2)}_{n\omega}$ for SH, TH, FH from Eq.~\eqref{5_eq:g_m_nw} generated from thermal light (shaded) and superbunched BSV (colored). The expected values are very high, moreover, they increase with the harmonic number and for more fluctuating pump. The second-order CF $g^{(2)}_{n\omega}$ is equal to $6$ for SH from thermal light~\cite{Qu1995,Allevi2015} and to $184$ for FH from superbunched BSV.

The experimental results are in good agreement with the theory. They show a huge increase of fluctuations in the generated optical harmonics. The observed CF $g^{(2)}=110\pm20$ for the FH from the superbunched BSV is close to the maximal values reported in the literature~\cite{Meuret2015,Meuret2017}. It is lower than the expected value of 184 due to the unwanted SHG discussed above, see section~\ref{5_sub:CFs_exp}; nevertheless, it is in agreement with Eq.~\eqref{5_eq:g_m_nw} if the measured values of $g^{(8)}_{\omega}$ and $g^{(4)}_{\omega}$ are used.

As discussed it is important to keep the probability of two-photon events per detector much lower than the one for single-photon events, especially for highly fluctuating light. At high mean photon number the probability of two-photon events within a single pulse is too high, which leads to the effect similar to the saturation of both detectors and to the resulting reduction of the measured CF $g^{(2)}$. At the same time, at small mean number of photons the count rates are too low for a precise measurement and the dark noise contribution (about $9\times10^{-4}$ counts per pulse), despite being subtracted, strongly affects the measurement.

\begin{figure}[!htb]
\begin{center}
\includegraphics[width=0.46\textwidth]{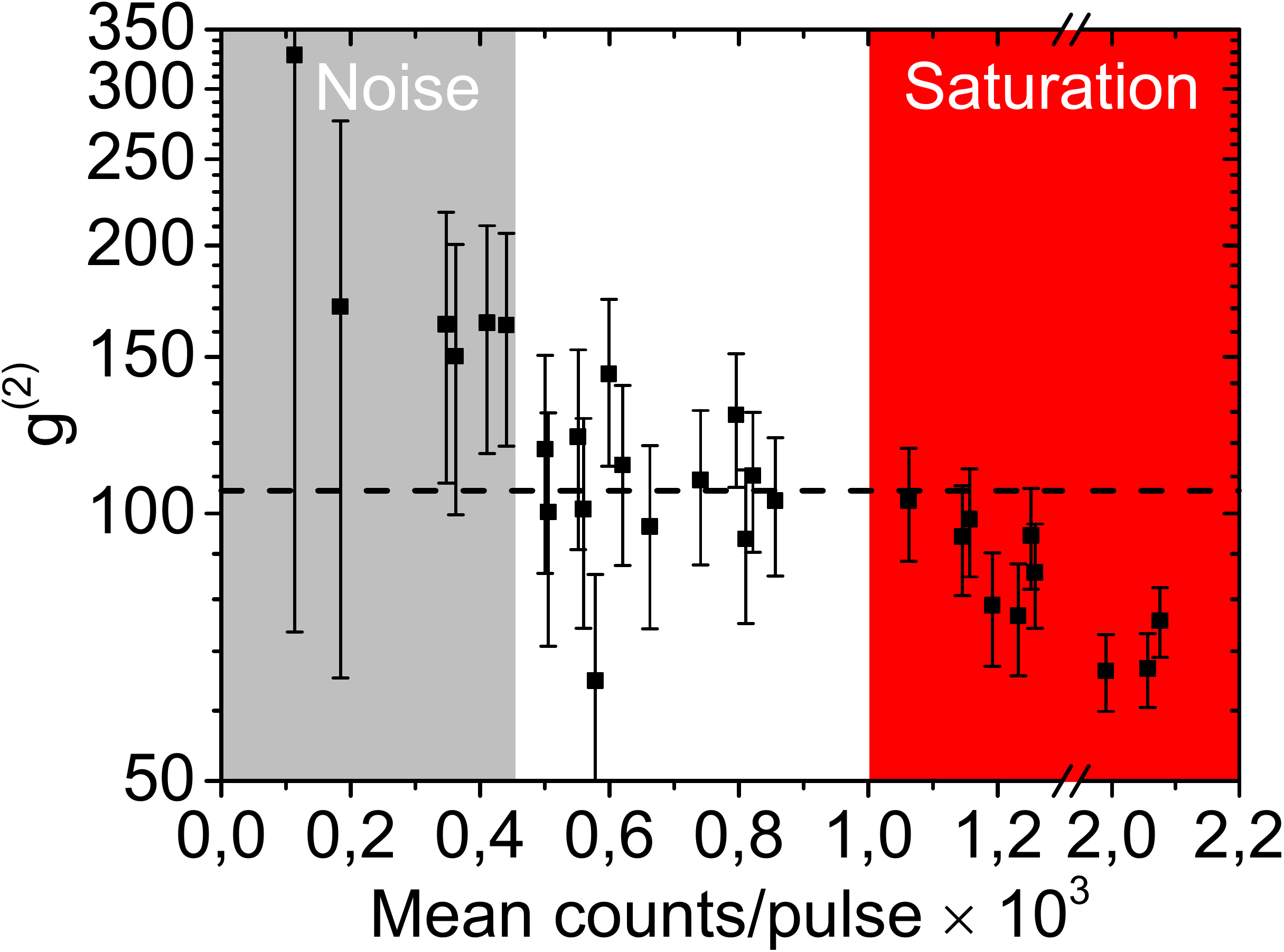}
\caption{Dependence of the measured CF $g^{(2)}$ for the FH on the mean number of counts per pulse (points). The expected value of $g^{(2)}$ calculated from Eq.~\eqref{5_eq:g_m_nw} for the measured BSV statistics is shown by dashed line.}
\label{5_fig:g2_vs_counts}
\end{center}
\end{figure}

Fig.~\ref{5_fig:g2_vs_counts} shows, as an example, the measurement of CF $g^{(2)}$ for the FH radiation versus the detected photon number, which is varied by attenuating the beam. This procedure should not change the value of $g^{(2)}$ due to its invariance to losses. The measured value is trustable only within a narrow range of mean photon numbers. During the measurements discussed above the mean photon number is kept within this range. Note that the `saturation border' moves to the higher mean photon numbers for the less fluctuating light while the `noise border' remains the same for the all measurements.

The probability distributions for the radiation of different harmonics are measured using the visible detector described in section~\ref{4_sub:setup_BS_interf}. Due to the low efficiency of the processes it was impossible to measure the FH and the TH from the thermal BSV. Moreover, for the same reason the TH is produced from the superbunched BSV with already a bit reduced fluctuations, similarly to the FH for the measurements with APD.

\begin{figure}[!htb]
\begin{center}
\includegraphics[width=0.65\textwidth]{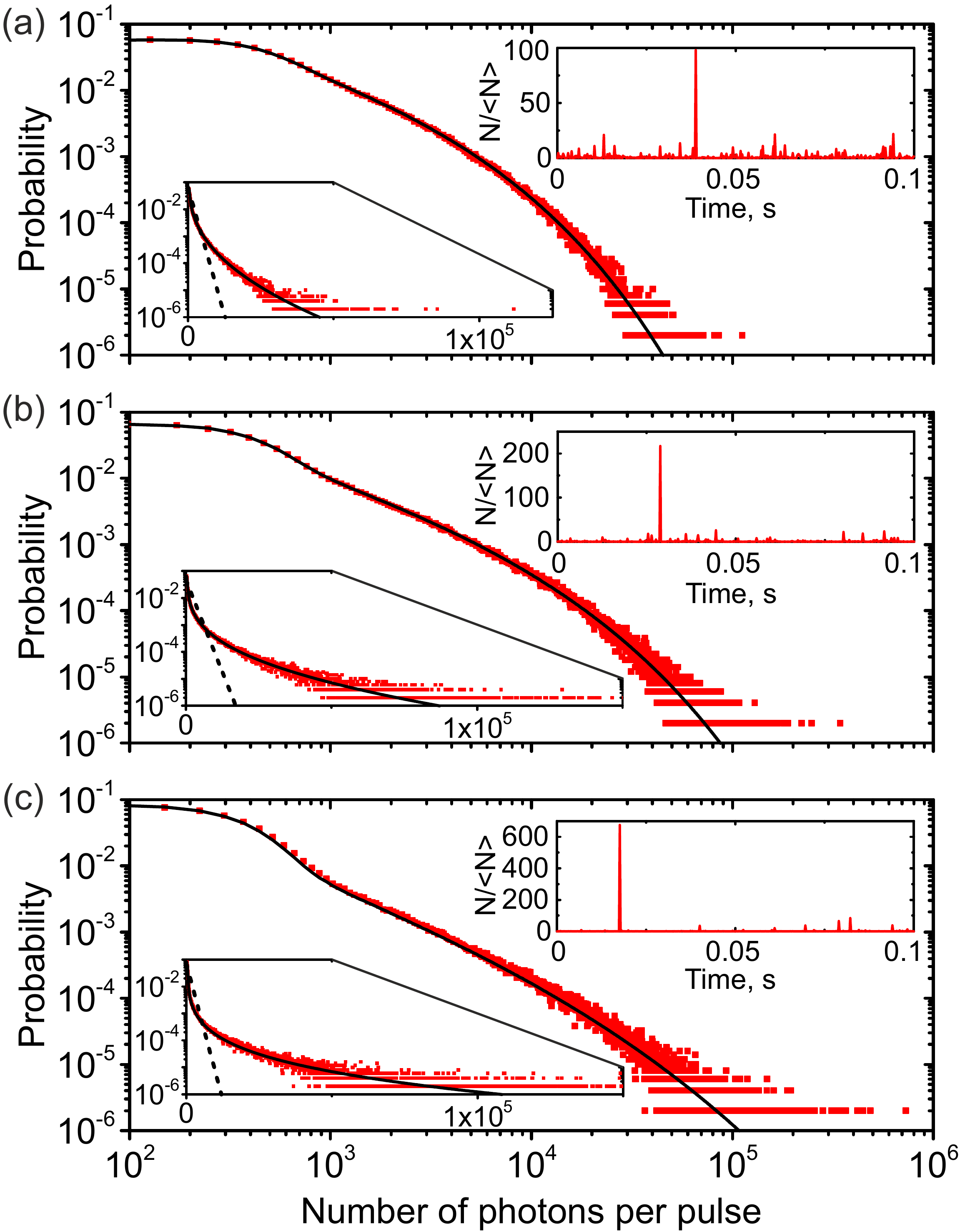}
\caption{Photon-number histograms for the SH generated from the thermal (a) and superbunched BSV (b), and the TH generated from the superbunched BSV (c). The experimental data are shown by red points in log-log and log-linear scales (bottom left insets). The theoretical distributions with an account for the photodetector noise are shown by black lines. To emphasize rogue-wave behavior, the thermal distribution~\eqref{5_eq:P_th} is shown by dashed lines in the insets. The top right insets show 0.1~s time traces with the largest extreme event.}
\label{5_fig:P_nw_log}
\end{center}
\end{figure}

The histograms for the measured harmonics are shown in Fig.~\ref{5_fig:P_nw_log} by red points. They are in a perfect agreement with the theoretical distributions [Eqs.~\eqref{5_eq:P_nw_th} and \eqref{5_eq:P_nw_sb}, solid lines] taking into account the noise distribution in a similar way as in the measurement of BSV photon-number distributions (section~\ref{5_sub:BSV_prob}). This time the detector's noise is equivalent to 270 photons per pulse. Again, it is reasonable to emphasize that the fitting parameters are not used, the mean values $\mean{N_{n\omega}}$ are taken from the experimental data and shown in Table~\ref{5_tab:HG_param} together with the measured CF $g^{(2)}$. 

\begin{table}[!htb]
\centering
\renewcommand{\arraystretch}{1.25}
\begin{tabular}{|c|c|c|}
\hline
&$\langle N_{n\omega}\rangle$, ph./pulse&$g^{(2)}$\\
\hline
$2\omega$ from thermal&$1.15\times10^3$&$6.08\pm0.07$\\
\hline
$2\omega$ from superb.&$1.59\times10^3$&$11.4\pm0.2$\\
\hline
$3\omega$ from superb.&$1.08\times10^3$&$33.1\pm1.6$\\
\hline
\end{tabular}
\caption{Measured characteristics of the SH and TH generated from the thermal and superbunched BSV.}
\label{5_tab:HG_param}
\renewcommand{\arraystretch}{1}
\end{table}

In log-linear scale the distributions demonstrate a large deviation from the thermal light scaling (lower insets, dashed lines). The top right insets show $0.1$~s time traces containing pulses with the highest photon numbers $N_{n\omega}^{\max}$ detected within 100~s of total measurement. The probability of such events is $10^{-6}$. The maximal values $N_{n\omega}^{\max}$ exceed the mean values $\langle N_{n\omega}\rangle$ by more than two orders of magnitude and differ from them by more than $40$ standard deviations.

This behavior is especially pronounced for the TH radiation,  $N_{3\omega}^{\max}=675\langle N_{3\omega}\rangle$. For comparison, the probability of events with $N>670\langle N\rangle$ for thermal light [Eq.~\eqref{5_eq:P_th}] is less than $10^{-290}$.

\begin{figure}[!htb]
\begin{center}
\includegraphics[width=0.65\textwidth]{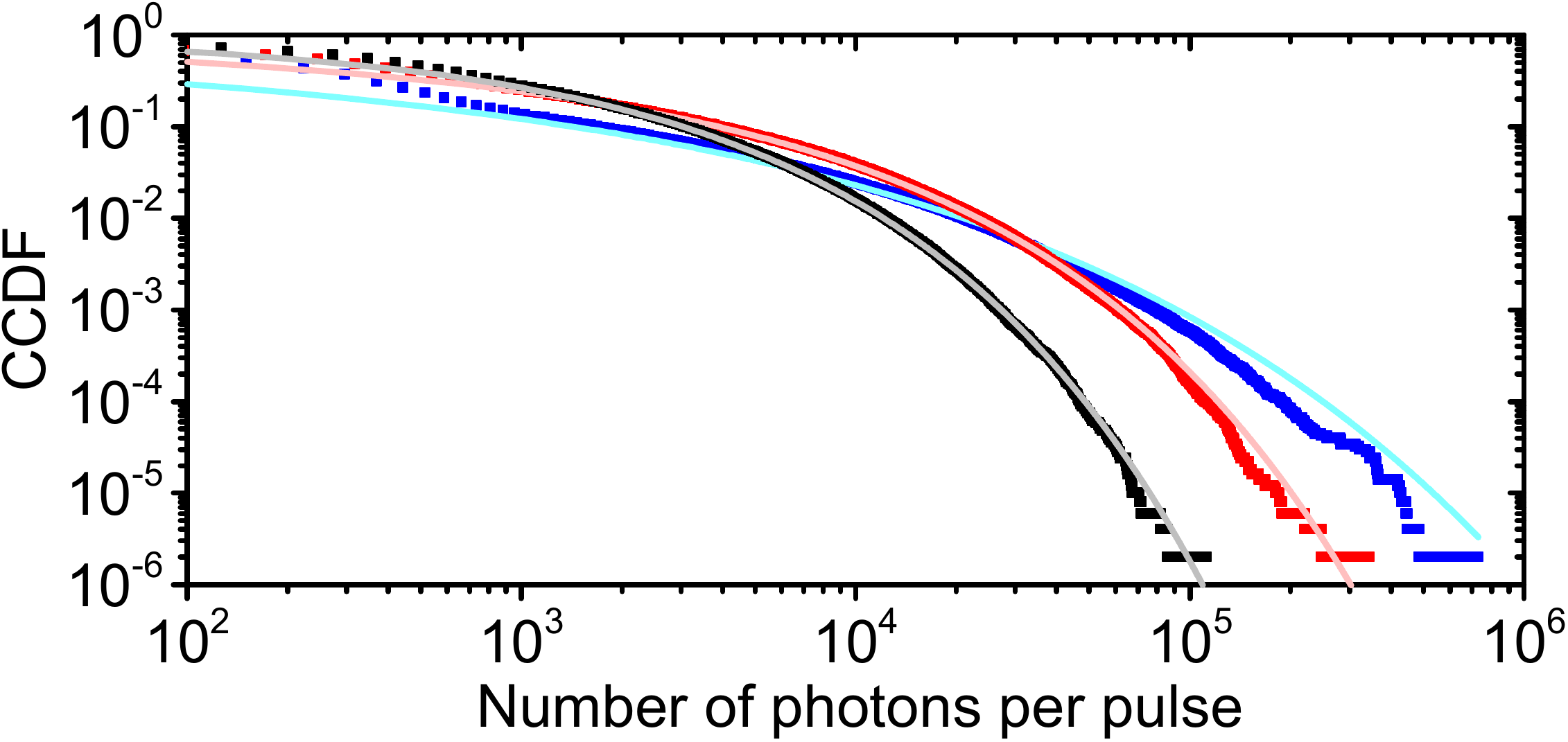}
\caption{CCDFs for the SH generated from the thermal (black points) and superbunched (red points) BSV and the TH generated from the superbunched BSV (blue points). The corresponding theoretical distributions are shown by gray, pink, and cyan lines, respectively.}
\label{5_fig:CCDF_nw}
\end{center}
\end{figure}

The experimental CCDFs are also perfectly consistent with Eqs.~\eqref{5_eq:CCDF_nw_th} and \eqref{5_eq:CCDF_nw_sb} for $n=2,3$, see Fig.~\ref{5_fig:CCDF_nw}. Again, the mean values $\langle N_{n\omega}\rangle$ are taken from the experimental data (Table~\ref{5_tab:HG_param}).

\subsection{Supercontinuum generation: Pareto photon distribution}
\label{5_sub:SC_extreme}

The experimental results presented here were obtained by Mathieu Manceau. Author's contribution was the data elaboration and the explanation of the observed effect.

\begin{figure}[!htb]
\begin{center}
\includegraphics[width=1\textwidth]{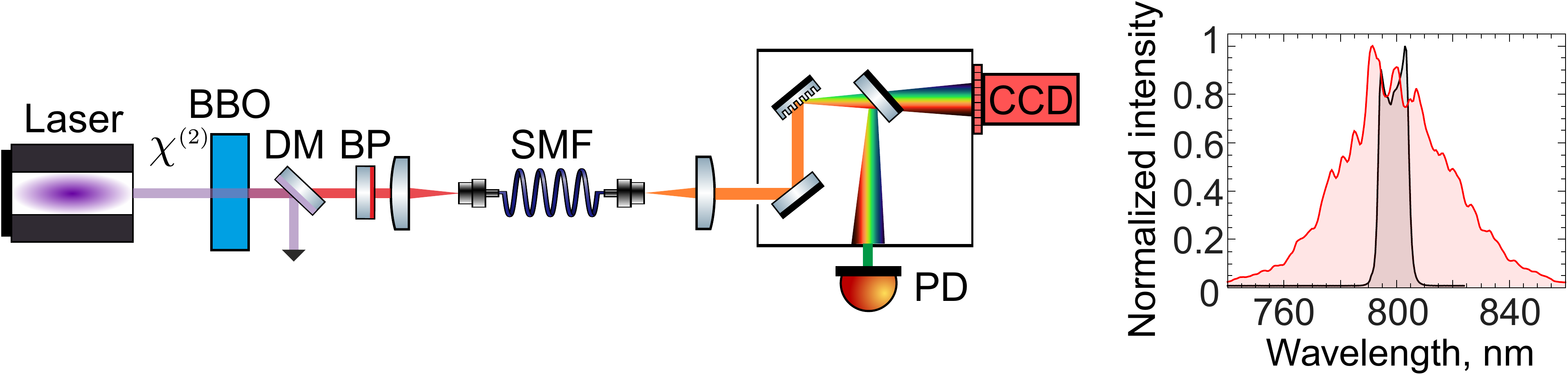}
\caption{Left: experimental setup for supercontinuum generation. Right: the averaged spectrum after the fiber under pumping with 0.2~nJ (grey) and 40~nJ BSV pulses (pink).}
\label{5_fig:setup_SC}
\end{center}
\end{figure}

The supercontinuum is pumped by the BSV centered at $800$ nm and generated through the type-I collinear degenerate PDC in two 3 mm BBO crystals~\cite{Manceau2017} from the laser used in the experiment on temporal walk-off matching (section~\ref{4_sub:setup_temp_walk}), see Fig.~\ref{5_fig:setup_SC} (left). After the laser is cut off by a dichroic mirror (DM), the BSV is filtered by a bandpass filter (BP) and coupled into a standard $5$~m single-mode fused silica fiber (SMF), where the supercontinuum is generated.

The fiber itself provides a single-mode spatial filtering~\cite{Perez2015_2}. The frequency filtering for the BSV is done by either 10~nm or 3~nm filter resulting in the number of modes $M=5\pm1$ or $2\pm0.2$, respectively. This number is obtained from the $g^{(2)}$ measurement, see section~\ref{5_sub:multimode}, performed at low BSV power, namely with the energy per pulse $\mathscr{E}=3$~pJ. The supercontinuum is then spectrally filtered with a monochromator with $1$~nm resolution and analyzed by the visible photodetector (PD) used for the study of optical harmonics\footnote{The detector has exactly the same design, but the calibration coefficient varies from detector to detector; for this one it is equal to 8.2~pV$\times$s per photon.}.

The spectrum after the fiber (Fig.~\ref{5_fig:setup_SC}, right) is almost unchanged (grey) if the BSV is weak, $\mathscr{E}=0.2$~nJ, but it is considerably broadened (pink) if the input energy per pulse $\mathscr{E}=40$~nJ.

\begin{figure}[!htb]
\begin{center}
\includegraphics[width=1\textwidth]{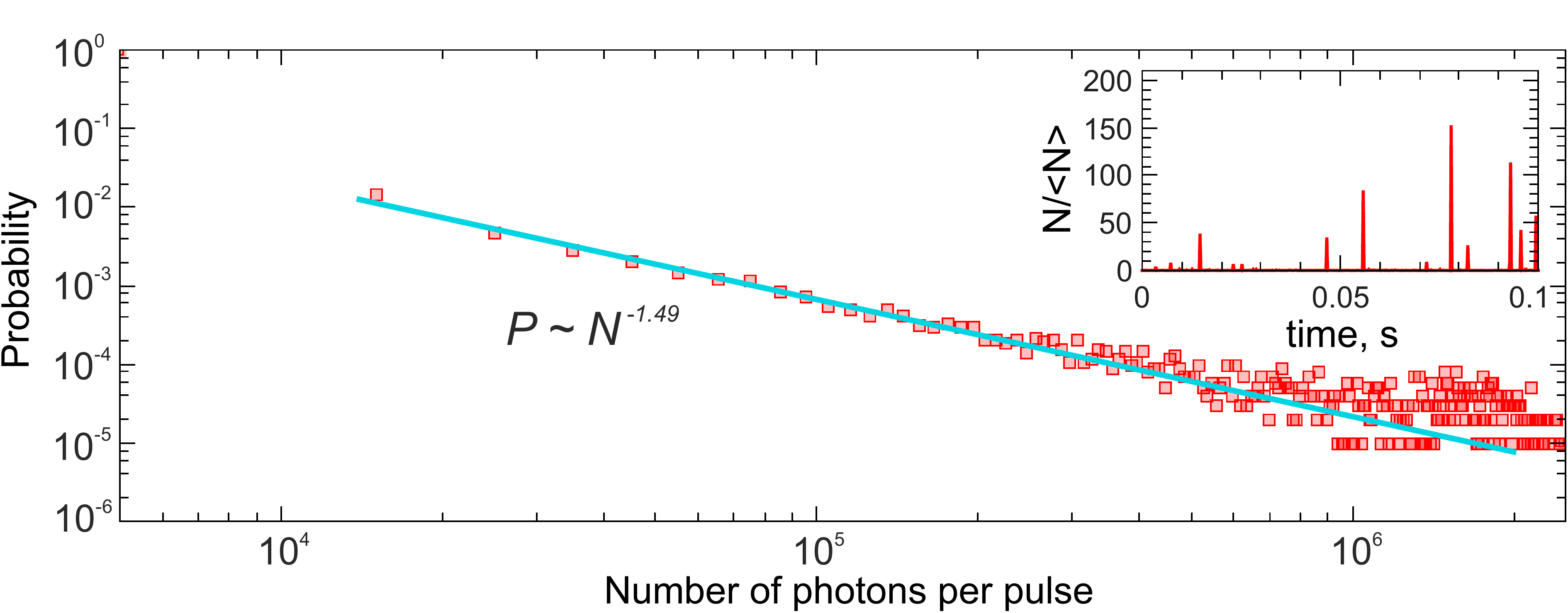}
\caption{Photon-number histogram for the supercontinuum pumped with $48$ nJ pulses of BSV (points), and the Pareto distribution~\eqref{5_eq:P_pareto} with $\alpha=0.49$ (solid line). The inset shows a 0.1~s time trace with the largest extreme event.}
\label{5_fig:P_sc}
\end{center}
\end{figure}

The supercontinuum is measured at wavelength $770$ nm under the pumping with $48$~nJ BSV pulses with 10~nm bandwidth; the obtained photon-number histogram is shown in Fig.~\ref{5_fig:P_sc}. The distribution demonstrates a linear behavior in log-log scale within two orders of magnitude. It has the Pareto scaling with $\alpha_e=0.49$ (solid line), which is found from the CCDF fit for more accurate determination.

\begin{figure}[!htb]
\begin{center}
\includegraphics[width=0.65\textwidth]{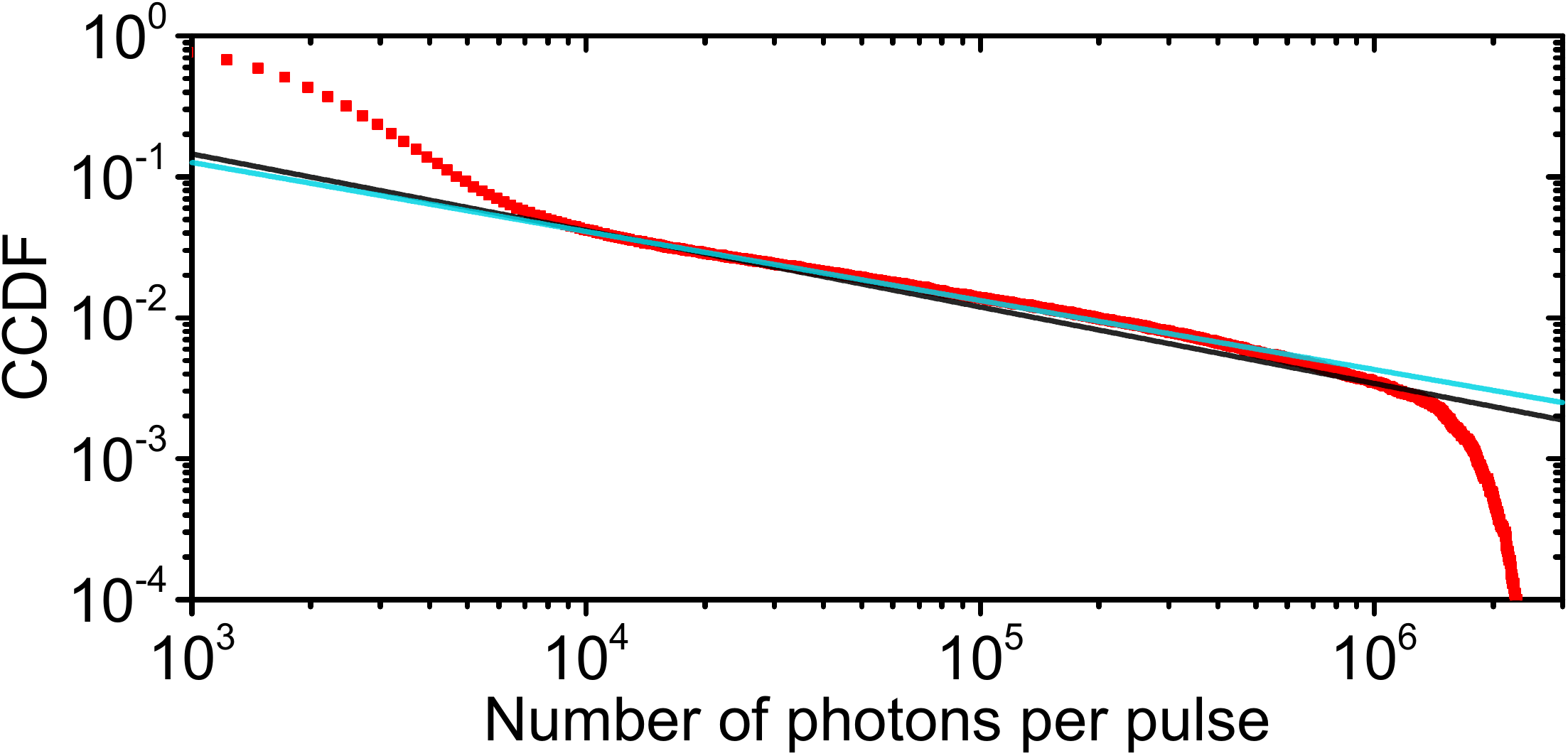}
\caption{CCDF for the supercontinuum from Fig.~\ref{5_fig:P_sc} (red points), its fit (blue line) with Eq.~\eqref{5_eq:CCDF_Pareto} leading to the exponent $\alpha_e=0.49$, and Eq.~\eqref{5_eq:CCDF_Pareto} with $\alpha=0.53$ (black line) obtained from the maximum likelihood estimator $\tilde{\alpha}$ [Eq.~\eqref{5_eq:tilde_alpha}].}
\label{5_fig:CCDF_sc}
\end{center}
\end{figure}

The experimental CCDF (Fig.~\ref{5_fig:CCDF_sc}) is fitted within its linear range, from $10^4$ to $8\times10^5$ photons per pulse. The uncertainty of the fit $\Delta \alpha_{e}=0.02$ is mainly caused by imprecise estimation of the start and end points of this range. The peak below $10^4$ photons, which follows the exponential distribution~\eqref{5_eq:P_th}, is attributed to the thermalization effect in supercontinuum~\cite{PICOZZI2014} that occurs due to the energy exchange between supercontinuum modes. At high photon numbers the scaling fails due to the detector saturation, which truncates the distribution for $N>10^6$ photons.

The other method for the exponent estimation is suggested in Refs.~\cite{Newman2005,Clauset2009}. It is so-called maximum likelihood estimator $\tilde{\alpha}$ given by
\begin{equation}\label{5_eq:tilde_alpha}
\tilde{\alpha}=1+s\left[\sum_{l=1}^s\ln\left(\frac{N_l}{N_{\min}}\right)\right]^{-1}.
\end{equation}
Here $N_{\min}$ is the photon number at which the Pareto scaling starts, $\{N_l\}$ are the observed photon numbers larger than $N_{\min}$, and $s$ is the dataset size after excluding $N<N_{\min}$. For the distribution from Fig.~\ref{5_fig:P_sc} it gives $\tilde{\alpha}=0.53$, which is similar to the obtained exponent $\alpha_e$. In Fig.~\ref{5_fig:CCDF_sc} the Pareto distribution with $\alpha=0.53$ is shown by black line.

\begin{figure}[!htb]
\begin{center}
\includegraphics[width=0.55\textwidth]{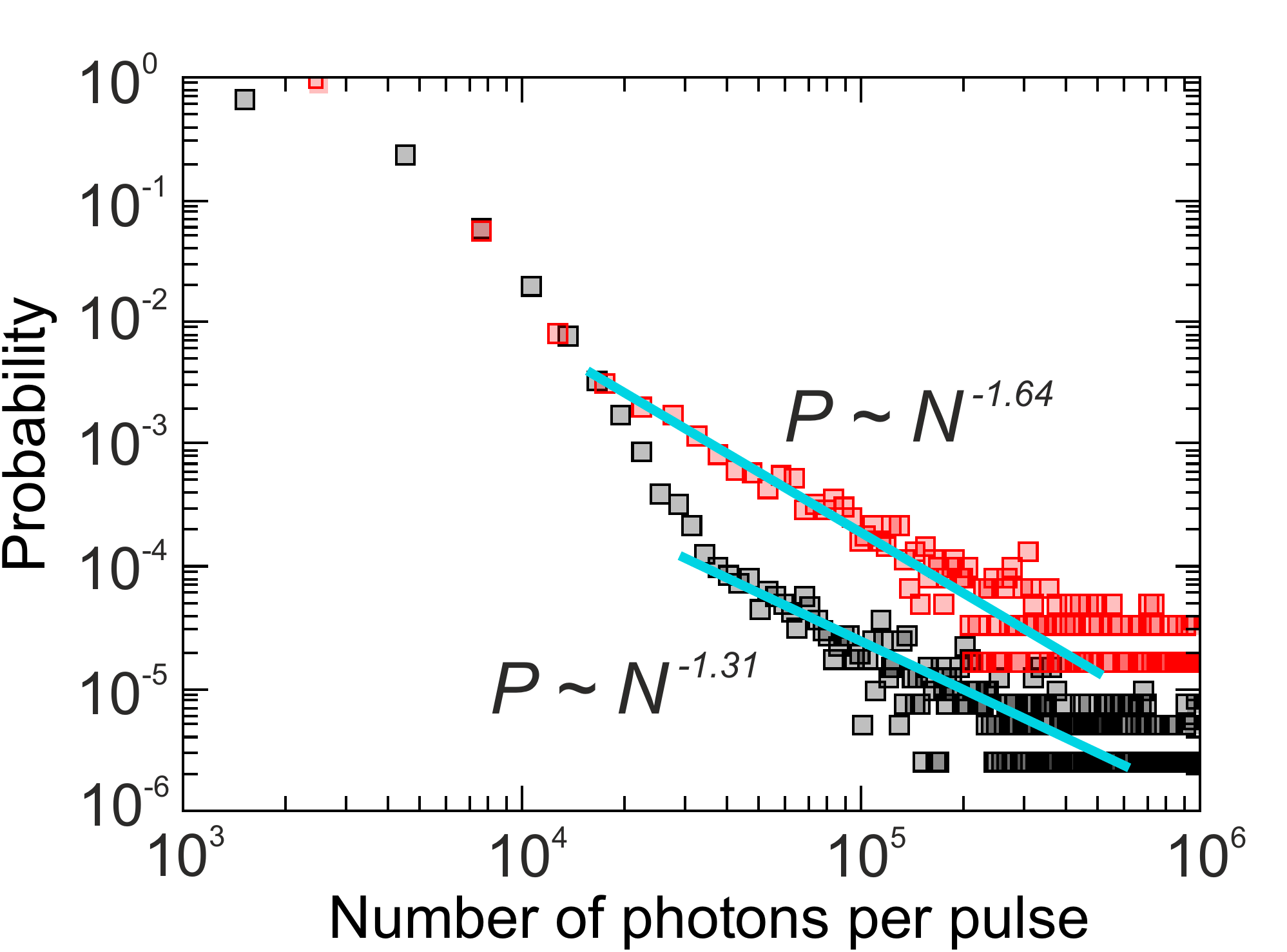}
\caption{Photon-number histograms for the supercontinuum pumped by 30~nJ BSV pulses with 10~nm (red points) and 3~nm (black points) bandwidth and their Pareto fits with $\alpha=0.64$ and 0.31, respectively.}
\label{5_fig:P_sc_2}
\end{center}
\end{figure}

In order to test that the exponent depends on the BSV power and its number of modes, we measure the probability distributions for the supercontinuum with the BSV energy per pulse reduced to $30$~nJ. The results are shown in Fig.~\ref{5_fig:P_sc_2}. For the pumping with 10 and 3~nm bandwidth the fits of CCDFs provide exponents $\alpha_e=0.64$ and $0.31$, respectively, meaning that the variation of BSV power and bandwidth provides a possibility to control the exponent of the Pareto distribution for supercontinuum.

Surprisingly, the simple FWM theory provides not only the Pareto scaling, it also gives the reasonable exponents. The theoretical exponents $\alpha_{t}$ are estimated from Eq.~\eqref{5_eq:alpha_scaling} using the corresponding number $M$ of modes and the gain $\kappa\langle N_\omega\rangle$ of the process. We get the gain $\kappa\langle N_\omega\rangle=4\pm0.5$ and $2.5\pm0.5$ for the energy per pulse $\mathscr{E}=48$ and 30~nJ, respectively, from the nonlinear dependence similar to the one obtained for BSV, see section~\ref{3_sec:broadening}. The uncertainties in both measurements lead to the relative error $\delta \alpha_{t}=25\%$.

\begin{table}[h]
\centering
\renewcommand{\arraystretch}{1.25}
\begin{tabular}{|c|c|c|c|c|}
\hline
& $\mathscr{E}$, $\Delta\lambda_\omega$&$\alpha_{e}$&$\alpha_{t}$&$\langle N_F\rangle$, ph./pulse\\
\hline
Fig.~\ref{5_fig:P_sc}&48~nJ, 10 nm&$0.49$&$0.31$&$1.18\times10^4$\\
\hline
Fig.~\ref{5_fig:P_sc_2}&30~nJ, 10 nm&$0.64$&$0.5$&$6.5\times10^3$\\
\cline{2-5}
&30~nJ, 3 nm&$0.31$&$0.2$&$5.6\times10^3$\\
\hline
\end{tabular}
\caption{Characteristics of the supercontinuum photon-number distributions. $\mathscr{E}$ and $\Delta\lambda_\omega$ are the BSV energy per pulse and bandwidth, respectively; $\alpha_{e}$ and $\alpha_{t}$ are the experimental and theoretical exponents; $\langle N_F\rangle$ is the mean number of photons.} \label{5_tab:Table_SC}
\renewcommand{\arraystretch}{1}
\end{table}

The theoretical Pareto exponents $\alpha_{t}$ are somewhat smaller than the experimental ones (Table~\ref{5_tab:Table_SC}). Such a difference could be caused by the approximations made, which are actually not true. For example, the undepleted-pump approximation; as one can see from Fig.~\ref{5_fig:setup_SC} (right) the intensities of the BSV at 800~nm and the obtained supercontinuum are comparable. Moreover, it is not taken into account that the supercontinuum is so strong that it also produces FWM sidebands. Finally, our gain estimation is very crude, because the mean value $\langle N_F\rangle$ is not defined for the exponent $\alpha<1$.

\subsection{Extreme bunching: outlook and possible applications}

The power-law probability distributions shown in Figs.~\ref{5_fig:P_sc} and \ref{5_fig:P_sc_2} have very unusual features. With the power exponent less than 2, $\alpha<1$, the mean number of photons is not defined and depends on the time of observation, making this distribution much different from the ones reported by the others~\cite{Borlaug2009,Kasparian2009,Alves2019}. Similar to the fractal `coastline paradox'~\cite{Mandelbrot1967}, where the coastline appears to be longer, the better one measures, the mean photon number will be the higher, the more data are used to determine it, $\langle N_F\rangle\propto s^{\nicefrac{1}{\alpha}-1}$, where $s$ is the dataset size~\cite{Newman2005}. This fractal-like behavior is typical for the Pareto distributions with indefinite mean values~\cite{Park1996}. Of course, in a real experiment the distribution gets always truncated through some physical mechanisms. For example, the coastline contains no details less than a centimeter. Therefore, the coastline length is also defined if it is measured with a centimeter accuracy.

Superbunching has interesting consequences for photon subtraction experiments~\cite{Zavatta2008}. In a photon subtraction experiment a quantum state of light $|\psi\rangle$ is fed to a BS, after which a single-photon detector can register a reflected photon. Provided that the detector registers a single photon, the state of light after the BS is the photon-subtracted one, $|\psi'\rangle\propto\op{a}|\psi\rangle$. Counterintuitively, the photon-subtracted state has the mean photon number increased by a factor $g^{(2)}$ compared to the initial state~\cite{Bogdanov2017,Hlouvsek2017}.

Furthermore, the $g^{(2)}$ measurement can be indeed considered as a photon subtraction experiment, because the mean number of photons in the state $|\psi'\rangle$ is equal to the mean number of coincidences. In our paper~\cite{Manceau2018} we perform such experiment for the supercontinuum and get a 140 times increase of the mean photon number. This drastic increase in the mean photon number shows that the optical harmonics or supercontinuum can be brought out of equilibrium by the subtraction process much more efficiently than thermal light or superbunched BSV. As a result, much more work could be, in principle, extracted, which makes the supercontinuum a useful resource for proof-of-principle tests of quantum thermodynamics~\cite{Hlouvsek2017}.

The radiation with strong fluctuations in the photon number can be also useful in the ghost imaging~\cite{Pittman1995,Gatti2004}, where the contrast of the image is given by the CF $g^{(2)}$~\cite{Agafonov2009,Chan2009}. In particular, recently the time-domain ghost imaging has been demonstrated with the incoherent supercontinuum~\cite{Amiot2018}, although without high photon-number fluctuations. It is significant that the use of supercontinuum with the Pareto photon-number distribution will drastically increase the contrast.

\chapter{Conclusion}

This thesis has presented the detailed research on different spectral and statistical properties of high-gain PDC and its potential applications.

Firstly, it was shown that in the anomalous range of group velocity dispersion the PDC spectrum is restricted in both angle and wavelength. Such a spectrum provides a possibility to collect all radiation without losses, which in its respect is an advantage for experiments with squeezed light. Moreover, the ring-shaped frequency-wavevector spectrum provides a new type of spatiotemporal coherence and entanglement of photon pairs. One can obtain an entangled photon pair with the photons being never at the same point at the same time. Such a pair could be used in the time- and space-resolved nonlinear spectroscopy with two-photon light.

Secondly, the broadening of the total and correlation spectral widths in the high-gain regime has been demonstrated. Apart from that, the asymmetry of PDC frequency spectra has been explained. Moreover, two methods for studying the mode structure and correlation properties of PDC have been developed. The first one is based on the PDC fluctuations, while the second one uses the spectral properties of sum frequency generation from PDC.

Thirdly, a macroscopic analogue of Hong-Ou-Mandel interference was measured and discussed. Although for high flux of entangled photons the standard measurement technique based on the second-order correlation function leads to a very low visibility, a modified technique based on the normalized variance of the difference signal provides a visibility close to unity. With this technique, the interference manifests itself as a peak, which additionally provides the information about the PDC spectral properties.

Fourthly, the photon statistics after the interference of twin beams on a 50\% beam splitter were analyzed. In this interference the intrinsic phase fluctuations between twin beams are converted into photon-number ones. This photon-number uncertainty is observed as a `U' shape in the output photon-number distribution. It is typical for the interference of high-order Fock states and could have an important application for quantum state engineering. The effect was also mimicked by classical beams with an artificially mixed phase.

Fifthly, it has been demonstrated that the spatial and temporal walk-off matching can be used for giant narrowband twin-beam generation. If one of the twin beams is emitted along the pump Poynting vector or its group velocity matches that of the pump, the generation of both beams is drastically enhanced. This enhancement leads to a considerable narrowing of the wavelength and angular spectrum; thus, one can use these effects for building a highly tunable optical parametric generator without any cavity.

Sixthly, the measured normalized correlation functions and probability distributions show that BSV produced via PDC has strong and ultrafast photon-number fluctuations. The correlation functions can be measured under narrowband filtering, since the filtering does not change their maximum values. Moreover, strong photon-number fluctuations enhance efficiency of multiphoton effects and this enhancement has been demonstrated for the generation of the second, third, and forth optical harmonics. For the latter one the generation efficiency is enhanced two orders of magnitude compared to the pumping with coherent light. This could be particularly useful for the multiphoton microscopy of fragile structures, including biological objects, in which the damage threshold plays an important role. Furthermore, BSV could be applied in the ultrafast spectroscopy and two-dimensional fluorescence spectroscopy.

Lastly, the generated harmonics fluctuate even more than BSV and have an exceptional statistics. The photon numbers are described by heavy-tailed photon-number distributions, events whose magnitude exceeds the mean value by more than 100 times were observed rather frequently. Furthermore, for the supercontinuum generated from BSV, the power-law photon-number distributions with the power exponents less than 2 were observed. As a result, these distributions decay so slowly, that the mean photon number is indefinite, to say nothing of higher moments. Such fluctuating statistics could be notably useful for quantum thermodynamics and ghost imaging.

In the end, the author believes that high-gain PDC will take an important place in science and industry, and therefore hopes that this thesis would be helpful for both communities.

\appendix
\chapter{List of acronyms and abbreviations}

\begin{tabular}{ll}
APD & avalanche photodiode\\
BBO & beta barium borate\\
BS & beam splitter\\
BSV & bright squeezed vacuum\\
CCDF & complementary cumulative distribution function\\
CF & correlation function\\
CW & continuous-wave\\
FH & fourth harmonic\\
FHG & fourth harmonic generation\\
FT & Fourier transform\\
FWM &four-wave mixing\\
GDD & group delay dispersion\\
GVD & group velocity dispersion\\
HOM & Hong-Ou-Mandel\\
IR & infrared\\
JSA & joint spectral amplitude\\
JSI & joint spectral intensity\\
OCT & optical coherence tomography\\
OPA & optical parametric amplification\\
OPG & optical parametric generator\\
PBS & polarizing beam splitter\\
PDC & parametric down-conversion\\
SFG & sum frequency generation\\
SH & second harmonic\\
SHG & second harmonic generation\\
SPDC & spontaneous parametric down-conversion\\
TH & third harmonic\\
THG & third harmonic generation\\
\end{tabular}

\chapter*{Acknowledgments}
\addcontentsline{toc}{chapter}{Acknowledgments}

First of all, I would like to thanks my supervisor, Maria, who introduced and guided me though the world of quantum optics, spent a lot of hours and days in the discussions about how the things should be done and arranged, tried to improve my presentations and writing skills, in particular the way how this thesis is written. Hopefully, in the end we always found a good solution. Thank you really a lot!

The second thanks goes to Timur and Olga, who were the first people I met in the quantum optics lab, my first co-supervisors, microchiefs. You explained me a lot of practical stuff. Thank you!

One more big thanks goes to Farid, who showed me how theoreticians work. With a lot of efforts I learned how to calculate analytically the difficult stuff. Thank you!

Thank you, Gerd, it was great to be part of MPL and the Leuchs division in particular.

Thanks 4piPAC Group for good-neighborhood in FAU offices and shared lab. Special thanks to Markus for helping with the German summary. 

Radim, Tatiana, Olga, Vlad, Ulrik, Eugene, Polina, Falk, Victor, and Denis, thank you a lot for the nice and productive collaboration!

Thanks the QuaRad -- Chekhova research group!  Andrea, Mathieu, Angela, Gaetano, Hani, Felix, Cameron, Sascha, Luo, Mickey, Tomas, and Paula, thank you a lot for long-term and short-term work cooperation, for the help with the equipment and guidance! Sorry for the criticism at group meetings, sometimes it was too harsh.

\begin{wrapfigure}{r}{0.38\textwidth}
\vspace{-7pt}
\includegraphics[width=0.38\textwidth]{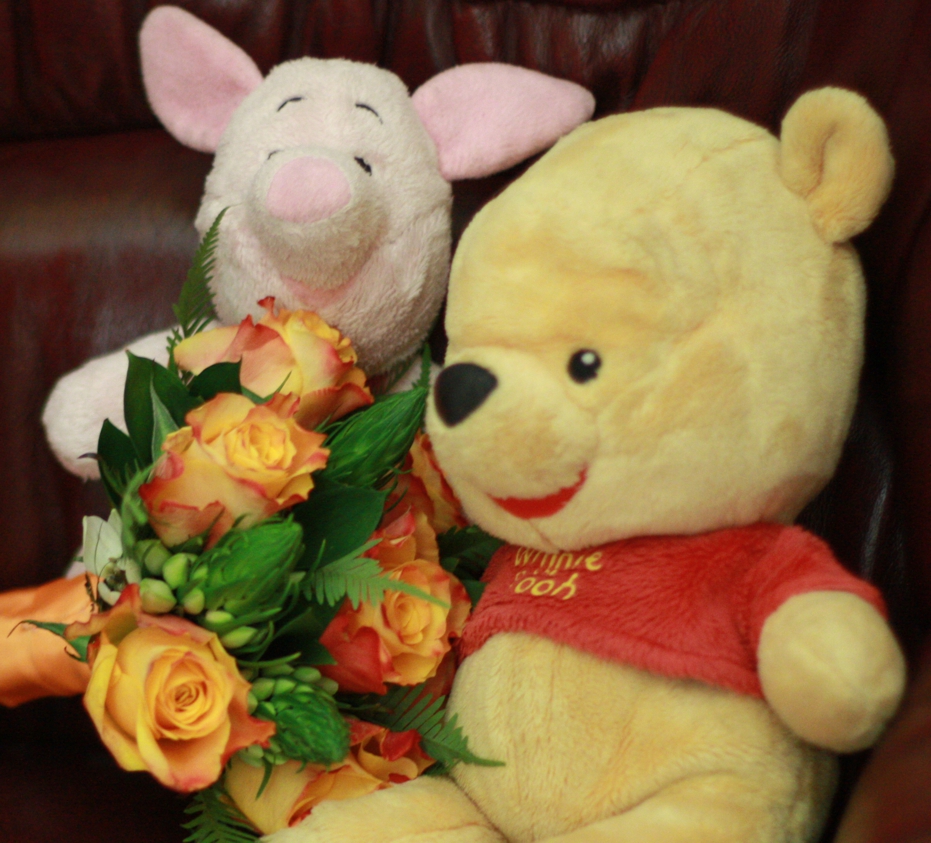}
\end{wrapfigure}
Finally, my biggest thanks goes to Vera and Sofia. We'll be friends Forever, won't we? My life is becoming much better with you. You are the best, I love you very very much!!!\\
\rule{9.6cm}{1 pt}\\
\noindent
Piglet sidled up to Pooh from behind.\\
`Pooh!' he whispered.\\
`Yes, Piglet?'\\
`Nothing,' said Piglet, taking Pooh's paw.\\`I just wanted to be sure of you.'

\clearpage

\phantomsection
\addcontentsline{toc}{chapter}{Bibliography}
\bibliographystyle{abbrv}
\bibliography{thesis_biblio}

\clearpage

\includepdf{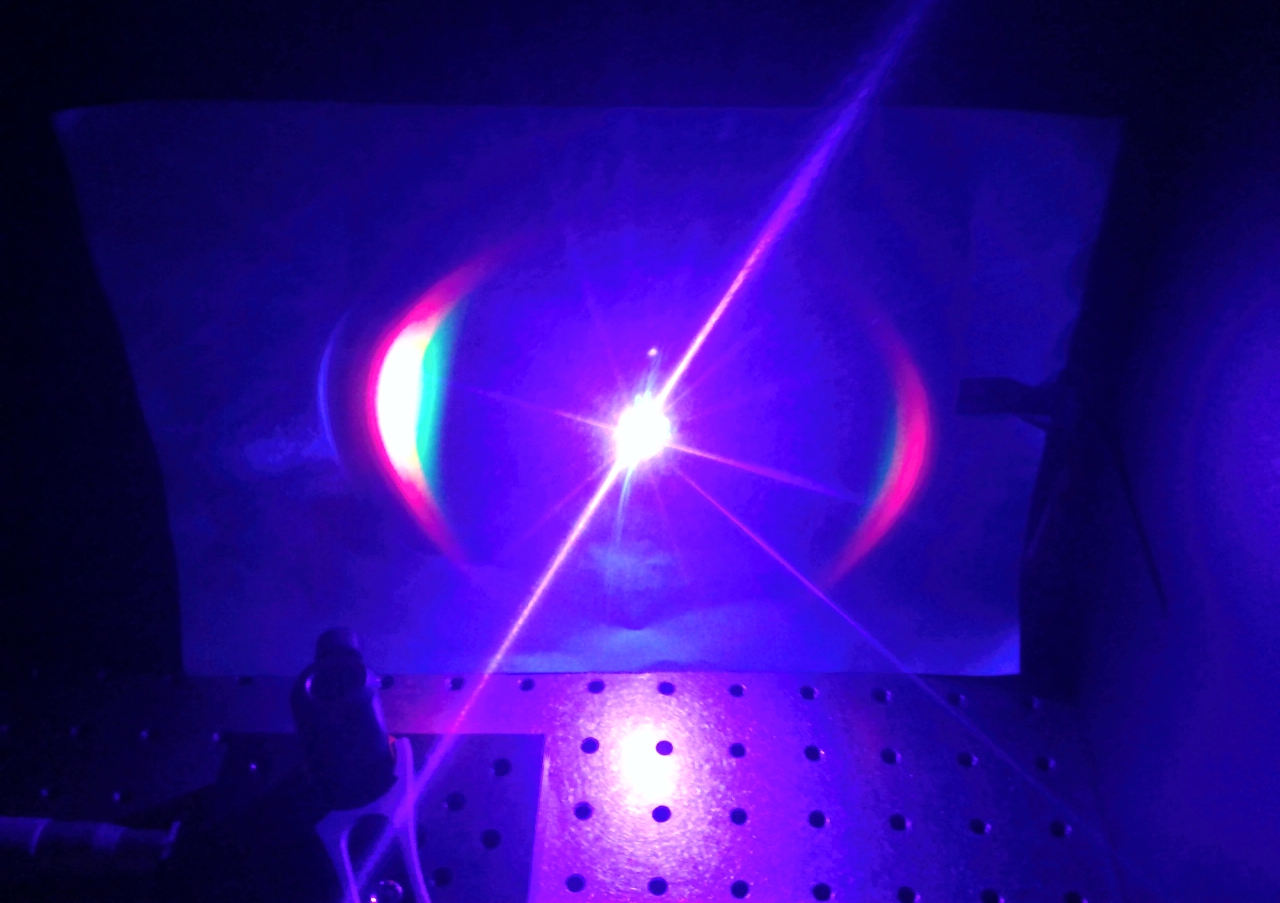}

\end{document}